\documentclass[11pt,a4paper]{article}
\usepackage{isabelle,isabellesym}
\isabellestyle{it}
 
\begin{document}

\title{Compositional properties of crypto-based components}
\author{Maria Spichkova}
\maketitle

\begin{abstract}
This paper presents an Isabelle/HOL+Isar~\cite{npw,IsabelleManual} set of theories which allows  to specify
crypto-based components and to verify their composition properties wrt. cryptographic aspects.  
We introduce a formalisation of the security property of data secrecy, 
the corresponding definitions and proofs. 
A part of these definitions is based on  \cite{sj_TB08}.\\
Please note that here we import here the Isabelle/HOL theory ListExtras.thy, presented in \cite{FocusStreamsCaseStudies-AFP}.
\end{abstract}

\tableofcontents
 
\newpage
\begin{isabellebody}%
\def\isabellecontext{Secrecy{\isacharunderscore}types}%
\isamarkupheader{Auxiliary data types%
}
\isamarkuptrue%
\isadelimtheory
\endisadelimtheory
\isatagtheory
\isacommand{theory}\isamarkupfalse%
\ Secrecy{\isacharunderscore}types\isanewline
\isakeyword{imports}\ Main\isanewline
\isakeyword{begin}\isanewline
\isanewline
\isamarkupcmt{We assume disjoint sets: Data of data values,%
}
\isanewline
\isamarkupcmt{Secrets of unguessable values, Keys - set of cryptographic  keys.%
}
\ \ \isanewline
\isamarkupcmt{Based on these sets, we specify the sets EncType of encryptors that may be%
}
\isanewline
\isamarkupcmt{used for encryption or decryption, and Expression of expression items.%
}
\isanewline
\isamarkupcmt{The specification (component) identifiers should be listed in the set specID,%
}
\isanewline
\isamarkupcmt{the channel indentifiers should be listed in the set chanID.%
}
\endisatagtheory
{\isafoldtheory}%
\isadelimtheory
\endisadelimtheory
\ \isanewline
\isanewline
\isacommand{datatype}\isamarkupfalse%
\ Keys\ {\isacharequal}\ CKey\ {\isacharbar}\ CKeyP\ {\isacharbar}\ SKey\ {\isacharbar}\ SKeyP\ {\isacharbar}\ genKey\ \isanewline
\isacommand{datatype}\isamarkupfalse%
\ Secrets\ {\isacharequal}\ secretD\ {\isacharbar}\ N\ {\isacharbar}\ NA\isanewline
\isacommand{type{\isacharunderscore}synonym}\isamarkupfalse%
\ Var\ \ {\isacharequal}\ {\isachardoublequoteopen}nat{\isachardoublequoteclose}\isanewline
\isacommand{type{\isacharunderscore}synonym}\isamarkupfalse%
\ Data\ {\isacharequal}\ {\isachardoublequoteopen}nat{\isachardoublequoteclose}\isanewline
\isacommand{datatype}\isamarkupfalse%
\ KS\ \ \ \ \ \ \ \ \ \ {\isacharequal}\ kKS\ Keys\ {\isacharbar}\ sKS\ Secrets\isanewline
\isacommand{datatype}\isamarkupfalse%
\ EncType\ \ {\isacharequal}\ kEnc\ Keys\ {\isacharbar}\ vEnc\ Var\isanewline
\isacommand{datatype}\isamarkupfalse%
\ specID\ {\isacharequal}\ sComp{\isadigit{1}}\ {\isacharbar}\ sComp{\isadigit{2}}\ {\isacharbar}\ sComp{\isadigit{3}}\ {\isacharbar}\ sComp{\isadigit{4}}\isanewline
\isacommand{datatype}\isamarkupfalse%
\ Expression\ {\isacharequal}\ kE\ Keys\ {\isacharbar}\ sE\ Secrets\ {\isacharbar}\ dE\ Data\ {\isacharbar}\ idE\ specID\ \isanewline
\isacommand{datatype}\isamarkupfalse%
\ chanID\ {\isacharequal}\ ch{\isadigit{1}}\ {\isacharbar}\ ch{\isadigit{2}}\ \ \ {\isacharbar}\ ch{\isadigit{3}}\ \ {\isacharbar}\ ch{\isadigit{4}}\isanewline
\isanewline
\isacommand{primrec}\isamarkupfalse%
\ Expression{\isadigit{2}}KSL{\isacharcolon}{\isacharcolon}\ {\isachardoublequoteopen}Expression\ list\ {\isasymRightarrow}\ KS\ list{\isachardoublequoteclose}\ \isanewline
\isakeyword{where}\isanewline
\ \ \ {\isachardoublequoteopen}Expression{\isadigit{2}}KSL\ {\isacharbrackleft}{\isacharbrackright}\ {\isacharequal}\ {\isacharbrackleft}{\isacharbrackright}{\isachardoublequoteclose}\ {\isacharbar}\isanewline
\ \ \ {\isachardoublequoteopen}Expression{\isadigit{2}}KSL\ {\isacharparenleft}x{\isacharhash}xs{\isacharparenright}\ {\isacharequal}\ \isanewline
\ \ \ \ \ {\isacharparenleft}{\isacharparenleft}case\ x\ of\ {\isacharparenleft}kE\ m{\isacharparenright}\ {\isasymRightarrow}\ {\isacharbrackleft}kKS\ m{\isacharbrackright}\ \isanewline
\ \ \ \ \ \ \ \ \ \ \ \ \ \ \ \ \ \ {\isacharbar}\ {\isacharparenleft}sE\ m{\isacharparenright}\ {\isasymRightarrow}\ {\isacharbrackleft}sKS\ m{\isacharbrackright}\ \isanewline
\ \ \ \ \ \ \ \ \ \ \ \ \ \ \ \ \ \ {\isacharbar}\ {\isacharparenleft}dE\ m{\isacharparenright}\ {\isasymRightarrow}\ {\isacharbrackleft}{\isacharbrackright}\ \isanewline
\ \ \ \ \ \ \ \ \ \ \ \ \ \ \ \ \ \ {\isacharbar}\ {\isacharparenleft}idE\ m{\isacharparenright}\ {\isasymRightarrow}\ {\isacharbrackleft}{\isacharbrackright}{\isacharparenright}\ {\isacharat}\ Expression{\isadigit{2}}KSL\ xs{\isacharparenright}\ {\isachardoublequoteclose}\isanewline
\isanewline
\isacommand{primrec}\isamarkupfalse%
\ KS{\isadigit{2}}Expression{\isacharcolon}{\isacharcolon}\ {\isachardoublequoteopen}KS\ {\isasymRightarrow}\ Expression{\isachardoublequoteclose}\ \isanewline
\isakeyword{where}\isanewline
\ \ {\isachardoublequoteopen}KS{\isadigit{2}}Expression\ {\isacharparenleft}kKS\ m{\isacharparenright}\ {\isacharequal}\ {\isacharparenleft}kE\ m{\isacharparenright}{\isachardoublequoteclose}\ \ {\isacharbar}\isanewline
\ \ {\isachardoublequoteopen}KS{\isadigit{2}}Expression\ {\isacharparenleft}sKS\ m{\isacharparenright}\ {\isacharequal}\ {\isacharparenleft}sE\ m{\isacharparenright}{\isachardoublequoteclose}\isanewline
\isadelimtheory
\isanewline
\endisadelimtheory
\isatagtheory
\isacommand{end}\isamarkupfalse%
\endisatagtheory
{\isafoldtheory}%
\isadelimtheory
\endisadelimtheory
\end{isabellebody}%

%
\begin{isabellebody}%
\def\isabellecontext{inout}%
\isamarkupheader{Correctness of the relations between sets of Input/Output channels%
}
\isamarkuptrue%
\isadelimtheory
\endisadelimtheory
\isatagtheory
\isacommand{theory}\isamarkupfalse%
\ \ inout\ \isanewline
\isakeyword{imports}\ Secrecy{\isacharunderscore}types\isanewline
\isakeyword{begin}%
\endisatagtheory
{\isafoldtheory}%
\isadelimtheory
\endisadelimtheory
\isanewline
\isanewline
\isacommand{consts}\isamarkupfalse%
\ \isanewline
\ \ subcomponents\ {\isacharcolon}{\isacharcolon}\ \ {\isachardoublequoteopen}specID\ {\isasymRightarrow}\ specID\ set{\isachardoublequoteclose}\isanewline
\isanewline
\isamarkupcmt{Mappings, defining sets of input, local, and output channels%
}
\isanewline
\isamarkupcmt{of a component%
}
\isanewline
\isacommand{consts}\isamarkupfalse%
\isanewline
\ \ ins\ {\isacharcolon}{\isacharcolon}\ {\isachardoublequoteopen}specID\ {\isasymRightarrow}\ chanID\ set{\isachardoublequoteclose}\isanewline
\ \ loc\ {\isacharcolon}{\isacharcolon}\ {\isachardoublequoteopen}specID\ {\isasymRightarrow}\ chanID\ set{\isachardoublequoteclose}\isanewline
\ \ out\ {\isacharcolon}{\isacharcolon}\ {\isachardoublequoteopen}specID\ {\isasymRightarrow}\ chanID\ set{\isachardoublequoteclose}\isanewline
\isanewline
\isamarkupcmt{Predicate insuring the correct mapping from the component identifier%
}
\isanewline
\isamarkupcmt{to the set of input channels of a component%
}
\isanewline
\isacommand{definition}\isamarkupfalse%
\isanewline
\ \ inStream\ {\isacharcolon}{\isacharcolon}\ {\isachardoublequoteopen}specID\ {\isasymRightarrow}\ chanID\ set\ {\isasymRightarrow}\ bool{\isachardoublequoteclose}\isanewline
\isakeyword{where}\isanewline
\ \ {\isachardoublequoteopen}inStream\ x\ y\ \ {\isasymequiv}\ {\isacharparenleft}ins\ x\ {\isacharequal}\ y{\isacharparenright}{\isachardoublequoteclose}\isanewline
\isanewline
\isamarkupcmt{Predicate insuring the correct mapping from the component identifier%
}
\isanewline
\isamarkupcmt{to the set of local channels of a component%
}
\isanewline
\isacommand{definition}\isamarkupfalse%
\isanewline
\ \ locStream\ {\isacharcolon}{\isacharcolon}\ {\isachardoublequoteopen}specID\ {\isasymRightarrow}\ chanID\ set\ {\isasymRightarrow}\ bool{\isachardoublequoteclose}\isanewline
\isakeyword{where}\isanewline
\ \ {\isachardoublequoteopen}locStream\ x\ y\ {\isasymequiv}\ {\isacharparenleft}loc\ x\ {\isacharequal}\ y{\isacharparenright}{\isachardoublequoteclose}\isanewline
\isanewline
\isamarkupcmt{Predicate insuring the correct mapping from the component identifier%
}
\isanewline
\isamarkupcmt{to the set of output channels of a component%
}
\isanewline
\isacommand{definition}\isamarkupfalse%
\isanewline
\ \ outStream\ {\isacharcolon}{\isacharcolon}\ {\isachardoublequoteopen}specID\ {\isasymRightarrow}\ chanID\ set\ {\isasymRightarrow}\ bool{\isachardoublequoteclose}\isanewline
\isakeyword{where}\isanewline
\ \ {\isachardoublequoteopen}outStream\ x\ y\ {\isasymequiv}\ {\isacharparenleft}out\ x\ {\isacharequal}\ y{\isacharparenright}{\isachardoublequoteclose}\isanewline
\isanewline
\isamarkupcmt{Predicate insuring the correct relations between%
}
\isanewline
\isamarkupcmt{to the set of input, output and local channels of a component%
}
\isanewline
\isacommand{definition}\isamarkupfalse%
\isanewline
\ \ correctInOutLoc\ {\isacharcolon}{\isacharcolon}\ {\isachardoublequoteopen}specID\ {\isasymRightarrow}\ bool{\isachardoublequoteclose}\isanewline
\isakeyword{where}\isanewline
\ \ {\isachardoublequoteopen}correctInOutLoc\ x\ {\isasymequiv}\ \isanewline
\ \ \ {\isacharparenleft}ins\ x{\isacharparenright}\ {\isasyminter}\ {\isacharparenleft}out\ x{\isacharparenright}\ {\isacharequal}\ {\isacharbraceleft}{\isacharbraceright}\ \isanewline
\ \ \ \ {\isasymand}\ {\isacharparenleft}ins\ x{\isacharparenright}\ {\isasyminter}\ {\isacharparenleft}loc\ x{\isacharparenright}\ {\isacharequal}\ {\isacharbraceleft}{\isacharbraceright}\ \isanewline
\ \ \ \ {\isasymand}\ {\isacharparenleft}loc\ x{\isacharparenright}\ {\isasyminter}\ {\isacharparenleft}out\ x{\isacharparenright}\ {\isacharequal}\ {\isacharbraceleft}{\isacharbraceright}\ {\isachardoublequoteclose}\ \isanewline
\isanewline
\isamarkupcmt{Predicate insuring the correct relations between%
}
\isanewline
\isamarkupcmt{sets of input channels within a composed component%
}
\isanewline
\isacommand{definition}\isamarkupfalse%
\isanewline
\ \ correctCompositionIn\ {\isacharcolon}{\isacharcolon}\ {\isachardoublequoteopen}specID\ {\isasymRightarrow}\ bool{\isachardoublequoteclose}\isanewline
\isakeyword{where}\isanewline
\ \ {\isachardoublequoteopen}correctCompositionIn\ x\ {\isasymequiv}\ \isanewline
\ \ {\isacharparenleft}ins\ x{\isacharparenright}\ {\isacharequal}\ {\isacharparenleft}{\isasymUnion}\ {\isacharparenleft}ins\ {\isacharbackquote}\ {\isacharparenleft}subcomponents\ x{\isacharparenright}{\isacharparenright}\ {\isacharminus}\ {\isacharparenleft}loc\ x{\isacharparenright}{\isacharparenright}\isanewline
\ \ {\isasymand}\ {\isacharparenleft}ins\ x{\isacharparenright}\ {\isasyminter}\ {\isacharparenleft}{\isasymUnion}\ {\isacharparenleft}out\ {\isacharbackquote}\ {\isacharparenleft}subcomponents\ x{\isacharparenright}{\isacharparenright}{\isacharparenright}\ {\isacharequal}\ {\isacharbraceleft}{\isacharbraceright}{\isachardoublequoteclose}\ \isanewline
\isanewline
\isamarkupcmt{Predicate insuring the correct relations between%
}
\isanewline
\isamarkupcmt{sets of output channels within a composed component%
}
\isanewline
\isacommand{definition}\isamarkupfalse%
\isanewline
\ \ correctCompositionOut\ {\isacharcolon}{\isacharcolon}\ {\isachardoublequoteopen}specID\ {\isasymRightarrow}\ bool{\isachardoublequoteclose}\isanewline
\isakeyword{where}\isanewline
\ \ {\isachardoublequoteopen}correctCompositionOut\ x\ {\isasymequiv}\ \isanewline
\ \ {\isacharparenleft}out\ x{\isacharparenright}\ {\isacharequal}\ {\isacharparenleft}{\isasymUnion}\ {\isacharparenleft}out\ {\isacharbackquote}\ {\isacharparenleft}subcomponents\ x{\isacharparenright}{\isacharparenright}{\isacharminus}\ {\isacharparenleft}loc\ x{\isacharparenright}{\isacharparenright}\isanewline
\ \ {\isasymand}\ {\isacharparenleft}out\ x{\isacharparenright}\ {\isasyminter}\ {\isacharparenleft}{\isasymUnion}\ {\isacharparenleft}ins\ {\isacharbackquote}\ {\isacharparenleft}subcomponents\ x{\isacharparenright}{\isacharparenright}{\isacharparenright}\ {\isacharequal}\ {\isacharbraceleft}{\isacharbraceright}\ {\isachardoublequoteclose}\ \isanewline
\isanewline
\isamarkupcmt{Predicate insuring the correct relations between%
}
\isanewline
\isamarkupcmt{sets of local channels within a composed component%
}
\isanewline
\isacommand{definition}\isamarkupfalse%
\isanewline
\ \ correctCompositionLoc\ {\isacharcolon}{\isacharcolon}\ {\isachardoublequoteopen}specID\ {\isasymRightarrow}\ bool{\isachardoublequoteclose}\isanewline
\isakeyword{where}\isanewline
\ \ {\isachardoublequoteopen}correctCompositionLoc\ x\ {\isasymequiv}\ \isanewline
\ \ \ {\isacharparenleft}loc\ x{\isacharparenright}\ {\isacharequal}\ {\isasymUnion}\ {\isacharparenleft}ins\ {\isacharbackquote}\ {\isacharparenleft}subcomponents\ x{\isacharparenright}{\isacharparenright}\isanewline
\ \ \ \ \ \ \ \ \ \ \ {\isasyminter}\ {\isasymUnion}\ {\isacharparenleft}out\ {\isacharbackquote}\ {\isacharparenleft}subcomponents\ x{\isacharparenright}{\isacharparenright}{\isachardoublequoteclose}\ \isanewline
\isanewline
\isamarkupcmt{If a component is an elementary one (has no subcomponents)%
}
\isanewline
\isamarkupcmt{its set of local channels should be empty%
}
\isanewline
\isacommand{lemma}\isamarkupfalse%
\ subcomponents{\isacharunderscore}loc{\isacharcolon}\isanewline
\isakeyword{assumes}\ {\isachardoublequoteopen}correctCompositionLoc\ x{\isachardoublequoteclose}\isanewline
\ \ \ \ \ \ \ \isakeyword{and}\ {\isachardoublequoteopen}subcomponents\ x\ {\isacharequal}\ {\isacharbraceleft}{\isacharbraceright}{\isachardoublequoteclose}\isanewline
\isakeyword{shows}\ {\isachardoublequoteopen}loc\ x\ {\isacharequal}\ {\isacharbraceleft}{\isacharbraceright}{\isachardoublequoteclose}\isanewline
\isadelimproof
\endisadelimproof
\isatagproof
\isacommand{using}\isamarkupfalse%
\ assms\ \isacommand{by}\isamarkupfalse%
\ {\isacharparenleft}simp\ add{\isacharcolon}\ correctCompositionLoc{\isacharunderscore}def{\isacharparenright}%
\endisatagproof
{\isafoldproof}%
\isadelimproof
\isanewline
\endisadelimproof
\isadelimtheory
\isanewline
\endisadelimtheory
\isatagtheory
\isacommand{end}\isamarkupfalse%
\endisatagtheory
{\isafoldtheory}%
\isadelimtheory
\endisadelimtheory
\end{isabellebody}%

%
\begin{isabellebody}%
\def\isabellecontext{Secrecy}%
\isamarkupheader{Secrecy: Definitions and properties%
}
\isamarkuptrue%
\isadelimtheory
\endisadelimtheory
\isatagtheory
\isacommand{theory}\isamarkupfalse%
\ Secrecy\isanewline
\isakeyword{imports}\ Secrecy{\isacharunderscore}types\ inout\ ListExtras\isanewline
\isakeyword{begin}\isanewline
\isanewline
\isamarkupcmt{Encryption, decryption, signature creation and signature verification functions%
}
\isanewline
\isamarkupcmt{For these functions we define only their signatures and general axioms,%
}
\isanewline
\isamarkupcmt{because in order to reason effectively, we view them as abstract functions and%
}
\isanewline
\isamarkupcmt{abstract from their implementation details%
}
\endisatagtheory
{\isafoldtheory}%
\isadelimtheory
\endisadelimtheory
\ \isanewline
\isacommand{consts}\isamarkupfalse%
\ \isanewline
\ \ Enc\ \ {\isacharcolon}{\isacharcolon}\ {\isachardoublequoteopen}Keys\ {\isasymRightarrow}\ Expression\ list\ {\isasymRightarrow}\ Expression\ list{\isachardoublequoteclose}\isanewline
\ \ Decr\ {\isacharcolon}{\isacharcolon}\ {\isachardoublequoteopen}Keys\ {\isasymRightarrow}\ Expression\ list\ {\isasymRightarrow}\ Expression\ list{\isachardoublequoteclose}\isanewline
\ \ Sign\ {\isacharcolon}{\isacharcolon}\ {\isachardoublequoteopen}Keys\ {\isasymRightarrow}\ Expression\ list\ {\isasymRightarrow}\ Expression\ list{\isachardoublequoteclose}\isanewline
\ \ Ext\ \ \ {\isacharcolon}{\isacharcolon}\ {\isachardoublequoteopen}Keys\ {\isasymRightarrow}\ Expression\ list\ {\isasymRightarrow}\ Expression\ list{\isachardoublequoteclose}\isanewline
\isanewline
\isamarkupcmt{Axioms on relations between encription and decription keys%
}
\isanewline
\isacommand{axiomatization}\isamarkupfalse%
\isanewline
\ \ \ EncrDecrKeys\ {\isacharcolon}{\isacharcolon}\ {\isachardoublequoteopen}Keys\ \ {\isasymRightarrow}\ Keys\ {\isasymRightarrow}\ bool{\isachardoublequoteclose}\isanewline
\isakeyword{where}\isanewline
ExtSign{\isacharcolon}\ \isanewline
\ {\isachardoublequoteopen}EncrDecrKeys\ K{\isadigit{1}}\ K{\isadigit{2}}\ {\isasymlongrightarrow}\ {\isacharparenleft}Ext\ K{\isadigit{1}}\ {\isacharparenleft}Sign\ K{\isadigit{2}}\ E{\isacharparenright}{\isacharparenright}\ {\isacharequal}\ E{\isachardoublequoteclose}\ \ \isakeyword{and}\isanewline
DecrEnc{\isacharcolon}\isanewline
\ {\isachardoublequoteopen}EncrDecrKeys\ K{\isadigit{1}}\ K{\isadigit{2}}\ {\isasymlongrightarrow}\ {\isacharparenleft}Decr\ K{\isadigit{2}}\ {\isacharparenleft}Enc\ K{\isadigit{1}}\ E{\isacharparenright}{\isacharparenright}\ {\isacharequal}\ E{\isachardoublequoteclose}\isanewline
\isanewline
\isamarkupcmt{Set of private keys of a component%
}
\isanewline
\isacommand{consts}\isamarkupfalse%
\isanewline
\ specKeys\ {\isacharcolon}{\isacharcolon}\ {\isachardoublequoteopen}specID\ {\isasymRightarrow}\ Keys\ set{\isachardoublequoteclose}\isanewline
\isamarkupcmt{Set of unguessable values used by a component%
}
\isanewline
\isacommand{consts}\isamarkupfalse%
\ \isanewline
\ specSecrets\ {\isacharcolon}{\isacharcolon}\ {\isachardoublequoteopen}specID\ {\isasymRightarrow}\ Secrets\ set{\isachardoublequoteclose}\isanewline
\isanewline
\isamarkupcmt{Join set of private keys and unguessable values used by a component%
}
\isanewline
\isacommand{definition}\isamarkupfalse%
\isanewline
\ \ specKeysSecrets\ {\isacharcolon}{\isacharcolon}\ {\isachardoublequoteopen}specID\ {\isasymRightarrow}\ KS\ set{\isachardoublequoteclose}\isanewline
\isakeyword{where}\isanewline
\ {\isachardoublequoteopen}specKeysSecrets\ C\ {\isasymequiv}\isanewline
\ \ {\isacharbraceleft}y\ {\isachardot}\ \ {\isasymexists}\ x{\isachardot}\ y\ {\isacharequal}\ {\isacharparenleft}kKS\ x{\isacharparenright}\ \ {\isasymand}\ {\isacharparenleft}x\ {\isasymin}\ {\isacharparenleft}specKeys\ C{\isacharparenright}{\isacharparenright}{\isacharbraceright}\ {\isasymunion}\isanewline
\ \ {\isacharbraceleft}z\ {\isachardot}\ \ {\isasymexists}\ s{\isachardot}\ z\ {\isacharequal}\ {\isacharparenleft}sKS\ s{\isacharparenright}\ \ {\isasymand}\ {\isacharparenleft}s\ {\isasymin}\ {\isacharparenleft}specSecrets\ C{\isacharparenright}{\isacharparenright}{\isacharbraceright}{\isachardoublequoteclose}\isanewline
\isanewline
\isamarkupcmt{Predicate defining that a list of expression items does not contain%
}
\isanewline
\isamarkupcmt{any private key  or unguessable value used by a component%
}
\isanewline
\isacommand{definition}\isamarkupfalse%
\isanewline
\ \ notSpecKeysSecretsExpr\ {\isacharcolon}{\isacharcolon}\ {\isachardoublequoteopen}specID\ {\isasymRightarrow}\ \ Expression\ list\ {\isasymRightarrow}\ bool{\isachardoublequoteclose}\isanewline
\isakeyword{where}\isanewline
\ \ \ \ \ {\isachardoublequoteopen}notSpecKeysSecretsExpr\ P\ e\ {\isasymequiv}\isanewline
\ \ \ \ \ {\isacharparenleft}{\isasymforall}\ x{\isachardot}\ {\isacharparenleft}kE\ x{\isacharparenright}\ mem\ e\ {\isasymlongrightarrow}\ {\isacharparenleft}kKS\ x{\isacharparenright}\ {\isasymnotin}\ specKeysSecrets\ P{\isacharparenright}\ {\isasymand}\isanewline
\ \ \ \ \ {\isacharparenleft}{\isasymforall}\ y{\isachardot}\ {\isacharparenleft}sE\ y{\isacharparenright}\ mem\ e\ {\isasymlongrightarrow}\ {\isacharparenleft}sKS\ y{\isacharparenright}\ {\isasymnotin}\ specKeysSecrets\ P{\isacharparenright}{\isachardoublequoteclose}\isanewline
\isanewline
\isamarkupcmt{If a component is a composite one, the set of its private keys%
}
\ \isanewline
\isamarkupcmt{is a union of the subcomponents' sets of the private keys%
}
\isanewline
\isacommand{definition}\isamarkupfalse%
\isanewline
\ \ correctCompositionKeys\ {\isacharcolon}{\isacharcolon}\ \ {\isachardoublequoteopen}specID\ {\isasymRightarrow}\ bool{\isachardoublequoteclose}\isanewline
\isakeyword{where}\isanewline
\ \ {\isachardoublequoteopen}correctCompositionKeys\ x\ {\isasymequiv}\isanewline
\ \ \ \ subcomponents\ x\ {\isasymnoteq}\ {\isacharbraceleft}{\isacharbraceright}\ {\isasymlongrightarrow}\ \isanewline
\ \ \ \ specKeys\ x\ {\isacharequal}\ \ {\isasymUnion}\ {\isacharparenleft}specKeys\ {\isacharbackquote}\ {\isacharparenleft}subcomponents\ x{\isacharparenright}{\isacharparenright}{\isachardoublequoteclose}\ \isanewline
\isanewline
\isamarkupcmt{If a component is a composite one, the set of its unguessable values%
}
\ \isanewline
\isamarkupcmt{is a union of the subcomponents' sets of the unguessable values%
}
\isanewline
\isacommand{definition}\isamarkupfalse%
\isanewline
\ \ correctCompositionSecrets\ {\isacharcolon}{\isacharcolon}\ \ {\isachardoublequoteopen}specID\ {\isasymRightarrow}\ bool{\isachardoublequoteclose}\isanewline
\isakeyword{where}\isanewline
\ \ {\isachardoublequoteopen}correctCompositionSecrets\ x\ {\isasymequiv}\isanewline
\ \ \ \ subcomponents\ x\ {\isasymnoteq}\ {\isacharbraceleft}{\isacharbraceright}\ {\isasymlongrightarrow}\ \isanewline
\ \ \ \ specSecrets\ x\ {\isacharequal}\ \ {\isasymUnion}\ {\isacharparenleft}specSecrets\ {\isacharbackquote}\ {\isacharparenleft}subcomponents\ x{\isacharparenright}{\isacharparenright}{\isachardoublequoteclose}\ \isanewline
\isanewline
\isamarkupcmt{If a component is a composite one, the set of its private keys and%
}
\ \isanewline
\isamarkupcmt{unguessable values is a union of the corresponding sets of its subcomponents%
}
\isanewline
\isacommand{definition}\isamarkupfalse%
\isanewline
\ \ correctCompositionKS\ {\isacharcolon}{\isacharcolon}\ \ {\isachardoublequoteopen}specID\ {\isasymRightarrow}\ bool{\isachardoublequoteclose}\isanewline
\isakeyword{where}\isanewline
\ \ {\isachardoublequoteopen}correctCompositionKS\ x\ {\isasymequiv}\isanewline
\ \ \ \ subcomponents\ x\ {\isasymnoteq}\ {\isacharbraceleft}{\isacharbraceright}\ {\isasymlongrightarrow}\ \isanewline
\ \ \ \ specKeysSecrets\ x\ {\isacharequal}\ \ {\isasymUnion}\ {\isacharparenleft}specKeysSecrets\ {\isacharbackquote}\ {\isacharparenleft}subcomponents\ x{\isacharparenright}{\isacharparenright}{\isachardoublequoteclose}\ \isanewline
\isanewline
\isamarkupcmt{Predicate defining set of correctness properties of the component's%
}
\isanewline
\isamarkupcmt{interface  and relations on its private keys and unguessable values%
}
\isanewline
\isacommand{definition}\isamarkupfalse%
\isanewline
\ \ correctComponentSecrecy\ \ {\isacharcolon}{\isacharcolon}\ \ {\isachardoublequoteopen}specID\ {\isasymRightarrow}\ bool{\isachardoublequoteclose}\isanewline
\isakeyword{where}\ \isanewline
\ \ {\isachardoublequoteopen}correctComponentSecrecy\ x\ {\isasymequiv}\ \isanewline
\ \ \ \ correctCompositionKS\ x\ {\isasymand}\ \isanewline
\ \ \ \ correctCompositionSecrets\ x\ {\isasymand}\ \isanewline
\ \ \ \ correctCompositionKeys\ x\ {\isasymand}\ \isanewline
\ \ \ \ correctCompositionLoc\ x\ {\isasymand}\isanewline
\ \ \ \ correctCompositionIn\ x\ {\isasymand}\isanewline
\ \ \ \ correctCompositionOut\ x\ {\isasymand}\ \isanewline
\ \ \ \ correctInOutLoc\ x{\isachardoublequoteclose}\isanewline
\isanewline
\isamarkupcmt{Predicate exprChannel I E defines whether the expression item E can be sent via the channel I%
}
\ \ \ \ \isanewline
\isacommand{consts}\isamarkupfalse%
\isanewline
\ exprChannel\ {\isacharcolon}{\isacharcolon}\ {\isachardoublequoteopen}chanID\ {\isasymRightarrow}\ Expression\ {\isasymRightarrow}\ bool{\isachardoublequoteclose}\isanewline
\isanewline
\isamarkupcmt{Predicate eoutM sP M E defines whether the component sP may eventually%
}
\isanewline
\isamarkupcmt{output an expression E if there exists a time interval t of%
}
\ \isanewline
\isamarkupcmt{an output channel which contains this expression E%
}
\isanewline
\isacommand{definition}\isamarkupfalse%
\isanewline
\ \ eout\ {\isacharcolon}{\isacharcolon}\ {\isachardoublequoteopen}specID\ \ {\isasymRightarrow}\ Expression\ {\isasymRightarrow}\ bool{\isachardoublequoteclose}\isanewline
\isakeyword{where}\isanewline
\ {\isachardoublequoteopen}eout\ sP\ E\ {\isasymequiv}\ \isanewline
\ \ {\isasymexists}\ {\isacharparenleft}ch\ {\isacharcolon}{\isacharcolon}\ chanID{\isacharparenright}{\isachardot}\ {\isacharparenleft}{\isacharparenleft}ch\ {\isasymin}\ {\isacharparenleft}out\ sP{\isacharparenright}{\isacharparenright}\ {\isasymand}\ {\isacharparenleft}exprChannel\ ch\ E{\isacharparenright}{\isacharparenright}{\isachardoublequoteclose}\isanewline
\isanewline
\isamarkupcmt{Predicate eout sP E defines whether the component sP may eventually%
}
\isanewline
\isamarkupcmt{output an expression E via subset of channels M,%
}
\isanewline
\isamarkupcmt{which is a subset of output channels of sP,%
}
\isanewline
\isamarkupcmt{and if there exists a time interval t of%
}
\ \isanewline
\isamarkupcmt{an output channel which contains this expression E%
}
\isanewline
\isacommand{definition}\isamarkupfalse%
\isanewline
\ \ eoutM\ {\isacharcolon}{\isacharcolon}\ {\isachardoublequoteopen}specID\ \ {\isasymRightarrow}\ chanID\ set\ {\isasymRightarrow}\ Expression\ {\isasymRightarrow}\ bool{\isachardoublequoteclose}\isanewline
\isakeyword{where}\isanewline
\ {\isachardoublequoteopen}eoutM\ sP\ M\ E\ {\isasymequiv}\ \isanewline
\ \ {\isasymexists}\ {\isacharparenleft}ch\ {\isacharcolon}{\isacharcolon}\ chanID{\isacharparenright}{\isachardot}\ {\isacharparenleft}{\isacharparenleft}ch\ {\isasymin}\ {\isacharparenleft}out\ sP{\isacharparenright}{\isacharparenright}\ {\isasymand}\ {\isacharparenleft}ch\ {\isasymin}\ M{\isacharparenright}\ {\isasymand}\ {\isacharparenleft}exprChannel\ ch\ E{\isacharparenright}{\isacharparenright}{\isachardoublequoteclose}\isanewline
\isanewline
\isamarkupcmt{Predicate ineM sP M E defines whether a component sP may eventually%
}
\isanewline
\isamarkupcmt{get an expression E  if there exists a time interval t of%
}
\ \isanewline
\isamarkupcmt{an input stream  which contains this expression E%
}
\isanewline
\isacommand{definition}\isamarkupfalse%
\isanewline
\ \ ine\ {\isacharcolon}{\isacharcolon}\ {\isachardoublequoteopen}specID\ \ {\isasymRightarrow}\ Expression\ {\isasymRightarrow}\ bool{\isachardoublequoteclose}\isanewline
\isakeyword{where}\isanewline
\ {\isachardoublequoteopen}ine\ sP\ E\ {\isasymequiv}\ \isanewline
\ \ {\isasymexists}\ {\isacharparenleft}ch\ {\isacharcolon}{\isacharcolon}\ chanID{\isacharparenright}{\isachardot}\ {\isacharparenleft}{\isacharparenleft}ch\ {\isasymin}\ {\isacharparenleft}ins\ sP{\isacharparenright}{\isacharparenright}\ {\isasymand}\ {\isacharparenleft}exprChannel\ ch\ E{\isacharparenright}{\isacharparenright}{\isachardoublequoteclose}\isanewline
\isanewline
\isamarkupcmt{Predicate ine sP E defines whether a component sP may eventually%
}
\isanewline
\isamarkupcmt{get an expression E via subset of channels M,%
}
\isanewline
\isamarkupcmt{which is a subset of input channels of sP,%
}
\isanewline
\isamarkupcmt{and if there exists a time interval t of%
}
\ \isanewline
\isamarkupcmt{an input stream  which contains this expression E%
}
\isanewline
\isacommand{definition}\isamarkupfalse%
\isanewline
\ \ ineM\ {\isacharcolon}{\isacharcolon}\ {\isachardoublequoteopen}specID\ \ {\isasymRightarrow}\ chanID\ set\ {\isasymRightarrow}\ Expression\ {\isasymRightarrow}\ bool{\isachardoublequoteclose}\isanewline
\isakeyword{where}\isanewline
\ {\isachardoublequoteopen}ineM\ sP\ M\ E\ {\isasymequiv}\ \isanewline
\ \ {\isasymexists}\ {\isacharparenleft}ch\ {\isacharcolon}{\isacharcolon}\ chanID{\isacharparenright}{\isachardot}\ {\isacharparenleft}{\isacharparenleft}ch\ {\isasymin}\ {\isacharparenleft}ins\ sP{\isacharparenright}{\isacharparenright}\ {\isasymand}\ {\isacharparenleft}ch\ {\isasymin}\ M{\isacharparenright}\ {\isasymand}\ {\isacharparenleft}exprChannel\ ch\ E{\isacharparenright}{\isacharparenright}{\isachardoublequoteclose}\isanewline
\isanewline
\isamarkupcmt{This predicate defines whether an input channel ch of a component sP%
}
\isanewline
\isamarkupcmt{is the only one input channel of this component%
}
\isanewline
\isamarkupcmt{via which it may eventually output an expression E%
}
\isanewline
\isacommand{definition}\isamarkupfalse%
\isanewline
\ \ out{\isacharunderscore}exprChannelSingle\ {\isacharcolon}{\isacharcolon}\ {\isachardoublequoteopen}specID\ \ {\isasymRightarrow}\ chanID\ {\isasymRightarrow}\ Expression\ {\isasymRightarrow}\ bool{\isachardoublequoteclose}\isanewline
\isakeyword{where}\isanewline
\ {\isachardoublequoteopen}out{\isacharunderscore}exprChannelSingle\ sP\ ch\ E\ {\isasymequiv}\ \isanewline
\ \ {\isacharparenleft}ch\ {\isasymin}\ {\isacharparenleft}out\ sP{\isacharparenright}{\isacharparenright}\ {\isasymand}\ \ \isanewline
\ \ {\isacharparenleft}exprChannel\ ch\ E{\isacharparenright}\ \ {\isasymand}\isanewline
\ \ {\isacharparenleft}{\isasymforall}\ {\isacharparenleft}x\ {\isacharcolon}{\isacharcolon}\ chanID{\isacharparenright}\ {\isacharparenleft}t\ {\isacharcolon}{\isacharcolon}\ nat{\isacharparenright}{\isachardot}\ {\isacharparenleft}{\isacharparenleft}x\ {\isasymin}\ {\isacharparenleft}out\ sP{\isacharparenright}{\isacharparenright}\ {\isasymand}\ {\isacharparenleft}x\ {\isasymnoteq}\ ch{\isacharparenright}\ {\isasymlongrightarrow}\ {\isasymnot}\ exprChannel\ x\ E{\isacharparenright}{\isacharparenright}{\isachardoublequoteclose}\isanewline
\isanewline
\isamarkupcmt{This predicate  yields true if only the channels from the set chSet,%
}
\isanewline
\isamarkupcmt{which is a subset of input channels of the  component sP,%
}
\isanewline
\isamarkupcmt{may eventually output an expression E%
}
\isanewline
\isacommand{definition}\isamarkupfalse%
\isanewline
\ out{\isacharunderscore}exprChannelSet\ {\isacharcolon}{\isacharcolon}\ {\isachardoublequoteopen}specID\ \ {\isasymRightarrow}\ chanID\ set\ {\isasymRightarrow}\ Expression\ {\isasymRightarrow}\ bool{\isachardoublequoteclose}\isanewline
\isakeyword{where}\isanewline
\ {\isachardoublequoteopen}out{\isacharunderscore}exprChannelSet\ sP\ chSet\ E\ {\isasymequiv}\ \isanewline
\ \ \ {\isacharparenleft}{\isacharparenleft}{\isasymforall}\ {\isacharparenleft}x\ {\isacharcolon}{\isacharcolon}chanID{\isacharparenright}{\isachardot}\ {\isacharparenleft}{\isacharparenleft}x\ {\isasymin}\ chSet{\isacharparenright}\ {\isasymlongrightarrow}\ {\isacharparenleft}{\isacharparenleft}x\ {\isasymin}\ {\isacharparenleft}out\ sP{\isacharparenright}{\isacharparenright}\ {\isasymand}\ {\isacharparenleft}exprChannel\ x\ E{\isacharparenright}{\isacharparenright}{\isacharparenright}{\isacharparenright}\isanewline
\ \ \ {\isasymand}\isanewline
\ \ \ {\isacharparenleft}{\isasymforall}\ {\isacharparenleft}x\ {\isacharcolon}{\isacharcolon}\ chanID{\isacharparenright}{\isachardot}\ {\isacharparenleft}{\isacharparenleft}x\ {\isasymnotin}\ chSet{\isacharparenright}\ {\isasymand}\ {\isacharparenleft}x\ {\isasymin}\ {\isacharparenleft}out\ sP{\isacharparenright}{\isacharparenright}\ {\isasymlongrightarrow}\ {\isasymnot}\ exprChannel\ x\ E{\isacharparenright}{\isacharparenright}{\isacharparenright}{\isachardoublequoteclose}\isanewline
\isanewline
\isamarkupcmt{This redicate defines whether%
}
\isanewline
\isamarkupcmt{an input channel ch of a component sP is the only one input channel%
}
\isanewline
\isamarkupcmt{of this component via which it may eventually get an expression E%
}
\isanewline
\isacommand{definition}\isamarkupfalse%
\isanewline
\ ine{\isacharunderscore}exprChannelSingle\ {\isacharcolon}{\isacharcolon}\ {\isachardoublequoteopen}specID\ \ {\isasymRightarrow}\ chanID\ {\isasymRightarrow}\ Expression\ {\isasymRightarrow}\ bool{\isachardoublequoteclose}\isanewline
\isakeyword{where}\isanewline
\ {\isachardoublequoteopen}ine{\isacharunderscore}exprChannelSingle\ sP\ ch\ E\ {\isasymequiv}\ \isanewline
\ \ {\isacharparenleft}ch\ {\isasymin}\ {\isacharparenleft}ins\ sP{\isacharparenright}{\isacharparenright}\ {\isasymand}\isanewline
\ \ {\isacharparenleft}exprChannel\ ch\ E{\isacharparenright}\ \ {\isasymand}\isanewline
\ \ {\isacharparenleft}{\isasymforall}\ {\isacharparenleft}x\ {\isacharcolon}{\isacharcolon}\ chanID{\isacharparenright}\ {\isacharparenleft}t\ {\isacharcolon}{\isacharcolon}\ nat{\isacharparenright}{\isachardot}\ {\isacharparenleft}{\isacharparenleft}\ x\ {\isasymin}\ {\isacharparenleft}ins\ sP{\isacharparenright}{\isacharparenright}\ {\isasymand}\ {\isacharparenleft}x\ {\isasymnoteq}\ ch{\isacharparenright}\ {\isasymlongrightarrow}\ {\isasymnot}\ exprChannel\ x\ E{\isacharparenright}{\isacharparenright}{\isachardoublequoteclose}\isanewline
\isanewline
\isamarkupcmt{This predicate yields true if the component sP may eventually%
}
\isanewline
\isamarkupcmt{get an expression E only via the channels from the set chSet,%
}
\isanewline
\isamarkupcmt{which is a subset of input channels of sP%
}
\isanewline
\isacommand{definition}\isamarkupfalse%
\isanewline
\ ine{\isacharunderscore}exprChannelSet\ {\isacharcolon}{\isacharcolon}\ {\isachardoublequoteopen}specID\ \ {\isasymRightarrow}\ chanID\ set\ {\isasymRightarrow}\ Expression\ {\isasymRightarrow}\ bool{\isachardoublequoteclose}\isanewline
\isakeyword{where}\isanewline
\ {\isachardoublequoteopen}ine{\isacharunderscore}exprChannelSet\ sP\ chSet\ E\ {\isasymequiv}\ \isanewline
\ \ \ {\isacharparenleft}{\isacharparenleft}{\isasymforall}\ {\isacharparenleft}x\ {\isacharcolon}{\isacharcolon}chanID{\isacharparenright}{\isachardot}\ {\isacharparenleft}{\isacharparenleft}x\ {\isasymin}\ chSet{\isacharparenright}\ {\isasymlongrightarrow}\ {\isacharparenleft}{\isacharparenleft}x\ {\isasymin}\ {\isacharparenleft}ins\ sP{\isacharparenright}{\isacharparenright}\ {\isasymand}\ {\isacharparenleft}exprChannel\ x\ E{\isacharparenright}{\isacharparenright}{\isacharparenright}{\isacharparenright}\isanewline
\ \ \ {\isasymand}\isanewline
\ \ \ {\isacharparenleft}{\isasymforall}\ {\isacharparenleft}x\ {\isacharcolon}{\isacharcolon}\ chanID{\isacharparenright}{\isachardot}\ {\isacharparenleft}{\isacharparenleft}x\ {\isasymnotin}\ chSet{\isacharparenright}\ {\isasymand}\ {\isacharparenleft}\ x\ {\isasymin}\ {\isacharparenleft}ins\ sP{\isacharparenright}{\isacharparenright}\ {\isasymlongrightarrow}\ {\isasymnot}\ exprChannel\ x\ E{\isacharparenright}{\isacharparenright}{\isacharparenright}{\isachardoublequoteclose}\isanewline
\isanewline
\isamarkupcmt{If a list of expression items does not contain any private key%
}
\isanewline
\isamarkupcmt{or unguessable value of a component P, then the first element%
}
\ \isanewline
\isamarkupcmt{of the list is neither a private key nor unguessable value of P%
}
\isanewline
\isacommand{lemma}\isamarkupfalse%
\ notSpecKeysSecretsExpr{\isacharunderscore}L{\isadigit{1}}{\isacharcolon}\isanewline
\isakeyword{assumes}\ {\isachardoublequoteopen}notSpecKeysSecretsExpr\ P\ {\isacharparenleft}a\ {\isacharhash}\ l{\isacharparenright}{\isachardoublequoteclose}\isanewline
\isakeyword{shows}\ \ \ \ {\isachardoublequoteopen}notSpecKeysSecretsExpr\ P\ {\isacharbrackleft}a{\isacharbrackright}{\isachardoublequoteclose}\isanewline
\isadelimproof
\endisadelimproof
\isatagproof
\isacommand{using}\isamarkupfalse%
\ assms\ \isacommand{by}\isamarkupfalse%
\ {\isacharparenleft}simp\ add{\isacharcolon}\ notSpecKeysSecretsExpr{\isacharunderscore}def{\isacharparenright}\isanewline
\isanewline
\isamarkupcmt{If a list of expression items does not contain any private key%
}
\isanewline
\isamarkupcmt{or unguessable value of a component P, then this list without its first%
}
\ \isanewline
\isamarkupcmt{element does not contain them too%
}
\endisatagproof
{\isafoldproof}%
\isadelimproof
\isanewline
\endisadelimproof
\isacommand{lemma}\isamarkupfalse%
\ notSpecKeysSecretsExpr{\isacharunderscore}L{\isadigit{2}}{\isacharcolon}\isanewline
\isakeyword{assumes}\ {\isachardoublequoteopen}notSpecKeysSecretsExpr\ P\ {\isacharparenleft}a\ {\isacharhash}\ l{\isacharparenright}{\isachardoublequoteclose}\isanewline
\isakeyword{shows}\ \ \ \ {\isachardoublequoteopen}notSpecKeysSecretsExpr\ P\ l{\isachardoublequoteclose}\ \isanewline
\isadelimproof
\endisadelimproof
\isatagproof
\isacommand{using}\isamarkupfalse%
\ assms\ \isacommand{by}\isamarkupfalse%
\ {\isacharparenleft}simp\ add{\isacharcolon}\ notSpecKeysSecretsExpr{\isacharunderscore}def{\isacharparenright}\isanewline
\isanewline
\isamarkupcmt{If a channel belongs to the set of input channels of a component P%
}
\isanewline
\isamarkupcmt{and does not belong to the set of local channels of the compositon of P and Q%
}
\ \isanewline
\isamarkupcmt{then it belongs to the set of input channels of this composition%
}
\endisatagproof
{\isafoldproof}%
\isadelimproof
\isanewline
\endisadelimproof
\isacommand{lemma}\isamarkupfalse%
\ correctCompositionIn{\isacharunderscore}L{\isadigit{1}}{\isacharcolon}\isanewline
\isakeyword{assumes}\ {\isachardoublequoteopen}subcomponents\ PQ\ {\isacharequal}\ {\isacharbraceleft}P{\isacharcomma}Q{\isacharbraceright}{\isachardoublequoteclose}\ \isanewline
\ \ \ \ \ \ \ \isakeyword{and}\ {\isachardoublequoteopen}correctCompositionIn\ PQ{\isachardoublequoteclose}\ \isanewline
\ \ \ \ \ \ \ \isakeyword{and}\ {\isachardoublequoteopen}ch\ {\isasymnotin}\ loc\ PQ{\isachardoublequoteclose}\isanewline
\ \ \ \ \ \ \ \isakeyword{and}\ {\isachardoublequoteopen}ch\ {\isasymin}\ ins\ P{\isachardoublequoteclose}\ \isanewline
\isakeyword{shows}\ \ \ \ {\isachardoublequoteopen}ch\ {\isasymin}\ ins\ PQ{\isachardoublequoteclose}\isanewline
\isadelimproof
\endisadelimproof
\isatagproof
\isacommand{using}\isamarkupfalse%
\ assms\ \isacommand{by}\isamarkupfalse%
\ {\isacharparenleft}simp\ add{\isacharcolon}\ correctCompositionIn{\isacharunderscore}def{\isacharparenright}\isanewline
\isanewline
\isamarkupcmt{If a channel belongs to the set of input channels of the compositon of P and Q%
}
\isanewline
\isamarkupcmt{then it belongs to the set of input channels either of P or of Q%
}
\endisatagproof
{\isafoldproof}%
\isadelimproof
\isanewline
\endisadelimproof
\isacommand{lemma}\isamarkupfalse%
\ correctCompositionIn{\isacharunderscore}L{\isadigit{2}}{\isacharcolon}\isanewline
\isakeyword{assumes}\ {\isachardoublequoteopen}subcomponents\ PQ\ {\isacharequal}\ {\isacharbraceleft}P{\isacharcomma}Q{\isacharbraceright}{\isachardoublequoteclose}\isanewline
\ \ \ \ \ \ \ \isakeyword{and}\ {\isachardoublequoteopen}correctCompositionIn\ PQ{\isachardoublequoteclose}\ \isanewline
\ \ \ \ \ \ \ \isakeyword{and}\ {\isachardoublequoteopen}ch\ {\isasymin}\ ins\ PQ{\isachardoublequoteclose}\ \isanewline
\isakeyword{shows}\ \ \ \ {\isachardoublequoteopen}{\isacharparenleft}ch\ {\isasymin}\ ins\ P{\isacharparenright}\ {\isasymor}\ {\isacharparenleft}ch\ {\isasymin}\ ins\ Q{\isacharparenright}{\isachardoublequoteclose}\ \isanewline
\isadelimproof
\endisadelimproof
\isatagproof
\isacommand{using}\isamarkupfalse%
\ assms\ \isacommand{by}\isamarkupfalse%
\ {\isacharparenleft}simp\ add{\isacharcolon}\ correctCompositionIn{\isacharunderscore}def{\isacharparenright}%
\endisatagproof
{\isafoldproof}%
\isadelimproof
\isanewline
\endisadelimproof
\isanewline
\isacommand{lemma}\isamarkupfalse%
\ ineM{\isacharunderscore}L{\isadigit{1}}{\isacharcolon}\isanewline
\isakeyword{assumes}\ {\isachardoublequoteopen}ch\ {\isasymin}\ M{\isachardoublequoteclose}\ \isanewline
\ \ \ \ \ \ \ \isakeyword{and}\ {\isachardoublequoteopen}ch\ {\isasymin}\ ins\ P{\isachardoublequoteclose}\isanewline
\ \ \ \ \ \ \ \isakeyword{and}\ {\isachardoublequoteopen}exprChannel\ ch\ E{\isachardoublequoteclose}\isanewline
\isakeyword{shows}\ \ \ \ {\isachardoublequoteopen}ineM\ P\ M\ E{\isachardoublequoteclose}\isanewline
\isadelimproof
\endisadelimproof
\isatagproof
\isacommand{using}\isamarkupfalse%
\ assms\ \isacommand{by}\isamarkupfalse%
\ {\isacharparenleft}simp\ add{\isacharcolon}\ ineM{\isacharunderscore}def{\isacharcomma}\ blast{\isacharparenright}%
\endisatagproof
{\isafoldproof}%
\isadelimproof
\isanewline
\endisadelimproof
\isanewline
\isacommand{lemma}\isamarkupfalse%
\ ineM{\isacharunderscore}ine{\isacharcolon}\isanewline
\isakeyword{assumes}\ {\isachardoublequoteopen}ineM\ P\ M\ E{\isachardoublequoteclose}\isanewline
\isakeyword{shows}\ \ \ \ {\isachardoublequoteopen}ine\ P\ E{\isachardoublequoteclose}\isanewline
\isadelimproof
\endisadelimproof
\isatagproof
\isacommand{using}\isamarkupfalse%
\ assms\ \isacommand{by}\isamarkupfalse%
\ {\isacharparenleft}simp\ add{\isacharcolon}\ ineM{\isacharunderscore}def\ ine{\isacharunderscore}def{\isacharcomma}\ blast{\isacharparenright}%
\endisatagproof
{\isafoldproof}%
\isadelimproof
\isanewline
\endisadelimproof
\isanewline
\isacommand{lemma}\isamarkupfalse%
\ not{\isacharunderscore}ine{\isacharunderscore}ineM{\isacharcolon}\isanewline
\isakeyword{assumes}\ {\isachardoublequoteopen}{\isasymnot}\ ine\ P\ E{\isachardoublequoteclose}\isanewline
\isakeyword{shows}\ \ \ \ {\isachardoublequoteopen}{\isasymnot}\ ineM\ P\ M\ E{\isachardoublequoteclose}\isanewline
\isadelimproof
\endisadelimproof
\isatagproof
\isacommand{using}\isamarkupfalse%
\ assms\ \isacommand{by}\isamarkupfalse%
\ {\isacharparenleft}simp\ add{\isacharcolon}\ ineM{\isacharunderscore}def\ ine{\isacharunderscore}def{\isacharparenright}%
\endisatagproof
{\isafoldproof}%
\isadelimproof
\isanewline
\endisadelimproof
\isanewline
\isacommand{lemma}\isamarkupfalse%
\ eoutM{\isacharunderscore}eout{\isacharcolon}\isanewline
\isakeyword{assumes}\ {\isachardoublequoteopen}eoutM\ P\ M\ E{\isachardoublequoteclose}\isanewline
\isakeyword{shows}\ \ \ \ {\isachardoublequoteopen}eout\ P\ E{\isachardoublequoteclose}\isanewline
\isadelimproof
\endisadelimproof
\isatagproof
\isacommand{using}\isamarkupfalse%
\ assms\ \isacommand{by}\isamarkupfalse%
\ {\isacharparenleft}simp\ add{\isacharcolon}\ eoutM{\isacharunderscore}def\ eout{\isacharunderscore}def{\isacharcomma}\ blast{\isacharparenright}%
\endisatagproof
{\isafoldproof}%
\isadelimproof
\isanewline
\endisadelimproof
\isanewline
\isacommand{lemma}\isamarkupfalse%
\ not{\isacharunderscore}eout{\isacharunderscore}eoutM{\isacharcolon}\isanewline
\isakeyword{assumes}\ {\isachardoublequoteopen}{\isasymnot}\ eout\ P\ E{\isachardoublequoteclose}\isanewline
\isakeyword{shows}\ \ \ \ {\isachardoublequoteopen}{\isasymnot}\ eoutM\ P\ M\ E{\isachardoublequoteclose}\isanewline
\isadelimproof
\endisadelimproof
\isatagproof
\isacommand{using}\isamarkupfalse%
\ assms\ \isacommand{by}\isamarkupfalse%
\ {\isacharparenleft}simp\ add{\isacharcolon}\ eoutM{\isacharunderscore}def\ eout{\isacharunderscore}def{\isacharparenright}%
\endisatagproof
{\isafoldproof}%
\isadelimproof
\isanewline
\endisadelimproof
\isanewline
\isacommand{lemma}\isamarkupfalse%
\ correctCompositionKeys{\isacharunderscore}subcomp{\isadigit{1}}{\isacharcolon}\isanewline
\isakeyword{assumes}\ {\isachardoublequoteopen}correctCompositionKeys\ C{\isachardoublequoteclose}\isanewline
\ \ \ \ \ \ \ \ \isakeyword{and}\ {\isachardoublequoteopen}x\ {\isasymin}\ subcomponents\ C{\isachardoublequoteclose}\ \isanewline
\ \ \ \ \ \ \ \ \isakeyword{and}\ {\isachardoublequoteopen}xb\ {\isasymin}\ specKeys\ C{\isachardoublequoteclose}\isanewline
\isakeyword{shows}\ {\isachardoublequoteopen}{\isasymexists}\ x\ {\isasymin}\ subcomponents\ C{\isachardot}\ {\isacharparenleft}xb\ {\isasymin}\ specKeys\ x{\isacharparenright}{\isachardoublequoteclose}\isanewline
\isadelimproof
\endisadelimproof
\isatagproof
\isacommand{using}\isamarkupfalse%
\ assms\ \isacommand{by}\isamarkupfalse%
\ {\isacharparenleft}simp\ add{\isacharcolon}\ correctCompositionKeys{\isacharunderscore}def{\isacharcomma}\ auto{\isacharparenright}%
\endisatagproof
{\isafoldproof}%
\isadelimproof
\isanewline
\endisadelimproof
\isanewline
\isacommand{lemma}\isamarkupfalse%
\ correctCompositionSecrets{\isacharunderscore}subcomp{\isadigit{1}}{\isacharcolon}\isanewline
\isakeyword{assumes}\ {\isachardoublequoteopen}correctCompositionSecrets\ C{\isachardoublequoteclose}\ \isanewline
\ \ \ \ \ \ \ \ \isakeyword{and}\ {\isachardoublequoteopen}x\ {\isasymin}\ subcomponents\ C{\isachardoublequoteclose}\isanewline
\ \ \ \ \ \ \ \ \isakeyword{and}\ {\isachardoublequoteopen}s\ {\isasymin}\ specSecrets\ C{\isachardoublequoteclose}\isanewline
\isakeyword{shows}\ \ {\isachardoublequoteopen}{\isasymexists}\ x\ {\isasymin}\ subcomponents\ C{\isachardot}\ {\isacharparenleft}s\ {\isasymin}\ specSecrets\ x{\isacharparenright}{\isachardoublequoteclose}\isanewline
\isadelimproof
\endisadelimproof
\isatagproof
\isacommand{using}\isamarkupfalse%
\ assms\ \isacommand{by}\isamarkupfalse%
\ {\isacharparenleft}simp\ add{\isacharcolon}\ correctCompositionSecrets{\isacharunderscore}def{\isacharcomma}\ auto{\isacharparenright}%
\endisatagproof
{\isafoldproof}%
\isadelimproof
\isanewline
\endisadelimproof
\isanewline
\isacommand{lemma}\isamarkupfalse%
\ correctCompositionKeys{\isacharunderscore}subcomp{\isadigit{2}}{\isacharcolon}\isanewline
\isakeyword{assumes}\ {\isachardoublequoteopen}correctCompositionKeys\ C{\isachardoublequoteclose}\isanewline
\ \ \ \ \ \ \ \isakeyword{and}\ {\isachardoublequoteopen}xb\ {\isasymin}\ subcomponents\ C{\isachardoublequoteclose}\isanewline
\ \ \ \ \ \ \ \isakeyword{and}\ {\isachardoublequoteopen}xc\ {\isasymin}\ specKeys\ xb{\isachardoublequoteclose}\isanewline
\isakeyword{shows}\ {\isachardoublequoteopen}xc\ {\isasymin}\ specKeys\ C{\isachardoublequoteclose}\isanewline
\isadelimproof
\endisadelimproof
\isatagproof
\isacommand{using}\isamarkupfalse%
\ assms\ \isacommand{by}\isamarkupfalse%
\ {\isacharparenleft}simp\ add{\isacharcolon}\ correctCompositionKeys{\isacharunderscore}def{\isacharcomma}\ auto{\isacharparenright}%
\endisatagproof
{\isafoldproof}%
\isadelimproof
\isanewline
\endisadelimproof
\isanewline
\isacommand{lemma}\isamarkupfalse%
\ correctCompositionSecrets{\isacharunderscore}subcomp{\isadigit{2}}{\isacharcolon}\isanewline
\isakeyword{assumes}\ {\isachardoublequoteopen}correctCompositionSecrets\ C{\isachardoublequoteclose}\isanewline
\ \ \ \ \ \ \ \ \isakeyword{and}\ {\isachardoublequoteopen}xb\ {\isasymin}\ subcomponents\ C{\isachardoublequoteclose}\isanewline
\ \ \ \ \ \ \ \ \isakeyword{and}\ {\isachardoublequoteopen}xc\ {\isasymin}\ specSecrets\ xb{\isachardoublequoteclose}\isanewline
\isakeyword{shows}\ \ \ \ \ {\isachardoublequoteopen}xc\ {\isasymin}\ specSecrets\ C{\isachardoublequoteclose}\isanewline
\isadelimproof
\endisadelimproof
\isatagproof
\isacommand{using}\isamarkupfalse%
\ assms\ \isacommand{by}\isamarkupfalse%
\ {\isacharparenleft}simp\ add{\isacharcolon}\ correctCompositionSecrets{\isacharunderscore}def{\isacharcomma}\ auto{\isacharparenright}%
\endisatagproof
{\isafoldproof}%
\isadelimproof
\isanewline
\endisadelimproof
\isanewline
\isacommand{lemma}\isamarkupfalse%
\ correctCompKS{\isacharunderscore}Keys{\isacharcolon}\isanewline
\isakeyword{assumes}\ {\isachardoublequoteopen}correctCompositionKS\ C{\isachardoublequoteclose}\isanewline
\isakeyword{shows}\ \ \ \ {\isachardoublequoteopen}correctCompositionKeys\ C{\isachardoublequoteclose}\isanewline
\isadelimproof
\endisadelimproof
\isatagproof
\isacommand{proof}\isamarkupfalse%
\ {\isacharparenleft}cases\ {\isachardoublequoteopen}subcomponents\ C\ {\isacharequal}\ {\isacharbraceleft}{\isacharbraceright}{\isachardoublequoteclose}{\isacharparenright}\isanewline
\ \ \isacommand{assume}\isamarkupfalse%
\ {\isachardoublequoteopen}subcomponents\ C\ {\isacharequal}\ {\isacharbraceleft}{\isacharbraceright}{\isachardoublequoteclose}\isanewline
\ \ \isacommand{from}\isamarkupfalse%
\ this\ \isakeyword{and}\ assms\ \isacommand{show}\isamarkupfalse%
\ {\isacharquery}thesis\isanewline
\ \ \isacommand{by}\isamarkupfalse%
\ {\isacharparenleft}simp\ add{\isacharcolon}\ correctCompositionKeys{\isacharunderscore}def{\isacharparenright}\isanewline
\isacommand{next}\isamarkupfalse%
\isanewline
\ \ \isacommand{assume}\isamarkupfalse%
\ {\isachardoublequoteopen}subcomponents\ C\ {\isasymnoteq}\ {\isacharbraceleft}{\isacharbraceright}{\isachardoublequoteclose}\isanewline
\ \ \isacommand{from}\isamarkupfalse%
\ this\ \isakeyword{and}\ assms\ \isacommand{show}\isamarkupfalse%
\ {\isacharquery}thesis\ \isanewline
\ \ \isacommand{by}\isamarkupfalse%
\ {\isacharparenleft}simp\ add{\isacharcolon}\ correctCompositionKS{\isacharunderscore}def\ \isanewline
\ \ \ \ \ \ \ \ \ \ \ \ \ \ \ \ correctCompositionKeys{\isacharunderscore}def\isanewline
\ \ \ \ \ \ \ \ \ \ \ \ \ \ \ \ specKeysSecrets{\isacharunderscore}def{\isacharcomma}\ blast{\isacharparenright}\isanewline
\isacommand{qed}\isamarkupfalse%
\endisatagproof
{\isafoldproof}%
\isadelimproof
\isanewline
\endisadelimproof
\isanewline
\isacommand{lemma}\isamarkupfalse%
\ correctCompKS{\isacharunderscore}Secrets{\isacharcolon}\isanewline
\isakeyword{assumes}\ {\isachardoublequoteopen}correctCompositionKS\ C{\isachardoublequoteclose}\isanewline
\isakeyword{shows}\ \ \ \ {\isachardoublequoteopen}correctCompositionSecrets\ C{\isachardoublequoteclose}\isanewline
\isadelimproof
\endisadelimproof
\isatagproof
\isacommand{proof}\isamarkupfalse%
\ {\isacharparenleft}cases\ {\isachardoublequoteopen}subcomponents\ C\ {\isacharequal}\ {\isacharbraceleft}{\isacharbraceright}{\isachardoublequoteclose}{\isacharparenright}\isanewline
\ \ \isacommand{assume}\isamarkupfalse%
\ {\isachardoublequoteopen}subcomponents\ C\ {\isacharequal}\ {\isacharbraceleft}{\isacharbraceright}{\isachardoublequoteclose}\isanewline
\ \ \isacommand{from}\isamarkupfalse%
\ this\ \isakeyword{and}\ assms\ \isacommand{show}\isamarkupfalse%
\ {\isacharquery}thesis\isanewline
\ \ \isacommand{by}\isamarkupfalse%
\ {\isacharparenleft}simp\ add{\isacharcolon}\ correctCompositionSecrets{\isacharunderscore}def{\isacharparenright}\isanewline
\isacommand{next}\isamarkupfalse%
\isanewline
\ \ \isacommand{assume}\isamarkupfalse%
\ {\isachardoublequoteopen}subcomponents\ C\ {\isasymnoteq}\ {\isacharbraceleft}{\isacharbraceright}{\isachardoublequoteclose}\isanewline
\ \ \isacommand{from}\isamarkupfalse%
\ this\ \isakeyword{and}\ assms\ \isacommand{show}\isamarkupfalse%
\ {\isacharquery}thesis\ \isanewline
\ \ \isacommand{by}\isamarkupfalse%
\ {\isacharparenleft}simp\ add{\isacharcolon}\ correctCompositionKS{\isacharunderscore}def\ \isanewline
\ \ \ \ \ \ \ \ \ \ \ \ \ \ \ \ correctCompositionSecrets{\isacharunderscore}def\isanewline
\ \ \ \ \ \ \ \ \ \ \ \ \ \ \ \ specKeysSecrets{\isacharunderscore}def{\isacharcomma}\ blast{\isacharparenright}\isanewline
\isacommand{qed}\isamarkupfalse%
\endisatagproof
{\isafoldproof}%
\isadelimproof
\isanewline
\endisadelimproof
\isanewline
\isacommand{lemma}\isamarkupfalse%
\ correctCompKS{\isacharunderscore}KeysSecrets{\isacharcolon}\isanewline
\isakeyword{assumes}\ {\isachardoublequoteopen}correctCompositionKeys\ C{\isachardoublequoteclose}\isanewline
\ \ \ \ \ \ \ \ \isakeyword{and}\ {\isachardoublequoteopen}correctCompositionSecrets\ C{\isachardoublequoteclose}\isanewline
\isakeyword{shows}\ \ \ \ {\isachardoublequoteopen}correctCompositionKS\ C{\isachardoublequoteclose}\isanewline
\isadelimproof
\endisadelimproof
\isatagproof
\isacommand{proof}\isamarkupfalse%
\ {\isacharparenleft}cases\ {\isachardoublequoteopen}subcomponents\ C\ {\isacharequal}\ {\isacharbraceleft}{\isacharbraceright}{\isachardoublequoteclose}{\isacharparenright}\isanewline
\ \ \isacommand{assume}\isamarkupfalse%
\ {\isachardoublequoteopen}subcomponents\ C\ {\isacharequal}\ {\isacharbraceleft}{\isacharbraceright}{\isachardoublequoteclose}\isanewline
\ \ \isacommand{from}\isamarkupfalse%
\ this\ \isakeyword{and}\ assms\ \isacommand{show}\isamarkupfalse%
\ {\isacharquery}thesis\isanewline
\ \ \isacommand{by}\isamarkupfalse%
\ {\isacharparenleft}simp\ add{\isacharcolon}\ correctCompositionKS{\isacharunderscore}def{\isacharparenright}\isanewline
\isacommand{next}\isamarkupfalse%
\isanewline
\ \ \isacommand{assume}\isamarkupfalse%
\ {\isachardoublequoteopen}subcomponents\ C\ {\isasymnoteq}\ {\isacharbraceleft}{\isacharbraceright}{\isachardoublequoteclose}\isanewline
\ \ \isacommand{from}\isamarkupfalse%
\ this\ \isakeyword{and}\ assms\ \isacommand{show}\isamarkupfalse%
\ {\isacharquery}thesis\ \isanewline
\ \ \isacommand{by}\isamarkupfalse%
\ {\isacharparenleft}simp\ add{\isacharcolon}\ correctCompositionKS{\isacharunderscore}def\ \isanewline
\ \ \ \ \ \ \ \ \ \ \ \ \ \ \ \ correctCompositionKeys{\isacharunderscore}def\ \isanewline
\ \ \ \ \ \ \ \ \ \ \ \ \ \ \ \ correctCompositionSecrets{\isacharunderscore}def\isanewline
\ \ \ \ \ \ \ \ \ \ \ \ \ \ \ \ specKeysSecrets{\isacharunderscore}def{\isacharcomma}\ blast{\isacharparenright}\isanewline
\isacommand{qed}\isamarkupfalse%
\endisatagproof
{\isafoldproof}%
\isadelimproof
\ \isanewline
\endisadelimproof
\isanewline
\isacommand{lemma}\isamarkupfalse%
\ correctCompositionKS{\isacharunderscore}subcomp{\isadigit{1}}{\isacharcolon}\isanewline
\isakeyword{assumes}\ {\isachardoublequoteopen}correctCompositionKS\ C{\isachardoublequoteclose}\isanewline
\ \ \ \ \ \ \ \isakeyword{and}\ h{\isadigit{1}}{\isacharcolon}{\isachardoublequoteopen}x\ {\isasymin}\ subcomponents\ C{\isachardoublequoteclose}\isanewline
\ \ \ \ \ \ \ \isakeyword{and}\ {\isachardoublequoteopen}xa\ {\isasymin}\ specKeys\ C{\isachardoublequoteclose}\isanewline
\isakeyword{shows}\ \ \ \ {\isachardoublequoteopen}{\isasymexists}\ y\ {\isasymin}\ subcomponents\ C{\isachardot}\ {\isacharparenleft}xa\ {\isasymin}\ specKeys\ y{\isacharparenright}{\isachardoublequoteclose}\isanewline
\isadelimproof
\endisadelimproof
\isatagproof
\isacommand{proof}\isamarkupfalse%
\ {\isacharparenleft}cases\ {\isachardoublequoteopen}subcomponents\ C\ {\isacharequal}\ {\isacharbraceleft}{\isacharbraceright}{\isachardoublequoteclose}{\isacharparenright}\isanewline
\ \ \isacommand{assume}\isamarkupfalse%
\ {\isachardoublequoteopen}subcomponents\ C\ {\isacharequal}\ {\isacharbraceleft}{\isacharbraceright}{\isachardoublequoteclose}\isanewline
\ \ \isacommand{from}\isamarkupfalse%
\ this\ \isakeyword{and}\ h{\isadigit{1}}\ \isacommand{show}\isamarkupfalse%
\ {\isacharquery}thesis\ \isacommand{by}\isamarkupfalse%
\ simp\ \isanewline
\isacommand{next}\isamarkupfalse%
\isanewline
\ \ \isacommand{assume}\isamarkupfalse%
\ {\isachardoublequoteopen}subcomponents\ C\ {\isasymnoteq}\ {\isacharbraceleft}{\isacharbraceright}{\isachardoublequoteclose}\isanewline
\ \ \isacommand{from}\isamarkupfalse%
\ this\ \isakeyword{and}\ assms\ \isacommand{show}\isamarkupfalse%
\ {\isacharquery}thesis\ \isanewline
\ \ \isacommand{by}\isamarkupfalse%
\ {\isacharparenleft}simp\ add{\isacharcolon}\ correctCompositionKS{\isacharunderscore}def\ specKeysSecrets{\isacharunderscore}def{\isacharcomma}\ blast{\isacharparenright}\ \isanewline
\isacommand{qed}\isamarkupfalse%
\endisatagproof
{\isafoldproof}%
\isadelimproof
\isanewline
\endisadelimproof
\isanewline
\isacommand{lemma}\isamarkupfalse%
\ correctCompositionKS{\isacharunderscore}subcomp{\isadigit{2}}{\isacharcolon}\isanewline
\isakeyword{assumes}\ {\isachardoublequoteopen}correctCompositionKS\ C{\isachardoublequoteclose}\isanewline
\ \ \ \ \ \ \ \ \isakeyword{and}\ h{\isadigit{1}}{\isacharcolon}{\isachardoublequoteopen}x\ {\isasymin}\ subcomponents\ C{\isachardoublequoteclose}\isanewline
\ \ \ \ \ \ \ \ \isakeyword{and}\ {\isachardoublequoteopen}xa\ {\isasymin}\ specSecrets\ C{\isachardoublequoteclose}\isanewline
\isakeyword{shows}\ \ \ \ {\isachardoublequoteopen}{\isasymexists}\ y\ {\isasymin}\ subcomponents\ C{\isachardot}\ xa\ {\isasymin}\ specSecrets\ y{\isachardoublequoteclose}\isanewline
\isadelimproof
\endisadelimproof
\isatagproof
\isacommand{proof}\isamarkupfalse%
\ {\isacharparenleft}cases\ {\isachardoublequoteopen}subcomponents\ C\ {\isacharequal}\ {\isacharbraceleft}{\isacharbraceright}{\isachardoublequoteclose}{\isacharparenright}\isanewline
\ \ \isacommand{assume}\isamarkupfalse%
\ {\isachardoublequoteopen}subcomponents\ C\ {\isacharequal}\ {\isacharbraceleft}{\isacharbraceright}{\isachardoublequoteclose}\isanewline
\ \ \isacommand{from}\isamarkupfalse%
\ this\ \isakeyword{and}\ h{\isadigit{1}}\ \isacommand{show}\isamarkupfalse%
\ {\isacharquery}thesis\ \isacommand{by}\isamarkupfalse%
\ simp\ \isanewline
\isacommand{next}\isamarkupfalse%
\isanewline
\ \ \isacommand{assume}\isamarkupfalse%
\ {\isachardoublequoteopen}subcomponents\ C\ {\isasymnoteq}\ {\isacharbraceleft}{\isacharbraceright}{\isachardoublequoteclose}\isanewline
\ \ \isacommand{from}\isamarkupfalse%
\ this\ \isakeyword{and}\ assms\ \isacommand{show}\isamarkupfalse%
\ {\isacharquery}thesis\ \isanewline
\ \ \isacommand{by}\isamarkupfalse%
\ {\isacharparenleft}simp\ add{\isacharcolon}\ correctCompositionKS{\isacharunderscore}def\ specKeysSecrets{\isacharunderscore}def{\isacharcomma}\ blast{\isacharparenright}\isanewline
\isacommand{qed}\isamarkupfalse%
\endisatagproof
{\isafoldproof}%
\isadelimproof
\isanewline
\endisadelimproof
\isanewline
\isacommand{lemma}\isamarkupfalse%
\ correctCompositionKS{\isacharunderscore}subcomp{\isadigit{3}}{\isacharcolon}\isanewline
\isakeyword{assumes}\ {\isachardoublequoteopen}correctCompositionKS\ C{\isachardoublequoteclose}\isanewline
\ \ \ \ \ \ \ \isakeyword{and}\ {\isachardoublequoteopen}x\ {\isasymin}\ subcomponents\ C{\isachardoublequoteclose}\isanewline
\ \ \ \ \ \ \ \isakeyword{and}\ {\isachardoublequoteopen}xa\ {\isasymin}\ specKeys\ x{\isachardoublequoteclose}\isanewline
\isakeyword{shows}\ \ \ \ {\isachardoublequoteopen}xa\ {\isasymin}\ specKeys\ C{\isachardoublequoteclose}\isanewline
\isadelimproof
\endisadelimproof
\isatagproof
\isacommand{using}\isamarkupfalse%
\ assms\ \isanewline
\isacommand{by}\isamarkupfalse%
\ {\isacharparenleft}simp\ add{\isacharcolon}\ correctCompositionKS{\isacharunderscore}def\ specKeysSecrets{\isacharunderscore}def{\isacharcomma}\ auto{\isacharparenright}%
\endisatagproof
{\isafoldproof}%
\isadelimproof
\isanewline
\endisadelimproof
\isanewline
\isacommand{lemma}\isamarkupfalse%
\ correctCompositionKS{\isacharunderscore}subcomp{\isadigit{4}}{\isacharcolon}\isanewline
\isakeyword{assumes}\ {\isachardoublequoteopen}correctCompositionKS\ C{\isachardoublequoteclose}\isanewline
\ \ \ \ \ \ \ \ \isakeyword{and}\ {\isachardoublequoteopen}x\ {\isasymin}\ subcomponents\ C{\isachardoublequoteclose}\isanewline
\ \ \ \ \ \ \ \ \isakeyword{and}\ {\isachardoublequoteopen}xa\ {\isasymin}\ specSecrets\ x{\isachardoublequoteclose}\ \isanewline
\isakeyword{shows}\ \ \ \ \ {\isachardoublequoteopen}xa\ {\isasymin}\ specSecrets\ C{\isachardoublequoteclose}\isanewline
\isadelimproof
\endisadelimproof
\isatagproof
\isacommand{using}\isamarkupfalse%
\ assms\ \isanewline
\isacommand{by}\isamarkupfalse%
\ {\isacharparenleft}simp\ add{\isacharcolon}\ correctCompositionKS{\isacharunderscore}def\ specKeysSecrets{\isacharunderscore}def{\isacharcomma}\ auto{\isacharparenright}%
\endisatagproof
{\isafoldproof}%
\isadelimproof
\isanewline
\endisadelimproof
\isanewline
\isacommand{lemma}\isamarkupfalse%
\ correctCompositionKS{\isacharunderscore}PQ{\isacharcolon}\isanewline
\isakeyword{assumes}\ {\isachardoublequoteopen}subcomponents\ PQ\ {\isacharequal}\ {\isacharbraceleft}P{\isacharcomma}\ Q{\isacharbraceright}{\isachardoublequoteclose}\isanewline
\ \ \ \ \ \ \ \isakeyword{and}\ {\isachardoublequoteopen}correctCompositionKS\ PQ{\isachardoublequoteclose}\ \isanewline
\ \ \ \ \ \ \ \isakeyword{and}\ {\isachardoublequoteopen}ks\ {\isasymin}\ specKeysSecrets\ PQ{\isachardoublequoteclose}\isanewline
\isakeyword{shows}\ \ \ \ {\isachardoublequoteopen}ks\ {\isasymin}\ specKeysSecrets\ P\ {\isasymor}\ ks\ {\isasymin}\ specKeysSecrets\ Q{\isachardoublequoteclose}\isanewline
\isadelimproof
\endisadelimproof
\isatagproof
\isacommand{using}\isamarkupfalse%
\ assms\ \isacommand{by}\isamarkupfalse%
\ {\isacharparenleft}simp\ add{\isacharcolon}\ correctCompositionKS{\isacharunderscore}def{\isacharparenright}%
\endisatagproof
{\isafoldproof}%
\isadelimproof
\isanewline
\endisadelimproof
\isanewline
\isacommand{lemma}\isamarkupfalse%
\ correctCompositionKS{\isacharunderscore}neg{\isadigit{1}}{\isacharcolon}\isanewline
\isakeyword{assumes}\ {\isachardoublequoteopen}subcomponents\ PQ\ {\isacharequal}\ {\isacharbraceleft}P{\isacharcomma}\ Q{\isacharbraceright}{\isachardoublequoteclose}\isanewline
\ \ \ \ \ \ \ \isakeyword{and}\ {\isachardoublequoteopen}correctCompositionKS\ PQ{\isachardoublequoteclose}\ \isanewline
\ \ \ \ \ \ \ \isakeyword{and}\ {\isachardoublequoteopen}ks\ {\isasymnotin}\ specKeysSecrets\ P{\isachardoublequoteclose}\isanewline
\ \ \ \ \ \ \ \isakeyword{and}\ {\isachardoublequoteopen}ks\ {\isasymnotin}\ specKeysSecrets\ Q{\isachardoublequoteclose}\isanewline
\isakeyword{shows}\ \ \ \ {\isachardoublequoteopen}ks\ {\isasymnotin}\ specKeysSecrets\ PQ{\isachardoublequoteclose}\isanewline
\isadelimproof
\endisadelimproof
\isatagproof
\isacommand{using}\isamarkupfalse%
\ assms\ \isacommand{by}\isamarkupfalse%
\ {\isacharparenleft}simp\ add{\isacharcolon}\ correctCompositionKS{\isacharunderscore}def{\isacharparenright}%
\endisatagproof
{\isafoldproof}%
\isadelimproof
\isanewline
\endisadelimproof
\isanewline
\isacommand{lemma}\isamarkupfalse%
\ correctCompositionKS{\isacharunderscore}negP{\isacharcolon}\isanewline
\isakeyword{assumes}\ {\isachardoublequoteopen}subcomponents\ PQ\ {\isacharequal}\ {\isacharbraceleft}P{\isacharcomma}\ Q{\isacharbraceright}{\isachardoublequoteclose}\isanewline
\ \ \ \ \ \ \ \ \isakeyword{and}\ {\isachardoublequoteopen}correctCompositionKS\ PQ{\isachardoublequoteclose}\ \isanewline
\ \ \ \ \ \ \ \ \isakeyword{and}\ {\isachardoublequoteopen}ks\ {\isasymnotin}\ specKeysSecrets\ PQ{\isachardoublequoteclose}\ \isanewline
\isakeyword{shows}\ \ \ \ \ {\isachardoublequoteopen}ks\ {\isasymnotin}\ specKeysSecrets\ P{\isachardoublequoteclose}\isanewline
\isadelimproof
\endisadelimproof
\isatagproof
\isacommand{using}\isamarkupfalse%
\ assms\ \isacommand{by}\isamarkupfalse%
\ {\isacharparenleft}simp\ add{\isacharcolon}\ correctCompositionKS{\isacharunderscore}def{\isacharparenright}%
\endisatagproof
{\isafoldproof}%
\isadelimproof
\isanewline
\endisadelimproof
\isanewline
\isacommand{lemma}\isamarkupfalse%
\ correctCompositionKS{\isacharunderscore}negQ{\isacharcolon}\isanewline
\isakeyword{assumes}\ {\isachardoublequoteopen}subcomponents\ PQ\ {\isacharequal}\ {\isacharbraceleft}P{\isacharcomma}\ Q{\isacharbraceright}{\isachardoublequoteclose}\isanewline
\ \ \ \ \ \ \ \ \isakeyword{and}\ {\isachardoublequoteopen}correctCompositionKS\ PQ{\isachardoublequoteclose}\ \isanewline
\ \ \ \ \ \ \ \ \isakeyword{and}\ {\isachardoublequoteopen}ks\ {\isasymnotin}\ specKeysSecrets\ PQ{\isachardoublequoteclose}\ \isanewline
\isakeyword{shows}\ \ \ \ \ {\isachardoublequoteopen}ks\ {\isasymnotin}\ specKeysSecrets\ Q{\isachardoublequoteclose}\isanewline
\isadelimproof
\endisadelimproof
\isatagproof
\isacommand{using}\isamarkupfalse%
\ assms\ \isacommand{by}\isamarkupfalse%
\ {\isacharparenleft}simp\ add{\isacharcolon}\ correctCompositionKS{\isacharunderscore}def{\isacharparenright}%
\endisatagproof
{\isafoldproof}%
\isadelimproof
\isanewline
\endisadelimproof
\isanewline
\isacommand{lemma}\isamarkupfalse%
\ out{\isacharunderscore}exprChannelSingle{\isacharunderscore}Set{\isacharcolon}\isanewline
\isakeyword{assumes}\ {\isachardoublequoteopen}out{\isacharunderscore}exprChannelSingle\ P\ ch\ E{\isachardoublequoteclose}\isanewline
\isakeyword{shows}\ \ \ \ {\isachardoublequoteopen}out{\isacharunderscore}exprChannelSet\ P\ {\isacharbraceleft}ch{\isacharbraceright}\ E{\isachardoublequoteclose}\isanewline
\isadelimproof
\endisadelimproof
\isatagproof
\isacommand{using}\isamarkupfalse%
\ assms\ \isanewline
\isacommand{by}\isamarkupfalse%
\ {\isacharparenleft}simp\ add{\isacharcolon}\ out{\isacharunderscore}exprChannelSingle{\isacharunderscore}def\ out{\isacharunderscore}exprChannelSet{\isacharunderscore}def{\isacharparenright}%
\endisatagproof
{\isafoldproof}%
\isadelimproof
\isanewline
\endisadelimproof
\isanewline
\isacommand{lemma}\isamarkupfalse%
\ out{\isacharunderscore}exprChannelSet{\isacharunderscore}Single{\isacharcolon}\isanewline
\isakeyword{assumes}\ {\isachardoublequoteopen}out{\isacharunderscore}exprChannelSet\ P\ {\isacharbraceleft}ch{\isacharbraceright}\ E{\isachardoublequoteclose}\isanewline
\isakeyword{shows}\ \ \ \ {\isachardoublequoteopen}out{\isacharunderscore}exprChannelSingle\ P\ ch\ E{\isachardoublequoteclose}\isanewline
\isadelimproof
\endisadelimproof
\isatagproof
\isacommand{using}\isamarkupfalse%
\ assms\isanewline
\isacommand{by}\isamarkupfalse%
\ {\isacharparenleft}simp\ add{\isacharcolon}\ out{\isacharunderscore}exprChannelSingle{\isacharunderscore}def\ out{\isacharunderscore}exprChannelSet{\isacharunderscore}def{\isacharparenright}%
\endisatagproof
{\isafoldproof}%
\isadelimproof
\isanewline
\endisadelimproof
\isanewline
\isacommand{lemma}\isamarkupfalse%
\ ine{\isacharunderscore}exprChannelSingle{\isacharunderscore}Set{\isacharcolon}\isanewline
\isakeyword{assumes}\ {\isachardoublequoteopen}ine{\isacharunderscore}exprChannelSingle\ P\ ch\ E{\isachardoublequoteclose}\isanewline
\ \ \isakeyword{shows}\ {\isachardoublequoteopen}ine{\isacharunderscore}exprChannelSet\ P\ {\isacharbraceleft}ch{\isacharbraceright}\ E{\isachardoublequoteclose}\isanewline
\isadelimproof
\endisadelimproof
\isatagproof
\isacommand{using}\isamarkupfalse%
\ assms\ \isanewline
\isacommand{by}\isamarkupfalse%
\ {\isacharparenleft}simp\ add{\isacharcolon}\ ine{\isacharunderscore}exprChannelSingle{\isacharunderscore}def\ ine{\isacharunderscore}exprChannelSet{\isacharunderscore}def{\isacharparenright}%
\endisatagproof
{\isafoldproof}%
\isadelimproof
\isanewline
\endisadelimproof
\isanewline
\isacommand{lemma}\isamarkupfalse%
\ ine{\isacharunderscore}exprChannelSet{\isacharunderscore}Single{\isacharcolon}\isanewline
\isakeyword{assumes}\ {\isachardoublequoteopen}ine{\isacharunderscore}exprChannelSet\ P\ {\isacharbraceleft}ch{\isacharbraceright}\ E{\isachardoublequoteclose}\isanewline
\isakeyword{shows}\ \ \ \ {\isachardoublequoteopen}ine{\isacharunderscore}exprChannelSingle\ P\ ch\ E{\isachardoublequoteclose}\isanewline
\isadelimproof
\endisadelimproof
\isatagproof
\isacommand{using}\isamarkupfalse%
\ assms\ \isanewline
\isacommand{by}\isamarkupfalse%
\ {\isacharparenleft}simp\ add{\isacharcolon}\ ine{\isacharunderscore}exprChannelSingle{\isacharunderscore}def\ ine{\isacharunderscore}exprChannelSet{\isacharunderscore}def{\isacharparenright}%
\endisatagproof
{\isafoldproof}%
\isadelimproof
\isanewline
\endisadelimproof
\isanewline
\isacommand{lemma}\isamarkupfalse%
\ ine{\isacharunderscore}ins{\isacharunderscore}neg{\isadigit{1}}{\isacharcolon}\isanewline
\isakeyword{assumes}\ {\isachardoublequoteopen}{\isasymnot}\ ine\ P\ m{\isachardoublequoteclose}\ \isanewline
\ \ \ \ \ \ \ \isakeyword{and}\ {\isachardoublequoteopen}exprChannel\ x\ m{\isachardoublequoteclose}\isanewline
\isakeyword{shows}\ \ \ \ {\isachardoublequoteopen}x\ {\isasymnotin}\ ins\ P{\isachardoublequoteclose}\isanewline
\isadelimproof
\endisadelimproof
\isatagproof
\isacommand{using}\isamarkupfalse%
\ assms\ \isacommand{by}\isamarkupfalse%
\ {\isacharparenleft}simp\ add{\isacharcolon}\ ine{\isacharunderscore}def{\isacharcomma}\ auto{\isacharparenright}%
\endisatagproof
{\isafoldproof}%
\isadelimproof
\isanewline
\endisadelimproof
\isanewline
\isacommand{theorem}\isamarkupfalse%
\ TBtheorem{\isadigit{1}}a{\isacharcolon}\isanewline
\isakeyword{assumes}\ {\isachardoublequoteopen}ine\ PQ\ E{\isachardoublequoteclose}\ \isanewline
\ \ \ \ \ \ \ \isakeyword{and}\ {\isachardoublequoteopen}subcomponents\ PQ\ {\isacharequal}\ {\isacharbraceleft}P{\isacharcomma}Q{\isacharbraceright}{\isachardoublequoteclose}\isanewline
\ \ \ \ \ \ \ \isakeyword{and}\ {\isachardoublequoteopen}correctCompositionIn\ PQ{\isachardoublequoteclose}\isanewline
\isakeyword{shows}\ {\isachardoublequoteopen}ine\ P\ E\ \ {\isasymor}\ ine\ Q\ E{\isachardoublequoteclose}\isanewline
\isadelimproof
\endisadelimproof
\isatagproof
\isacommand{using}\isamarkupfalse%
\ assms\ \isacommand{by}\isamarkupfalse%
\ {\isacharparenleft}simp\ add{\isacharcolon}\ ine{\isacharunderscore}def\ correctCompositionIn{\isacharunderscore}def{\isacharcomma}\ auto{\isacharparenright}%
\endisatagproof
{\isafoldproof}%
\isadelimproof
\isanewline
\endisadelimproof
\isanewline
\isacommand{theorem}\isamarkupfalse%
\ TBtheorem{\isadigit{1}}b{\isacharcolon}\isanewline
\isakeyword{assumes}\ {\isachardoublequoteopen}ineM\ PQ\ M\ E{\isachardoublequoteclose}\isanewline
\ \ \ \ \ \ \ \isakeyword{and}\ {\isachardoublequoteopen}subcomponents\ PQ\ {\isacharequal}\ {\isacharbraceleft}P{\isacharcomma}Q{\isacharbraceright}{\isachardoublequoteclose}\isanewline
\ \ \ \ \ \ \ \isakeyword{and}\ {\isachardoublequoteopen}correctCompositionIn\ PQ{\isachardoublequoteclose}\ \isanewline
\isakeyword{shows}\ \ \ \ {\isachardoublequoteopen}ineM\ P\ M\ E\ {\isasymor}\ ineM\ Q\ M\ E{\isachardoublequoteclose}\isanewline
\isadelimproof
\endisadelimproof
\isatagproof
\isacommand{using}\isamarkupfalse%
\ assms\ \isacommand{by}\isamarkupfalse%
\ {\isacharparenleft}simp\ add{\isacharcolon}\ ineM{\isacharunderscore}def\ correctCompositionIn{\isacharunderscore}def{\isacharcomma}\ auto{\isacharparenright}%
\endisatagproof
{\isafoldproof}%
\isadelimproof
\isanewline
\endisadelimproof
\isanewline
\isacommand{theorem}\isamarkupfalse%
\ TBtheorem{\isadigit{2}}a{\isacharcolon}\isanewline
\isakeyword{assumes}\ {\isachardoublequoteopen}eout\ PQ\ E{\isachardoublequoteclose}\isanewline
\ \ \ \ \ \ \ \isakeyword{and}\ {\isachardoublequoteopen}subcomponents\ PQ\ {\isacharequal}\ {\isacharbraceleft}P{\isacharcomma}Q{\isacharbraceright}{\isachardoublequoteclose}\isanewline
\ \ \ \ \ \ \ \isakeyword{and}\ {\isachardoublequoteopen}correctCompositionOut\ PQ{\isachardoublequoteclose}\isanewline
\isakeyword{shows}\ \ \ \ {\isachardoublequoteopen}eout\ P\ E\ {\isasymor}\ eout\ Q\ E{\isachardoublequoteclose}\isanewline
\isadelimproof
\endisadelimproof
\isatagproof
\isacommand{using}\isamarkupfalse%
\ assms\ \isacommand{by}\isamarkupfalse%
\ {\isacharparenleft}simp\ add{\isacharcolon}\ eout{\isacharunderscore}def\ correctCompositionOut{\isacharunderscore}def{\isacharcomma}\ auto{\isacharparenright}%
\endisatagproof
{\isafoldproof}%
\isadelimproof
\isanewline
\endisadelimproof
\isanewline
\isacommand{theorem}\isamarkupfalse%
\ TBtheorem{\isadigit{2}}b{\isacharcolon}\isanewline
\isakeyword{assumes}\ {\isachardoublequoteopen}eoutM\ PQ\ M\ E{\isachardoublequoteclose}\isanewline
\ \ \ \ \ \ \ \isakeyword{and}\ {\isachardoublequoteopen}subcomponents\ PQ\ {\isacharequal}\ {\isacharbraceleft}P{\isacharcomma}Q{\isacharbraceright}{\isachardoublequoteclose}\isanewline
\ \ \ \ \ \ \ \isakeyword{and}\ {\isachardoublequoteopen}correctCompositionOut\ PQ{\isachardoublequoteclose}\isanewline
\isakeyword{shows}\ \ \ \ {\isachardoublequoteopen}eoutM\ P\ M\ E\ {\isasymor}\ eoutM\ Q\ M\ E{\isachardoublequoteclose}\isanewline
\isadelimproof
\endisadelimproof
\isatagproof
\isacommand{using}\isamarkupfalse%
\ assms\ \isacommand{by}\isamarkupfalse%
\ {\isacharparenleft}simp\ add{\isacharcolon}\ eoutM{\isacharunderscore}def\ correctCompositionOut{\isacharunderscore}def{\isacharcomma}\ auto{\isacharparenright}%
\endisatagproof
{\isafoldproof}%
\isadelimproof
\isanewline
\endisadelimproof
\isanewline
\isacommand{lemma}\isamarkupfalse%
\ correctCompositionIn{\isacharunderscore}prop{\isadigit{1}}{\isacharcolon}\isanewline
\isakeyword{assumes}\ {\isachardoublequoteopen}subcomponents\ PQ\ {\isacharequal}\ {\isacharbraceleft}P{\isacharcomma}Q{\isacharbraceright}{\isachardoublequoteclose}\isanewline
\ \ \ \ \ \ \ \isakeyword{and}\ {\isachardoublequoteopen}correctCompositionIn\ PQ{\isachardoublequoteclose}\isanewline
\ \ \ \ \ \ \ \isakeyword{and}\ {\isachardoublequoteopen}x\ {\isasymin}\ {\isacharparenleft}ins\ PQ{\isacharparenright}{\isachardoublequoteclose}\isanewline
\isakeyword{shows}\ \ \ {\isachardoublequoteopen}{\isacharparenleft}x\ {\isasymin}\ {\isacharparenleft}ins\ P{\isacharparenright}{\isacharparenright}\ {\isasymor}\ {\isacharparenleft}x\ {\isasymin}\ {\isacharparenleft}ins\ Q{\isacharparenright}{\isacharparenright}{\isachardoublequoteclose}\ \isanewline
\isadelimproof
\endisadelimproof
\isatagproof
\isacommand{using}\isamarkupfalse%
\ assms\ \isacommand{by}\isamarkupfalse%
\ {\isacharparenleft}simp\ add{\isacharcolon}\ correctCompositionIn{\isacharunderscore}def{\isacharparenright}%
\endisatagproof
{\isafoldproof}%
\isadelimproof
\isanewline
\endisadelimproof
\isanewline
\isacommand{lemma}\isamarkupfalse%
\ correctCompositionOut{\isacharunderscore}prop{\isadigit{1}}{\isacharcolon}\isanewline
\isakeyword{assumes}\ {\isachardoublequoteopen}subcomponents\ PQ\ {\isacharequal}\ {\isacharbraceleft}P{\isacharcomma}Q{\isacharbraceright}{\isachardoublequoteclose}\isanewline
\ \ \ \ \ \ \ \isakeyword{and}\ {\isachardoublequoteopen}correctCompositionOut\ PQ{\isachardoublequoteclose}\isanewline
\ \ \ \ \ \ \ \isakeyword{and}\ {\isachardoublequoteopen}x\ {\isasymin}\ {\isacharparenleft}out\ PQ{\isacharparenright}{\isachardoublequoteclose}\isanewline
\isakeyword{shows}\ \ \ \ {\isachardoublequoteopen}{\isacharparenleft}x\ {\isasymin}\ {\isacharparenleft}out\ P{\isacharparenright}{\isacharparenright}\ {\isasymor}\ {\isacharparenleft}x\ {\isasymin}\ {\isacharparenleft}out\ Q{\isacharparenright}{\isacharparenright}{\isachardoublequoteclose}\ \isanewline
\isadelimproof
\endisadelimproof
\isatagproof
\isacommand{using}\isamarkupfalse%
\ assms\ \isacommand{by}\isamarkupfalse%
\ {\isacharparenleft}simp\ add{\isacharcolon}\ correctCompositionOut{\isacharunderscore}def{\isacharparenright}%
\endisatagproof
{\isafoldproof}%
\isadelimproof
\isanewline
\endisadelimproof
\isanewline
\isacommand{theorem}\isamarkupfalse%
\ TBtheorem{\isadigit{3}}a{\isacharcolon}\isanewline
\isakeyword{assumes}\ {\isachardoublequoteopen}{\isasymnot}\ {\isacharparenleft}ine\ P\ E{\isacharparenright}{\isachardoublequoteclose}\isanewline
\ \ \ \ \ \ \ \isakeyword{and}\ {\isachardoublequoteopen}{\isasymnot}\ {\isacharparenleft}ine\ Q\ E{\isacharparenright}{\isachardoublequoteclose}\isanewline
\ \ \ \ \ \ \ \isakeyword{and}\ {\isachardoublequoteopen}subcomponents\ PQ\ {\isacharequal}\ {\isacharbraceleft}P{\isacharcomma}Q{\isacharbraceright}{\isachardoublequoteclose}\isanewline
\ \ \ \ \ \ \ \isakeyword{and}\ {\isachardoublequoteopen}correctCompositionIn\ PQ{\isachardoublequoteclose}\isanewline
\isakeyword{shows}\ \ \ \ {\isachardoublequoteopen}{\isasymnot}\ {\isacharparenleft}ine\ PQ\ E{\isacharparenright}{\isachardoublequoteclose}\isanewline
\isadelimproof
\endisadelimproof
\isatagproof
\isacommand{using}\isamarkupfalse%
\ assms\ \isacommand{by}\isamarkupfalse%
\ {\isacharparenleft}simp\ add{\isacharcolon}\ ine{\isacharunderscore}def\ correctCompositionIn{\isacharunderscore}def{\isacharcomma}\ auto\ {\isacharparenright}%
\endisatagproof
{\isafoldproof}%
\isadelimproof
\isanewline
\endisadelimproof
\isanewline
\isacommand{theorem}\isamarkupfalse%
\ TBlemma{\isadigit{3}}b{\isacharcolon}\isanewline
\isakeyword{assumes}\ h{\isadigit{1}}{\isacharcolon}{\isachardoublequoteopen}{\isasymnot}\ {\isacharparenleft}ineM\ P\ M\ E{\isacharparenright}{\isachardoublequoteclose}\isanewline
\ \ \ \ \ \ \ \isakeyword{and}\ h{\isadigit{2}}{\isacharcolon}{\isachardoublequoteopen}{\isasymnot}\ {\isacharparenleft}ineM\ Q\ M\ E{\isacharparenright}{\isachardoublequoteclose}\isanewline
\ \ \ \ \ \ \ \isakeyword{and}\ subPQ{\isacharcolon}{\isachardoublequoteopen}subcomponents\ PQ\ {\isacharequal}\ {\isacharbraceleft}P{\isacharcomma}Q{\isacharbraceright}{\isachardoublequoteclose}\isanewline
\ \ \ \ \ \ \ \isakeyword{and}\ cCompI{\isacharcolon}{\isachardoublequoteopen}correctCompositionIn\ PQ{\isachardoublequoteclose}\isanewline
\ \ \ \ \ \ \ \isakeyword{and}\ chM{\isacharcolon}{\isachardoublequoteopen}ch\ {\isasymin}\ M{\isachardoublequoteclose}\ \isanewline
\ \ \ \ \ \ \ \isakeyword{and}\ chPQ{\isacharcolon}{\isachardoublequoteopen}ch\ {\isasymin}\ ins\ PQ{\isachardoublequoteclose}\isanewline
\ \ \ \ \ \ \ \isakeyword{and}\ eCh{\isacharcolon}{\isachardoublequoteopen}exprChannel\ ch\ E{\isachardoublequoteclose}\isanewline
\isakeyword{shows}\ {\isachardoublequoteopen}False{\isachardoublequoteclose}\isanewline
\isadelimproof
\endisadelimproof
\isatagproof
\isacommand{proof}\isamarkupfalse%
\ {\isacharparenleft}cases\ {\isachardoublequoteopen}ch\ {\isasymin}\ ins\ P{\isachardoublequoteclose}{\isacharparenright}\isanewline
\ \ \isacommand{assume}\isamarkupfalse%
\ a{\isadigit{1}}{\isacharcolon}{\isachardoublequoteopen}ch\ {\isasymin}\ ins\ P{\isachardoublequoteclose}\isanewline
\ \ \isacommand{from}\isamarkupfalse%
\ a{\isadigit{1}}\ \isakeyword{and}\ chM\ \isakeyword{and}\ eCh\ \isacommand{have}\isamarkupfalse%
\ {\isachardoublequoteopen}ineM\ P\ M\ E{\isachardoublequoteclose}\ \isacommand{by}\isamarkupfalse%
\ {\isacharparenleft}simp\ add{\isacharcolon}\ ineM{\isacharunderscore}L{\isadigit{1}}{\isacharparenright}\isanewline
\ \ \isacommand{from}\isamarkupfalse%
\ this\ \isakeyword{and}\ h{\isadigit{1}}\ \isacommand{show}\isamarkupfalse%
\ {\isacharquery}thesis\ \isacommand{by}\isamarkupfalse%
\ simp\isanewline
\isacommand{next}\isamarkupfalse%
\isanewline
\ \ \isacommand{assume}\isamarkupfalse%
\ a{\isadigit{2}}{\isacharcolon}{\isachardoublequoteopen}ch\ {\isasymnotin}\ ins\ P{\isachardoublequoteclose}\ \isanewline
\ \ \isacommand{from}\isamarkupfalse%
\ subPQ\ \isakeyword{and}\ cCompI\ \isakeyword{and}\ chPQ\ \isacommand{have}\isamarkupfalse%
\ {\isachardoublequoteopen}{\isacharparenleft}ch\ {\isasymin}\ ins\ P{\isacharparenright}\ {\isasymor}\ {\isacharparenleft}ch\ {\isasymin}\ ins\ Q{\isacharparenright}{\isachardoublequoteclose}\isanewline
\ \ \ \ \isacommand{by}\isamarkupfalse%
\ {\isacharparenleft}simp\ add{\isacharcolon}\ correctCompositionIn{\isacharunderscore}L{\isadigit{2}}{\isacharparenright}\isanewline
\ \ \isacommand{from}\isamarkupfalse%
\ this\ \isakeyword{and}\ a{\isadigit{2}}\ \isacommand{have}\isamarkupfalse%
\ {\isachardoublequoteopen}ch\ {\isasymin}\ ins\ Q{\isachardoublequoteclose}\ \isacommand{by}\isamarkupfalse%
\ simp\ \isanewline
\ \ \isacommand{from}\isamarkupfalse%
\ this\ \isakeyword{and}\ chM\ \isakeyword{and}\ eCh\ \isacommand{have}\isamarkupfalse%
\ {\isachardoublequoteopen}ineM\ Q\ M\ E{\isachardoublequoteclose}\ \isacommand{by}\isamarkupfalse%
\ {\isacharparenleft}simp\ add{\isacharcolon}\ ineM{\isacharunderscore}L{\isadigit{1}}{\isacharparenright}\isanewline
\ \ \isacommand{from}\isamarkupfalse%
\ this\ \isakeyword{and}\ h{\isadigit{2}}\ \isacommand{show}\isamarkupfalse%
\ {\isacharquery}thesis\ \isacommand{by}\isamarkupfalse%
\ simp\isanewline
\isacommand{qed}\isamarkupfalse%
\endisatagproof
{\isafoldproof}%
\isadelimproof
\isanewline
\endisadelimproof
\isanewline
\isacommand{theorem}\isamarkupfalse%
\ TBtheorem{\isadigit{3}}b{\isacharcolon}\isanewline
\isakeyword{assumes}\ {\isachardoublequoteopen}{\isasymnot}\ {\isacharparenleft}ineM\ P\ M\ E{\isacharparenright}{\isachardoublequoteclose}\isanewline
\ \ \ \ \ \ \ \isakeyword{and}\ {\isachardoublequoteopen}{\isasymnot}\ {\isacharparenleft}ineM\ Q\ M\ E{\isacharparenright}{\isachardoublequoteclose}\isanewline
\ \ \ \ \ \ \ \isakeyword{and}\ {\isachardoublequoteopen}subcomponents\ PQ\ {\isacharequal}\ {\isacharbraceleft}P{\isacharcomma}Q{\isacharbraceright}{\isachardoublequoteclose}\isanewline
\ \ \ \ \ \ \ \isakeyword{and}\ {\isachardoublequoteopen}correctCompositionIn\ PQ{\isachardoublequoteclose}\isanewline
\isakeyword{shows}\ \ \ \ {\isachardoublequoteopen}{\isasymnot}\ {\isacharparenleft}ineM\ PQ\ M\ E{\isacharparenright}{\isachardoublequoteclose}\isanewline
\isadelimproof
\endisadelimproof
\isatagproof
\isacommand{using}\isamarkupfalse%
\ assms\ \isacommand{by}\isamarkupfalse%
\ {\isacharparenleft}metis\ TBtheorem{\isadigit{1}}b{\isacharparenright}%
\endisatagproof
{\isafoldproof}%
\isadelimproof
\ \ \ \ \isanewline
\endisadelimproof
\isanewline
\isacommand{theorem}\isamarkupfalse%
\ TBtheorem{\isadigit{4}}a{\isacharunderscore}empty{\isacharcolon}\isanewline
\isakeyword{assumes}\ {\isachardoublequoteopen}{\isacharparenleft}ine\ P\ E{\isacharparenright}\ {\isasymor}\ {\isacharparenleft}ine\ Q\ E{\isacharparenright}{\isachardoublequoteclose}\isanewline
\ \ \ \ \ \ \ \isakeyword{and}\ {\isachardoublequoteopen}subcomponents\ PQ\ {\isacharequal}\ {\isacharbraceleft}P{\isacharcomma}Q{\isacharbraceright}{\isachardoublequoteclose}\isanewline
\ \ \ \ \ \ \ \isakeyword{and}\ {\isachardoublequoteopen}correctCompositionIn\ PQ{\isachardoublequoteclose}\isanewline
\ \ \ \ \ \ \ \isakeyword{and}\ {\isachardoublequoteopen}loc\ PQ\ {\isacharequal}\ {\isacharbraceleft}{\isacharbraceright}{\isachardoublequoteclose}\isanewline
\isakeyword{shows}\ \ \ \ {\isachardoublequoteopen}ine\ PQ\ E{\isachardoublequoteclose}\isanewline
\isadelimproof
\endisadelimproof
\isatagproof
\isacommand{using}\isamarkupfalse%
\ assms\ \isacommand{by}\isamarkupfalse%
\ {\isacharparenleft}simp\ add{\isacharcolon}\ ine{\isacharunderscore}def\ correctCompositionIn{\isacharunderscore}def{\isacharcomma}\ auto{\isacharparenright}%
\endisatagproof
{\isafoldproof}%
\isadelimproof
\isanewline
\endisadelimproof
\isanewline
\isacommand{theorem}\isamarkupfalse%
\ TBtheorem{\isadigit{4}}a{\isacharunderscore}P{\isacharcolon}\isanewline
\isakeyword{assumes}\ {\isachardoublequoteopen}ine\ P\ E{\isachardoublequoteclose}\isanewline
\ \ \ \ \ \ \ \isakeyword{and}\ {\isachardoublequoteopen}subcomponents\ PQ\ {\isacharequal}\ {\isacharbraceleft}P{\isacharcomma}Q{\isacharbraceright}{\isachardoublequoteclose}\isanewline
\ \ \ \ \ \ \ \isakeyword{and}\ {\isachardoublequoteopen}correctCompositionIn\ PQ{\isachardoublequoteclose}\isanewline
\ \ \ \ \ \ \ \isakeyword{and}\ {\isachardoublequoteopen}{\isasymexists}\ ch{\isachardot}\ {\isacharparenleft}ch\ {\isasymin}\ {\isacharparenleft}ins\ P{\isacharparenright}\ {\isasymand}\ exprChannel\ ch\ E\ {\isasymand}\ ch\ {\isasymnotin}\ {\isacharparenleft}loc\ PQ{\isacharparenright}{\isacharparenright}{\isachardoublequoteclose}\isanewline
\isakeyword{shows}\ \ \ \ {\isachardoublequoteopen}ine\ PQ\ E{\isachardoublequoteclose}\isanewline
\isadelimproof
\endisadelimproof
\isatagproof
\isacommand{using}\isamarkupfalse%
\ assms\ \isacommand{by}\isamarkupfalse%
\ {\isacharparenleft}simp\ add{\isacharcolon}\ ine{\isacharunderscore}def\ correctCompositionIn{\isacharunderscore}def{\isacharcomma}\ auto{\isacharparenright}%
\endisatagproof
{\isafoldproof}%
\isadelimproof
\ \isanewline
\endisadelimproof
\isanewline
\isacommand{theorem}\isamarkupfalse%
\ TBtheorem{\isadigit{4}}b{\isacharunderscore}P{\isacharcolon}\isanewline
\isakeyword{assumes}\ {\isachardoublequoteopen}ineM\ P\ M\ E{\isachardoublequoteclose}\isanewline
\ \ \ \ \ \ \ \isakeyword{and}\ {\isachardoublequoteopen}subcomponents\ PQ\ {\isacharequal}\ {\isacharbraceleft}P{\isacharcomma}Q{\isacharbraceright}{\isachardoublequoteclose}\isanewline
\ \ \ \ \ \ \ \isakeyword{and}\ {\isachardoublequoteopen}correctCompositionIn\ PQ{\isachardoublequoteclose}\isanewline
\ \ \ \ \ \ \ \isakeyword{and}\ {\isachardoublequoteopen}{\isasymexists}\ ch{\isachardot}\ {\isacharparenleft}{\isacharparenleft}ch\ {\isasymin}\ {\isacharparenleft}ins\ Q{\isacharparenright}{\isacharparenright}\ {\isasymand}\ {\isacharparenleft}exprChannel\ ch\ E{\isacharparenright}\ {\isasymand}\ \isanewline
\ \ \ \ \ \ \ \ \ \ \ \ \ \ \ \ \ \ \ \ \ \ \ \ {\isacharparenleft}ch\ {\isasymnotin}\ {\isacharparenleft}loc\ PQ{\isacharparenright}{\isacharparenright}\ {\isasymand}\ {\isacharparenleft}ch\ {\isasymin}\ M{\isacharparenright}{\isacharparenright}{\isachardoublequoteclose}\isanewline
\isakeyword{shows}\ \ \ \ {\isachardoublequoteopen}ineM\ PQ\ M\ E{\isachardoublequoteclose}\isanewline
\isadelimproof
\endisadelimproof
\isatagproof
\isacommand{using}\isamarkupfalse%
\ assms\ \isacommand{by}\isamarkupfalse%
\ {\isacharparenleft}simp\ add{\isacharcolon}\ ineM{\isacharunderscore}def\ correctCompositionIn{\isacharunderscore}def{\isacharcomma}\ auto{\isacharparenright}%
\endisatagproof
{\isafoldproof}%
\isadelimproof
\ \isanewline
\endisadelimproof
\isanewline
\isacommand{theorem}\isamarkupfalse%
\ TBtheorem{\isadigit{4}}a{\isacharunderscore}PQ{\isacharcolon}\isanewline
\isakeyword{assumes}\ {\isachardoublequoteopen}{\isacharparenleft}ine\ P\ E{\isacharparenright}\ {\isasymor}\ {\isacharparenleft}ine\ Q\ E{\isacharparenright}{\isachardoublequoteclose}\isanewline
\ \ \ \ \ \ \ \isakeyword{and}\ {\isachardoublequoteopen}subcomponents\ PQ\ {\isacharequal}\ {\isacharbraceleft}P{\isacharcomma}Q{\isacharbraceright}{\isachardoublequoteclose}\isanewline
\ \ \ \ \ \ \ \isakeyword{and}\ {\isachardoublequoteopen}correctCompositionIn\ PQ{\isachardoublequoteclose}\isanewline
\ \ \ \ \ \ \ \isakeyword{and}\ {\isachardoublequoteopen}{\isasymexists}\ ch{\isachardot}\ {\isacharparenleft}{\isacharparenleft}{\isacharparenleft}ch\ {\isasymin}\ {\isacharparenleft}ins\ P{\isacharparenright}{\isacharparenright}\ {\isasymor}\ {\isacharparenleft}ch\ {\isasymin}\ {\isacharparenleft}ins\ Q{\isacharparenright}\ {\isacharparenright}{\isacharparenright}\ {\isasymand}\ \isanewline
\ \ \ \ \ \ \ \ \ \ \ \ \ \ \ \ \ \ \ \ \ \ \ \ \ {\isacharparenleft}exprChannel\ ch\ E{\isacharparenright}\ {\isasymand}\ \ {\isacharparenleft}ch\ {\isasymnotin}\ {\isacharparenleft}loc\ PQ{\isacharparenright}{\isacharparenright}{\isacharparenright}{\isachardoublequoteclose}\isanewline
\isakeyword{shows}\ \ \ \ {\isachardoublequoteopen}ine\ PQ\ E{\isachardoublequoteclose}\isanewline
\isadelimproof
\endisadelimproof
\isatagproof
\isacommand{using}\isamarkupfalse%
\ assms\ \isacommand{by}\isamarkupfalse%
\ {\isacharparenleft}simp\ add{\isacharcolon}\ ine{\isacharunderscore}def\ correctCompositionIn{\isacharunderscore}def{\isacharcomma}\ auto{\isacharparenright}%
\endisatagproof
{\isafoldproof}%
\isadelimproof
\ \isanewline
\endisadelimproof
\isanewline
\isacommand{theorem}\isamarkupfalse%
\ TBtheorem{\isadigit{4}}b{\isacharunderscore}PQ{\isacharcolon}\isanewline
\isakeyword{assumes}\ {\isachardoublequoteopen}{\isacharparenleft}ineM\ P\ M\ E{\isacharparenright}\ {\isasymor}\ {\isacharparenleft}ineM\ Q\ M\ E{\isacharparenright}{\isachardoublequoteclose}\ \isanewline
\ \ \ \ \ \ \ \isakeyword{and}\ {\isachardoublequoteopen}subcomponents\ PQ\ {\isacharequal}\ {\isacharbraceleft}P{\isacharcomma}Q{\isacharbraceright}{\isachardoublequoteclose}\isanewline
\ \ \ \ \ \ \ \isakeyword{and}\ {\isachardoublequoteopen}correctCompositionIn\ PQ{\isachardoublequoteclose}\isanewline
\ \ \ \ \ \ \ \isakeyword{and}\ {\isachardoublequoteopen}{\isasymexists}\ ch{\isachardot}\ {\isacharparenleft}{\isacharparenleft}{\isacharparenleft}ch\ {\isasymin}\ {\isacharparenleft}ins\ P{\isacharparenright}{\isacharparenright}\ {\isasymor}\ {\isacharparenleft}ch\ {\isasymin}\ {\isacharparenleft}ins\ Q{\isacharparenright}\ {\isacharparenright}{\isacharparenright}\ {\isasymand}\ \isanewline
\ \ \ \ \ \ \ \ \ \ \ \ \ \ \ \ \ \ \ \ \ \ \ \ \ {\isacharparenleft}ch\ {\isasymin}\ M{\isacharparenright}\ {\isasymand}\ {\isacharparenleft}exprChannel\ ch\ E{\isacharparenright}\ {\isasymand}\ \ {\isacharparenleft}ch\ {\isasymnotin}\ {\isacharparenleft}loc\ PQ{\isacharparenright}{\isacharparenright}{\isacharparenright}{\isachardoublequoteclose}\isanewline
\isakeyword{shows}\ \ \ \ \ {\isachardoublequoteopen}ineM\ PQ\ M\ E{\isachardoublequoteclose}\isanewline
\isadelimproof
\endisadelimproof
\isatagproof
\isacommand{using}\isamarkupfalse%
\ assms\ \isacommand{by}\isamarkupfalse%
\ {\isacharparenleft}simp\ add{\isacharcolon}\ ineM{\isacharunderscore}def\ correctCompositionIn{\isacharunderscore}def{\isacharcomma}\ auto{\isacharparenright}%
\endisatagproof
{\isafoldproof}%
\isadelimproof
\ \isanewline
\endisadelimproof
\isanewline
\isacommand{theorem}\isamarkupfalse%
\ TBtheorem{\isadigit{4}}a{\isacharunderscore}notP{\isadigit{1}}{\isacharcolon}\isanewline
\isakeyword{assumes}\ {\isachardoublequoteopen}ine\ P\ E{\isachardoublequoteclose}\isanewline
\ \ \ \ \ \ \ \isakeyword{and}\ {\isachardoublequoteopen}{\isasymnot}\ ine\ Q\ E{\isachardoublequoteclose}\isanewline
\ \ \ \ \ \ \ \isakeyword{and}\ {\isachardoublequoteopen}subcomponents\ PQ\ {\isacharequal}\ {\isacharbraceleft}P{\isacharcomma}Q{\isacharbraceright}{\isachardoublequoteclose}\isanewline
\ \ \ \ \ \ \ \isakeyword{and}\ {\isachardoublequoteopen}correctCompositionIn\ PQ{\isachardoublequoteclose}\ \isanewline
\ \ \ \ \ \ \ \isakeyword{and}\ {\isachardoublequoteopen}{\isasymexists}\ ch{\isachardot}\ {\isacharparenleft}{\isacharparenleft}ine{\isacharunderscore}exprChannelSingle\ P\ ch\ E{\isacharparenright}\ {\isasymand}\ {\isacharparenleft}ch\ {\isasymin}\ {\isacharparenleft}loc\ PQ{\isacharparenright}{\isacharparenright}{\isacharparenright}{\isachardoublequoteclose}\isanewline
\isakeyword{shows}\ \ \ \ {\isachardoublequoteopen}{\isasymnot}\ ine\ PQ\ E{\isachardoublequoteclose}\isanewline
\isadelimproof
\endisadelimproof
\isatagproof
\isacommand{using}\isamarkupfalse%
\ assms\ \isanewline
\isacommand{by}\isamarkupfalse%
\ {\isacharparenleft}simp\ add{\isacharcolon}\ ine{\isacharunderscore}def\ correctCompositionIn{\isacharunderscore}def\ \isanewline
\ \ \ \ \ \ \ \ \ \ \ \ \ \ \ \ \ \ \ \ \ ine{\isacharunderscore}exprChannelSingle{\isacharunderscore}def{\isacharcomma}\ auto{\isacharparenright}%
\endisatagproof
{\isafoldproof}%
\isadelimproof
\ \isanewline
\endisadelimproof
\isanewline
\isacommand{theorem}\isamarkupfalse%
\ TBtheorem{\isadigit{4}}b{\isacharunderscore}notP{\isadigit{1}}{\isacharcolon}\isanewline
\isakeyword{assumes}\ {\isachardoublequoteopen}ineM\ P\ M\ E{\isachardoublequoteclose}\isanewline
\ \ \ \ \ \ \ \isakeyword{and}\ {\isachardoublequoteopen}{\isasymnot}\ ineM\ Q\ M\ E{\isachardoublequoteclose}\isanewline
\ \ \ \ \ \ \ \isakeyword{and}\ {\isachardoublequoteopen}subcomponents\ PQ\ {\isacharequal}\ {\isacharbraceleft}P{\isacharcomma}Q{\isacharbraceright}{\isachardoublequoteclose}\isanewline
\ \ \ \ \ \ \ \isakeyword{and}\ {\isachardoublequoteopen}correctCompositionIn\ PQ{\isachardoublequoteclose}\ \ \isanewline
\ \ \ \ \ \ \ \isakeyword{and}\ {\isachardoublequoteopen}{\isasymexists}\ ch{\isachardot}\ {\isacharparenleft}{\isacharparenleft}ine{\isacharunderscore}exprChannelSingle\ P\ ch\ E{\isacharparenright}\ {\isasymand}\ {\isacharparenleft}ch\ {\isasymin}\ M{\isacharparenright}\ \isanewline
\ \ \ \ \ \ \ \ \ \ \ \ \ \ \ \ \ \ \ \ \ {\isasymand}\ {\isacharparenleft}ch\ {\isasymin}\ {\isacharparenleft}loc\ PQ{\isacharparenright}{\isacharparenright}{\isacharparenright}{\isachardoublequoteclose}\isanewline
\isakeyword{shows}\ \ \ \ {\isachardoublequoteopen}{\isasymnot}\ ineM\ PQ\ M\ E{\isachardoublequoteclose}\isanewline
\isadelimproof
\endisadelimproof
\isatagproof
\isacommand{using}\isamarkupfalse%
\ assms\isanewline
\isacommand{by}\isamarkupfalse%
\ {\isacharparenleft}simp\ add{\isacharcolon}\ ineM{\isacharunderscore}def\ correctCompositionIn{\isacharunderscore}def\ \isanewline
\ \ \ \ \ \ \ \ \ \ \ \ \ \ \ \ \ \ \ \ \ ine{\isacharunderscore}exprChannelSingle{\isacharunderscore}def{\isacharcomma}\ auto{\isacharparenright}%
\endisatagproof
{\isafoldproof}%
\isadelimproof
\ \isanewline
\endisadelimproof
\isanewline
\isacommand{theorem}\isamarkupfalse%
\ TBtheorem{\isadigit{4}}a{\isacharunderscore}notP{\isadigit{2}}{\isacharcolon}\isanewline
\isakeyword{assumes}\ {\isachardoublequoteopen}{\isasymnot}\ ine\ Q\ E{\isachardoublequoteclose}\isanewline
\ \ \ \ \ \ \ \isakeyword{and}\ {\isachardoublequoteopen}subcomponents\ PQ\ {\isacharequal}\ {\isacharbraceleft}P{\isacharcomma}Q{\isacharbraceright}{\isachardoublequoteclose}\isanewline
\ \ \ \ \ \ \ \isakeyword{and}\ {\isachardoublequoteopen}correctCompositionIn\ PQ{\isachardoublequoteclose}\ \isanewline
\ \ \ \ \ \ \ \isakeyword{and}\ {\isachardoublequoteopen}ine{\isacharunderscore}exprChannelSet\ P\ ChSet\ E{\isachardoublequoteclose}\isanewline
\ \ \ \ \ \ \ \isakeyword{and}\ {\isachardoublequoteopen}{\isasymforall}\ {\isacharparenleft}x\ {\isacharcolon}{\isacharcolon}chanID{\isacharparenright}{\isachardot}\ {\isacharparenleft}{\isacharparenleft}x\ {\isasymin}\ ChSet{\isacharparenright}\ {\isasymlongrightarrow}\ {\isacharparenleft}x\ {\isasymin}\ {\isacharparenleft}loc\ PQ{\isacharparenright}{\isacharparenright}{\isacharparenright}{\isachardoublequoteclose}\ \isanewline
\isakeyword{shows}\ \ \ \ {\isachardoublequoteopen}{\isasymnot}\ ine\ PQ\ E{\isachardoublequoteclose}\isanewline
\isadelimproof
\endisadelimproof
\isatagproof
\isacommand{using}\isamarkupfalse%
\ assms\ \isanewline
\isacommand{by}\isamarkupfalse%
\ {\isacharparenleft}simp\ add{\isacharcolon}\ ine{\isacharunderscore}def\ correctCompositionIn{\isacharunderscore}def\ \isanewline
\ \ \ \ \ \ \ \ \ \ \ \ \ \ \ \ \ \ \ \ \ ine{\isacharunderscore}exprChannelSet{\isacharunderscore}def{\isacharcomma}\ auto{\isacharparenright}%
\endisatagproof
{\isafoldproof}%
\isadelimproof
\ \isanewline
\endisadelimproof
\isanewline
\isacommand{theorem}\isamarkupfalse%
\ TBtheorem{\isadigit{4}}b{\isacharunderscore}notP{\isadigit{2}}{\isacharcolon}\isanewline
\isakeyword{assumes}\ {\isachardoublequoteopen}{\isasymnot}\ ineM\ Q\ M\ E{\isachardoublequoteclose}\isanewline
\ \ \ \ \ \ \ \isakeyword{and}\ {\isachardoublequoteopen}subcomponents\ PQ\ {\isacharequal}\ {\isacharbraceleft}P{\isacharcomma}Q{\isacharbraceright}{\isachardoublequoteclose}\isanewline
\ \ \ \ \ \ \ \isakeyword{and}\ {\isachardoublequoteopen}correctCompositionIn\ PQ{\isachardoublequoteclose}\isanewline
\ \ \ \ \ \ \ \isakeyword{and}\ {\isachardoublequoteopen}ine{\isacharunderscore}exprChannelSet\ P\ ChSet\ E{\isachardoublequoteclose}\isanewline
\ \ \ \ \ \ \ \isakeyword{and}\ {\isachardoublequoteopen}{\isasymforall}\ {\isacharparenleft}x\ {\isacharcolon}{\isacharcolon}chanID{\isacharparenright}{\isachardot}\ {\isacharparenleft}{\isacharparenleft}x\ {\isasymin}\ ChSet{\isacharparenright}\ {\isasymlongrightarrow}\ {\isacharparenleft}x\ {\isasymin}\ {\isacharparenleft}loc\ PQ{\isacharparenright}{\isacharparenright}{\isacharparenright}{\isachardoublequoteclose}\isanewline
\isakeyword{shows}\ \ \ \ {\isachardoublequoteopen}{\isasymnot}\ ineM\ PQ\ M\ E{\isachardoublequoteclose}\isanewline
\isadelimproof
\endisadelimproof
\isatagproof
\isacommand{using}\isamarkupfalse%
\ assms\ \isanewline
\isacommand{by}\isamarkupfalse%
\ {\isacharparenleft}simp\ add{\isacharcolon}\ ineM{\isacharunderscore}def\ correctCompositionIn{\isacharunderscore}def\ \isanewline
\ \ \ \ \ \ \ \ \ \ \ \ \ \ \ \ \ \ \ \ \ ine{\isacharunderscore}exprChannelSet{\isacharunderscore}def{\isacharcomma}\ auto{\isacharparenright}%
\endisatagproof
{\isafoldproof}%
\isadelimproof
\ \isanewline
\endisadelimproof
\isanewline
\isacommand{theorem}\isamarkupfalse%
\ TBtheorem{\isadigit{4}}a{\isacharunderscore}notPQ{\isacharcolon}\isanewline
\isakeyword{assumes}\ {\isachardoublequoteopen}subcomponents\ PQ\ {\isacharequal}\ {\isacharbraceleft}P{\isacharcomma}Q{\isacharbraceright}{\isachardoublequoteclose}\isanewline
\ \ \ \ \ \ \ \isakeyword{and}\ {\isachardoublequoteopen}correctCompositionIn\ PQ{\isachardoublequoteclose}\isanewline
\ \ \ \ \ \ \ \isakeyword{and}\ {\isachardoublequoteopen}ine{\isacharunderscore}exprChannelSet\ P\ ChSetP\ E{\isachardoublequoteclose}\isanewline
\ \ \ \ \ \ \ \isakeyword{and}\ {\isachardoublequoteopen}ine{\isacharunderscore}exprChannelSet\ Q\ ChSetQ\ E{\isachardoublequoteclose}\isanewline
\ \ \ \ \ \ \ \isakeyword{and}\ {\isachardoublequoteopen}{\isasymforall}\ {\isacharparenleft}x\ {\isacharcolon}{\isacharcolon}chanID{\isacharparenright}{\isachardot}\ {\isacharparenleft}{\isacharparenleft}x\ {\isasymin}\ ChSetP{\isacharparenright}\ {\isasymlongrightarrow}\ {\isacharparenleft}x\ {\isasymin}\ {\isacharparenleft}loc\ PQ{\isacharparenright}{\isacharparenright}{\isacharparenright}{\isachardoublequoteclose}\isanewline
\ \ \ \ \ \ \ \isakeyword{and}\ {\isachardoublequoteopen}{\isasymforall}\ {\isacharparenleft}x\ {\isacharcolon}{\isacharcolon}chanID{\isacharparenright}{\isachardot}\ {\isacharparenleft}{\isacharparenleft}x\ {\isasymin}\ ChSetQ{\isacharparenright}\ {\isasymlongrightarrow}\ {\isacharparenleft}x\ {\isasymin}\ {\isacharparenleft}loc\ PQ{\isacharparenright}{\isacharparenright}{\isacharparenright}{\isachardoublequoteclose}\isanewline
\isakeyword{shows}\ \ \ \ {\isachardoublequoteopen}{\isasymnot}\ ine\ PQ\ E{\isachardoublequoteclose}\isanewline
\isadelimproof
\endisadelimproof
\isatagproof
\isacommand{using}\isamarkupfalse%
\ assms\ \isanewline
\isacommand{by}\isamarkupfalse%
\ {\isacharparenleft}simp\ add{\isacharcolon}\ ine{\isacharunderscore}def\ correctCompositionIn{\isacharunderscore}def\ \isanewline
\ \ \ \ \ \ \ \ \ \ \ \ \ \ \ \ \ \ \ \ \ ine{\isacharunderscore}exprChannelSet{\isacharunderscore}def{\isacharcomma}\ auto{\isacharparenright}%
\endisatagproof
{\isafoldproof}%
\isadelimproof
\isanewline
\endisadelimproof
\isanewline
\isacommand{lemma}\isamarkupfalse%
\ ineM{\isacharunderscore}Un{\isadigit{1}}{\isacharcolon}\isanewline
\isakeyword{assumes}\ {\isachardoublequoteopen}ineM\ P\ A\ E{\isachardoublequoteclose}\isanewline
\isakeyword{shows}\ \ \ \ {\isachardoublequoteopen}ineM\ P\ {\isacharparenleft}A\ Un\ B{\isacharparenright}\ E{\isachardoublequoteclose}\isanewline
\isadelimproof
\endisadelimproof
\isatagproof
\isacommand{using}\isamarkupfalse%
\ assms\ \isacommand{by}\isamarkupfalse%
\ {\isacharparenleft}simp\ add{\isacharcolon}\ ineM{\isacharunderscore}def{\isacharcomma}\ auto{\isacharparenright}%
\endisatagproof
{\isafoldproof}%
\isadelimproof
\isanewline
\endisadelimproof
\isanewline
\isacommand{theorem}\isamarkupfalse%
\ TBtheorem{\isadigit{4}}b{\isacharunderscore}notPQ{\isacharcolon}\isanewline
\isakeyword{assumes}\ {\isachardoublequoteopen}subcomponents\ PQ\ {\isacharequal}\ {\isacharbraceleft}P{\isacharcomma}Q{\isacharbraceright}{\isachardoublequoteclose}\isanewline
\ \ \ \ \ \ \ \isakeyword{and}\ {\isachardoublequoteopen}correctCompositionIn\ PQ{\isachardoublequoteclose}\ \isanewline
\ \ \ \ \ \ \ \isakeyword{and}\ {\isachardoublequoteopen}ine{\isacharunderscore}exprChannelSet\ P\ ChSetP\ E{\isachardoublequoteclose}\isanewline
\ \ \ \ \ \ \ \isakeyword{and}\ {\isachardoublequoteopen}ine{\isacharunderscore}exprChannelSet\ Q\ ChSetQ\ E{\isachardoublequoteclose}\isanewline
\ \ \ \ \ \ \ \isakeyword{and}\ {\isachardoublequoteopen}{\isasymforall}\ {\isacharparenleft}x\ {\isacharcolon}{\isacharcolon}chanID{\isacharparenright}{\isachardot}\ {\isacharparenleft}{\isacharparenleft}x\ {\isasymin}\ ChSetP{\isacharparenright}\ {\isasymlongrightarrow}\ {\isacharparenleft}x\ {\isasymin}\ {\isacharparenleft}loc\ PQ{\isacharparenright}{\isacharparenright}{\isacharparenright}{\isachardoublequoteclose}\isanewline
\ \ \ \ \ \ \ \isakeyword{and}\ {\isachardoublequoteopen}{\isasymforall}\ {\isacharparenleft}x\ {\isacharcolon}{\isacharcolon}chanID{\isacharparenright}{\isachardot}\ {\isacharparenleft}{\isacharparenleft}x\ {\isasymin}\ ChSetQ{\isacharparenright}\ {\isasymlongrightarrow}\ {\isacharparenleft}x\ {\isasymin}\ {\isacharparenleft}loc\ PQ{\isacharparenright}{\isacharparenright}{\isacharparenright}{\isachardoublequoteclose}\isanewline
\isakeyword{shows}\ \ \ \ {\isachardoublequoteopen}\ {\isasymnot}\ ineM\ PQ\ M\ E{\isachardoublequoteclose}\isanewline
\isadelimproof
\endisadelimproof
\isatagproof
\isacommand{using}\isamarkupfalse%
\ assms\ \isanewline
\isacommand{by}\isamarkupfalse%
\ {\isacharparenleft}simp\ add{\isacharcolon}\ ineM{\isacharunderscore}def\ correctCompositionIn{\isacharunderscore}def\ \isanewline
\ \ \ \ \ \ \ \ \ \ \ \ \ \ \ \ \ \ \ \ \ ine{\isacharunderscore}exprChannelSet{\isacharunderscore}def{\isacharcomma}\ auto{\isacharparenright}%
\endisatagproof
{\isafoldproof}%
\isadelimproof
\ \isanewline
\endisadelimproof
\isanewline
\isacommand{lemma}\isamarkupfalse%
\ ine{\isacharunderscore}nonempty{\isacharunderscore}exprChannelSet{\isacharcolon}\isanewline
\isakeyword{assumes}\ {\isachardoublequoteopen}ine{\isacharunderscore}exprChannelSet\ P\ ChSet\ E{\isachardoublequoteclose}\isanewline
\ \ \ \ \ \ \ \isakeyword{and}\ {\isachardoublequoteopen}ChSet\ {\isasymnoteq}\ {\isacharbraceleft}{\isacharbraceright}{\isachardoublequoteclose}\isanewline
\isakeyword{shows}\ \ \ \ {\isachardoublequoteopen}ine\ P\ E\ {\isachardoublequoteclose}\isanewline
\isadelimproof
\endisadelimproof
\isatagproof
\isacommand{using}\isamarkupfalse%
\ assms\ \isacommand{by}\isamarkupfalse%
\ {\isacharparenleft}simp\ add{\isacharcolon}\ ine{\isacharunderscore}def\ ine{\isacharunderscore}exprChannelSet{\isacharunderscore}def{\isacharcomma}\ auto{\isacharparenright}%
\endisatagproof
{\isafoldproof}%
\isadelimproof
\isanewline
\endisadelimproof
\isanewline
\isacommand{lemma}\isamarkupfalse%
\ ine{\isacharunderscore}empty{\isacharunderscore}exprChannelSet{\isacharcolon}\isanewline
\isakeyword{assumes}\ {\isachardoublequoteopen}ine{\isacharunderscore}exprChannelSet\ P\ ChSet\ E{\isachardoublequoteclose}\isanewline
\ \ \ \ \ \ \ \isakeyword{and}\ {\isachardoublequoteopen}ChSet\ {\isacharequal}\ {\isacharbraceleft}{\isacharbraceright}{\isachardoublequoteclose}\isanewline
\isakeyword{shows}\ \ \ \ {\isachardoublequoteopen}{\isasymnot}\ ine\ P\ E{\isachardoublequoteclose}\isanewline
\isadelimproof
\endisadelimproof
\isatagproof
\isacommand{using}\isamarkupfalse%
\ assms\ \isacommand{by}\isamarkupfalse%
\ {\isacharparenleft}simp\ add{\isacharcolon}\ ine{\isacharunderscore}def\ ine{\isacharunderscore}exprChannelSet{\isacharunderscore}def{\isacharparenright}%
\endisatagproof
{\isafoldproof}%
\isadelimproof
\isanewline
\endisadelimproof
\isanewline
\isacommand{theorem}\isamarkupfalse%
\ TBtheorem{\isadigit{5}}a{\isacharunderscore}empty{\isacharcolon}\isanewline
\isakeyword{assumes}\ {\isachardoublequoteopen}{\isacharparenleft}eout\ P\ E{\isacharparenright}\ {\isasymor}\ {\isacharparenleft}eout\ Q\ E{\isacharparenright}{\isachardoublequoteclose}\isanewline
\ \ \ \ \ \ \ \isakeyword{and}\ {\isachardoublequoteopen}subcomponents\ PQ\ {\isacharequal}\ {\isacharbraceleft}P{\isacharcomma}Q{\isacharbraceright}{\isachardoublequoteclose}\isanewline
\ \ \ \ \ \ \ \isakeyword{and}\ {\isachardoublequoteopen}correctCompositionOut\ PQ{\isachardoublequoteclose}\isanewline
\ \ \ \ \ \ \ \isakeyword{and}\ {\isachardoublequoteopen}loc\ PQ\ {\isacharequal}\ {\isacharbraceleft}{\isacharbraceright}{\isachardoublequoteclose}\isanewline
\isakeyword{shows}\ \ \ \ {\isachardoublequoteopen}eout\ PQ\ E{\isachardoublequoteclose}\isanewline
\isadelimproof
\endisadelimproof
\isatagproof
\isacommand{using}\isamarkupfalse%
\ assms\ \isacommand{by}\isamarkupfalse%
\ {\isacharparenleft}simp\ add{\isacharcolon}\ eout{\isacharunderscore}def\ correctCompositionOut{\isacharunderscore}def{\isacharcomma}\ auto{\isacharparenright}%
\endisatagproof
{\isafoldproof}%
\isadelimproof
\isanewline
\endisadelimproof
\isanewline
\isacommand{theorem}\isamarkupfalse%
\ TBtheorem{\isadigit{4}}{\isadigit{5}}a{\isacharunderscore}P{\isacharcolon}\isanewline
\isakeyword{assumes}\ {\isachardoublequoteopen}eout\ P\ E{\isachardoublequoteclose}\isanewline
\ \ \ \ \ \ \ \isakeyword{and}\ {\isachardoublequoteopen}subcomponents\ PQ\ {\isacharequal}\ {\isacharbraceleft}P{\isacharcomma}Q{\isacharbraceright}{\isachardoublequoteclose}\isanewline
\ \ \ \ \ \ \ \isakeyword{and}\ {\isachardoublequoteopen}correctCompositionOut\ PQ{\isachardoublequoteclose}\isanewline
\ \ \ \ \ \ \ \isakeyword{and}\ {\isachardoublequoteopen}{\isasymexists}\ ch{\isachardot}\ {\isacharparenleft}{\isacharparenleft}ch\ {\isasymin}\ {\isacharparenleft}out\ P{\isacharparenright}{\isacharparenright}\ {\isasymand}\ {\isacharparenleft}exprChannel\ ch\ E{\isacharparenright}\ {\isasymand}\ \isanewline
\ \ \ \ \ \ \ \ \ \ \ \ \ \ \ \ \ \ \ \ \ \ \ \ {\isacharparenleft}ch\ {\isasymnotin}\ {\isacharparenleft}loc\ PQ{\isacharparenright}{\isacharparenright}{\isacharparenright}{\isachardoublequoteclose}\isanewline
\isakeyword{shows}\ \ \ \ {\isachardoublequoteopen}eout\ PQ\ E{\isachardoublequoteclose}\isanewline
\isadelimproof
\endisadelimproof
\isatagproof
\isacommand{using}\isamarkupfalse%
\ assms\ \isacommand{by}\isamarkupfalse%
\ {\isacharparenleft}simp\ add{\isacharcolon}\ eout{\isacharunderscore}def\ correctCompositionOut{\isacharunderscore}def{\isacharcomma}\ auto{\isacharparenright}%
\endisatagproof
{\isafoldproof}%
\isadelimproof
\isanewline
\endisadelimproof
\isanewline
\isacommand{theorem}\isamarkupfalse%
\ TBtheore{\isadigit{5}}{\isadigit{4}}b{\isacharunderscore}P{\isacharcolon}\isanewline
\isakeyword{assumes}\ {\isachardoublequoteopen}eoutM\ P\ M\ E{\isachardoublequoteclose}\isanewline
\ \ \ \ \ \ \ \isakeyword{and}\ {\isachardoublequoteopen}subcomponents\ PQ\ {\isacharequal}\ {\isacharbraceleft}P{\isacharcomma}Q{\isacharbraceright}{\isachardoublequoteclose}\isanewline
\ \ \ \ \ \ \ \isakeyword{and}\ {\isachardoublequoteopen}correctCompositionOut\ PQ{\isachardoublequoteclose}\ \isanewline
\ \ \ \ \ \ \ \isakeyword{and}\ {\isachardoublequoteopen}{\isasymexists}\ ch{\isachardot}\ {\isacharparenleft}{\isacharparenleft}ch\ {\isasymin}\ {\isacharparenleft}out\ Q{\isacharparenright}{\isacharparenright}\ {\isasymand}\ {\isacharparenleft}exprChannel\ ch\ E{\isacharparenright}\ {\isasymand}\ \isanewline
\ \ \ \ \ \ \ \ \ \ \ \ \ \ \ \ \ \ \ \ \ \ \ \ {\isacharparenleft}ch\ {\isasymnotin}\ {\isacharparenleft}loc\ PQ{\isacharparenright}{\isacharparenright}\ {\isasymand}\ {\isacharparenleft}ch\ {\isasymin}\ M{\isacharparenright}\ {\isacharparenright}{\isachardoublequoteclose}\isanewline
\isakeyword{shows}\ \ \ \ {\isachardoublequoteopen}eoutM\ PQ\ M\ E{\isachardoublequoteclose}\isanewline
\isadelimproof
\endisadelimproof
\isatagproof
\isacommand{using}\isamarkupfalse%
\ assms\ \isacommand{by}\isamarkupfalse%
\ {\isacharparenleft}simp\ add{\isacharcolon}\ eoutM{\isacharunderscore}def\ correctCompositionOut{\isacharunderscore}def{\isacharcomma}\ auto{\isacharparenright}%
\endisatagproof
{\isafoldproof}%
\isadelimproof
\isanewline
\endisadelimproof
\isanewline
\isacommand{theorem}\isamarkupfalse%
\ TBtheorem{\isadigit{5}}a{\isacharunderscore}PQ{\isacharcolon}\isanewline
\isakeyword{assumes}\ {\isachardoublequoteopen}{\isacharparenleft}eout\ P\ E{\isacharparenright}\ {\isasymor}\ {\isacharparenleft}eout\ Q\ E{\isacharparenright}{\isachardoublequoteclose}\isanewline
\ \ \ \ \ \ \ \isakeyword{and}\ {\isachardoublequoteopen}subcomponents\ PQ\ {\isacharequal}\ {\isacharbraceleft}P{\isacharcomma}Q{\isacharbraceright}{\isachardoublequoteclose}\isanewline
\ \ \ \ \ \ \ \isakeyword{and}\ {\isachardoublequoteopen}correctCompositionOut\ PQ{\isachardoublequoteclose}\isanewline
\ \ \ \ \ \ \ \isakeyword{and}\ {\isachardoublequoteopen}{\isasymexists}\ ch{\isachardot}\ {\isacharparenleft}{\isacharparenleft}{\isacharparenleft}ch\ {\isasymin}\ {\isacharparenleft}out\ P{\isacharparenright}{\isacharparenright}\ {\isasymor}\ {\isacharparenleft}ch\ {\isasymin}\ {\isacharparenleft}out\ Q{\isacharparenright}\ {\isacharparenright}{\isacharparenright}\ {\isasymand}\ \isanewline
\ \ \ \ \ \ \ \ \ \ \ \ \ \ \ \ \ \ \ \ \ \ \ \ {\isacharparenleft}exprChannel\ ch\ E{\isacharparenright}\ {\isasymand}\ \ {\isacharparenleft}ch\ {\isasymnotin}\ {\isacharparenleft}loc\ PQ{\isacharparenright}{\isacharparenright}{\isacharparenright}{\isachardoublequoteclose}\isanewline
\isakeyword{shows}\ \ \ \ {\isachardoublequoteopen}eout\ PQ\ E{\isachardoublequoteclose}\isanewline
\isadelimproof
\endisadelimproof
\isatagproof
\isacommand{using}\isamarkupfalse%
\ assms\ \isacommand{by}\isamarkupfalse%
\ {\isacharparenleft}simp\ add{\isacharcolon}\ eout{\isacharunderscore}def\ correctCompositionOut{\isacharunderscore}def{\isacharcomma}\ auto{\isacharparenright}%
\endisatagproof
{\isafoldproof}%
\isadelimproof
\isanewline
\endisadelimproof
\isanewline
\isacommand{theorem}\isamarkupfalse%
\ TBtheorem{\isadigit{5}}b{\isacharunderscore}PQ{\isacharcolon}\isanewline
\isakeyword{assumes}\ {\isachardoublequoteopen}{\isacharparenleft}eoutM\ P\ M\ E{\isacharparenright}\ {\isasymor}\ {\isacharparenleft}eoutM\ Q\ M\ E{\isacharparenright}{\isachardoublequoteclose}\ \isanewline
\ \ \ \ \ \ \ \isakeyword{and}\ {\isachardoublequoteopen}subcomponents\ PQ\ {\isacharequal}\ {\isacharbraceleft}P{\isacharcomma}Q{\isacharbraceright}{\isachardoublequoteclose}\isanewline
\ \ \ \ \ \ \ \isakeyword{and}\ {\isachardoublequoteopen}correctCompositionOut\ PQ{\isachardoublequoteclose}\isanewline
\ \ \ \ \ \ \ \isakeyword{and}\ {\isachardoublequoteopen}{\isasymexists}\ ch{\isachardot}\ {\isacharparenleft}{\isacharparenleft}{\isacharparenleft}ch\ {\isasymin}\ {\isacharparenleft}out\ P{\isacharparenright}{\isacharparenright}\ {\isasymor}\ {\isacharparenleft}ch\ {\isasymin}\ {\isacharparenleft}out\ Q{\isacharparenright}\ {\isacharparenright}{\isacharparenright}\ {\isasymand}\ {\isacharparenleft}ch\ {\isasymin}\ M{\isacharparenright}\ \isanewline
\ \ \ \ \ \ \ \ \ \ \ \ \ \ \ \ \ \ \ \ \ \ {\isasymand}\ {\isacharparenleft}exprChannel\ ch\ E{\isacharparenright}\ {\isasymand}\ \ {\isacharparenleft}ch\ {\isasymnotin}\ {\isacharparenleft}loc\ PQ{\isacharparenright}{\isacharparenright}{\isacharparenright}{\isachardoublequoteclose}\isanewline
\isakeyword{shows}\ \ \ \ {\isachardoublequoteopen}eoutM\ PQ\ M\ E{\isachardoublequoteclose}\isanewline
\isadelimproof
\endisadelimproof
\isatagproof
\isacommand{using}\isamarkupfalse%
\ assms\ \isacommand{by}\isamarkupfalse%
\ {\isacharparenleft}simp\ add{\isacharcolon}\ eoutM{\isacharunderscore}def\ correctCompositionOut{\isacharunderscore}def{\isacharcomma}\ auto{\isacharparenright}%
\endisatagproof
{\isafoldproof}%
\isadelimproof
\ \isanewline
\endisadelimproof
\isanewline
\isacommand{theorem}\isamarkupfalse%
\ TBtheorem{\isadigit{5}}a{\isacharunderscore}notP{\isadigit{1}}{\isacharcolon}\isanewline
\isakeyword{assumes}\ {\isachardoublequoteopen}eout\ P\ E{\isachardoublequoteclose}\isanewline
\ \ \ \ \ \ \ \isakeyword{and}\ {\isachardoublequoteopen}{\isasymnot}\ eout\ Q\ E{\isachardoublequoteclose}\isanewline
\ \ \ \ \ \ \ \isakeyword{and}\ {\isachardoublequoteopen}subcomponents\ PQ\ {\isacharequal}\ {\isacharbraceleft}P{\isacharcomma}Q{\isacharbraceright}{\isachardoublequoteclose}\isanewline
\ \ \ \ \ \ \ \isakeyword{and}\ {\isachardoublequoteopen}correctCompositionOut\ PQ{\isachardoublequoteclose}\isanewline
\ \ \ \ \ \ \ \isakeyword{and}\ {\isachardoublequoteopen}{\isasymexists}\ ch{\isachardot}\ {\isacharparenleft}{\isacharparenleft}out{\isacharunderscore}exprChannelSingle\ P\ ch\ E{\isacharparenright}\ {\isasymand}\ {\isacharparenleft}ch\ {\isasymin}\ {\isacharparenleft}loc\ PQ{\isacharparenright}{\isacharparenright}{\isacharparenright}{\isachardoublequoteclose}\isanewline
\isakeyword{shows}\ \ \ \ {\isachardoublequoteopen}{\isasymnot}\ eout\ PQ\ E{\isachardoublequoteclose}\isanewline
\isadelimproof
\endisadelimproof
\isatagproof
\isacommand{using}\isamarkupfalse%
\ assms\ \isanewline
\isacommand{by}\isamarkupfalse%
\ {\isacharparenleft}simp\ add{\isacharcolon}\ eout{\isacharunderscore}def\ correctCompositionOut{\isacharunderscore}def\ \isanewline
\ \ \ \ \ \ \ \ \ \ \ \ \ \ \ \ \ \ \ \ \ \ out{\isacharunderscore}exprChannelSingle{\isacharunderscore}def{\isacharcomma}\ auto{\isacharparenright}%
\endisatagproof
{\isafoldproof}%
\isadelimproof
\ \isanewline
\endisadelimproof
\isanewline
\isacommand{theorem}\isamarkupfalse%
\ TBtheorem{\isadigit{5}}b{\isacharunderscore}notP{\isadigit{1}}{\isacharcolon}\isanewline
\isakeyword{assumes}\ {\isachardoublequoteopen}eoutM\ P\ M\ E{\isachardoublequoteclose}\isanewline
\ \ \ \ \ \ \ \isakeyword{and}\ {\isachardoublequoteopen}{\isasymnot}\ eoutM\ Q\ M\ E{\isachardoublequoteclose}\isanewline
\ \ \ \ \ \ \ \isakeyword{and}\ {\isachardoublequoteopen}subcomponents\ PQ\ {\isacharequal}\ {\isacharbraceleft}P{\isacharcomma}Q{\isacharbraceright}{\isachardoublequoteclose}\isanewline
\ \ \ \ \ \ \ \isakeyword{and}\ {\isachardoublequoteopen}correctCompositionOut\ PQ{\isachardoublequoteclose}\isanewline
\ \ \ \ \ \ \ \isakeyword{and}\ {\isachardoublequoteopen}{\isasymexists}\ ch{\isachardot}\ {\isacharparenleft}{\isacharparenleft}out{\isacharunderscore}exprChannelSingle\ P\ ch\ E{\isacharparenright}\ {\isasymand}\ {\isacharparenleft}ch\ {\isasymin}\ M{\isacharparenright}\ \isanewline
\ \ \ \ \ \ \ \ \ \ \ \ \ \ \ \ \ \ \ {\isasymand}\ {\isacharparenleft}ch\ {\isasymin}\ {\isacharparenleft}loc\ PQ{\isacharparenright}{\isacharparenright}{\isacharparenright}{\isachardoublequoteclose}\isanewline
\isakeyword{shows}\ \ \ \ {\isachardoublequoteopen}{\isasymnot}\ eoutM\ PQ\ M\ E{\isachardoublequoteclose}\isanewline
\isadelimproof
\endisadelimproof
\isatagproof
\isacommand{using}\isamarkupfalse%
\ assms\ \isanewline
\isacommand{by}\isamarkupfalse%
\ {\isacharparenleft}simp\ add{\isacharcolon}\ eoutM{\isacharunderscore}def\ correctCompositionOut{\isacharunderscore}def\ \isanewline
\ \ \ \ \ \ \ \ \ \ \ \ \ \ \ \ \ \ \ \ \ out{\isacharunderscore}exprChannelSingle{\isacharunderscore}def{\isacharcomma}\ auto{\isacharparenright}%
\endisatagproof
{\isafoldproof}%
\isadelimproof
\ \isanewline
\endisadelimproof
\isanewline
\isacommand{theorem}\isamarkupfalse%
\ TBtheorem{\isadigit{5}}a{\isacharunderscore}notP{\isadigit{2}}{\isacharcolon}\isanewline
\isakeyword{assumes}\ {\isachardoublequoteopen}{\isasymnot}\ eout\ Q\ E{\isachardoublequoteclose}\isanewline
\ \ \ \ \ \ \ \isakeyword{and}\ {\isachardoublequoteopen}subcomponents\ PQ\ {\isacharequal}\ {\isacharbraceleft}P{\isacharcomma}Q{\isacharbraceright}{\isachardoublequoteclose}\isanewline
\ \ \ \ \ \ \ \isakeyword{and}\ {\isachardoublequoteopen}correctCompositionOut\ PQ{\isachardoublequoteclose}\ \isanewline
\ \ \ \ \ \ \ \isakeyword{and}\ {\isachardoublequoteopen}out{\isacharunderscore}exprChannelSet\ P\ ChSet\ E{\isachardoublequoteclose}\isanewline
\ \ \ \ \ \ \ \isakeyword{and}\ {\isachardoublequoteopen}{\isasymforall}\ {\isacharparenleft}x\ {\isacharcolon}{\isacharcolon}chanID{\isacharparenright}{\isachardot}\ {\isacharparenleft}{\isacharparenleft}x\ {\isasymin}\ ChSet{\isacharparenright}\ {\isasymlongrightarrow}\ {\isacharparenleft}x\ {\isasymin}\ {\isacharparenleft}loc\ PQ{\isacharparenright}{\isacharparenright}{\isacharparenright}{\isachardoublequoteclose}\isanewline
\isakeyword{shows}\ \ \ \ {\isachardoublequoteopen}{\isasymnot}\ eout\ PQ\ E{\isachardoublequoteclose}\isanewline
\isadelimproof
\endisadelimproof
\isatagproof
\isacommand{using}\isamarkupfalse%
\ assms\isanewline
\isacommand{by}\isamarkupfalse%
\ {\isacharparenleft}simp\ add{\isacharcolon}\ eout{\isacharunderscore}def\ correctCompositionOut{\isacharunderscore}def\ \isanewline
\ \ \ \ \ \ \ \ \ \ \ \ \ \ \ \ \ \ \ \ \ out{\isacharunderscore}exprChannelSet{\isacharunderscore}def{\isacharcomma}\ auto{\isacharparenright}%
\endisatagproof
{\isafoldproof}%
\isadelimproof
\isanewline
\endisadelimproof
\isanewline
\isacommand{theorem}\isamarkupfalse%
\ TBtheorem{\isadigit{5}}b{\isacharunderscore}notP{\isadigit{2}}{\isacharcolon}\isanewline
\isakeyword{assumes}\ {\isachardoublequoteopen}{\isasymnot}\ eoutM\ Q\ M\ E{\isachardoublequoteclose}\isanewline
\ \ \ \ \ \ \ \isakeyword{and}\ {\isachardoublequoteopen}subcomponents\ PQ\ {\isacharequal}\ {\isacharbraceleft}P{\isacharcomma}Q{\isacharbraceright}{\isachardoublequoteclose}\isanewline
\ \ \ \ \ \ \ \isakeyword{and}\ {\isachardoublequoteopen}correctCompositionOut\ PQ{\isachardoublequoteclose}\isanewline
\ \ \ \ \ \ \ \isakeyword{and}\ {\isachardoublequoteopen}out{\isacharunderscore}exprChannelSet\ P\ ChSet\ E{\isachardoublequoteclose}\isanewline
\ \ \ \ \ \ \ \isakeyword{and}\ {\isachardoublequoteopen}{\isasymforall}\ {\isacharparenleft}x\ {\isacharcolon}{\isacharcolon}chanID{\isacharparenright}{\isachardot}\ {\isacharparenleft}{\isacharparenleft}x\ {\isasymin}\ ChSet{\isacharparenright}\ {\isasymlongrightarrow}\ {\isacharparenleft}x\ {\isasymin}\ {\isacharparenleft}loc\ PQ{\isacharparenright}{\isacharparenright}{\isacharparenright}{\isachardoublequoteclose}\ \isanewline
\isakeyword{shows}\ \ \ \ {\isachardoublequoteopen}{\isasymnot}\ eoutM\ PQ\ M\ E{\isachardoublequoteclose}\isanewline
\isadelimproof
\endisadelimproof
\isatagproof
\isacommand{using}\isamarkupfalse%
\ assms\isanewline
\isacommand{by}\isamarkupfalse%
\ {\isacharparenleft}simp\ add{\isacharcolon}\ eoutM{\isacharunderscore}def\ correctCompositionOut{\isacharunderscore}def\ \isanewline
\ \ \ \ \ \ \ \ \ \ \ \ \ \ \ \ \ \ \ \ \ out{\isacharunderscore}exprChannelSet{\isacharunderscore}def{\isacharcomma}\ auto{\isacharparenright}%
\endisatagproof
{\isafoldproof}%
\isadelimproof
\isanewline
\endisadelimproof
\isanewline
\isacommand{theorem}\isamarkupfalse%
\ TBtheorem{\isadigit{5}}a{\isacharunderscore}notPQ{\isacharcolon}\isanewline
\isakeyword{assumes}\ {\isachardoublequoteopen}subcomponents\ PQ\ {\isacharequal}\ {\isacharbraceleft}P{\isacharcomma}Q{\isacharbraceright}{\isachardoublequoteclose}\isanewline
\ \ \ \ \ \ \ \isakeyword{and}\ {\isachardoublequoteopen}correctCompositionOut\ PQ{\isachardoublequoteclose}\isanewline
\ \ \ \ \ \ \ \isakeyword{and}\ {\isachardoublequoteopen}out{\isacharunderscore}exprChannelSet\ P\ ChSetP\ E{\isachardoublequoteclose}\isanewline
\ \ \ \ \ \ \ \isakeyword{and}\ {\isachardoublequoteopen}out{\isacharunderscore}exprChannelSet\ Q\ ChSetQ\ E{\isachardoublequoteclose}\isanewline
\ \ \ \ \ \ \ \isakeyword{and}\ {\isachardoublequoteopen}{\isasymforall}\ {\isacharparenleft}x\ {\isacharcolon}{\isacharcolon}chanID{\isacharparenright}{\isachardot}\ {\isacharparenleft}{\isacharparenleft}x\ {\isasymin}\ ChSetP{\isacharparenright}\ {\isasymlongrightarrow}\ {\isacharparenleft}x\ {\isasymin}\ {\isacharparenleft}loc\ PQ{\isacharparenright}{\isacharparenright}{\isacharparenright}{\isachardoublequoteclose}\isanewline
\ \ \ \ \ \ \ \isakeyword{and}\ {\isachardoublequoteopen}{\isasymforall}\ {\isacharparenleft}x\ {\isacharcolon}{\isacharcolon}chanID{\isacharparenright}{\isachardot}\ {\isacharparenleft}{\isacharparenleft}x\ {\isasymin}\ ChSetQ{\isacharparenright}\ {\isasymlongrightarrow}\ {\isacharparenleft}x\ {\isasymin}\ {\isacharparenleft}loc\ PQ{\isacharparenright}{\isacharparenright}{\isacharparenright}{\isachardoublequoteclose}\isanewline
\isakeyword{shows}\ \ \ \ {\isachardoublequoteopen}{\isasymnot}\ eout\ PQ\ E{\isachardoublequoteclose}\isanewline
\isadelimproof
\endisadelimproof
\isatagproof
\isacommand{using}\isamarkupfalse%
\ assms\isanewline
\isacommand{by}\isamarkupfalse%
\ {\isacharparenleft}simp\ add{\isacharcolon}\ eout{\isacharunderscore}def\ correctCompositionOut{\isacharunderscore}def\ \isanewline
\ \ \ \ \ \ \ \ \ \ \ \ \ \ \ \ \ \ \ \ \ out{\isacharunderscore}exprChannelSet{\isacharunderscore}def{\isacharcomma}\ auto{\isacharparenright}%
\endisatagproof
{\isafoldproof}%
\isadelimproof
\ \isanewline
\endisadelimproof
\isanewline
\isacommand{theorem}\isamarkupfalse%
\ TBtheorem{\isadigit{5}}b{\isacharunderscore}notPQ{\isacharcolon}\isanewline
\isakeyword{assumes}\ {\isachardoublequoteopen}subcomponents\ PQ\ {\isacharequal}\ {\isacharbraceleft}P{\isacharcomma}Q{\isacharbraceright}{\isachardoublequoteclose}\isanewline
\ \ \ \ \ \ \ \isakeyword{and}\ {\isachardoublequoteopen}correctCompositionOut\ PQ{\isachardoublequoteclose}\isanewline
\ \ \ \ \ \ \ \isakeyword{and}\ {\isachardoublequoteopen}out{\isacharunderscore}exprChannelSet\ P\ ChSetP\ E{\isachardoublequoteclose}\isanewline
\ \ \ \ \ \ \ \isakeyword{and}\ {\isachardoublequoteopen}out{\isacharunderscore}exprChannelSet\ Q\ ChSetQ\ E{\isachardoublequoteclose}\isanewline
\ \ \ \ \ \ \ \isakeyword{and}\ {\isachardoublequoteopen}M\ {\isacharequal}\ ChSetP\ {\isasymunion}\ ChSetQ{\isachardoublequoteclose}\isanewline
\ \ \ \ \ \ \ \isakeyword{and}\ {\isachardoublequoteopen}{\isasymforall}\ {\isacharparenleft}x\ {\isacharcolon}{\isacharcolon}chanID{\isacharparenright}{\isachardot}\ {\isacharparenleft}{\isacharparenleft}x\ {\isasymin}\ ChSetP{\isacharparenright}\ {\isasymlongrightarrow}\ {\isacharparenleft}x\ {\isasymin}\ {\isacharparenleft}loc\ PQ{\isacharparenright}{\isacharparenright}{\isacharparenright}{\isachardoublequoteclose}\isanewline
\ \ \ \ \ \ \ \isakeyword{and}\ {\isachardoublequoteopen}{\isasymforall}\ {\isacharparenleft}x\ {\isacharcolon}{\isacharcolon}chanID{\isacharparenright}{\isachardot}\ {\isacharparenleft}{\isacharparenleft}x\ {\isasymin}\ ChSetQ{\isacharparenright}\ {\isasymlongrightarrow}\ {\isacharparenleft}x\ {\isasymin}\ {\isacharparenleft}loc\ PQ{\isacharparenright}{\isacharparenright}{\isacharparenright}{\isachardoublequoteclose}\isanewline
\isakeyword{shows}\ \ \ \ {\isachardoublequoteopen}{\isasymnot}\ eoutM\ PQ\ M\ E{\isachardoublequoteclose}\isanewline
\isadelimproof
\endisadelimproof
\isatagproof
\isacommand{using}\isamarkupfalse%
\ assms\ \isanewline
\isacommand{by}\isamarkupfalse%
\ {\isacharparenleft}simp\ add{\isacharcolon}\ eoutM{\isacharunderscore}def\ correctCompositionOut{\isacharunderscore}def\ \isanewline
\ \ \ \ \ \ \ \ \ \ \ \ \ \ \ \ \ \ \ \ \ out{\isacharunderscore}exprChannelSet{\isacharunderscore}def{\isacharcomma}\ auto{\isacharparenright}%
\endisatagproof
{\isafoldproof}%
\isadelimproof
\ \isanewline
\endisadelimproof
\isadelimtheory
\isanewline
\endisadelimtheory
\isatagtheory
\isacommand{end}\isamarkupfalse%
\endisatagtheory
{\isafoldtheory}%
\isadelimtheory
\endisadelimtheory
\ \end{isabellebody}%

%
\begin{isabellebody}%
\def\isabellecontext{CompLocalSecrets}%
\isamarkupheader{Local Secrets of a component%
}
\isamarkuptrue%
\isadelimtheory
\endisadelimtheory
\isatagtheory
\isacommand{theory}\isamarkupfalse%
\ CompLocalSecrets\isanewline
\isakeyword{imports}\ Secrecy\ \isanewline
\isakeyword{begin}\isanewline
\isanewline
\isamarkupcmt{Set of local secrets: the set of secrets which does not belong to%
}
\isanewline
\isamarkupcmt{the set of private keys and unguessable values, but are transmitted%
}
\isanewline
\isamarkupcmt{via local channels or belongs to the local secrets of its subcomponents%
}
\endisatagtheory
{\isafoldtheory}%
\isadelimtheory
\endisadelimtheory
\isanewline
\isacommand{axiomatization}\isamarkupfalse%
\isanewline
\ \ LocalSecrets\ {\isacharcolon}{\isacharcolon}\ {\isachardoublequoteopen}specID\ \ {\isasymRightarrow}\ KS\ set{\isachardoublequoteclose}\isanewline
\isakeyword{where}\isanewline
LocalSecretsDef{\isacharcolon}\isanewline
\ {\isachardoublequoteopen}LocalSecrets\ A\ {\isacharequal}\isanewline
\ \ {\isacharbraceleft}{\isacharparenleft}m\ {\isacharcolon}{\isacharcolon}\ KS{\isacharparenright}{\isachardot}\ m\ {\isasymnotin}\ specKeysSecrets\ A\ \ {\isasymand}\ \isanewline
\ \ \ \ \ \ \ \ \ \ \ \ \ \ {\isacharparenleft}{\isacharparenleft}{\isasymexists}\ x\ y{\isachardot}\ {\isacharparenleft}{\isacharparenleft}x\ {\isasymin}\ loc\ A{\isacharparenright}\ {\isasymand}\ m\ {\isacharequal}\ {\isacharparenleft}kKS\ y{\isacharparenright}\ {\isasymand}\ {\isacharparenleft}exprChannel\ x\ {\isacharparenleft}kE\ y{\isacharparenright}{\isacharparenright}{\isacharparenright}{\isacharparenright}\ \isanewline
\ \ \ \ \ \ \ \ \ \ \ \ \ \ {\isacharbar}{\isacharparenleft}{\isasymexists}\ x\ z{\isachardot}\ {\isacharparenleft}{\isacharparenleft}x\ {\isasymin}\ loc\ A{\isacharparenright}\ {\isasymand}\ m\ {\isacharequal}\ {\isacharparenleft}sKS\ z{\isacharparenright}\ {\isasymand}\ {\isacharparenleft}exprChannel\ x\ {\isacharparenleft}sE\ z{\isacharparenright}{\isacharparenright}\ {\isacharparenright}{\isacharparenright}\ {\isacharparenright}{\isacharbraceright}\ \isanewline
\ \ \ {\isasymunion}\ \ {\isacharparenleft}{\isasymUnion}\ {\isacharparenleft}LocalSecrets\ {\isacharbackquote}\ {\isacharparenleft}subcomponents\ A{\isacharparenright}\ {\isacharparenright}{\isacharparenright}{\isachardoublequoteclose}\isanewline
\isanewline
\isacommand{lemma}\isamarkupfalse%
\ LocalSecretsComposition{\isadigit{1}}{\isacharcolon}\isanewline
\isakeyword{assumes}\ {\isachardoublequoteopen}ls\ {\isasymin}\ LocalSecrets\ P{\isachardoublequoteclose}\isanewline
\ \ \ \ \ \ \ \isakeyword{and}\ {\isachardoublequoteopen}subcomponents\ PQ\ {\isacharequal}\ {\isacharbraceleft}P{\isacharcomma}\ Q{\isacharbraceright}{\isachardoublequoteclose}\isanewline
\isakeyword{shows}\ \ \ \ {\isachardoublequoteopen}ls\ {\isasymin}\ LocalSecrets\ PQ{\isachardoublequoteclose}\isanewline
\isadelimproof
\endisadelimproof
\isatagproof
\isacommand{using}\isamarkupfalse%
\ assms\ \isacommand{by}\isamarkupfalse%
\ {\isacharparenleft}simp\ {\isacharparenleft}no{\isacharunderscore}asm{\isacharparenright}\ only{\isacharcolon}\ LocalSecretsDef{\isacharcomma}\ auto{\isacharparenright}%
\endisatagproof
{\isafoldproof}%
\isadelimproof
\isanewline
\endisadelimproof
\isanewline
\isacommand{lemma}\isamarkupfalse%
\ \ LocalSecretsComposition{\isacharunderscore}exprChannel{\isacharunderscore}k{\isacharcolon}\isanewline
\isakeyword{assumes}\ {\isachardoublequoteopen}exprChannel\ x\ {\isacharparenleft}kE\ Keys{\isacharparenright}{\isachardoublequoteclose}\isanewline
\ \ \ \ \ \ \ \isakeyword{and}\ {\isachardoublequoteopen}{\isasymnot}\ ine\ P\ {\isacharparenleft}kE\ Keys{\isacharparenright}{\isachardoublequoteclose}\isanewline
\ \ \ \ \ \ \ \isakeyword{and}\ {\isachardoublequoteopen}{\isasymnot}\ ine\ Q\ {\isacharparenleft}kE\ Keys{\isacharparenright}{\isachardoublequoteclose}\isanewline
\ \ \ \ \ \ \ \isakeyword{and}\ {\isachardoublequoteopen}{\isasymnot}\ {\isacharparenleft}x\ {\isasymnotin}\ ins\ P\ {\isasymand}\ x\ {\isasymnotin}\ ins\ Q{\isacharparenright}{\isachardoublequoteclose}\isanewline
\isakeyword{shows}\ {\isachardoublequoteopen}False{\isachardoublequoteclose}\isanewline
\isadelimproof
\endisadelimproof
\isatagproof
\isacommand{using}\isamarkupfalse%
\ assms\ \isacommand{by}\isamarkupfalse%
\ {\isacharparenleft}metis\ ine{\isacharunderscore}def{\isacharparenright}%
\endisatagproof
{\isafoldproof}%
\isadelimproof
\isanewline
\endisadelimproof
\isanewline
\isacommand{lemma}\isamarkupfalse%
\ \ LocalSecretsComposition{\isacharunderscore}exprChannel{\isacharunderscore}s{\isacharcolon}\isanewline
\isakeyword{assumes}\ {\isachardoublequoteopen}exprChannel\ x\ {\isacharparenleft}sE\ Secrets{\isacharparenright}{\isachardoublequoteclose}\isanewline
\ \ \ \ \ \ \ \isakeyword{and}\ {\isachardoublequoteopen}{\isasymnot}\ ine\ P\ {\isacharparenleft}sE\ Secrets{\isacharparenright}{\isachardoublequoteclose}\isanewline
\ \ \ \ \ \ \ \isakeyword{and}\ {\isachardoublequoteopen}{\isasymnot}\ ine\ Q\ {\isacharparenleft}sE\ Secrets{\isacharparenright}{\isachardoublequoteclose}\isanewline
\ \ \ \ \ \ \ \isakeyword{and}\ {\isachardoublequoteopen}{\isasymnot}\ {\isacharparenleft}x\ {\isasymnotin}\ ins\ P\ {\isasymand}\ x\ {\isasymnotin}\ ins\ Q{\isacharparenright}{\isachardoublequoteclose}\isanewline
\isakeyword{shows}\ {\isachardoublequoteopen}False{\isachardoublequoteclose}\isanewline
\isadelimproof
\endisadelimproof
\isatagproof
\isacommand{using}\isamarkupfalse%
\ assms\ \isacommand{by}\isamarkupfalse%
\ {\isacharparenleft}metis\ ine{\isacharunderscore}ins{\isacharunderscore}neg{\isadigit{1}}{\isacharparenright}%
\endisatagproof
{\isafoldproof}%
\isadelimproof
\isanewline
\endisadelimproof
\isanewline
\isacommand{lemma}\isamarkupfalse%
\ LocalSecretsComposition{\isacharunderscore}neg{\isadigit{1}}{\isacharunderscore}k{\isacharcolon}\isanewline
\isakeyword{assumes}\ {\isachardoublequoteopen}subcomponents\ PQ\ {\isacharequal}\ {\isacharbraceleft}P{\isacharcomma}\ Q{\isacharbraceright}{\isachardoublequoteclose}\isanewline
\ \ \ \ \ \ \ \isakeyword{and}\ {\isachardoublequoteopen}correctCompositionLoc\ PQ{\isachardoublequoteclose}\isanewline
\ \ \ \ \ \ \ \isakeyword{and}\ {\isachardoublequoteopen}{\isasymnot}\ ine\ P\ {\isacharparenleft}kE\ Keys{\isacharparenright}{\isachardoublequoteclose}\isanewline
\ \ \ \ \ \ \ \isakeyword{and}\ {\isachardoublequoteopen}{\isasymnot}\ ine\ Q\ {\isacharparenleft}kE\ Keys{\isacharparenright}{\isachardoublequoteclose}\isanewline
\ \ \ \ \ \ \ \isakeyword{and}\ {\isachardoublequoteopen}kKS\ Keys\ {\isasymnotin}\ LocalSecrets\ P{\isachardoublequoteclose}\isanewline
\ \ \ \ \ \ \ \isakeyword{and}\ {\isachardoublequoteopen}kKS\ Keys\ {\isasymnotin}\ LocalSecrets\ Q{\isachardoublequoteclose}\isanewline
\isakeyword{shows}\ \ \ \ {\isachardoublequoteopen}kKS\ Keys\ {\isasymnotin}\ LocalSecrets\ PQ{\isachardoublequoteclose}\isanewline
\isadelimproof
\endisadelimproof
\isatagproof
\isacommand{proof}\isamarkupfalse%
\ {\isacharminus}\ \isanewline
\ \ \isacommand{from}\isamarkupfalse%
\ assms\ \isacommand{show}\isamarkupfalse%
\ {\isacharquery}thesis\ \isanewline
\ \ \ \ \isacommand{apply}\isamarkupfalse%
\ {\isacharparenleft}simp\ {\isacharparenleft}no{\isacharunderscore}asm{\isacharparenright}\ only{\isacharcolon}\ LocalSecretsDef{\isacharcomma}\ \isanewline
\ \ \ \ \ \ \ \ \ \ \ simp\ add{\isacharcolon}\ correctCompositionLoc{\isacharunderscore}def{\isacharcomma}\ clarify{\isacharparenright}\isanewline
\ \ \ \ \isacommand{by}\isamarkupfalse%
\ {\isacharparenleft}rule\ LocalSecretsComposition{\isacharunderscore}exprChannel{\isacharunderscore}k{\isacharcomma}\ auto{\isacharparenright}\isanewline
\isacommand{qed}\isamarkupfalse%
\endisatagproof
{\isafoldproof}%
\isadelimproof
\isanewline
\endisadelimproof
\isanewline
\isacommand{lemma}\isamarkupfalse%
\ LocalSecretsComposition{\isacharunderscore}neg{\isacharunderscore}k{\isacharcolon}\isanewline
\isakeyword{assumes}\ {\isachardoublequoteopen}subcomponents\ PQ\ {\isacharequal}\ {\isacharbraceleft}P{\isacharcomma}Q{\isacharbraceright}{\isachardoublequoteclose}\isanewline
\ \ \ \ \ \ \ \isakeyword{and}\ {\isachardoublequoteopen}correctCompositionLoc\ PQ{\isachardoublequoteclose}\isanewline
\ \ \ \ \ \ \ \isakeyword{and}\ {\isachardoublequoteopen}correctCompositionKS\ PQ{\isachardoublequoteclose}\isanewline
\ \ \ \ \ \ \ \isakeyword{and}\ {\isachardoublequoteopen}{\isacharparenleft}kKS\ m{\isacharparenright}\ {\isasymnotin}\ specKeysSecrets\ P{\isachardoublequoteclose}\isanewline
\ \ \ \ \ \ \ \isakeyword{and}\ {\isachardoublequoteopen}{\isacharparenleft}kKS\ m{\isacharparenright}\ {\isasymnotin}\ specKeysSecrets\ Q{\isachardoublequoteclose}\isanewline
\ \ \ \ \ \ \ \isakeyword{and}\ {\isachardoublequoteopen}{\isasymnot}\ ine\ P\ {\isacharparenleft}kE\ m{\isacharparenright}{\isachardoublequoteclose}\isanewline
\ \ \ \ \ \ \ \isakeyword{and}\ {\isachardoublequoteopen}{\isasymnot}\ ine\ Q\ {\isacharparenleft}kE\ m{\isacharparenright}{\isachardoublequoteclose}\isanewline
\ \ \ \ \ \ \ \isakeyword{and}\ {\isachardoublequoteopen}{\isacharparenleft}kKS\ m{\isacharparenright}\ {\isasymnotin}\ {\isacharparenleft}{\isacharparenleft}LocalSecrets\ P{\isacharparenright}\ {\isasymunion}\ {\isacharparenleft}LocalSecrets\ Q{\isacharparenright}{\isacharparenright}{\isachardoublequoteclose}\isanewline
\isakeyword{shows}\ \ \ \ {\isachardoublequoteopen}{\isacharparenleft}kKS\ m{\isacharparenright}\ {\isasymnotin}\ {\isacharparenleft}LocalSecrets\ PQ{\isacharparenright}{\isachardoublequoteclose}\isanewline
\isadelimproof
\endisadelimproof
\isatagproof
\isacommand{proof}\isamarkupfalse%
\ {\isacharminus}\isanewline
\ \ \isacommand{from}\isamarkupfalse%
\ assms\ \isacommand{show}\isamarkupfalse%
\ {\isacharquery}thesis\ \isanewline
\ \ \ \ \isacommand{apply}\isamarkupfalse%
\ {\isacharparenleft}simp\ {\isacharparenleft}no{\isacharunderscore}asm{\isacharparenright}\ only{\isacharcolon}\ LocalSecretsDef{\isacharcomma}\ \isanewline
\ \ \ \ \ \ \ \ \ \ \ simp\ add{\isacharcolon}\ correctCompositionLoc{\isacharunderscore}def{\isacharcomma}\ clarify{\isacharparenright}\isanewline
\ \ \ \ \isacommand{by}\isamarkupfalse%
\ {\isacharparenleft}rule\ LocalSecretsComposition{\isacharunderscore}exprChannel{\isacharunderscore}k{\isacharcomma}\ auto{\isacharparenright}\isanewline
\isacommand{qed}\isamarkupfalse%
\endisatagproof
{\isafoldproof}%
\isadelimproof
\ \ \isanewline
\endisadelimproof
\isanewline
\isacommand{lemma}\isamarkupfalse%
\ LocalSecretsComposition{\isacharunderscore}neg{\isacharunderscore}s{\isacharcolon}\isanewline
\isakeyword{assumes}\ subPQ{\isacharcolon}{\isachardoublequoteopen}subcomponents\ PQ\ {\isacharequal}\ {\isacharbraceleft}P{\isacharcomma}Q{\isacharbraceright}{\isachardoublequoteclose}\isanewline
\ \ \ \ \ \ \ \isakeyword{and}\ cCompLoc{\isacharcolon}{\isachardoublequoteopen}correctCompositionLoc\ PQ{\isachardoublequoteclose}\isanewline
\ \ \ \ \ \ \ \isakeyword{and}\ cCompKS{\isacharcolon}{\isachardoublequoteopen}correctCompositionKS\ PQ{\isachardoublequoteclose}\isanewline
\ \ \ \ \ \ \ \isakeyword{and}\ notKSP{\isacharcolon}{\isachardoublequoteopen}{\isacharparenleft}sKS\ m{\isacharparenright}\ {\isasymnotin}\ specKeysSecrets\ P{\isachardoublequoteclose}\isanewline
\ \ \ \ \ \ \ \isakeyword{and}\ notKSQ{\isacharcolon}{\isachardoublequoteopen}{\isacharparenleft}sKS\ m{\isacharparenright}\ {\isasymnotin}\ specKeysSecrets\ Q{\isachardoublequoteclose}\isanewline
\ \ \ \ \ \ \ \isakeyword{and}\ {\isachardoublequoteopen}{\isasymnot}\ ine\ P\ {\isacharparenleft}sE\ m{\isacharparenright}{\isachardoublequoteclose}\isanewline
\ \ \ \ \ \ \ \isakeyword{and}\ {\isachardoublequoteopen}{\isasymnot}\ ine\ Q\ {\isacharparenleft}sE\ m{\isacharparenright}{\isachardoublequoteclose}\isanewline
\ \ \ \ \ \ \ \isakeyword{and}\ notLocSeqPQ{\isacharcolon}{\isachardoublequoteopen}{\isacharparenleft}sKS\ m{\isacharparenright}\ {\isasymnotin}\ {\isacharparenleft}{\isacharparenleft}LocalSecrets\ P{\isacharparenright}\ {\isasymunion}\ {\isacharparenleft}LocalSecrets\ Q{\isacharparenright}{\isacharparenright}{\isachardoublequoteclose}\isanewline
\isakeyword{shows}\ \ \ {\isachardoublequoteopen}{\isacharparenleft}sKS\ m{\isacharparenright}\ {\isasymnotin}\ {\isacharparenleft}LocalSecrets\ PQ{\isacharparenright}{\isachardoublequoteclose}\isanewline
\isadelimproof
\endisadelimproof
\isatagproof
\isacommand{proof}\isamarkupfalse%
\ {\isacharminus}\isanewline
\ \ \isacommand{from}\isamarkupfalse%
\ subPQ\ \isakeyword{and}\ cCompKS\ \isakeyword{and}\ notKSP\ \isakeyword{and}\ notKSQ\isanewline
\ \ \isacommand{have}\isamarkupfalse%
\ sg{\isadigit{1}}{\isacharcolon}{\isachardoublequoteopen}sKS\ m\ {\isasymnotin}\ specKeysSecrets\ PQ{\isachardoublequoteclose}\isanewline
\ \ \ \ \isacommand{by}\isamarkupfalse%
\ {\isacharparenleft}simp\ add{\isacharcolon}\ correctCompositionKS{\isacharunderscore}neg{\isadigit{1}}{\isacharparenright}\ \isanewline
\ \ \isacommand{from}\isamarkupfalse%
\ subPQ\ \isakeyword{and}\ cCompLoc\ \isakeyword{and}\ notLocSeqPQ\ \isacommand{have}\isamarkupfalse%
\ sg{\isadigit{2}}{\isacharcolon}\isanewline
\ \ \ {\isachardoublequoteopen}sKS\ m\ {\isasymnotin}\ \ {\isasymUnion}\ {\isacharparenleft}LocalSecrets\ {\isacharbackquote}\ subcomponents\ PQ{\isacharparenright}{\isachardoublequoteclose}\isanewline
\ \ \ \ \isacommand{by}\isamarkupfalse%
\ simp\isanewline
\ \ \isacommand{from}\isamarkupfalse%
\ sg{\isadigit{1}}\ \isakeyword{and}\ sg{\isadigit{2}}\ \isakeyword{and}\ assms\ \isacommand{show}\isamarkupfalse%
\ {\isacharquery}thesis\ \isanewline
\ \ \ \ \isacommand{apply}\isamarkupfalse%
\ {\isacharparenleft}simp\ {\isacharparenleft}no{\isacharunderscore}asm{\isacharparenright}\ only{\isacharcolon}\ LocalSecretsDef{\isacharcomma}\ \isanewline
\ \ \ \ \ \ \ \ \ \ \ simp\ add{\isacharcolon}\ correctCompositionLoc{\isacharunderscore}def{\isacharcomma}\ clarify{\isacharparenright}\isanewline
\ \ \ \ \isacommand{by}\isamarkupfalse%
\ {\isacharparenleft}rule\ LocalSecretsComposition{\isacharunderscore}exprChannel{\isacharunderscore}s{\isacharcomma}\ auto{\isacharparenright}\isanewline
\isacommand{qed}\isamarkupfalse%
\endisatagproof
{\isafoldproof}%
\isadelimproof
\ \ \isanewline
\endisadelimproof
\isanewline
\isacommand{lemma}\isamarkupfalse%
\ LocalSecretsComposition{\isacharunderscore}neg{\isacharcolon}\isanewline
\isakeyword{assumes}\ {\isachardoublequoteopen}subcomponents\ PQ\ {\isacharequal}\ {\isacharbraceleft}P{\isacharcomma}Q{\isacharbraceright}{\isachardoublequoteclose}\ \isanewline
\ \ \ \ \ \ \ \isakeyword{and}\ {\isachardoublequoteopen}correctCompositionLoc\ PQ{\isachardoublequoteclose}\ \isanewline
\ \ \ \ \ \ \ \isakeyword{and}\ {\isachardoublequoteopen}correctCompositionKS\ PQ{\isachardoublequoteclose}\isanewline
\ \ \ \ \ \ \ \isakeyword{and}\ {\isachardoublequoteopen}ks\ {\isasymnotin}\ specKeysSecrets\ P{\isachardoublequoteclose}\isanewline
\ \ \ \ \ \ \ \isakeyword{and}\ {\isachardoublequoteopen}ks\ {\isasymnotin}\ specKeysSecrets\ Q{\isachardoublequoteclose}\isanewline
\ \ \ \ \ \ \ \isakeyword{and}\ h{\isadigit{1}}{\isacharcolon}{\isachardoublequoteopen}{\isasymforall}\ m{\isachardot}\ ks\ {\isacharequal}\ kKS\ m\ {\isasymlongrightarrow}\ {\isacharparenleft}{\isasymnot}\ ine\ P\ {\isacharparenleft}kE\ m{\isacharparenright}\ {\isasymand}\ {\isasymnot}\ ine\ Q\ {\isacharparenleft}kE\ m{\isacharparenright}{\isacharparenright}{\isachardoublequoteclose}\isanewline
\ \ \ \ \ \ \ \isakeyword{and}\ h{\isadigit{2}}{\isacharcolon}{\isachardoublequoteopen}{\isasymforall}\ m{\isachardot}\ ks\ {\isacharequal}\ sKS\ m\ {\isasymlongrightarrow}\ {\isacharparenleft}{\isasymnot}\ ine\ P\ {\isacharparenleft}sE\ m{\isacharparenright}\ {\isasymand}\ {\isasymnot}\ ine\ Q\ {\isacharparenleft}sE\ m{\isacharparenright}{\isacharparenright}{\isachardoublequoteclose}\isanewline
\ \ \ \ \ \ \ \isakeyword{and}\ {\isachardoublequoteopen}ks\ {\isasymnotin}\ {\isacharparenleft}{\isacharparenleft}LocalSecrets\ P{\isacharparenright}\ {\isasymunion}\ {\isacharparenleft}LocalSecrets\ Q{\isacharparenright}{\isacharparenright}{\isachardoublequoteclose}\isanewline
\isakeyword{shows}\ \ \ {\isachardoublequoteopen}ks\ {\isasymnotin}\ {\isacharparenleft}LocalSecrets\ PQ{\isacharparenright}{\isachardoublequoteclose}\isanewline
\isadelimproof
\endisadelimproof
\isatagproof
\isacommand{proof}\isamarkupfalse%
\ {\isacharparenleft}cases\ {\isachardoublequoteopen}ks{\isachardoublequoteclose}{\isacharparenright}\isanewline
\ \ \isacommand{fix}\isamarkupfalse%
\ m\isanewline
\ \ \isacommand{assume}\isamarkupfalse%
\ a{\isadigit{1}}{\isacharcolon}{\isachardoublequoteopen}ks\ {\isacharequal}\ kKS\ m{\isachardoublequoteclose}\isanewline
\ \ \isacommand{from}\isamarkupfalse%
\ this\ \isakeyword{and}\ h{\isadigit{1}}\ \isacommand{have}\isamarkupfalse%
\ {\isachardoublequoteopen}{\isasymnot}\ ine\ P\ {\isacharparenleft}kE\ m{\isacharparenright}\ {\isasymand}\ {\isasymnot}\ ine\ Q\ {\isacharparenleft}kE\ m{\isacharparenright}{\isachardoublequoteclose}\ \isacommand{by}\isamarkupfalse%
\ simp\isanewline
\ \ \isacommand{from}\isamarkupfalse%
\ this\ \isakeyword{and}\ a{\isadigit{1}}\ \isakeyword{and}\ assms\ \isacommand{show}\isamarkupfalse%
\ {\isacharquery}thesis\isanewline
\ \ \ \ \isacommand{by}\isamarkupfalse%
\ {\isacharparenleft}simp\ add{\isacharcolon}\ LocalSecretsComposition{\isacharunderscore}neg{\isacharunderscore}k{\isacharparenright}\isanewline
\isacommand{next}\isamarkupfalse%
\isanewline
\ \ \isacommand{fix}\isamarkupfalse%
\ m\isanewline
\ \ \isacommand{assume}\isamarkupfalse%
\ a{\isadigit{2}}{\isacharcolon}{\isachardoublequoteopen}ks\ {\isacharequal}\ sKS\ m{\isachardoublequoteclose}\isanewline
\ \ \isacommand{from}\isamarkupfalse%
\ this\ \isakeyword{and}\ h{\isadigit{2}}\ \isacommand{have}\isamarkupfalse%
\ {\isachardoublequoteopen}{\isasymnot}\ ine\ P\ {\isacharparenleft}sE\ m{\isacharparenright}\ {\isasymand}\ {\isasymnot}\ ine\ Q\ {\isacharparenleft}sE\ m{\isacharparenright}{\isachardoublequoteclose}\ \isacommand{by}\isamarkupfalse%
\ simp\isanewline
\ \ \isacommand{from}\isamarkupfalse%
\ this\ \isakeyword{and}\ a{\isadigit{2}}\ \isakeyword{and}\ assms\ \isacommand{show}\isamarkupfalse%
\ {\isacharquery}thesis\isanewline
\ \ \ \ \isacommand{by}\isamarkupfalse%
\ {\isacharparenleft}simp\ add{\isacharcolon}\ LocalSecretsComposition{\isacharunderscore}neg{\isacharunderscore}s{\isacharparenright}\isanewline
\isacommand{qed}\isamarkupfalse%
\endisatagproof
{\isafoldproof}%
\isadelimproof
\isanewline
\endisadelimproof
\isanewline
\isacommand{lemma}\isamarkupfalse%
\ LocalSecretsComposition{\isacharunderscore}neg{\isadigit{1}}{\isacharunderscore}s{\isacharcolon}\isanewline
\isakeyword{assumes}\ {\isachardoublequoteopen}subcomponents\ PQ\ {\isacharequal}\ {\isacharbraceleft}P{\isacharcomma}\ Q{\isacharbraceright}{\isachardoublequoteclose}\isanewline
\ \ \ \ \ \ \ \isakeyword{and}\ {\isachardoublequoteopen}correctCompositionLoc\ PQ{\isachardoublequoteclose}\isanewline
\ \ \ \ \ \ \ \isakeyword{and}\ {\isachardoublequoteopen}{\isasymnot}\ ine\ P\ {\isacharparenleft}sE\ s{\isacharparenright}{\isachardoublequoteclose}\isanewline
\ \ \ \ \ \ \ \isakeyword{and}\ {\isachardoublequoteopen}{\isasymnot}\ ine\ Q\ {\isacharparenleft}sE\ s{\isacharparenright}{\isachardoublequoteclose}\isanewline
\ \ \ \ \ \ \ \isakeyword{and}\ {\isachardoublequoteopen}sKS\ s\ {\isasymnotin}\ LocalSecrets\ P{\isachardoublequoteclose}\ \isanewline
\ \ \ \ \ \ \ \isakeyword{and}\ {\isachardoublequoteopen}sKS\ s\ {\isasymnotin}\ LocalSecrets\ Q{\isachardoublequoteclose}\isanewline
\isakeyword{shows}\ \ \ \ {\isachardoublequoteopen}sKS\ s\ {\isasymnotin}\ LocalSecrets\ PQ{\isachardoublequoteclose}\isanewline
\isadelimproof
\endisadelimproof
\isatagproof
\isacommand{proof}\isamarkupfalse%
\ {\isacharminus}\ \isanewline
\ \ \isacommand{from}\isamarkupfalse%
\ assms\ \isacommand{have}\isamarkupfalse%
\ \isanewline
\ \ \ {\isachardoublequoteopen}sKS\ s\ {\isasymnotin}\ \ {\isasymUnion}\ {\isacharparenleft}LocalSecrets\ {\isacharbackquote}\ subcomponents\ PQ{\isacharparenright}{\isachardoublequoteclose}\isanewline
\ \ \ \ \isacommand{by}\isamarkupfalse%
\ simp\isanewline
\ \ \ \isacommand{from}\isamarkupfalse%
\ \ assms\ \isakeyword{and}\ this\ \isacommand{show}\isamarkupfalse%
\ {\isacharquery}thesis\ \isanewline
\ \ \ \ \isacommand{apply}\isamarkupfalse%
\ {\isacharparenleft}simp\ {\isacharparenleft}no{\isacharunderscore}asm{\isacharparenright}\ only{\isacharcolon}\ LocalSecretsDef{\isacharcomma}\ \isanewline
\ \ \ \ \ \ \ \ \ \ \ simp\ add{\isacharcolon}\ correctCompositionLoc{\isacharunderscore}def{\isacharcomma}\ clarify{\isacharparenright}\isanewline
\ \ \ \ \isacommand{by}\isamarkupfalse%
\ {\isacharparenleft}rule\ LocalSecretsComposition{\isacharunderscore}exprChannel{\isacharunderscore}s{\isacharcomma}\ auto{\isacharparenright}\isanewline
\isacommand{qed}\isamarkupfalse%
\endisatagproof
{\isafoldproof}%
\isadelimproof
\ \ \isanewline
\endisadelimproof
\isanewline
\isacommand{lemma}\isamarkupfalse%
\ LocalSecretsComposition{\isacharunderscore}neg{\isadigit{1}}{\isacharcolon}\isanewline
\isakeyword{assumes}\ {\isachardoublequoteopen}subcomponents\ PQ\ {\isacharequal}\ {\isacharbraceleft}P{\isacharcomma}\ Q{\isacharbraceright}{\isachardoublequoteclose}\isanewline
\ \ \ \ \ \ \ \isakeyword{and}\ {\isachardoublequoteopen}correctCompositionLoc\ PQ{\isachardoublequoteclose}\isanewline
\ \ \ \ \ \ \ \isakeyword{and}\ h{\isadigit{1}}{\isacharcolon}{\isachardoublequoteopen}{\isasymforall}\ m{\isachardot}\ ks\ {\isacharequal}\ kKS\ m\ {\isasymlongrightarrow}\ {\isacharparenleft}{\isasymnot}\ ine\ P\ {\isacharparenleft}kE\ m{\isacharparenright}\ {\isasymand}\ {\isasymnot}\ ine\ Q\ {\isacharparenleft}kE\ m{\isacharparenright}{\isacharparenright}{\isachardoublequoteclose}\ \isanewline
\ \ \ \ \ \ \ \isakeyword{and}\ h{\isadigit{2}}{\isacharcolon}{\isachardoublequoteopen}{\isasymforall}\ m{\isachardot}\ ks\ {\isacharequal}\ sKS\ m\ {\isasymlongrightarrow}\ {\isacharparenleft}{\isasymnot}\ ine\ P\ {\isacharparenleft}sE\ m{\isacharparenright}\ {\isasymand}\ {\isasymnot}\ ine\ Q\ {\isacharparenleft}sE\ m{\isacharparenright}{\isacharparenright}{\isachardoublequoteclose}\isanewline
\ \ \ \ \ \ \ \isakeyword{and}\ {\isachardoublequoteopen}ks\ {\isasymnotin}\ LocalSecrets\ P{\isachardoublequoteclose}\isanewline
\ \ \ \ \ \ \ \isakeyword{and}\ {\isachardoublequoteopen}ks\ {\isasymnotin}\ LocalSecrets\ Q{\isachardoublequoteclose}\isanewline
\isakeyword{shows}\ \ \ \ {\isachardoublequoteopen}ks\ {\isasymnotin}\ LocalSecrets\ PQ{\isachardoublequoteclose}\isanewline
\isadelimproof
\endisadelimproof
\isatagproof
\isacommand{proof}\isamarkupfalse%
\ {\isacharparenleft}cases\ {\isachardoublequoteopen}ks{\isachardoublequoteclose}{\isacharparenright}\isanewline
\ \ \isacommand{fix}\isamarkupfalse%
\ m\isanewline
\ \ \isacommand{assume}\isamarkupfalse%
\ a{\isadigit{1}}{\isacharcolon}{\isachardoublequoteopen}ks\ {\isacharequal}\ kKS\ m{\isachardoublequoteclose}\isanewline
\ \ \isacommand{from}\isamarkupfalse%
\ this\ \isakeyword{and}\ h{\isadigit{1}}\ \isacommand{have}\isamarkupfalse%
\ {\isachardoublequoteopen}{\isasymnot}\ ine\ P\ {\isacharparenleft}kE\ m{\isacharparenright}\ {\isasymand}\ {\isasymnot}\ ine\ Q\ {\isacharparenleft}kE\ m{\isacharparenright}{\isachardoublequoteclose}\ \isacommand{by}\isamarkupfalse%
\ simp\isanewline
\ \ \isacommand{from}\isamarkupfalse%
\ this\ \isakeyword{and}\ a{\isadigit{1}}\ \isakeyword{and}\ assms\ \isacommand{show}\isamarkupfalse%
\ {\isacharquery}thesis\ \isanewline
\ \ \ \ \isacommand{by}\isamarkupfalse%
\ {\isacharparenleft}simp\ add{\isacharcolon}\ LocalSecretsComposition{\isacharunderscore}neg{\isadigit{1}}{\isacharunderscore}k{\isacharparenright}\isanewline
\isacommand{next}\isamarkupfalse%
\isanewline
\ \ \isacommand{fix}\isamarkupfalse%
\ m\isanewline
\ \ \isacommand{assume}\isamarkupfalse%
\ a{\isadigit{2}}{\isacharcolon}{\isachardoublequoteopen}ks\ {\isacharequal}\ sKS\ m{\isachardoublequoteclose}\isanewline
\ \ \isacommand{from}\isamarkupfalse%
\ this\ \isakeyword{and}\ h{\isadigit{2}}\ \isacommand{have}\isamarkupfalse%
\ {\isachardoublequoteopen}{\isasymnot}\ ine\ P\ {\isacharparenleft}sE\ m{\isacharparenright}\ {\isasymand}\ {\isasymnot}\ ine\ Q\ {\isacharparenleft}sE\ m{\isacharparenright}{\isachardoublequoteclose}\ \isacommand{by}\isamarkupfalse%
\ simp\isanewline
\ \ \isacommand{from}\isamarkupfalse%
\ this\ \isakeyword{and}\ a{\isadigit{2}}\ \isakeyword{and}\ assms\ \isacommand{show}\isamarkupfalse%
\ {\isacharquery}thesis\ \isanewline
\ \ \ \ \isacommand{by}\isamarkupfalse%
\ {\isacharparenleft}simp\ add{\isacharcolon}\ LocalSecretsComposition{\isacharunderscore}neg{\isadigit{1}}{\isacharunderscore}s{\isacharparenright}\isanewline
\isacommand{qed}\isamarkupfalse%
\endisatagproof
{\isafoldproof}%
\isadelimproof
\isanewline
\endisadelimproof
\isanewline
\isacommand{lemma}\isamarkupfalse%
\ LocalSecretsComposition{\isacharunderscore}ine{\isadigit{1}}{\isacharunderscore}k{\isacharcolon}\isanewline
\isakeyword{assumes}\ {\isachardoublequoteopen}kKS\ k\ {\isasymin}\ LocalSecrets\ PQ{\isachardoublequoteclose}\ \isanewline
\ \ \ \ \ \ \ \isakeyword{and}\ {\isachardoublequoteopen}subcomponents\ PQ\ {\isacharequal}\ {\isacharbraceleft}P{\isacharcomma}\ Q{\isacharbraceright}{\isachardoublequoteclose}\isanewline
\ \ \ \ \ \ \ \isakeyword{and}\ {\isachardoublequoteopen}correctCompositionLoc\ PQ{\isachardoublequoteclose}\ \isanewline
\ \ \ \ \ \ \ \isakeyword{and}\ {\isachardoublequoteopen}{\isasymnot}\ ine\ Q\ {\isacharparenleft}kE\ k{\isacharparenright}{\isachardoublequoteclose}\isanewline
\ \ \ \ \ \ \ \isakeyword{and}\ {\isachardoublequoteopen}kKS\ k\ {\isasymnotin}\ LocalSecrets\ P{\isachardoublequoteclose}\isanewline
\ \ \ \ \ \ \ \isakeyword{and}\ {\isachardoublequoteopen}kKS\ k\ {\isasymnotin}\ LocalSecrets\ Q{\isachardoublequoteclose}\isanewline
\isakeyword{shows}\ \ \ \ {\isachardoublequoteopen}ine\ P\ {\isacharparenleft}kE\ k{\isacharparenright}{\isachardoublequoteclose}\isanewline
\isadelimproof
\endisadelimproof
\isatagproof
\isacommand{using}\isamarkupfalse%
\ assms\ \isacommand{by}\isamarkupfalse%
\ {\isacharparenleft}metis\ LocalSecretsComposition{\isacharunderscore}neg{\isadigit{1}}{\isacharunderscore}k{\isacharparenright}%
\endisatagproof
{\isafoldproof}%
\isadelimproof
\isanewline
\endisadelimproof
\isanewline
\isacommand{lemma}\isamarkupfalse%
\ LocalSecretsComposition{\isacharunderscore}ine{\isadigit{1}}{\isacharunderscore}s{\isacharcolon}\isanewline
\isakeyword{assumes}\ {\isachardoublequoteopen}sKS\ s\ {\isasymin}\ LocalSecrets\ PQ{\isachardoublequoteclose}\ \isanewline
\ \ \ \ \ \ \ \isakeyword{and}\ {\isachardoublequoteopen}subcomponents\ PQ\ {\isacharequal}\ {\isacharbraceleft}P{\isacharcomma}\ Q{\isacharbraceright}{\isachardoublequoteclose}\isanewline
\ \ \ \ \ \ \ \isakeyword{and}\ {\isachardoublequoteopen}correctCompositionLoc\ PQ{\isachardoublequoteclose}\ \isanewline
\ \ \ \ \ \ \ \isakeyword{and}\ {\isachardoublequoteopen}{\isasymnot}\ ine\ Q\ {\isacharparenleft}sE\ s{\isacharparenright}{\isachardoublequoteclose}\isanewline
\ \ \ \ \ \ \ \isakeyword{and}\ {\isachardoublequoteopen}sKS\ s\ {\isasymnotin}\ LocalSecrets\ P{\isachardoublequoteclose}\isanewline
\ \ \ \ \ \ \ \isakeyword{and}\ {\isachardoublequoteopen}sKS\ s\ {\isasymnotin}\ LocalSecrets\ Q{\isachardoublequoteclose}\isanewline
\isakeyword{shows}\ \ \ \ {\isachardoublequoteopen}ine\ P\ {\isacharparenleft}sE\ s{\isacharparenright}{\isachardoublequoteclose}\isanewline
\isadelimproof
\endisadelimproof
\isatagproof
\isacommand{using}\isamarkupfalse%
\ assms\ \isacommand{by}\isamarkupfalse%
\ {\isacharparenleft}metis\ LocalSecretsComposition{\isacharunderscore}neg{\isadigit{1}}{\isacharunderscore}s{\isacharparenright}%
\endisatagproof
{\isafoldproof}%
\isadelimproof
\isanewline
\endisadelimproof
\isanewline
\isacommand{lemma}\isamarkupfalse%
\ LocalSecretsComposition{\isacharunderscore}ine{\isadigit{2}}{\isacharunderscore}k{\isacharcolon}\isanewline
\isakeyword{assumes}\ {\isachardoublequoteopen}kKS\ k\ {\isasymin}\ LocalSecrets\ PQ{\isachardoublequoteclose}\isanewline
\ \ \ \ \ \ \ \isakeyword{and}\ {\isachardoublequoteopen}subcomponents\ PQ\ {\isacharequal}\ {\isacharbraceleft}P{\isacharcomma}\ Q{\isacharbraceright}{\isachardoublequoteclose}\isanewline
\ \ \ \ \ \ \ \isakeyword{and}\ {\isachardoublequoteopen}correctCompositionLoc\ PQ{\isachardoublequoteclose}\isanewline
\ \ \ \ \ \ \ \isakeyword{and}\ {\isachardoublequoteopen}{\isasymnot}\ ine\ P\ {\isacharparenleft}kE\ k{\isacharparenright}{\isachardoublequoteclose}\isanewline
\ \ \ \ \ \ \ \isakeyword{and}\ {\isachardoublequoteopen}kKS\ k\ {\isasymnotin}\ LocalSecrets\ P{\isachardoublequoteclose}\isanewline
\ \ \ \ \ \ \ \isakeyword{and}\ {\isachardoublequoteopen}kKS\ k\ {\isasymnotin}\ LocalSecrets\ Q{\isachardoublequoteclose}\isanewline
\isakeyword{shows}\ \ \ {\isachardoublequoteopen}ine\ Q\ {\isacharparenleft}kE\ k{\isacharparenright}{\isachardoublequoteclose}\ \isanewline
\isadelimproof
\endisadelimproof
\isatagproof
\isacommand{using}\isamarkupfalse%
\ assms\ \ \isacommand{by}\isamarkupfalse%
\ {\isacharparenleft}metis\ LocalSecretsComposition{\isacharunderscore}ine{\isadigit{1}}{\isacharunderscore}k{\isacharparenright}%
\endisatagproof
{\isafoldproof}%
\isadelimproof
\isanewline
\endisadelimproof
\isanewline
\isacommand{lemma}\isamarkupfalse%
\ LocalSecretsComposition{\isacharunderscore}ine{\isadigit{2}}{\isacharunderscore}s{\isacharcolon}\isanewline
\isakeyword{assumes}\ {\isachardoublequoteopen}sKS\ s\ {\isasymin}\ LocalSecrets\ PQ{\isachardoublequoteclose}\ \isanewline
\ \ \ \ \ \ \ \isakeyword{and}\ {\isachardoublequoteopen}subcomponents\ PQ\ {\isacharequal}\ {\isacharbraceleft}P{\isacharcomma}\ Q{\isacharbraceright}{\isachardoublequoteclose}\isanewline
\ \ \ \ \ \ \ \isakeyword{and}\ {\isachardoublequoteopen}correctCompositionLoc\ PQ{\isachardoublequoteclose}\isanewline
\ \ \ \ \ \ \ \isakeyword{and}\ {\isachardoublequoteopen}{\isasymnot}\ ine\ P\ {\isacharparenleft}sE\ s{\isacharparenright}{\isachardoublequoteclose}\isanewline
\ \ \ \ \ \ \ \isakeyword{and}\ {\isachardoublequoteopen}sKS\ s\ {\isasymnotin}\ LocalSecrets\ P{\isachardoublequoteclose}\isanewline
\ \ \ \ \ \ \ \isakeyword{and}\ {\isachardoublequoteopen}sKS\ s\ {\isasymnotin}\ LocalSecrets\ Q{\isachardoublequoteclose}\isanewline
\isakeyword{shows}\ \ \ \ {\isachardoublequoteopen}ine\ Q\ {\isacharparenleft}sE\ s{\isacharparenright}{\isachardoublequoteclose}\isanewline
\isadelimproof
\endisadelimproof
\isatagproof
\isacommand{using}\isamarkupfalse%
\ assms\ \isacommand{by}\isamarkupfalse%
\ {\isacharparenleft}metis\ LocalSecretsComposition{\isacharunderscore}ine{\isadigit{1}}{\isacharunderscore}s{\isacharparenright}%
\endisatagproof
{\isafoldproof}%
\isadelimproof
\isanewline
\endisadelimproof
\isanewline
\isacommand{lemma}\isamarkupfalse%
\ LocalSecretsComposition{\isacharunderscore}neg{\isacharunderscore}loc{\isacharunderscore}k{\isacharcolon}\isanewline
\isakeyword{assumes}\ {\isachardoublequoteopen}kKS\ key\ {\isasymnotin}\ LocalSecrets\ P{\isachardoublequoteclose}\isanewline
\ \ \ \ \ \ \ \isakeyword{and}\ {\isachardoublequoteopen}exprChannel\ ch\ {\isacharparenleft}kE\ key{\isacharparenright}{\isachardoublequoteclose}\isanewline
\ \ \ \ \ \ \ \isakeyword{and}\ {\isachardoublequoteopen}kKS\ key\ {\isasymnotin}\ specKeysSecrets\ P{\isachardoublequoteclose}\isanewline
\isakeyword{shows}\ \ \ \ {\isachardoublequoteopen}ch\ {\isasymnotin}\ loc\ P{\isachardoublequoteclose}\isanewline
\isadelimproof
\endisadelimproof
\isatagproof
\isacommand{using}\isamarkupfalse%
\ assms\ \isacommand{by}\isamarkupfalse%
\ {\isacharparenleft}simp\ only{\isacharcolon}\ LocalSecretsDef{\isacharcomma}\ auto{\isacharparenright}%
\endisatagproof
{\isafoldproof}%
\isadelimproof
\isanewline
\endisadelimproof
\isanewline
\isacommand{lemma}\isamarkupfalse%
\ LocalSecretsComposition{\isacharunderscore}neg{\isacharunderscore}loc{\isacharunderscore}s{\isacharcolon}\isanewline
\isakeyword{assumes}\ {\isachardoublequoteopen}sKS\ secret\ {\isasymnotin}\ LocalSecrets\ P{\isachardoublequoteclose}\isanewline
\ \ \ \ \ \ \ \isakeyword{and}\ {\isachardoublequoteopen}exprChannel\ ch\ {\isacharparenleft}sE\ secret{\isacharparenright}{\isachardoublequoteclose}\isanewline
\ \ \ \ \ \ \ \isakeyword{and}\ {\isachardoublequoteopen}sKS\ secret\ {\isasymnotin}\ specKeysSecrets\ P{\isachardoublequoteclose}\isanewline
\isakeyword{shows}\ \ \ \ {\isachardoublequoteopen}ch\ {\isasymnotin}\ loc\ P{\isachardoublequoteclose}\isanewline
\isadelimproof
\endisadelimproof
\isatagproof
\isacommand{using}\isamarkupfalse%
\ assms\ \isacommand{by}\isamarkupfalse%
\ {\isacharparenleft}simp\ only{\isacharcolon}\ LocalSecretsDef{\isacharcomma}\ auto{\isacharparenright}%
\endisatagproof
{\isafoldproof}%
\isadelimproof
\isanewline
\endisadelimproof
\isanewline
\isacommand{lemma}\isamarkupfalse%
\ correctCompositionKS{\isacharunderscore}exprChannel{\isacharunderscore}k{\isacharunderscore}P{\isacharcolon}\isanewline
\isakeyword{assumes}\ {\isachardoublequoteopen}subcomponents\ PQ\ {\isacharequal}\ {\isacharbraceleft}P{\isacharcomma}Q{\isacharbraceright}{\isachardoublequoteclose}\ \isanewline
\ \ \ \ \ \ \ \isakeyword{and}\ {\isachardoublequoteopen}correctCompositionKS\ PQ{\isachardoublequoteclose}\isanewline
\ \ \ \ \ \ \ \isakeyword{and}\ {\isachardoublequoteopen}kKS\ key\ {\isasymnotin}\ LocalSecrets\ PQ{\isachardoublequoteclose}\isanewline
\ \ \ \ \ \ \ \isakeyword{and}\ {\isachardoublequoteopen}ch\ {\isasymin}\ ins\ P{\isachardoublequoteclose}\isanewline
\ \ \ \ \ \ \ \isakeyword{and}\ {\isachardoublequoteopen}exprChannel\ ch\ {\isacharparenleft}kE\ key{\isacharparenright}{\isachardoublequoteclose}\isanewline
\ \ \ \ \ \ \ \isakeyword{and}\ {\isachardoublequoteopen}kKS\ key\ {\isasymnotin}\ specKeysSecrets\ PQ{\isachardoublequoteclose}\isanewline
\ \ \ \ \ \ \ \isakeyword{and}\ {\isachardoublequoteopen}correctCompositionIn\ PQ{\isachardoublequoteclose}\isanewline
\isakeyword{shows}\ \ \ \ {\isachardoublequoteopen}ch\ {\isasymin}\ ins\ PQ\ {\isasymand}\ exprChannel\ ch\ {\isacharparenleft}kE\ key{\isacharparenright}{\isachardoublequoteclose}\isanewline
\isadelimproof
\endisadelimproof
\isatagproof
\isacommand{using}\isamarkupfalse%
\ assms\isanewline
\isacommand{by}\isamarkupfalse%
\ {\isacharparenleft}metis\ LocalSecretsComposition{\isacharunderscore}neg{\isacharunderscore}loc{\isacharunderscore}k\ correctCompositionIn{\isacharunderscore}L{\isadigit{1}}{\isacharparenright}%
\endisatagproof
{\isafoldproof}%
\isadelimproof
\isanewline
\endisadelimproof
\isanewline
\isacommand{lemma}\isamarkupfalse%
\ correctCompositionKS{\isacharunderscore}exprChannel{\isacharunderscore}k{\isacharunderscore}Pex{\isacharcolon}\isanewline
\isakeyword{assumes}\ {\isachardoublequoteopen}subcomponents\ PQ\ {\isacharequal}\ {\isacharbraceleft}P{\isacharcomma}Q{\isacharbraceright}{\isachardoublequoteclose}\ \isanewline
\ \ \ \ \ \ \ \isakeyword{and}\ {\isachardoublequoteopen}correctCompositionKS\ PQ{\isachardoublequoteclose}\isanewline
\ \ \ \ \ \ \ \isakeyword{and}\ {\isachardoublequoteopen}kKS\ key\ {\isasymnotin}\ LocalSecrets\ PQ{\isachardoublequoteclose}\isanewline
\ \ \ \ \ \ \ \isakeyword{and}\ {\isachardoublequoteopen}ch\ {\isasymin}\ ins\ P{\isachardoublequoteclose}\isanewline
\ \ \ \ \ \ \ \isakeyword{and}\ {\isachardoublequoteopen}exprChannel\ ch\ {\isacharparenleft}kE\ key{\isacharparenright}{\isachardoublequoteclose}\isanewline
\ \ \ \ \ \ \ \isakeyword{and}\ {\isachardoublequoteopen}kKS\ key\ {\isasymnotin}\ specKeysSecrets\ PQ{\isachardoublequoteclose}\isanewline
\ \ \ \ \ \ \ \isakeyword{and}\ {\isachardoublequoteopen}correctCompositionIn\ PQ{\isachardoublequoteclose}\isanewline
\isakeyword{shows}\ \ \ \ {\isachardoublequoteopen}{\isasymexists}ch{\isachardot}\ ch\ {\isasymin}\ ins\ PQ\ {\isasymand}\ exprChannel\ ch\ {\isacharparenleft}kE\ key{\isacharparenright}{\isachardoublequoteclose}\isanewline
\isadelimproof
\endisadelimproof
\isatagproof
\isacommand{using}\isamarkupfalse%
\ assms\isanewline
\isacommand{by}\isamarkupfalse%
\ {\isacharparenleft}metis\ correctCompositionKS{\isacharunderscore}exprChannel{\isacharunderscore}k{\isacharunderscore}P{\isacharparenright}%
\endisatagproof
{\isafoldproof}%
\isadelimproof
\isanewline
\endisadelimproof
\isanewline
\isacommand{lemma}\isamarkupfalse%
\ correctCompositionKS{\isacharunderscore}exprChannel{\isacharunderscore}k{\isacharunderscore}Q{\isacharcolon}\isanewline
\isakeyword{assumes}\ {\isachardoublequoteopen}subcomponents\ PQ\ {\isacharequal}\ {\isacharbraceleft}P{\isacharcomma}Q{\isacharbraceright}{\isachardoublequoteclose}\ \isanewline
\ \ \ \ \ \ \ \isakeyword{and}\ {\isachardoublequoteopen}correctCompositionKS\ PQ{\isachardoublequoteclose}\isanewline
\ \ \ \ \ \ \ \isakeyword{and}\ {\isachardoublequoteopen}kKS\ key\ {\isasymnotin}\ LocalSecrets\ PQ{\isachardoublequoteclose}\isanewline
\ \ \ \ \ \ \ \isakeyword{and}\ {\isachardoublequoteopen}ch\ {\isasymin}\ ins\ Q{\isachardoublequoteclose}\isanewline
\ \ \ \ \ \ \ \isakeyword{and}\ h{\isadigit{1}}{\isacharcolon}{\isachardoublequoteopen}exprChannel\ ch\ {\isacharparenleft}kE\ key{\isacharparenright}{\isachardoublequoteclose}\isanewline
\ \ \ \ \ \ \ \isakeyword{and}\ {\isachardoublequoteopen}kKS\ key\ {\isasymnotin}\ specKeysSecrets\ PQ{\isachardoublequoteclose}\isanewline
\ \ \ \ \ \ \ \isakeyword{and}\ {\isachardoublequoteopen}correctCompositionIn\ PQ{\isachardoublequoteclose}\isanewline
\isakeyword{shows}\ \ \ \ {\isachardoublequoteopen}ch\ {\isasymin}\ ins\ PQ\ {\isasymand}\ exprChannel\ ch\ {\isacharparenleft}kE\ key{\isacharparenright}{\isachardoublequoteclose}\isanewline
\isadelimproof
\endisadelimproof
\isatagproof
\isacommand{proof}\isamarkupfalse%
\ {\isacharminus}\ \isanewline
\ \ \isacommand{from}\isamarkupfalse%
\ assms\ \isacommand{have}\isamarkupfalse%
\ {\isachardoublequoteopen}ch\ {\isasymnotin}\ loc\ PQ{\isachardoublequoteclose}\ \isanewline
\ \ \ \ \isacommand{by}\isamarkupfalse%
\ {\isacharparenleft}simp\ add{\isacharcolon}\ LocalSecretsComposition{\isacharunderscore}neg{\isacharunderscore}loc{\isacharunderscore}k{\isacharparenright}\isanewline
\ \ \isacommand{from}\isamarkupfalse%
\ this\ \isakeyword{and}\ assms\ \isacommand{have}\isamarkupfalse%
\ {\isachardoublequoteopen}ch\ {\isasymin}\ ins\ PQ{\isachardoublequoteclose}\ \isanewline
\ \ \ \ \isacommand{by}\isamarkupfalse%
\ {\isacharparenleft}simp\ add{\isacharcolon}\ correctCompositionIn{\isacharunderscore}def{\isacharparenright}\ \isanewline
\ \ \isacommand{from}\isamarkupfalse%
\ this\ \isakeyword{and}\ h{\isadigit{1}}\ \isacommand{show}\isamarkupfalse%
\ {\isacharquery}thesis\ \isacommand{by}\isamarkupfalse%
\ simp\isanewline
\isacommand{qed}\isamarkupfalse%
\endisatagproof
{\isafoldproof}%
\isadelimproof
\isanewline
\endisadelimproof
\isanewline
\isacommand{lemma}\isamarkupfalse%
\ correctCompositionKS{\isacharunderscore}exprChannel{\isacharunderscore}k{\isacharunderscore}Qex{\isacharcolon}\isanewline
\isakeyword{assumes}\ {\isachardoublequoteopen}subcomponents\ PQ\ {\isacharequal}\ {\isacharbraceleft}P{\isacharcomma}Q{\isacharbraceright}{\isachardoublequoteclose}\ \isanewline
\ \ \ \ \ \ \ \ \isakeyword{and}\ {\isachardoublequoteopen}correctCompositionKS\ PQ{\isachardoublequoteclose}\isanewline
\ \ \ \ \ \ \ \ \isakeyword{and}\ {\isachardoublequoteopen}kKS\ key\ {\isasymnotin}\ LocalSecrets\ PQ{\isachardoublequoteclose}\isanewline
\ \ \ \ \ \ \ \ \isakeyword{and}\ {\isachardoublequoteopen}ch\ {\isasymin}\ ins\ Q{\isachardoublequoteclose}\isanewline
\ \ \ \ \ \ \ \ \isakeyword{and}\ {\isachardoublequoteopen}exprChannel\ ch\ {\isacharparenleft}kE\ key{\isacharparenright}{\isachardoublequoteclose}\isanewline
\ \ \ \ \ \ \ \ \isakeyword{and}\ {\isachardoublequoteopen}kKS\ key\ {\isasymnotin}\ specKeysSecrets\ PQ{\isachardoublequoteclose}\isanewline
\ \ \ \ \ \ \ \ \isakeyword{and}\ {\isachardoublequoteopen}correctCompositionIn\ PQ{\isachardoublequoteclose}\isanewline
\isakeyword{shows}\ \ \ \ {\isachardoublequoteopen}{\isasymexists}ch{\isachardot}\ ch\ {\isasymin}\ ins\ PQ\ {\isasymand}\ exprChannel\ ch\ {\isacharparenleft}kE\ key{\isacharparenright}{\isachardoublequoteclose}\isanewline
\isadelimproof
\endisadelimproof
\isatagproof
\isacommand{using}\isamarkupfalse%
\ assms\isanewline
\isacommand{by}\isamarkupfalse%
\ {\isacharparenleft}metis\ correctCompositionKS{\isacharunderscore}exprChannel{\isacharunderscore}k{\isacharunderscore}Q{\isacharparenright}%
\endisatagproof
{\isafoldproof}%
\isadelimproof
\isanewline
\endisadelimproof
\isanewline
\isacommand{lemma}\isamarkupfalse%
\ correctCompositionKS{\isacharunderscore}exprChannel{\isacharunderscore}s{\isacharunderscore}P{\isacharcolon}\isanewline
\isakeyword{assumes}\ {\isachardoublequoteopen}subcomponents\ PQ\ {\isacharequal}\ {\isacharbraceleft}P{\isacharcomma}Q{\isacharbraceright}{\isachardoublequoteclose}\ \isanewline
\ \ \ \ \ \ \ \isakeyword{and}\ {\isachardoublequoteopen}correctCompositionKS\ PQ{\isachardoublequoteclose}\isanewline
\ \ \ \ \ \ \ \isakeyword{and}\ {\isachardoublequoteopen}sKS\ secret\ {\isasymnotin}\ LocalSecrets\ PQ{\isachardoublequoteclose}\isanewline
\ \ \ \ \ \ \ \isakeyword{and}\ {\isachardoublequoteopen}ch\ {\isasymin}\ ins\ P{\isachardoublequoteclose}\isanewline
\ \ \ \ \ \ \ \isakeyword{and}\ {\isachardoublequoteopen}exprChannel\ ch\ {\isacharparenleft}sE\ secret{\isacharparenright}{\isachardoublequoteclose}\isanewline
\ \ \ \ \ \ \ \isakeyword{and}\ {\isachardoublequoteopen}sKS\ secret\ {\isasymnotin}\ specKeysSecrets\ PQ{\isachardoublequoteclose}\isanewline
\ \ \ \ \ \ \ \isakeyword{and}\ {\isachardoublequoteopen}correctCompositionIn\ PQ{\isachardoublequoteclose}\isanewline
\isakeyword{shows}\ \ \ \ {\isachardoublequoteopen}ch\ {\isasymin}\ ins\ PQ\ {\isasymand}\ exprChannel\ ch\ {\isacharparenleft}sE\ secret{\isacharparenright}{\isachardoublequoteclose}\isanewline
\isadelimproof
\endisadelimproof
\isatagproof
\isacommand{using}\isamarkupfalse%
\ assms\isanewline
\isacommand{by}\isamarkupfalse%
\ {\isacharparenleft}metis\ LocalSecretsComposition{\isacharunderscore}neg{\isacharunderscore}loc{\isacharunderscore}s\ correctCompositionIn{\isacharunderscore}L{\isadigit{1}}{\isacharparenright}%
\endisatagproof
{\isafoldproof}%
\isadelimproof
\isanewline
\endisadelimproof
\isanewline
\isacommand{lemma}\isamarkupfalse%
\ correctCompositionKS{\isacharunderscore}exprChannel{\isacharunderscore}s{\isacharunderscore}Pex{\isacharcolon}\isanewline
\isakeyword{assumes}\ {\isachardoublequoteopen}subcomponents\ PQ\ {\isacharequal}\ {\isacharbraceleft}P{\isacharcomma}Q{\isacharbraceright}{\isachardoublequoteclose}\ \isanewline
\ \ \ \ \ \ \ \isakeyword{and}\ {\isachardoublequoteopen}correctCompositionKS\ PQ{\isachardoublequoteclose}\isanewline
\ \ \ \ \ \ \ \isakeyword{and}\ {\isachardoublequoteopen}sKS\ secret\ {\isasymnotin}\ LocalSecrets\ PQ{\isachardoublequoteclose}\isanewline
\ \ \ \ \ \ \ \isakeyword{and}\ {\isachardoublequoteopen}ch\ {\isasymin}\ ins\ P{\isachardoublequoteclose}\isanewline
\ \ \ \ \ \ \ \isakeyword{and}\ {\isachardoublequoteopen}exprChannel\ ch\ {\isacharparenleft}sE\ secret{\isacharparenright}{\isachardoublequoteclose}\isanewline
\ \ \ \ \ \ \ \isakeyword{and}\ {\isachardoublequoteopen}sKS\ secret\ {\isasymnotin}\ specKeysSecrets\ PQ{\isachardoublequoteclose}\isanewline
\ \ \ \ \ \ \ \isakeyword{and}\ {\isachardoublequoteopen}correctCompositionIn\ PQ{\isachardoublequoteclose}\isanewline
\isakeyword{shows}\ \ \ \ {\isachardoublequoteopen}{\isasymexists}ch{\isachardot}\ ch\ {\isasymin}\ ins\ PQ\ {\isasymand}\ exprChannel\ ch\ {\isacharparenleft}sE\ secret{\isacharparenright}{\isachardoublequoteclose}\isanewline
\isadelimproof
\endisadelimproof
\isatagproof
\isacommand{using}\isamarkupfalse%
\ assms\ \ \isanewline
\isacommand{by}\isamarkupfalse%
\ {\isacharparenleft}metis\ correctCompositionKS{\isacharunderscore}exprChannel{\isacharunderscore}s{\isacharunderscore}P{\isacharparenright}%
\endisatagproof
{\isafoldproof}%
\isadelimproof
\isanewline
\endisadelimproof
\isanewline
\isacommand{lemma}\isamarkupfalse%
\ correctCompositionKS{\isacharunderscore}exprChannel{\isacharunderscore}s{\isacharunderscore}Q{\isacharcolon}\isanewline
\isakeyword{assumes}\ {\isachardoublequoteopen}subcomponents\ PQ\ {\isacharequal}\ {\isacharbraceleft}P{\isacharcomma}Q{\isacharbraceright}{\isachardoublequoteclose}\ \isanewline
\ \ \ \ \ \ \ \isakeyword{and}\ {\isachardoublequoteopen}correctCompositionKS\ PQ{\isachardoublequoteclose}\isanewline
\ \ \ \ \ \ \ \isakeyword{and}\ {\isachardoublequoteopen}sKS\ secret\ {\isasymnotin}\ LocalSecrets\ PQ{\isachardoublequoteclose}\isanewline
\ \ \ \ \ \ \ \isakeyword{and}\ {\isachardoublequoteopen}ch\ {\isasymin}\ ins\ Q{\isachardoublequoteclose}\isanewline
\ \ \ \ \ \ \ \isakeyword{and}\ h{\isadigit{1}}{\isacharcolon}{\isachardoublequoteopen}exprChannel\ ch\ {\isacharparenleft}sE\ secret{\isacharparenright}{\isachardoublequoteclose}\isanewline
\ \ \ \ \ \ \ \isakeyword{and}\ {\isachardoublequoteopen}sKS\ secret\ {\isasymnotin}\ specKeysSecrets\ PQ{\isachardoublequoteclose}\isanewline
\ \ \ \ \ \ \ \isakeyword{and}\ {\isachardoublequoteopen}correctCompositionIn\ PQ{\isachardoublequoteclose}\isanewline
\isakeyword{shows}\ \ \ \ {\isachardoublequoteopen}ch\ {\isasymin}\ ins\ PQ\ {\isasymand}\ exprChannel\ ch\ {\isacharparenleft}sE\ secret{\isacharparenright}{\isachardoublequoteclose}\isanewline
\isadelimproof
\endisadelimproof
\isatagproof
\isacommand{proof}\isamarkupfalse%
\ {\isacharminus}\ \isanewline
\ \ \isacommand{from}\isamarkupfalse%
\ assms\ \isacommand{have}\isamarkupfalse%
\ {\isachardoublequoteopen}ch\ {\isasymnotin}\ loc\ PQ{\isachardoublequoteclose}\ \isanewline
\ \ \ \ \isacommand{by}\isamarkupfalse%
\ {\isacharparenleft}simp\ add{\isacharcolon}\ LocalSecretsComposition{\isacharunderscore}neg{\isacharunderscore}loc{\isacharunderscore}s{\isacharparenright}\isanewline
\ \ \isacommand{from}\isamarkupfalse%
\ this\ \isakeyword{and}\ assms\ \isacommand{have}\isamarkupfalse%
\ {\isachardoublequoteopen}ch\ {\isasymin}\ ins\ PQ{\isachardoublequoteclose}\ \isanewline
\ \ \ \ \isacommand{by}\isamarkupfalse%
\ {\isacharparenleft}simp\ add{\isacharcolon}\ correctCompositionIn{\isacharunderscore}def{\isacharparenright}\ \isanewline
\ \ \isacommand{from}\isamarkupfalse%
\ this\ \isakeyword{and}\ h{\isadigit{1}}\ \isacommand{show}\isamarkupfalse%
\ {\isacharquery}thesis\ \isacommand{by}\isamarkupfalse%
\ simp\isanewline
\isacommand{qed}\isamarkupfalse%
\endisatagproof
{\isafoldproof}%
\isadelimproof
\isanewline
\endisadelimproof
\isanewline
\isacommand{lemma}\isamarkupfalse%
\ correctCompositionKS{\isacharunderscore}exprChannel{\isacharunderscore}s{\isacharunderscore}Qex{\isacharcolon}\isanewline
\isakeyword{assumes}\ {\isachardoublequoteopen}subcomponents\ PQ\ {\isacharequal}\ {\isacharbraceleft}P{\isacharcomma}Q{\isacharbraceright}{\isachardoublequoteclose}\ \isanewline
\ \ \ \ \ \ \ \isakeyword{and}\ {\isachardoublequoteopen}correctCompositionKS\ PQ{\isachardoublequoteclose}\isanewline
\ \ \ \ \ \ \ \isakeyword{and}\ {\isachardoublequoteopen}sKS\ secret\ {\isasymnotin}\ LocalSecrets\ PQ{\isachardoublequoteclose}\isanewline
\ \ \ \ \ \ \ \isakeyword{and}\ {\isachardoublequoteopen}ch\ {\isasymin}\ ins\ Q{\isachardoublequoteclose}\isanewline
\ \ \ \ \ \ \ \isakeyword{and}\ {\isachardoublequoteopen}exprChannel\ ch\ {\isacharparenleft}sE\ secret{\isacharparenright}{\isachardoublequoteclose}\isanewline
\ \ \ \ \ \ \ \isakeyword{and}\ {\isachardoublequoteopen}sKS\ secret\ {\isasymnotin}\ specKeysSecrets\ PQ{\isachardoublequoteclose}\isanewline
\ \ \ \ \ \ \ \isakeyword{and}\ {\isachardoublequoteopen}correctCompositionIn\ PQ{\isachardoublequoteclose}\isanewline
\isakeyword{shows}\ \ \ \ {\isachardoublequoteopen}{\isasymexists}ch{\isachardot}\ ch\ {\isasymin}\ ins\ PQ\ {\isasymand}\ exprChannel\ ch\ {\isacharparenleft}sE\ secret{\isacharparenright}{\isachardoublequoteclose}\isanewline
\isadelimproof
\endisadelimproof
\isatagproof
\isacommand{using}\isamarkupfalse%
\ assms\isanewline
\isacommand{by}\isamarkupfalse%
\ {\isacharparenleft}metis\ correctCompositionKS{\isacharunderscore}exprChannel{\isacharunderscore}s{\isacharunderscore}Q{\isacharparenright}%
\endisatagproof
{\isafoldproof}%
\isadelimproof
\isanewline
\endisadelimproof
\isadelimtheory
\isanewline
\endisadelimtheory
\isatagtheory
\isacommand{end}\isamarkupfalse%
\endisatagtheory
{\isafoldtheory}%
\isadelimtheory
\endisadelimtheory
\end{isabellebody}%

%
\begin{isabellebody}%
\def\isabellecontext{KnowledgeKeysSecrets}%
\isamarkupheader{Knowledge of Keys and Secrets%
}
\isamarkuptrue%
\isadelimtheory
\endisadelimtheory
\isatagtheory
\isacommand{theory}\isamarkupfalse%
\ KnowledgeKeysSecrets\isanewline
\isakeyword{imports}\ CompLocalSecrets\ \isanewline
\isakeyword{begin}%
\endisatagtheory
{\isafoldtheory}%
\isadelimtheory
\endisadelimtheory
~\\
An component $A$ knows a secret $m$ (or some secret expression $m$)  that does not belong to its local secrets , if
\begin{itemize}
	\item %
	$A$  may eventually get the secret $m$,
	\item 
	$m$ belongs to the set $LS_A$ of its local secrets, 
	\item %
	$A$ knows some list of expressions $m_2$ which is an concatenations of $m$ and some list of expressions $m_1$,
	\item %
	$m$ is a concatenation of some lists of secrets $m_1$ and $m_2$, and $A$ knows both these secrets,
	\item %
	$A$ knows some secret key $k^{-1}$ and the result of the encryption of the $m$ with the corresponding public key,
	\item %
	$A$ knows some public key $k$ and the result of the signature creation of the $m$ with the corresponding private key,%
	\item %
	$m$ is an encryption of some secret $m_1$ with a public key $k$, and $A$ knows both $m_1$ and $k$,
	\item %
	$m$ is the result of the signature creation of the $m_1$ with the key $k$, and $A$ knows both $m_1$ and $k$.
\end{itemize}
\isacommand{primrec}\isamarkupfalse%
\isanewline
\ \ know\ {\isacharcolon}{\isacharcolon}\ {\isachardoublequoteopen}specID\ {\isasymRightarrow}\ KS\ {\isasymRightarrow}\ bool{\isachardoublequoteclose}\isanewline
\isakeyword{where}\ \isanewline
\ {\isachardoublequoteopen}know\ A\ {\isacharparenleft}kKS\ m{\isacharparenright}\ {\isacharequal}\ \isanewline
\ \ {\isacharparenleft}{\isacharparenleft}ine\ A\ {\isacharparenleft}kE\ m{\isacharparenright}{\isacharparenright}\ {\isasymor}\ {\isacharparenleft}{\isacharparenleft}kKS\ m{\isacharparenright}\ {\isasymin}\ {\isacharparenleft}LocalSecrets\ A{\isacharparenright}{\isacharparenright}{\isacharparenright}{\isachardoublequoteclose}\ {\isacharbar}\ \isanewline
\ {\isachardoublequoteopen}know\ A\ {\isacharparenleft}sKS\ m{\isacharparenright}\ {\isacharequal}\ \isanewline
\ \ {\isacharparenleft}{\isacharparenleft}ine\ A\ {\isacharparenleft}sE\ m{\isacharparenright}{\isacharparenright}\ {\isasymor}\ {\isacharparenleft}{\isacharparenleft}sKS\ m{\isacharparenright}\ {\isasymin}\ {\isacharparenleft}LocalSecrets\ A{\isacharparenright}{\isacharparenright}{\isacharparenright}{\isachardoublequoteclose}\isanewline
\isanewline
\isacommand{axiomatization}\isamarkupfalse%
\isanewline
\ \ knows\ {\isacharcolon}{\isacharcolon}\ {\isachardoublequoteopen}specID\ {\isasymRightarrow}\ Expression\ list\ {\isasymRightarrow}\ bool{\isachardoublequoteclose}\isanewline
\isakeyword{where}\isanewline
knows{\isacharunderscore}emptyexpression{\isacharcolon}\isanewline
\ \ {\isachardoublequoteopen}knows\ C\ {\isacharbrackleft}{\isacharbrackright}\ {\isacharequal}\ True{\isachardoublequoteclose}\ \isakeyword{and}\isanewline
know{\isadigit{1}}k{\isacharcolon}\ \isanewline
\ \ {\isachardoublequoteopen}knows\ C\ {\isacharbrackleft}KS{\isadigit{2}}Expression\ {\isacharparenleft}kKS\ m{\isadigit{1}}{\isacharparenright}{\isacharbrackright}\ {\isacharequal}\ know\ C\ {\isacharparenleft}kKS\ m{\isadigit{1}}{\isacharparenright}{\isachardoublequoteclose}\ \isakeyword{and}\isanewline
know{\isadigit{1}}s{\isacharcolon}\isanewline
\ \ {\isachardoublequoteopen}knows\ C\ {\isacharbrackleft}KS{\isadigit{2}}Expression\ {\isacharparenleft}sKS\ m{\isadigit{2}}{\isacharparenright}{\isacharbrackright}\ {\isacharequal}\ know\ C\ {\isacharparenleft}sKS\ m{\isadigit{2}}{\isacharparenright}{\isachardoublequoteclose}\ \isakeyword{and}\isanewline
knows{\isadigit{2}}a{\isacharcolon}\ \isanewline
\ \ {\isachardoublequoteopen}knows\ A\ {\isacharparenleft}e{\isadigit{1}}\ {\isacharat}\ e{\isacharparenright}\ {\isasymlongrightarrow}\ knows\ A\ e{\isachardoublequoteclose}\ \isakeyword{and}\isanewline
knows{\isadigit{2}}b{\isacharcolon}\ \isanewline
\ \ {\isachardoublequoteopen}knows\ A\ {\isacharparenleft}e\ {\isacharat}\ e{\isadigit{1}}{\isacharparenright}\ {\isasymlongrightarrow}\ knows\ A\ e{\isachardoublequoteclose}\ \isakeyword{and}\isanewline
knows{\isadigit{3}}{\isacharcolon}\ \isanewline
\ \ {\isachardoublequoteopen}{\isacharparenleft}knows\ A\ e{\isadigit{1}}{\isacharparenright}\ {\isasymand}\ {\isacharparenleft}knows\ A\ e{\isadigit{2}}{\isacharparenright}\ {\isasymlongrightarrow}\ knows\ A\ {\isacharparenleft}e{\isadigit{1}}\ {\isacharat}\ e{\isadigit{2}}{\isacharparenright}{\isachardoublequoteclose}\ \isakeyword{and}\isanewline
knows{\isadigit{4}}{\isacharcolon}\ \isanewline
\ \ {\isachardoublequoteopen}{\isacharparenleft}IncrDecrKeys\ k{\isadigit{1}}\ k{\isadigit{2}}{\isacharparenright}\ {\isasymand}\ {\isacharparenleft}know\ A\ {\isacharparenleft}kKS\ k{\isadigit{2}}{\isacharparenright}{\isacharparenright}\ {\isasymand}\ {\isacharparenleft}knows\ A\ {\isacharparenleft}Enc\ k{\isadigit{1}}\ e{\isacharparenright}{\isacharparenright}\isanewline
\ \ \ {\isasymlongrightarrow}\ knows\ A\ e{\isachardoublequoteclose}\ \isanewline
\isakeyword{and}\isanewline
knows{\isadigit{5}}{\isacharcolon}\ \isanewline
\ \ {\isachardoublequoteopen}{\isacharparenleft}IncrDecrKeys\ k{\isadigit{1}}\ k{\isadigit{2}}{\isacharparenright}\ {\isasymand}\ {\isacharparenleft}know\ A\ {\isacharparenleft}kKS\ k{\isadigit{1}}{\isacharparenright}{\isacharparenright}\ {\isasymand}\ {\isacharparenleft}knows\ A\ {\isacharparenleft}Sign\ k{\isadigit{2}}\ e{\isacharparenright}{\isacharparenright}\isanewline
\ \ \ {\isasymlongrightarrow}\ knows\ A\ e{\isachardoublequoteclose}\isanewline
\isakeyword{and}\isanewline
knows{\isadigit{6}}{\isacharcolon}\ \isanewline
\ \ {\isachardoublequoteopen}{\isacharparenleft}know\ A\ {\isacharparenleft}kKS\ k{\isacharparenright}{\isacharparenright}\ {\isasymand}\ {\isacharparenleft}knows\ A\ e{\isadigit{1}}{\isacharparenright}\ {\isasymlongrightarrow}\ knows\ A\ {\isacharparenleft}Enc\ k\ e{\isadigit{1}}{\isacharparenright}{\isachardoublequoteclose}\isanewline
\isakeyword{and}\isanewline
knows{\isadigit{7}}{\isacharcolon}\ \isanewline
\ \ {\isachardoublequoteopen}{\isacharparenleft}know\ A\ {\isacharparenleft}kKS\ k{\isacharparenright}{\isacharparenright}\ {\isasymand}\ {\isacharparenleft}knows\ A\ e{\isadigit{1}}{\isacharparenright}\ {\isasymlongrightarrow}\ knows\ A\ {\isacharparenleft}Sign\ k\ e{\isadigit{1}}{\isacharparenright}{\isachardoublequoteclose}\isanewline
\isanewline
\isacommand{primrec}\isamarkupfalse%
\ \ eoutKnowCorrect\ {\isacharcolon}{\isacharcolon}\ {\isachardoublequoteopen}specID\ {\isasymRightarrow}\ KS\ {\isasymRightarrow}\ bool{\isachardoublequoteclose}\isanewline
\isakeyword{where}\isanewline
eout{\isacharunderscore}know{\isacharunderscore}k{\isacharcolon}\ \isanewline
\ \ {\isachardoublequoteopen}eoutKnowCorrect\ C\ {\isacharparenleft}kKS\ m{\isacharparenright}\ {\isacharequal}\ \isanewline
\ \ {\isacharparenleft}{\isacharparenleft}eout\ \ C\ {\isacharparenleft}kE\ m{\isacharparenright}{\isacharparenright}\ {\isasymlongleftrightarrow}\ {\isacharparenleft}m\ {\isasymin}\ {\isacharparenleft}specKeys\ C{\isacharparenright}\ {\isasymor}\ {\isacharparenleft}know\ C\ {\isacharparenleft}kKS\ m{\isacharparenright}{\isacharparenright}{\isacharparenright}\ {\isacharparenright}{\isachardoublequoteclose}\ \ {\isacharbar}\isanewline
eout{\isacharunderscore}know{\isacharunderscore}s{\isacharcolon}\ \isanewline
\ \ \ {\isachardoublequoteopen}eoutKnowCorrect\ C\ {\isacharparenleft}sKS\ m{\isacharparenright}\ {\isacharequal}\ \isanewline
\ \ {\isacharparenleft}{\isacharparenleft}eout\ \ C\ {\isacharparenleft}sE\ m{\isacharparenright}{\isacharparenright}\ {\isasymlongleftrightarrow}\ \ {\isacharparenleft}m\ {\isasymin}\ {\isacharparenleft}specSecrets\ C{\isacharparenright}\ {\isasymor}\ {\isacharparenleft}know\ C\ {\isacharparenleft}sKS\ m{\isacharparenright}{\isacharparenright}{\isacharparenright}\ {\isacharparenright}{\isachardoublequoteclose}\isanewline
\isanewline
\isacommand{definition}\isamarkupfalse%
\ eoutKnowsECorrect\ {\isacharcolon}{\isacharcolon}\ {\isachardoublequoteopen}specID\ {\isasymRightarrow}\ Expression\ {\isasymRightarrow}\ bool{\isachardoublequoteclose}\isanewline
\isakeyword{where}\isanewline
\ \ {\isachardoublequoteopen}eoutKnowsECorrect\ C\ e\ {\isasymequiv}\isanewline
\ \ \ {\isacharparenleft}{\isacharparenleft}eout\ \ C\ e{\isacharparenright}\ {\isasymlongleftrightarrow}\isanewline
\ \ \ {\isacharparenleft}{\isacharparenleft}{\isasymexists}\ k{\isachardot}\ e\ {\isacharequal}\ {\isacharparenleft}kE\ k{\isacharparenright}\ {\isasymand}\ {\isacharparenleft}k\ {\isasymin}\ specKeys\ C{\isacharparenright}{\isacharparenright}\ {\isasymor}\ \isanewline
\ \ \ \ {\isacharparenleft}{\isasymexists}\ s{\isachardot}\ e\ {\isacharequal}\ {\isacharparenleft}sE\ s{\isacharparenright}\ {\isasymand}\ {\isacharparenleft}s\ {\isasymin}\ specSecrets\ C{\isacharparenright}{\isacharparenright}\ {\isasymor}\isanewline
\ \ \ \ {\isacharparenleft}knows\ C\ {\isacharbrackleft}e{\isacharbrackright}{\isacharparenright}{\isacharparenright}{\isacharparenright}{\isachardoublequoteclose}\isanewline
\isanewline
\isacommand{lemma}\isamarkupfalse%
\ eoutKnowCorrect{\isacharunderscore}L{\isadigit{1}}k{\isacharcolon}\isanewline
\isakeyword{assumes}\ {\isachardoublequoteopen}eoutKnowCorrect\ C\ {\isacharparenleft}kKS\ m{\isacharparenright}{\isachardoublequoteclose}\ \ \isanewline
\ \ \ \ \ \ \ \isakeyword{and}\ {\isachardoublequoteopen}eout\ \ C\ {\isacharparenleft}kE\ m{\isacharparenright}{\isachardoublequoteclose}\isanewline
\isakeyword{shows}\ \ \ \ {\isachardoublequoteopen}m\ {\isasymin}\ {\isacharparenleft}specKeys\ C{\isacharparenright}\ {\isasymor}\ {\isacharparenleft}know\ C\ {\isacharparenleft}kKS\ m{\isacharparenright}{\isacharparenright}{\isachardoublequoteclose}\ \isanewline
\isadelimproof
\endisadelimproof
\isatagproof
\isacommand{using}\isamarkupfalse%
\ assms\ \isacommand{by}\isamarkupfalse%
\ {\isacharparenleft}metis\ eout{\isacharunderscore}know{\isacharunderscore}k{\isacharparenright}%
\endisatagproof
{\isafoldproof}%
\isadelimproof
\isanewline
\endisadelimproof
\isanewline
\isacommand{lemma}\isamarkupfalse%
\ eoutKnowCorrect{\isacharunderscore}L{\isadigit{1}}s{\isacharcolon}\isanewline
\isakeyword{assumes}\ {\isachardoublequoteopen}eoutKnowCorrect\ C\ {\isacharparenleft}sKS\ m{\isacharparenright}{\isachardoublequoteclose}\ \isanewline
\ \ \ \ \ \ \ \isakeyword{and}\ {\isachardoublequoteopen}eout\ \ C\ {\isacharparenleft}sE\ m{\isacharparenright}{\isachardoublequoteclose}\isanewline
\isakeyword{shows}\ \ \ \ {\isachardoublequoteopen}m\ {\isasymin}\ {\isacharparenleft}specSecrets\ C{\isacharparenright}\ {\isasymor}\ {\isacharparenleft}know\ C\ {\isacharparenleft}sKS\ m{\isacharparenright}{\isacharparenright}{\isachardoublequoteclose}\ \isanewline
\isadelimproof
\endisadelimproof
\isatagproof
\isacommand{using}\isamarkupfalse%
\ assms\ \isacommand{by}\isamarkupfalse%
\ {\isacharparenleft}metis\ eout{\isacharunderscore}know{\isacharunderscore}s{\isacharparenright}%
\endisatagproof
{\isafoldproof}%
\isadelimproof
\isanewline
\endisadelimproof
\isanewline
\isacommand{lemma}\isamarkupfalse%
\ eoutKnowsECorrect{\isacharunderscore}L{\isadigit{1}}{\isacharcolon}\isanewline
\isakeyword{assumes}\ {\isachardoublequoteopen}eoutKnowsECorrect\ C\ e{\isachardoublequoteclose}\ \isanewline
\ \ \ \ \ \ \ \isakeyword{and}\ {\isachardoublequoteopen}eout\ \ C\ e{\isachardoublequoteclose}\isanewline
\isakeyword{shows}\ {\isachardoublequoteopen}{\isacharparenleft}{\isasymexists}\ k{\isachardot}\ e\ {\isacharequal}\ {\isacharparenleft}kE\ k{\isacharparenright}\ {\isasymand}\ {\isacharparenleft}k\ {\isasymin}\ specKeys\ C{\isacharparenright}{\isacharparenright}\ {\isasymor}\ \isanewline
\ \ \ \ \ \ \ \ \ \ \ \ {\isacharparenleft}{\isasymexists}\ s{\isachardot}\ e\ {\isacharequal}\ {\isacharparenleft}sE\ s{\isacharparenright}\ {\isasymand}\ {\isacharparenleft}s\ {\isasymin}\ specSecrets\ C{\isacharparenright}{\isacharparenright}\ {\isasymor}\isanewline
\ \ \ \ \ \ \ \ \ \ \ \ {\isacharparenleft}knows\ C\ {\isacharbrackleft}e{\isacharbrackright}{\isacharparenright}{\isachardoublequoteclose}\ \isanewline
\isadelimproof
\endisadelimproof
\isatagproof
\isacommand{using}\isamarkupfalse%
\ assms\ \isacommand{by}\isamarkupfalse%
\ {\isacharparenleft}metis\ eoutKnowsECorrect{\isacharunderscore}def{\isacharparenright}%
\endisatagproof
{\isafoldproof}%
\isadelimproof
\isanewline
\endisadelimproof
\ \isanewline
\isacommand{lemma}\isamarkupfalse%
\ know{\isadigit{2}}knows{\isacharunderscore}k{\isacharcolon}\ \isanewline
\isakeyword{assumes}\ {\isachardoublequoteopen}know\ A\ {\isacharparenleft}kKS\ m{\isacharparenright}{\isachardoublequoteclose}\isanewline
\isakeyword{shows}\ \ \ \ {\isachardoublequoteopen}knows\ A\ {\isacharbrackleft}kE\ m{\isacharbrackright}{\isachardoublequoteclose}\ \isanewline
\isadelimproof
\endisadelimproof
\isatagproof
\isacommand{using}\isamarkupfalse%
\ assms\isanewline
\isacommand{by}\isamarkupfalse%
\ {\isacharparenleft}metis\ KS{\isadigit{2}}Expression{\isachardot}simps{\isacharparenleft}{\isadigit{1}}{\isacharparenright}\ know{\isadigit{1}}k{\isacharparenright}%
\endisatagproof
{\isafoldproof}%
\isadelimproof
\isanewline
\endisadelimproof
\isanewline
\isacommand{lemma}\isamarkupfalse%
\ knows{\isadigit{2}}know{\isacharunderscore}k{\isacharcolon}\ \isanewline
\isakeyword{assumes}\ {\isachardoublequoteopen}knows\ A\ {\isacharbrackleft}kE\ m{\isacharbrackright}{\isachardoublequoteclose}\ \isanewline
\isakeyword{shows}\ \ \ \ {\isachardoublequoteopen}know\ A\ {\isacharparenleft}kKS\ m{\isacharparenright}{\isachardoublequoteclose}\isanewline
\isadelimproof
\endisadelimproof
\isatagproof
\isacommand{using}\isamarkupfalse%
\ assms\isanewline
\isacommand{by}\isamarkupfalse%
\ {\isacharparenleft}metis\ KS{\isadigit{2}}Expression{\isachardot}simps{\isacharparenleft}{\isadigit{1}}{\isacharparenright}\ know{\isadigit{1}}k{\isacharparenright}%
\endisatagproof
{\isafoldproof}%
\isadelimproof
\isanewline
\endisadelimproof
\isanewline
\isacommand{lemma}\isamarkupfalse%
\ know{\isadigit{2}}knowsPQ{\isacharunderscore}k{\isacharcolon}\ \isanewline
\isakeyword{assumes}\ {\isachardoublequoteopen}know\ P\ {\isacharparenleft}kKS\ m{\isacharparenright}\ {\isasymor}\ know\ Q\ {\isacharparenleft}kKS\ m{\isacharparenright}{\isachardoublequoteclose}\isanewline
\isakeyword{shows}\ \ \ \ {\isachardoublequoteopen}knows\ P\ {\isacharbrackleft}kE\ m{\isacharbrackright}\ {\isasymor}\ knows\ Q\ {\isacharbrackleft}kE\ m{\isacharbrackright}{\isachardoublequoteclose}\ \isanewline
\isadelimproof
\endisadelimproof
\isatagproof
\isacommand{using}\isamarkupfalse%
\ assms\ \isacommand{by}\isamarkupfalse%
\ {\isacharparenleft}metis\ know{\isadigit{2}}knows{\isacharunderscore}k{\isacharparenright}%
\endisatagproof
{\isafoldproof}%
\isadelimproof
\isanewline
\endisadelimproof
\isanewline
\isacommand{lemma}\isamarkupfalse%
\ knows{\isadigit{2}}knowPQ{\isacharunderscore}k{\isacharcolon}\ \isanewline
\isakeyword{assumes}\ {\isachardoublequoteopen}knows\ P\ {\isacharbrackleft}kE\ m{\isacharbrackright}\ {\isasymor}\ knows\ Q\ {\isacharbrackleft}kE\ m{\isacharbrackright}{\isachardoublequoteclose}\isanewline
\isakeyword{shows}\ \ \ \ \ {\isachardoublequoteopen}know\ P\ {\isacharparenleft}kKS\ m{\isacharparenright}\ {\isasymor}\ know\ Q\ {\isacharparenleft}kKS\ m{\isacharparenright}{\isachardoublequoteclose}\isanewline
\isadelimproof
\endisadelimproof
\isatagproof
\isacommand{using}\isamarkupfalse%
\ assms\ \ \isacommand{by}\isamarkupfalse%
\ {\isacharparenleft}metis\ knows{\isadigit{2}}know{\isacharunderscore}k{\isacharparenright}%
\endisatagproof
{\isafoldproof}%
\isadelimproof
\isanewline
\endisadelimproof
\isanewline
\isacommand{lemma}\isamarkupfalse%
\ knows{\isadigit{1}}k{\isacharcolon}\ \isanewline
\ {\isachardoublequoteopen}know\ A\ {\isacharparenleft}kKS\ m{\isacharparenright}\ {\isacharequal}\ knows\ A\ {\isacharbrackleft}kE\ m{\isacharbrackright}{\isachardoublequoteclose}\isanewline
\isadelimproof
\endisadelimproof
\isatagproof
\isacommand{by}\isamarkupfalse%
\ {\isacharparenleft}metis\ know{\isadigit{2}}knows{\isacharunderscore}k\ knows{\isadigit{2}}know{\isacharunderscore}k{\isacharparenright}%
\endisatagproof
{\isafoldproof}%
\isadelimproof
\ \isanewline
\endisadelimproof
\isanewline
\isacommand{lemma}\isamarkupfalse%
\ know{\isadigit{2}}knows{\isacharunderscore}neg{\isacharunderscore}k{\isacharcolon}\ \isanewline
\isakeyword{assumes}\ \ {\isachardoublequoteopen}{\isasymnot}\ know\ A\ {\isacharparenleft}kKS\ m{\isacharparenright}{\isachardoublequoteclose}\isanewline
\isakeyword{shows}\ \ \ \ \ {\isachardoublequoteopen}{\isasymnot}\ knows\ A\ {\isacharbrackleft}kE\ m{\isacharbrackright}{\isachardoublequoteclose}\isanewline
\isadelimproof
\endisadelimproof
\isatagproof
\isacommand{using}\isamarkupfalse%
\ assms\ \isacommand{by}\isamarkupfalse%
\ {\isacharparenleft}metis\ knows{\isadigit{1}}k{\isacharparenright}%
\endisatagproof
{\isafoldproof}%
\isadelimproof
\ \isanewline
\endisadelimproof
\isanewline
\isacommand{lemma}\isamarkupfalse%
\ knows{\isadigit{2}}know{\isacharunderscore}neg{\isacharunderscore}k{\isacharcolon}\ \isanewline
\isakeyword{assumes}\ {\isachardoublequoteopen}{\isasymnot}\ knows\ A\ {\isacharbrackleft}kE\ m{\isacharbrackright}{\isachardoublequoteclose}\ \isanewline
\isakeyword{shows}\ \ \ \ {\isachardoublequoteopen}{\isasymnot}\ know\ A\ {\isacharparenleft}kKS\ m{\isacharparenright}{\isachardoublequoteclose}\isanewline
\isadelimproof
\endisadelimproof
\isatagproof
\isacommand{using}\isamarkupfalse%
\ assms\ \isacommand{by}\isamarkupfalse%
\ {\isacharparenleft}metis\ know{\isadigit{2}}knowsPQ{\isacharunderscore}k{\isacharparenright}%
\endisatagproof
{\isafoldproof}%
\isadelimproof
\isanewline
\endisadelimproof
\isanewline
\isacommand{lemma}\isamarkupfalse%
\ know{\isadigit{2}}knows{\isacharunderscore}s{\isacharcolon}\ \isanewline
\isakeyword{assumes}\ {\isachardoublequoteopen}know\ A\ {\isacharparenleft}sKS\ m{\isacharparenright}{\isachardoublequoteclose}\isanewline
\isakeyword{shows}\ \ \ \ {\isachardoublequoteopen}knows\ A\ {\isacharbrackleft}sE\ m{\isacharbrackright}{\isachardoublequoteclose}\isanewline
\isadelimproof
\endisadelimproof
\isatagproof
\isacommand{using}\isamarkupfalse%
\ assms\isanewline
\isacommand{by}\isamarkupfalse%
\ {\isacharparenleft}metis\ KS{\isadigit{2}}Expression{\isachardot}simps{\isacharparenleft}{\isadigit{2}}{\isacharparenright}\ know{\isadigit{1}}s{\isacharparenright}%
\endisatagproof
{\isafoldproof}%
\isadelimproof
\ \isanewline
\endisadelimproof
\isanewline
\isacommand{lemma}\isamarkupfalse%
\ knows{\isadigit{2}}know{\isacharunderscore}s{\isacharcolon}\ \isanewline
\isakeyword{assumes}\ {\isachardoublequoteopen}knows\ A\ {\isacharbrackleft}sE\ m{\isacharbrackright}{\isachardoublequoteclose}\ \isanewline
\isakeyword{shows}\ \ \ \ {\isachardoublequoteopen}know\ A\ {\isacharparenleft}sKS\ m{\isacharparenright}{\isachardoublequoteclose}\isanewline
\isadelimproof
\endisadelimproof
\isatagproof
\isacommand{using}\isamarkupfalse%
\ assms\isanewline
\isacommand{by}\isamarkupfalse%
\ {\isacharparenleft}metis\ KS{\isadigit{2}}Expression{\isachardot}simps{\isacharparenleft}{\isadigit{2}}{\isacharparenright}\ know{\isadigit{1}}s{\isacharparenright}%
\endisatagproof
{\isafoldproof}%
\isadelimproof
\ \isanewline
\endisadelimproof
\isanewline
\isacommand{lemma}\isamarkupfalse%
\ know{\isadigit{2}}knowsPQ{\isacharunderscore}s{\isacharcolon}\ \isanewline
\isakeyword{assumes}\ {\isachardoublequoteopen}know\ P\ {\isacharparenleft}sKS\ m{\isacharparenright}\ {\isasymor}\ know\ Q\ {\isacharparenleft}sKS\ m{\isacharparenright}{\isachardoublequoteclose}\isanewline
\isakeyword{shows}\ \ \ \ {\isachardoublequoteopen}knows\ P\ {\isacharbrackleft}sE\ m{\isacharbrackright}\ {\isasymor}\ knows\ Q\ {\isacharbrackleft}sE\ m{\isacharbrackright}{\isachardoublequoteclose}\ \isanewline
\isadelimproof
\endisadelimproof
\isatagproof
\isacommand{using}\isamarkupfalse%
\ assms\ \isacommand{by}\isamarkupfalse%
\ {\isacharparenleft}metis\ know{\isadigit{2}}knows{\isacharunderscore}s{\isacharparenright}%
\endisatagproof
{\isafoldproof}%
\isadelimproof
\isanewline
\endisadelimproof
\isanewline
\isacommand{lemma}\isamarkupfalse%
\ knows{\isadigit{2}}knowPQ{\isacharunderscore}s{\isacharcolon}\ \isanewline
\isakeyword{assumes}\ {\isachardoublequoteopen}knows\ P\ {\isacharbrackleft}sE\ m{\isacharbrackright}\ {\isasymor}\ knows\ Q\ {\isacharbrackleft}sE\ m{\isacharbrackright}{\isachardoublequoteclose}\isanewline
\isakeyword{shows}\ \ \ \ {\isachardoublequoteopen}know\ P\ {\isacharparenleft}sKS\ m{\isacharparenright}\ {\isasymor}\ know\ Q\ {\isacharparenleft}sKS\ m{\isacharparenright}{\isachardoublequoteclose}\isanewline
\isadelimproof
\endisadelimproof
\isatagproof
\isacommand{using}\isamarkupfalse%
\ assms\ \isacommand{by}\isamarkupfalse%
\ {\isacharparenleft}metis\ knows{\isadigit{2}}know{\isacharunderscore}s{\isacharparenright}%
\endisatagproof
{\isafoldproof}%
\isadelimproof
\isanewline
\endisadelimproof
\isanewline
\isacommand{lemma}\isamarkupfalse%
\ knows{\isadigit{1}}s{\isacharcolon}\isanewline
\ \ {\isachardoublequoteopen}know\ A\ {\isacharparenleft}sKS\ m{\isacharparenright}\ {\isacharequal}\ knows\ A\ {\isacharbrackleft}sE\ m{\isacharbrackright}{\isachardoublequoteclose}\isanewline
\isadelimproof
\endisadelimproof
\isatagproof
\isacommand{by}\isamarkupfalse%
\ {\isacharparenleft}metis\ know{\isadigit{2}}knows{\isacharunderscore}s\ knows{\isadigit{2}}know{\isacharunderscore}s{\isacharparenright}%
\endisatagproof
{\isafoldproof}%
\isadelimproof
\ \isanewline
\endisadelimproof
\isanewline
\isacommand{lemma}\isamarkupfalse%
\ know{\isadigit{2}}knows{\isacharunderscore}neg{\isacharunderscore}s{\isacharcolon}\ \isanewline
\isakeyword{assumes}\ {\isachardoublequoteopen}{\isasymnot}\ know\ A\ {\isacharparenleft}sKS\ m{\isacharparenright}{\isachardoublequoteclose}\isanewline
\isakeyword{shows}\ \ \ \ {\isachardoublequoteopen}{\isasymnot}\ knows\ A\ {\isacharbrackleft}sE\ m{\isacharbrackright}{\isachardoublequoteclose}\ \isanewline
\isadelimproof
\endisadelimproof
\isatagproof
\isacommand{using}\isamarkupfalse%
\ assms\ \isacommand{by}\isamarkupfalse%
\ {\isacharparenleft}metis\ knows{\isadigit{2}}know{\isacharunderscore}s{\isacharparenright}%
\endisatagproof
{\isafoldproof}%
\isadelimproof
\ \isanewline
\endisadelimproof
\isanewline
\isacommand{lemma}\isamarkupfalse%
\ knows{\isadigit{2}}know{\isacharunderscore}neg{\isacharunderscore}s{\isacharcolon}\ \isanewline
\isakeyword{assumes}\ {\isachardoublequoteopen}{\isasymnot}\ knows\ A\ {\isacharbrackleft}sE\ m{\isacharbrackright}{\isachardoublequoteclose}\ \isanewline
\isakeyword{shows}\ \ \ \ {\isachardoublequoteopen}{\isasymnot}\ know\ A\ {\isacharparenleft}sKS\ m{\isacharparenright}{\isachardoublequoteclose}\isanewline
\isadelimproof
\endisadelimproof
\isatagproof
\isacommand{using}\isamarkupfalse%
\ assms\ \isacommand{by}\isamarkupfalse%
\ {\isacharparenleft}metis\ \ know{\isadigit{2}}knows{\isacharunderscore}s{\isacharparenright}%
\endisatagproof
{\isafoldproof}%
\isadelimproof
\isanewline
\endisadelimproof
\isanewline
\isacommand{lemma}\isamarkupfalse%
\ knows{\isadigit{2}}{\isacharcolon}\ \isanewline
\isakeyword{assumes}\ {\isachardoublequoteopen}e{\isadigit{2}}\ {\isacharequal}\ e{\isadigit{1}}\ {\isacharat}\ e\ {\isasymor}\ e{\isadigit{2}}\ {\isacharequal}\ e\ {\isacharat}\ e{\isadigit{1}}{\isachardoublequoteclose}\ \isanewline
\ \ \ \ \ \ \ \isakeyword{and}\ {\isachardoublequoteopen}knows\ A\ e{\isadigit{2}}{\isachardoublequoteclose}\ \isanewline
\isakeyword{shows}\ \ \ \ {\isachardoublequoteopen}knows\ A\ e{\isachardoublequoteclose}\isanewline
\isadelimproof
\endisadelimproof
\isatagproof
\isacommand{using}\isamarkupfalse%
\ assms\ \isacommand{by}\isamarkupfalse%
\ {\isacharparenleft}metis\ knows{\isadigit{2}}a\ knows{\isadigit{2}}b{\isacharparenright}%
\endisatagproof
{\isafoldproof}%
\isadelimproof
\isanewline
\endisadelimproof
\ \isanewline
\isacommand{lemma}\isamarkupfalse%
\ correctCompositionInLoc{\isacharunderscore}exprChannel{\isacharcolon}\isanewline
\isakeyword{assumes}\ {\isachardoublequoteopen}subcomponents\ PQ\ {\isacharequal}\ {\isacharbraceleft}P{\isacharcomma}\ Q{\isacharbraceright}{\isachardoublequoteclose}\isanewline
\ \ \ \ \ \ \ \isakeyword{and}\ {\isachardoublequoteopen}correctCompositionIn\ PQ{\isachardoublequoteclose}\isanewline
\ \ \ \ \ \ \ \isakeyword{and}\ {\isachardoublequoteopen}ch\ {\isacharcolon}\ ins\ P{\isachardoublequoteclose}\isanewline
\ \ \ \ \ \ \ \isakeyword{and}\ {\isachardoublequoteopen}exprChannel\ ch\ m{\isachardoublequoteclose}\isanewline
\ \ \ \ \ \ \ \isakeyword{and}\ {\isachardoublequoteopen}{\isasymforall}\ x{\isachardot}\ x\ {\isasymin}\ ins\ PQ\ {\isasymlongrightarrow}\ {\isasymnot}\ exprChannel\ x\ m{\isachardoublequoteclose}\isanewline
\isakeyword{shows}\ \ \ \ {\isachardoublequoteopen}ch\ {\isacharcolon}\ loc\ PQ{\isachardoublequoteclose}\isanewline
\isadelimproof
\endisadelimproof
\isatagproof
\isacommand{using}\isamarkupfalse%
\ assms\ \isacommand{by}\isamarkupfalse%
\ {\isacharparenleft}simp\ add{\isacharcolon}\ correctCompositionIn{\isacharunderscore}def{\isacharcomma}\ auto{\isacharparenright}%
\endisatagproof
{\isafoldproof}%
\isadelimproof
\isanewline
\endisadelimproof
\isanewline
\isacommand{lemma}\isamarkupfalse%
\ eout{\isacharunderscore}know{\isacharunderscore}nonKS{\isacharunderscore}k{\isacharcolon}\ \isanewline
\isakeyword{assumes}\ {\isachardoublequoteopen}m\ {\isasymnotin}\ specKeys\ A{\isachardoublequoteclose}\isanewline
\ \ \ \ \ \ \ \ \isakeyword{and}\ {\isachardoublequoteopen}eout\ A\ {\isacharparenleft}kE\ m{\isacharparenright}{\isachardoublequoteclose}\isanewline
\ \ \ \ \ \ \ \ \isakeyword{and}\ {\isachardoublequoteopen}eoutKnowCorrect\ A\ {\isacharparenleft}kKS\ m{\isacharparenright}{\isachardoublequoteclose}\isanewline
\isakeyword{shows}\ \ \ \ \ {\isachardoublequoteopen}know\ A\ {\isacharparenleft}kKS\ m{\isacharparenright}{\isachardoublequoteclose}\isanewline
\isadelimproof
\endisadelimproof
\isatagproof
\isacommand{using}\isamarkupfalse%
\ assms\ \isacommand{by}\isamarkupfalse%
\ {\isacharparenleft}metis\ eoutKnowCorrect{\isacharunderscore}L{\isadigit{1}}k{\isacharparenright}%
\endisatagproof
{\isafoldproof}%
\isadelimproof
\isanewline
\endisadelimproof
\isanewline
\isacommand{lemma}\isamarkupfalse%
\ \ eout{\isacharunderscore}know{\isacharunderscore}nonKS{\isacharunderscore}s{\isacharcolon}\isanewline
\isakeyword{assumes}\ {\isachardoublequoteopen}m\ {\isasymnotin}\ specSecrets\ A{\isachardoublequoteclose}\isanewline
\ \ \ \ \ \ \ \ \isakeyword{and}\ {\isachardoublequoteopen}eout\ A\ {\isacharparenleft}sE\ m{\isacharparenright}{\isachardoublequoteclose}\isanewline
\ \ \ \ \ \ \ \ \isakeyword{and}\ {\isachardoublequoteopen}eoutKnowCorrect\ A\ {\isacharparenleft}sKS\ m{\isacharparenright}{\isachardoublequoteclose}\isanewline
\isakeyword{shows}\ \ \ \ {\isachardoublequoteopen}know\ A\ {\isacharparenleft}sKS\ m{\isacharparenright}{\isachardoublequoteclose}\isanewline
\isadelimproof
\endisadelimproof
\isatagproof
\isacommand{using}\isamarkupfalse%
\ assms\ \isacommand{by}\isamarkupfalse%
\ {\isacharparenleft}metis\ eoutKnowCorrect{\isacharunderscore}L{\isadigit{1}}s{\isacharparenright}%
\endisatagproof
{\isafoldproof}%
\isadelimproof
\isanewline
\endisadelimproof
\isanewline
\isacommand{lemma}\isamarkupfalse%
\ not{\isacharunderscore}know{\isacharunderscore}k{\isacharunderscore}not{\isacharunderscore}ine{\isacharcolon}\isanewline
\isakeyword{assumes}\ {\isachardoublequoteopen}{\isasymnot}\ know\ A\ {\isacharparenleft}kKS\ m{\isacharparenright}{\isachardoublequoteclose}\isanewline
\isakeyword{shows}\ \ \ \ {\isachardoublequoteopen}{\isasymnot}\ ine\ A\ {\isacharparenleft}kE\ m{\isacharparenright}{\isachardoublequoteclose}\isanewline
\isadelimproof
\endisadelimproof
\isatagproof
\isacommand{using}\isamarkupfalse%
\ assms\ \isacommand{by}\isamarkupfalse%
\ simp%
\endisatagproof
{\isafoldproof}%
\isadelimproof
\isanewline
\endisadelimproof
\isanewline
\isacommand{lemma}\isamarkupfalse%
\ not{\isacharunderscore}know{\isacharunderscore}s{\isacharunderscore}not{\isacharunderscore}ine{\isacharcolon}\isanewline
\isakeyword{assumes}\ {\isachardoublequoteopen}{\isasymnot}\ know\ A\ {\isacharparenleft}sKS\ m{\isacharparenright}{\isachardoublequoteclose}\isanewline
\isakeyword{shows}\ \ \ \ {\isachardoublequoteopen}{\isasymnot}\ ine\ A\ {\isacharparenleft}sE\ m{\isacharparenright}{\isachardoublequoteclose}\isanewline
\isadelimproof
\endisadelimproof
\isatagproof
\isacommand{using}\isamarkupfalse%
\ assms\ \isacommand{by}\isamarkupfalse%
\ simp%
\endisatagproof
{\isafoldproof}%
\isadelimproof
\isanewline
\endisadelimproof
\isanewline
\isacommand{lemma}\isamarkupfalse%
\ not{\isacharunderscore}know{\isacharunderscore}k{\isacharunderscore}not{\isacharunderscore}eout{\isacharcolon}\isanewline
\isakeyword{assumes}\ {\isachardoublequoteopen}m\ {\isasymnotin}\ specKeys\ A{\isachardoublequoteclose}\isanewline
\ \ \ \ \ \ \ \ \isakeyword{and}\ {\isachardoublequoteopen}{\isasymnot}\ know\ A\ {\isacharparenleft}kKS\ m{\isacharparenright}{\isachardoublequoteclose}\ \isanewline
\ \ \ \ \ \ \ \ \isakeyword{and}\ {\isachardoublequoteopen}eoutKnowCorrect\ A\ {\isacharparenleft}kKS\ m{\isacharparenright}{\isachardoublequoteclose}\isanewline
\isakeyword{shows}\ \ \ \ \ {\isachardoublequoteopen}{\isasymnot}\ eout\ A\ {\isacharparenleft}kE\ m{\isacharparenright}{\isachardoublequoteclose}\isanewline
\isadelimproof
\endisadelimproof
\isatagproof
\isacommand{using}\isamarkupfalse%
\ assms\ \isacommand{by}\isamarkupfalse%
\ {\isacharparenleft}metis\ eout{\isacharunderscore}know{\isacharunderscore}k{\isacharparenright}%
\endisatagproof
{\isafoldproof}%
\isadelimproof
\isanewline
\endisadelimproof
\isanewline
\isacommand{lemma}\isamarkupfalse%
\ not{\isacharunderscore}know{\isacharunderscore}s{\isacharunderscore}not{\isacharunderscore}eout{\isacharcolon}\isanewline
\isakeyword{assumes}\ {\isachardoublequoteopen}m\ {\isasymnotin}\ specSecrets\ A{\isachardoublequoteclose}\isanewline
\ \ \ \ \ \ \ \ \isakeyword{and}\ {\isachardoublequoteopen}{\isasymnot}\ know\ A\ {\isacharparenleft}sKS\ m{\isacharparenright}{\isachardoublequoteclose}\isanewline
\ \ \ \ \ \ \ \ \isakeyword{and}\ {\isachardoublequoteopen}eoutKnowCorrect\ A\ {\isacharparenleft}sKS\ m{\isacharparenright}{\isachardoublequoteclose}\isanewline
\isakeyword{shows}\ \ \ \ \ {\isachardoublequoteopen}{\isasymnot}\ eout\ A\ {\isacharparenleft}sE\ m{\isacharparenright}{\isachardoublequoteclose}\isanewline
\isadelimproof
\endisadelimproof
\isatagproof
\isacommand{using}\isamarkupfalse%
\ assms\ \isacommand{by}\isamarkupfalse%
\ {\isacharparenleft}metis\ eout{\isacharunderscore}know{\isacharunderscore}nonKS{\isacharunderscore}s{\isacharparenright}%
\endisatagproof
{\isafoldproof}%
\isadelimproof
\isanewline
\endisadelimproof
\isanewline
\isacommand{lemma}\isamarkupfalse%
\ adv{\isacharunderscore}not{\isacharunderscore}know{\isadigit{1}}{\isacharcolon}\isanewline
\isakeyword{assumes}\ {\isachardoublequoteopen}out\ P\ {\isasymsubseteq}\ ins\ A{\isachardoublequoteclose}\isanewline
\ \ \ \ \ \ \ \ \isakeyword{and}\ {\isachardoublequoteopen}{\isasymnot}\ know\ A\ {\isacharparenleft}kKS\ m{\isacharparenright}{\isachardoublequoteclose}\ \isanewline
\isakeyword{shows}\ \ \ \ {\isachardoublequoteopen}{\isasymnot}\ eout\ P\ {\isacharparenleft}kE\ m{\isacharparenright}{\isachardoublequoteclose}\ \isanewline
\isadelimproof
\endisadelimproof
\isatagproof
\isacommand{using}\isamarkupfalse%
\ assms\isanewline
\isacommand{by}\isamarkupfalse%
\ {\isacharparenleft}metis\ {\isacharparenleft}full{\isacharunderscore}types{\isacharparenright}\ eout{\isacharunderscore}def\ ine{\isacharunderscore}ins{\isacharunderscore}neg{\isadigit{1}}\ not{\isacharunderscore}know{\isacharunderscore}k{\isacharunderscore}not{\isacharunderscore}ine\ set{\isacharunderscore}rev{\isacharunderscore}mp{\isacharparenright}%
\endisatagproof
{\isafoldproof}%
\isadelimproof
\isanewline
\endisadelimproof
\isanewline
\isacommand{lemma}\isamarkupfalse%
\ \ adv{\isacharunderscore}not{\isacharunderscore}know{\isadigit{2}}{\isacharcolon}\isanewline
\isakeyword{assumes}\ {\isachardoublequoteopen}out\ P\ {\isasymsubseteq}\ ins\ A{\isachardoublequoteclose}\isanewline
\ \ \ \ \ \ \ \isakeyword{and}\ {\isachardoublequoteopen}{\isasymnot}\ know\ A\ {\isacharparenleft}sKS\ m{\isacharparenright}{\isachardoublequoteclose}\isanewline
\isakeyword{shows}\ \ \ \ {\isachardoublequoteopen}{\isasymnot}\ eout\ P\ {\isacharparenleft}sE\ m{\isacharparenright}{\isachardoublequoteclose}\isanewline
\isadelimproof
\endisadelimproof
\isatagproof
\isacommand{using}\isamarkupfalse%
\ assms\isanewline
\isacommand{by}\isamarkupfalse%
\ {\isacharparenleft}metis\ {\isacharparenleft}full{\isacharunderscore}types{\isacharparenright}\ eout{\isacharunderscore}def\ ine{\isacharunderscore}ins{\isacharunderscore}neg{\isadigit{1}}\ not{\isacharunderscore}know{\isacharunderscore}s{\isacharunderscore}not{\isacharunderscore}ine\ set{\isacharunderscore}rev{\isacharunderscore}mp{\isacharparenright}%
\endisatagproof
{\isafoldproof}%
\isadelimproof
\isanewline
\endisadelimproof
\isanewline
\isacommand{lemma}\isamarkupfalse%
\ LocalSecrets{\isacharunderscore}L{\isadigit{1}}{\isacharcolon}\isanewline
\isakeyword{assumes}\ {\isachardoublequoteopen}{\isacharparenleft}kKS{\isacharparenright}\ key\ {\isasymin}\ LocalSecrets\ P{\isachardoublequoteclose}\ \ \isanewline
\ \ \ \ \ \ \ \isakeyword{and}\ {\isachardoublequoteopen}{\isacharparenleft}kKS\ key{\isacharparenright}\ {\isasymnotin}\ {\isasymUnion}{\isacharparenleft}LocalSecrets\ {\isacharbackquote}\ subcomponents\ P{\isacharparenright}{\isachardoublequoteclose}\isanewline
\isakeyword{shows}\ \ \ \ {\isachardoublequoteopen}kKS\ key\ {\isasymnotin}\ specKeysSecrets\ P{\isachardoublequoteclose}\isanewline
\isadelimproof
\endisadelimproof
\isatagproof
\isacommand{using}\isamarkupfalse%
\ assms\ \isacommand{by}\isamarkupfalse%
\ {\isacharparenleft}simp\ only{\isacharcolon}\ LocalSecretsDef{\isacharcomma}\ auto{\isacharparenright}%
\endisatagproof
{\isafoldproof}%
\isadelimproof
\isanewline
\endisadelimproof
\isanewline
\isacommand{lemma}\isamarkupfalse%
\ LocalSecrets{\isacharunderscore}L{\isadigit{2}}{\isacharcolon}\isanewline
\isakeyword{assumes}\ {\isachardoublequoteopen}kKS\ key\ {\isasymin}\ LocalSecrets\ P{\isachardoublequoteclose}\ \ \isanewline
\ \ \ \ \ \ \ \isakeyword{and}\ {\isachardoublequoteopen}kKS\ key\ {\isasymin}\ specKeysSecrets\ P{\isachardoublequoteclose}\isanewline
\isakeyword{shows}\ \ \ \ {\isachardoublequoteopen}kKS\ key\ {\isasymin}\ {\isasymUnion}{\isacharparenleft}LocalSecrets\ {\isacharbackquote}\ subcomponents\ P{\isacharparenright}{\isachardoublequoteclose}\isanewline
\isadelimproof
\endisadelimproof
\isatagproof
\isacommand{using}\isamarkupfalse%
\ assms\ \isacommand{by}\isamarkupfalse%
\ {\isacharparenleft}simp\ only{\isacharcolon}\ LocalSecretsDef{\isacharcomma}\ auto{\isacharparenright}%
\endisatagproof
{\isafoldproof}%
\isadelimproof
\isanewline
\endisadelimproof
\isanewline
\isacommand{lemma}\isamarkupfalse%
\ know{\isacharunderscore}composition{\isadigit{1}}{\isacharcolon}\isanewline
\isakeyword{assumes}\ notKSP{\isacharcolon}{\isachardoublequoteopen}m\ {\isasymnotin}\ specKeysSecrets\ P{\isachardoublequoteclose}\isanewline
\ \ \ \ \ \ \ \isakeyword{and}\ notKSQ{\isacharcolon}{\isachardoublequoteopen}m\ {\isasymnotin}\ specKeysSecrets\ Q{\isachardoublequoteclose}\isanewline
\ \ \ \ \ \ \ \isakeyword{and}\ {\isachardoublequoteopen}know\ P\ m{\isachardoublequoteclose}\isanewline
\ \ \ \ \ \ \ \isakeyword{and}\ subPQ{\isacharcolon}{\isachardoublequoteopen}subcomponents\ PQ\ {\isacharequal}\ {\isacharbraceleft}P{\isacharcomma}Q{\isacharbraceright}{\isachardoublequoteclose}\ \isanewline
\ \ \ \ \ \ \ \isakeyword{and}\ cCompI{\isacharcolon}{\isachardoublequoteopen}correctCompositionIn\ PQ{\isachardoublequoteclose}\isanewline
\ \ \ \ \ \ \ \isakeyword{and}\ cCompKS{\isacharcolon}{\isachardoublequoteopen}correctCompositionKS\ PQ{\isachardoublequoteclose}\isanewline
\isakeyword{shows}\ \ \ \ {\isachardoublequoteopen}know\ PQ\ m{\isachardoublequoteclose}\isanewline
\isadelimproof
\endisadelimproof
\isatagproof
\isacommand{proof}\isamarkupfalse%
\ {\isacharparenleft}cases\ m{\isacharparenright}\isanewline
\ \ \isacommand{fix}\isamarkupfalse%
\ key\isanewline
\ \ \isacommand{assume}\isamarkupfalse%
\ a{\isadigit{1}}{\isacharcolon}{\isachardoublequoteopen}m\ {\isacharequal}\ kKS\ key{\isachardoublequoteclose}\isanewline
\ \ \isacommand{show}\isamarkupfalse%
\ {\isacharquery}thesis\isanewline
\ \ \isacommand{proof}\isamarkupfalse%
\ {\isacharparenleft}cases\ \ {\isachardoublequoteopen}ine\ P\ {\isacharparenleft}kE\ key{\isacharparenright}{\isachardoublequoteclose}{\isacharparenright}\ \isanewline
\ \ \ \ \ \isacommand{assume}\isamarkupfalse%
\ a{\isadigit{1}}{\isadigit{1}}{\isacharcolon}{\isachardoublequoteopen}ine\ P\ {\isacharparenleft}kE\ key{\isacharparenright}{\isachardoublequoteclose}\ \isanewline
\ \ \ \ \ \isacommand{from}\isamarkupfalse%
\ this\ \isacommand{have}\isamarkupfalse%
\ a{\isadigit{1}}{\isadigit{1}}ext{\isacharcolon}{\isachardoublequoteopen}ine\ P\ {\isacharparenleft}kE\ key{\isacharparenright}\ {\isacharbar}\ ine\ Q\ {\isacharparenleft}kE\ key{\isacharparenright}{\isachardoublequoteclose}\ \isacommand{by}\isamarkupfalse%
\ simp\isanewline
\ \ \ \ \ \isacommand{from}\isamarkupfalse%
\ subPQ\ \isakeyword{and}\ cCompKS\ \isakeyword{and}\ notKSP\ \isakeyword{and}\ notKSQ\ \isacommand{have}\isamarkupfalse%
\ {\isachardoublequoteopen}m\ {\isasymnotin}\ specKeysSecrets\ PQ{\isachardoublequoteclose}\ \isanewline
\ \ \ \ \ \ \ \isacommand{by}\isamarkupfalse%
\ {\isacharparenleft}rule\ correctCompositionKS{\isacharunderscore}neg{\isadigit{1}}{\isacharparenright}\ \isanewline
\ \ \ \ \ \isacommand{from}\isamarkupfalse%
\ this\ \isakeyword{and}\ a{\isadigit{1}}\ \isacommand{have}\isamarkupfalse%
\ sg{\isadigit{1}}{\isacharcolon}{\isachardoublequoteopen}kKS\ key\ {\isasymnotin}\ specKeysSecrets\ PQ{\isachardoublequoteclose}\ \isacommand{by}\isamarkupfalse%
\ simp\isanewline
\ \ \ \ \ \isacommand{from}\isamarkupfalse%
\ a{\isadigit{1}}\ \isakeyword{and}\ a{\isadigit{1}}{\isadigit{1}}ext\ \isakeyword{and}\ cCompKS\ \ \isacommand{show}\isamarkupfalse%
\ {\isacharquery}thesis\isanewline
\ \ \ \ \ \isacommand{proof}\isamarkupfalse%
\ {\isacharparenleft}cases\ {\isachardoublequoteopen}loc\ PQ\ {\isacharequal}\ {\isacharbraceleft}{\isacharbraceright}{\isachardoublequoteclose}{\isacharparenright}\isanewline
\ \ \ \ \ \ \ \isacommand{assume}\isamarkupfalse%
\ a{\isadigit{1}}{\isadigit{1}}locE{\isacharcolon}{\isachardoublequoteopen}loc\ PQ\ {\isacharequal}\ {\isacharbraceleft}{\isacharbraceright}{\isachardoublequoteclose}\isanewline
\ \ \ \ \ \ \ \isacommand{from}\isamarkupfalse%
\ a{\isadigit{1}}{\isadigit{1}}ext\ \isakeyword{and}\ subPQ\ \isakeyword{and}\ cCompI\ \isakeyword{and}\ a{\isadigit{1}}{\isadigit{1}}locE\ \isacommand{have}\isamarkupfalse%
\ {\isachardoublequoteopen}ine\ PQ\ {\isacharparenleft}kE\ key{\isacharparenright}{\isachardoublequoteclose}\ \isanewline
\ \ \ \ \ \ \ \ \ \isacommand{by}\isamarkupfalse%
\ {\isacharparenleft}rule\ TBtheorem{\isadigit{4}}a{\isacharunderscore}empty{\isacharparenright}\ \isanewline
\ \ \ \ \ \ \ \isacommand{from}\isamarkupfalse%
\ this\ \isakeyword{and}\ a{\isadigit{1}}\ \isacommand{show}\isamarkupfalse%
\ {\isacharquery}thesis\ \isacommand{by}\isamarkupfalse%
\ auto\isanewline
\ \ \ \ \ \isacommand{next}\isamarkupfalse%
\ \isanewline
\ \ \ \ \ \ \ \isacommand{assume}\isamarkupfalse%
\ a{\isadigit{1}}{\isadigit{1}}locNE{\isacharcolon}{\isachardoublequoteopen}loc\ PQ\ {\isasymnoteq}\ {\isacharbraceleft}{\isacharbraceright}{\isachardoublequoteclose}\isanewline
\ \ \ \ \ \ \ \isacommand{from}\isamarkupfalse%
\ a{\isadigit{1}}\ \isakeyword{and}\ a{\isadigit{1}}{\isadigit{1}}\ \isakeyword{and}\ sg{\isadigit{1}}\ \isakeyword{and}\ assms\ \isacommand{show}\isamarkupfalse%
\ {\isacharquery}thesis\isanewline
\ \ \ \ \ \ \ \ \ \isacommand{apply}\isamarkupfalse%
\ {\isacharparenleft}simp\ add{\isacharcolon}\ ine{\isacharunderscore}def{\isacharcomma}\ auto{\isacharparenright}\isanewline
\ \ \ \ \ \ \ \ \ \isacommand{by}\isamarkupfalse%
\ {\isacharparenleft}simp\ add{\isacharcolon}\ correctCompositionKS{\isacharunderscore}exprChannel{\isacharunderscore}k{\isacharunderscore}Pex{\isacharparenright}\ \isanewline
\ \ \ \ \ \isacommand{qed}\isamarkupfalse%
\isanewline
\ \ \ \isacommand{next}\isamarkupfalse%
\isanewline
\ \ \ \ \ \isacommand{assume}\isamarkupfalse%
\ a{\isadigit{1}}{\isadigit{2}}{\isacharcolon}{\isachardoublequoteopen}{\isasymnot}\ ine\ P\ {\isacharparenleft}kE\ key{\isacharparenright}{\isachardoublequoteclose}\isanewline
\ \ \ \ \ \isacommand{from}\isamarkupfalse%
\ this\ \isakeyword{and}\ a{\isadigit{1}}\ \isakeyword{and}\ assms\ \isacommand{show}\isamarkupfalse%
\ {\isacharquery}thesis\isanewline
\ \ \ \ \ \ \ \isacommand{by}\isamarkupfalse%
\ {\isacharparenleft}auto{\isacharcomma}\ simp\ add{\isacharcolon}\ \ LocalSecretsComposition{\isadigit{1}}{\isacharparenright}\isanewline
\ \ \ \isacommand{qed}\isamarkupfalse%
\isanewline
\ \isacommand{next}\isamarkupfalse%
\isanewline
\ \ \isacommand{fix}\isamarkupfalse%
\ secret\isanewline
\ \ \isacommand{assume}\isamarkupfalse%
\ a{\isadigit{2}}{\isacharcolon}{\isachardoublequoteopen}m\ {\isacharequal}\ sKS\ secret{\isachardoublequoteclose}\isanewline
\ \ \isacommand{show}\isamarkupfalse%
\ {\isacharquery}thesis\isanewline
\ \ \isacommand{proof}\isamarkupfalse%
\ {\isacharparenleft}cases\ \ {\isachardoublequoteopen}ine\ P\ {\isacharparenleft}sE\ secret{\isacharparenright}{\isachardoublequoteclose}{\isacharparenright}\ \isanewline
\ \ \ \ \ \isacommand{assume}\isamarkupfalse%
\ a{\isadigit{2}}{\isadigit{1}}{\isacharcolon}{\isachardoublequoteopen}ine\ P\ {\isacharparenleft}sE\ secret{\isacharparenright}{\isachardoublequoteclose}\ \isanewline
\ \ \ \ \ \isacommand{from}\isamarkupfalse%
\ this\ \isacommand{have}\isamarkupfalse%
\ a{\isadigit{2}}{\isadigit{1}}ext{\isacharcolon}{\isachardoublequoteopen}ine\ P\ {\isacharparenleft}sE\ secret{\isacharparenright}\ {\isacharbar}\ ine\ Q\ {\isacharparenleft}sE\ secret{\isacharparenright}{\isachardoublequoteclose}\ \isacommand{by}\isamarkupfalse%
\ simp\isanewline
\ \ \ \ \ \isacommand{from}\isamarkupfalse%
\ subPQ\ \isakeyword{and}\ cCompKS\ \isakeyword{and}\ notKSP\ \isakeyword{and}\ notKSQ\ \isacommand{have}\isamarkupfalse%
\ {\isachardoublequoteopen}m\ {\isasymnotin}\ specKeysSecrets\ PQ{\isachardoublequoteclose}\ \isanewline
\ \ \ \ \ \ \ \isacommand{by}\isamarkupfalse%
\ {\isacharparenleft}rule\ correctCompositionKS{\isacharunderscore}neg{\isadigit{1}}{\isacharparenright}\ \isanewline
\ \ \ \ \ \isacommand{from}\isamarkupfalse%
\ this\ \isakeyword{and}\ a{\isadigit{2}}\ \isacommand{have}\isamarkupfalse%
\ sg{\isadigit{2}}{\isacharcolon}{\isachardoublequoteopen}sKS\ secret\ {\isasymnotin}\ specKeysSecrets\ PQ{\isachardoublequoteclose}\ \isacommand{by}\isamarkupfalse%
\ simp\isanewline
\ \ \ \ \ \isacommand{from}\isamarkupfalse%
\ a{\isadigit{2}}\ \isakeyword{and}\ a{\isadigit{2}}{\isadigit{1}}ext\ \isakeyword{and}\ cCompKS\ \ \isacommand{show}\isamarkupfalse%
\ {\isacharquery}thesis\isanewline
\ \ \ \ \ \isacommand{proof}\isamarkupfalse%
\ {\isacharparenleft}cases\ {\isachardoublequoteopen}loc\ PQ\ {\isacharequal}\ {\isacharbraceleft}{\isacharbraceright}{\isachardoublequoteclose}{\isacharparenright}\isanewline
\ \ \ \ \ \ \ \isacommand{assume}\isamarkupfalse%
\ a{\isadigit{2}}{\isadigit{1}}locE{\isacharcolon}{\isachardoublequoteopen}loc\ PQ\ {\isacharequal}\ {\isacharbraceleft}{\isacharbraceright}{\isachardoublequoteclose}\isanewline
\ \ \ \ \ \ \ \isacommand{from}\isamarkupfalse%
\ a{\isadigit{2}}{\isadigit{1}}ext\ \isakeyword{and}\ subPQ\ \isakeyword{and}\ cCompI\ \isakeyword{and}\ a{\isadigit{2}}{\isadigit{1}}locE\ \isacommand{have}\isamarkupfalse%
\ {\isachardoublequoteopen}ine\ PQ\ {\isacharparenleft}sE\ secret{\isacharparenright}{\isachardoublequoteclose}\ \isanewline
\ \ \ \ \ \ \ \ \ \isacommand{by}\isamarkupfalse%
\ {\isacharparenleft}rule\ TBtheorem{\isadigit{4}}a{\isacharunderscore}empty{\isacharparenright}\ \isanewline
\ \ \ \ \ \ \ \isacommand{from}\isamarkupfalse%
\ this\ \isakeyword{and}\ a{\isadigit{2}}\ \isacommand{show}\isamarkupfalse%
\ {\isacharquery}thesis\ \isacommand{by}\isamarkupfalse%
\ auto\isanewline
\ \ \ \ \ \isacommand{next}\isamarkupfalse%
\ \isanewline
\ \ \ \ \ \ \ \isacommand{assume}\isamarkupfalse%
\ a{\isadigit{2}}{\isadigit{1}}locNE{\isacharcolon}{\isachardoublequoteopen}loc\ PQ\ {\isasymnoteq}\ {\isacharbraceleft}{\isacharbraceright}{\isachardoublequoteclose}\isanewline
\ \ \ \ \ \ \ \isacommand{from}\isamarkupfalse%
\ a{\isadigit{2}}\ \isakeyword{and}\ a{\isadigit{2}}{\isadigit{1}}\ \isakeyword{and}\ sg{\isadigit{2}}\ \isakeyword{and}\ assms\ \isacommand{show}\isamarkupfalse%
\ {\isacharquery}thesis\isanewline
\ \ \ \ \ \ \ \ \ \isacommand{apply}\isamarkupfalse%
\ {\isacharparenleft}simp\ add{\isacharcolon}\ ine{\isacharunderscore}def{\isacharcomma}\ auto{\isacharparenright}\isanewline
\ \ \ \ \ \ \ \ \ \isacommand{by}\isamarkupfalse%
\ {\isacharparenleft}simp\ add{\isacharcolon}\ correctCompositionKS{\isacharunderscore}exprChannel{\isacharunderscore}s{\isacharunderscore}Pex{\isacharparenright}\ \isanewline
\ \ \ \ \ \isacommand{qed}\isamarkupfalse%
\isanewline
\ \ \ \isacommand{next}\isamarkupfalse%
\isanewline
\ \ \ \ \ \isacommand{assume}\isamarkupfalse%
\ a{\isadigit{1}}{\isadigit{2}}{\isacharcolon}{\isachardoublequoteopen}{\isasymnot}\ ine\ P\ {\isacharparenleft}sE\ secret{\isacharparenright}{\isachardoublequoteclose}\isanewline
\ \ \ \ \ \isacommand{from}\isamarkupfalse%
\ this\ \isakeyword{and}\ a{\isadigit{2}}\ \isakeyword{and}\ assms\ \isacommand{show}\isamarkupfalse%
\ {\isacharquery}thesis\isanewline
\ \ \ \ \ \isacommand{by}\isamarkupfalse%
\ {\isacharparenleft}metis\ LocalSecretsComposition{\isadigit{1}}\ know{\isachardot}simps{\isacharparenleft}{\isadigit{2}}{\isacharparenright}{\isacharparenright}\isanewline
\ \ \ \isacommand{qed}\isamarkupfalse%
\isanewline
\isacommand{qed}\isamarkupfalse%
\endisatagproof
{\isafoldproof}%
\isadelimproof
\isanewline
\endisadelimproof
\isanewline
\isacommand{lemma}\isamarkupfalse%
\ know{\isacharunderscore}composition{\isadigit{2}}{\isacharcolon}\isanewline
\isakeyword{assumes}\ {\isachardoublequoteopen}m\ {\isasymnotin}\ specKeysSecrets\ P{\isachardoublequoteclose}\isanewline
\ \ \ \ \ \ \ \isakeyword{and}\ {\isachardoublequoteopen}m\ {\isasymnotin}\ specKeysSecrets\ Q{\isachardoublequoteclose}\isanewline
\ \ \ \ \ \ \ \isakeyword{and}\ {\isachardoublequoteopen}know\ Q\ m{\isachardoublequoteclose}\isanewline
\ \ \ \ \ \ \ \isakeyword{and}\ {\isachardoublequoteopen}subcomponents\ PQ\ {\isacharequal}\ {\isacharbraceleft}P{\isacharcomma}Q{\isacharbraceright}{\isachardoublequoteclose}\isanewline
\ \ \ \ \ \ \ \isakeyword{and}\ {\isachardoublequoteopen}correctCompositionIn\ PQ{\isachardoublequoteclose}\isanewline
\ \ \ \ \ \ \ \isakeyword{and}\ {\isachardoublequoteopen}correctCompositionKS\ PQ{\isachardoublequoteclose}\isanewline
\isakeyword{shows}\ \ \ \ {\isachardoublequoteopen}know\ PQ\ m{\isachardoublequoteclose}\isanewline
\isadelimproof
\endisadelimproof
\isatagproof
\isacommand{using}\isamarkupfalse%
\ assms\ \isacommand{by}\isamarkupfalse%
\ {\isacharparenleft}metis\ insert{\isacharunderscore}commute\ know{\isacharunderscore}composition{\isadigit{1}}{\isacharparenright}%
\endisatagproof
{\isafoldproof}%
\isadelimproof
\isanewline
\endisadelimproof
\isanewline
\isacommand{lemma}\isamarkupfalse%
\ know{\isacharunderscore}composition{\isacharcolon}\isanewline
\isakeyword{assumes}\ {\isachardoublequoteopen}m\ {\isasymnotin}\ specKeysSecrets\ P{\isachardoublequoteclose}\isanewline
\ \ \ \ \ \ \ \ \isakeyword{and}\ {\isachardoublequoteopen}m\ {\isasymnotin}\ specKeysSecrets\ Q{\isachardoublequoteclose}\isanewline
\ \ \ \ \ \ \ \ \isakeyword{and}\ {\isachardoublequoteopen}know\ P\ m\ {\isasymor}\ know\ Q\ m{\isachardoublequoteclose}\isanewline
\ \ \ \ \ \ \ \ \isakeyword{and}\ {\isachardoublequoteopen}subcomponents\ PQ\ {\isacharequal}\ {\isacharbraceleft}P{\isacharcomma}Q{\isacharbraceright}{\isachardoublequoteclose}\ \isanewline
\ \ \ \ \ \ \ \ \isakeyword{and}\ {\isachardoublequoteopen}correctCompositionIn\ PQ{\isachardoublequoteclose}\isanewline
\ \ \ \ \ \ \ \ \isakeyword{and}\ {\isachardoublequoteopen}correctCompositionKS\ PQ{\isachardoublequoteclose}\isanewline
\isakeyword{shows}\ \ \ \ {\isachardoublequoteopen}know\ PQ\ m{\isachardoublequoteclose}\isanewline
\isadelimproof
\endisadelimproof
\isatagproof
\isacommand{using}\isamarkupfalse%
\ assms\ \isacommand{by}\isamarkupfalse%
\ {\isacharparenleft}metis\ know{\isacharunderscore}composition{\isadigit{1}}\ know{\isacharunderscore}composition{\isadigit{2}}{\isacharparenright}%
\endisatagproof
{\isafoldproof}%
\isadelimproof
\isanewline
\endisadelimproof
\isanewline
\isacommand{theorem}\isamarkupfalse%
\ know{\isacharunderscore}composition{\isacharunderscore}neg{\isacharunderscore}ine{\isacharunderscore}k{\isacharcolon}\isanewline
\isakeyword{assumes}\ {\isachardoublequoteopen}{\isasymnot}\ know\ P\ {\isacharparenleft}kKS\ key{\isacharparenright}{\isachardoublequoteclose}\isanewline
\ \ \ \ \ \ \ \isakeyword{and}\ {\isachardoublequoteopen}{\isasymnot}\ know\ Q\ {\isacharparenleft}kKS\ key{\isacharparenright}{\isachardoublequoteclose}\isanewline
\ \ \ \ \ \ \ \isakeyword{and}\ {\isachardoublequoteopen}subcomponents\ PQ\ {\isacharequal}\ {\isacharbraceleft}P{\isacharcomma}Q{\isacharbraceright}{\isachardoublequoteclose}\isanewline
\ \ \ \ \ \ \ \isakeyword{and}\ {\isachardoublequoteopen}correctCompositionIn\ PQ{\isachardoublequoteclose}\isanewline
\isakeyword{shows}\ \ \ \ {\isachardoublequoteopen}{\isasymnot}\ {\isacharparenleft}ine\ PQ\ {\isacharparenleft}kE\ key{\isacharparenright}{\isacharparenright}{\isachardoublequoteclose}\isanewline
\isadelimproof
\endisadelimproof
\isatagproof
\isacommand{using}\isamarkupfalse%
\ assms\ \isacommand{by}\isamarkupfalse%
\ {\isacharparenleft}metis\ TBtheorem{\isadigit{3}}a\ not{\isacharunderscore}know{\isacharunderscore}k{\isacharunderscore}not{\isacharunderscore}ine{\isacharparenright}%
\endisatagproof
{\isafoldproof}%
\isadelimproof
\isanewline
\endisadelimproof
\isanewline
\isacommand{theorem}\isamarkupfalse%
\ know{\isacharunderscore}composition{\isacharunderscore}neg{\isacharunderscore}ine{\isacharunderscore}s{\isacharcolon}\isanewline
\isakeyword{assumes}\ {\isachardoublequoteopen}{\isasymnot}\ know\ P\ {\isacharparenleft}sKS\ secret{\isacharparenright}{\isachardoublequoteclose}\isanewline
\ \ \ \ \ \ \ \isakeyword{and}\ {\isachardoublequoteopen}{\isasymnot}\ know\ Q\ {\isacharparenleft}sKS\ secret{\isacharparenright}{\isachardoublequoteclose}\isanewline
\ \ \ \ \ \ \ \isakeyword{and}\ {\isachardoublequoteopen}subcomponents\ PQ\ {\isacharequal}\ {\isacharbraceleft}P{\isacharcomma}Q{\isacharbraceright}{\isachardoublequoteclose}\isanewline
\ \ \ \ \ \ \ \isakeyword{and}\ {\isachardoublequoteopen}correctCompositionIn\ PQ{\isachardoublequoteclose}\isanewline
\isakeyword{shows}\ \ \ \ {\isachardoublequoteopen}{\isasymnot}\ {\isacharparenleft}ine\ PQ\ {\isacharparenleft}sE\ secret{\isacharparenright}{\isacharparenright}{\isachardoublequoteclose}\isanewline
\isadelimproof
\endisadelimproof
\isatagproof
\isacommand{using}\isamarkupfalse%
\ assms\ \isacommand{by}\isamarkupfalse%
\ {\isacharparenleft}metis\ TBtheorem{\isadigit{3}}a\ not{\isacharunderscore}know{\isacharunderscore}s{\isacharunderscore}not{\isacharunderscore}ine{\isacharparenright}%
\endisatagproof
{\isafoldproof}%
\isadelimproof
\isanewline
\endisadelimproof
\isanewline
\isacommand{lemma}\isamarkupfalse%
\ know{\isacharunderscore}composition{\isacharunderscore}neg{\isadigit{1}}{\isacharcolon}\isanewline
\isakeyword{assumes}\ notknowP{\isacharcolon}{\isachardoublequoteopen}{\isasymnot}\ know\ P\ m{\isachardoublequoteclose}\isanewline
\ \ \ \ \ \ \ \isakeyword{and}\ notknowQ{\isacharcolon}{\isachardoublequoteopen}{\isasymnot}\ know\ Q\ m{\isachardoublequoteclose}\isanewline
\ \ \ \ \ \ \ \isakeyword{and}\ subPQ{\isacharcolon}{\isachardoublequoteopen}subcomponents\ PQ\ {\isacharequal}\ {\isacharbraceleft}P{\isacharcomma}Q{\isacharbraceright}{\isachardoublequoteclose}\isanewline
\ \ \ \ \ \ \ \isakeyword{and}\ cCompLoc{\isacharcolon}{\isachardoublequoteopen}correctCompositionLoc\ PQ{\isachardoublequoteclose}\isanewline
\ \ \ \ \ \ \ \isakeyword{and}\ cCompI{\isacharcolon}{\isachardoublequoteopen}correctCompositionIn\ PQ{\isachardoublequoteclose}\isanewline
\isakeyword{shows}\ \ \ \ {\isachardoublequoteopen}{\isasymnot}\ know\ PQ\ m{\isachardoublequoteclose}\isanewline
\isadelimproof
\endisadelimproof
\isatagproof
\isacommand{proof}\isamarkupfalse%
\ {\isacharparenleft}cases\ m{\isacharparenright}\isanewline
\ \ \isacommand{fix}\isamarkupfalse%
\ key\isanewline
\ \ \isacommand{assume}\isamarkupfalse%
\ a{\isadigit{1}}{\isacharcolon}{\isachardoublequoteopen}m\ {\isacharequal}\ kKS\ key{\isachardoublequoteclose}\isanewline
\ \ \isacommand{from}\isamarkupfalse%
\ notknowP\ \isakeyword{and}\ a{\isadigit{1}}\ \isacommand{have}\isamarkupfalse%
\ sg{\isadigit{1}}{\isacharcolon}{\isachardoublequoteopen}{\isasymnot}\ know\ P\ {\isacharparenleft}kKS\ key{\isacharparenright}{\isachardoublequoteclose}\ \isacommand{by}\isamarkupfalse%
\ simp\isanewline
\ \ \isacommand{then}\isamarkupfalse%
\ \isacommand{have}\isamarkupfalse%
\ sg{\isadigit{1}}a{\isacharcolon}{\isachardoublequoteopen}{\isasymnot}\ ine\ P\ {\isacharparenleft}kE\ key{\isacharparenright}{\isachardoublequoteclose}\ \isacommand{by}\isamarkupfalse%
\ simp\isanewline
\ \ \isacommand{from}\isamarkupfalse%
\ sg{\isadigit{1}}\ \isacommand{have}\isamarkupfalse%
\ sg{\isadigit{1}}b{\isacharcolon}{\isachardoublequoteopen}kKS\ key\ {\isasymnotin}\ LocalSecrets\ P{\isachardoublequoteclose}\ \isacommand{by}\isamarkupfalse%
\ simp\isanewline
\ \ \isacommand{from}\isamarkupfalse%
\ notknowQ\ \isakeyword{and}\ a{\isadigit{1}}\ \isacommand{have}\isamarkupfalse%
\ sg{\isadigit{2}}{\isacharcolon}{\isachardoublequoteopen}{\isasymnot}\ know\ Q\ {\isacharparenleft}kKS\ key{\isacharparenright}{\isachardoublequoteclose}\ \isacommand{by}\isamarkupfalse%
\ simp\isanewline
\ \ \isacommand{then}\isamarkupfalse%
\ \ \isacommand{have}\isamarkupfalse%
\ sg{\isadigit{2}}a{\isacharcolon}{\isachardoublequoteopen}{\isasymnot}\ ine\ Q\ {\isacharparenleft}kE\ key{\isacharparenright}{\isachardoublequoteclose}\ \isacommand{by}\isamarkupfalse%
\ simp\isanewline
\ \ \isacommand{from}\isamarkupfalse%
\ sg{\isadigit{2}}\ \isacommand{have}\isamarkupfalse%
\ sg{\isadigit{2}}b{\isacharcolon}{\isachardoublequoteopen}kKS\ key\ {\isasymnotin}\ LocalSecrets\ Q{\isachardoublequoteclose}\ \isacommand{by}\isamarkupfalse%
\ simp\isanewline
\ \ \isacommand{from}\isamarkupfalse%
\ sg{\isadigit{1}}\ \isakeyword{and}\ sg{\isadigit{2}}\ \isakeyword{and}\ subPQ\ \isakeyword{and}\ cCompI\ \isacommand{have}\isamarkupfalse%
\ sg{\isadigit{3}}{\isacharcolon}{\isachardoublequoteopen}{\isasymnot}\ ine\ PQ\ {\isacharparenleft}kE\ key{\isacharparenright}{\isachardoublequoteclose}\isanewline
\ \ \ \ \isacommand{by}\isamarkupfalse%
\ {\isacharparenleft}rule\ know{\isacharunderscore}composition{\isacharunderscore}neg{\isacharunderscore}ine{\isacharunderscore}k{\isacharparenright}\isanewline
\ \ \isacommand{from}\isamarkupfalse%
\ subPQ\ \isakeyword{and}\ cCompLoc\ \isakeyword{and}\ sg{\isadigit{1}}a\ \isakeyword{and}\ sg{\isadigit{2}}a\ \isakeyword{and}\ sg{\isadigit{1}}b\ \isakeyword{and}\ sg{\isadigit{2}}b\ \isacommand{have}\isamarkupfalse%
\ sg{\isadigit{4}}{\isacharcolon}\isanewline
\ \ {\isachardoublequoteopen}kKS\ key\ {\isasymnotin}\ LocalSecrets\ PQ{\isachardoublequoteclose}\ \isanewline
\ \ \ \ \isacommand{by}\isamarkupfalse%
\ {\isacharparenleft}rule\ LocalSecretsComposition{\isacharunderscore}neg{\isadigit{1}}{\isacharunderscore}k{\isacharparenright}\isanewline
\ \ \isacommand{from}\isamarkupfalse%
\ sg{\isadigit{3}}\ \isakeyword{and}\ sg{\isadigit{4}}\ \isakeyword{and}\ a{\isadigit{1}}\ \isacommand{show}\isamarkupfalse%
\ {\isacharquery}thesis\ \isacommand{by}\isamarkupfalse%
\ simp\isanewline
\isacommand{next}\isamarkupfalse%
\ \isanewline
\ \ \isacommand{fix}\isamarkupfalse%
\ secret\isanewline
\ \ \isacommand{assume}\isamarkupfalse%
\ a{\isadigit{2}}{\isacharcolon}{\isachardoublequoteopen}m\ {\isacharequal}\ sKS\ secret{\isachardoublequoteclose}\isanewline
\ \ \isacommand{from}\isamarkupfalse%
\ notknowP\ \isakeyword{and}\ a{\isadigit{2}}\ \isacommand{have}\isamarkupfalse%
\ sg{\isadigit{1}}{\isacharcolon}{\isachardoublequoteopen}{\isasymnot}\ know\ P\ {\isacharparenleft}sKS\ secret{\isacharparenright}{\isachardoublequoteclose}\ \isacommand{by}\isamarkupfalse%
\ simp\isanewline
\ \ \isacommand{then}\isamarkupfalse%
\ \isacommand{have}\isamarkupfalse%
\ sg{\isadigit{1}}a{\isacharcolon}{\isachardoublequoteopen}{\isasymnot}\ ine\ P\ {\isacharparenleft}sE\ secret{\isacharparenright}{\isachardoublequoteclose}\ \isacommand{by}\isamarkupfalse%
\ simp\isanewline
\ \ \isacommand{from}\isamarkupfalse%
\ sg{\isadigit{1}}\ \isacommand{have}\isamarkupfalse%
\ sg{\isadigit{1}}b{\isacharcolon}{\isachardoublequoteopen}sKS\ secret\ {\isasymnotin}\ LocalSecrets\ P{\isachardoublequoteclose}\ \isacommand{by}\isamarkupfalse%
\ simp\isanewline
\ \ \isacommand{from}\isamarkupfalse%
\ notknowQ\ \isakeyword{and}\ a{\isadigit{2}}\ \isacommand{have}\isamarkupfalse%
\ sg{\isadigit{2}}{\isacharcolon}{\isachardoublequoteopen}{\isasymnot}\ know\ Q\ {\isacharparenleft}sKS\ secret{\isacharparenright}{\isachardoublequoteclose}\ \isacommand{by}\isamarkupfalse%
\ simp\isanewline
\ \ \isacommand{then}\isamarkupfalse%
\ \isacommand{have}\isamarkupfalse%
\ sg{\isadigit{2}}a{\isacharcolon}{\isachardoublequoteopen}{\isasymnot}\ ine\ Q\ {\isacharparenleft}sE\ secret{\isacharparenright}{\isachardoublequoteclose}\ \isacommand{by}\isamarkupfalse%
\ simp\isanewline
\ \ \isacommand{from}\isamarkupfalse%
\ sg{\isadigit{2}}\ \isacommand{have}\isamarkupfalse%
\ sg{\isadigit{2}}b{\isacharcolon}{\isachardoublequoteopen}sKS\ secret\ {\isasymnotin}\ LocalSecrets\ Q{\isachardoublequoteclose}\ \isacommand{by}\isamarkupfalse%
\ simp\isanewline
\ \ \isacommand{from}\isamarkupfalse%
\ sg{\isadigit{1}}\ \isakeyword{and}\ sg{\isadigit{2}}\ \isakeyword{and}\ subPQ\ \isakeyword{and}\ cCompI\ \isacommand{have}\isamarkupfalse%
\ sg{\isadigit{3}}{\isacharcolon}{\isachardoublequoteopen}{\isasymnot}\ ine\ PQ\ {\isacharparenleft}sE\ secret{\isacharparenright}{\isachardoublequoteclose}\isanewline
\ \ \ \ \isacommand{by}\isamarkupfalse%
\ {\isacharparenleft}rule\ know{\isacharunderscore}composition{\isacharunderscore}neg{\isacharunderscore}ine{\isacharunderscore}s{\isacharparenright}\ \isanewline
\ \ \isacommand{from}\isamarkupfalse%
\ subPQ\ \isakeyword{and}\ cCompLoc\ \isakeyword{and}\ sg{\isadigit{1}}a\ \isakeyword{and}\ sg{\isadigit{2}}a\ \isakeyword{and}\ sg{\isadigit{1}}b\ \isakeyword{and}\ sg{\isadigit{2}}b\ \isacommand{have}\isamarkupfalse%
\ sg{\isadigit{4}}{\isacharcolon}\isanewline
\ \ {\isachardoublequoteopen}sKS\ secret\ {\isasymnotin}\ LocalSecrets\ PQ{\isachardoublequoteclose}\ \ \isanewline
\ \ \ \ \isacommand{by}\isamarkupfalse%
\ {\isacharparenleft}rule\ LocalSecretsComposition{\isacharunderscore}neg{\isadigit{1}}{\isacharunderscore}s{\isacharparenright}\isanewline
\ \ \isacommand{from}\isamarkupfalse%
\ sg{\isadigit{3}}\ \isakeyword{and}\ sg{\isadigit{4}}\ \isakeyword{and}\ a{\isadigit{2}}\ \isacommand{show}\isamarkupfalse%
\ {\isacharquery}thesis\ \isacommand{by}\isamarkupfalse%
\ simp\isanewline
\isacommand{qed}\isamarkupfalse%
\endisatagproof
{\isafoldproof}%
\isadelimproof
\isanewline
\endisadelimproof
\isanewline
\isacommand{lemma}\isamarkupfalse%
\ know{\isacharunderscore}decomposition{\isacharcolon}\isanewline
\isakeyword{assumes}\ knowPQ{\isacharcolon}{\isachardoublequoteopen}know\ PQ\ m{\isachardoublequoteclose}\isanewline
\ \ \ \ \ \ \ \isakeyword{and}\ subPQ{\isacharcolon}{\isachardoublequoteopen}subcomponents\ PQ\ {\isacharequal}\ {\isacharbraceleft}P{\isacharcomma}Q{\isacharbraceright}{\isachardoublequoteclose}\ \isanewline
\ \ \ \ \ \ \ \isakeyword{and}\ cCompI{\isacharcolon}{\isachardoublequoteopen}correctCompositionIn\ PQ{\isachardoublequoteclose}\isanewline
\ \ \ \ \ \ \ \isakeyword{and}\ cCompLoc{\isacharcolon}{\isachardoublequoteopen}correctCompositionLoc\ PQ{\isachardoublequoteclose}\isanewline
\isakeyword{shows}\ {\isachardoublequoteopen}know\ P\ m\ {\isasymor}\ know\ Q\ m{\isachardoublequoteclose}\isanewline
\isadelimproof
\endisadelimproof
\isatagproof
\isacommand{proof}\isamarkupfalse%
\ {\isacharparenleft}cases\ m{\isacharparenright}\isanewline
\ \ \isacommand{fix}\isamarkupfalse%
\ key\isanewline
\ \ \isacommand{assume}\isamarkupfalse%
\ a{\isadigit{1}}{\isacharcolon}{\isachardoublequoteopen}m\ {\isacharequal}\ kKS\ key{\isachardoublequoteclose}\isanewline
\ \ \isacommand{from}\isamarkupfalse%
\ this\ \isacommand{show}\isamarkupfalse%
\ {\isacharquery}thesis\isanewline
\ \ \isacommand{proof}\isamarkupfalse%
\ {\isacharparenleft}cases\ {\isachardoublequoteopen}ine\ PQ\ {\isacharparenleft}kE\ key{\isacharparenright}{\isachardoublequoteclose}{\isacharparenright}\isanewline
\ \ \ \ \isacommand{assume}\isamarkupfalse%
\ a{\isadigit{1}}{\isadigit{1}}{\isacharcolon}{\isachardoublequoteopen}ine\ PQ\ {\isacharparenleft}kE\ key{\isacharparenright}{\isachardoublequoteclose}\isanewline
\ \ \ \ \isacommand{from}\isamarkupfalse%
\ this\ \isakeyword{and}\ subPQ\ \isakeyword{and}\ cCompI\ \isakeyword{and}\ a{\isadigit{1}}\ \isacommand{have}\isamarkupfalse%
\ \isanewline
\ \ \ \ \ {\isachardoublequoteopen}ine\ P\ {\isacharparenleft}kE\ key{\isacharparenright}\ \ {\isasymor}\ ine\ Q\ {\isacharparenleft}kE\ key{\isacharparenright}{\isachardoublequoteclose}\isanewline
\ \ \ \ \ \ \isacommand{by}\isamarkupfalse%
\ {\isacharparenleft}simp\ add{\isacharcolon}\ TBtheorem{\isadigit{1}}a{\isacharparenright}\isanewline
\ \ \ \ \isacommand{from}\isamarkupfalse%
\ this\ \isakeyword{and}\ a{\isadigit{1}}\ \isacommand{show}\isamarkupfalse%
\ {\isacharquery}thesis\ \isacommand{by}\isamarkupfalse%
\ auto\isanewline
\ \ \isacommand{next}\isamarkupfalse%
\isanewline
\ \ \ \ \isacommand{assume}\isamarkupfalse%
\ a{\isadigit{1}}{\isadigit{2}}{\isacharcolon}{\isachardoublequoteopen}{\isasymnot}\ ine\ PQ\ {\isacharparenleft}kE\ key{\isacharparenright}{\isachardoublequoteclose}\isanewline
\ \ \ \ \isacommand{from}\isamarkupfalse%
\ this\ \isakeyword{and}\ knowPQ\ \isakeyword{and}\ a{\isadigit{1}}\ \isacommand{have}\isamarkupfalse%
\ sg{\isadigit{2}}{\isacharcolon}{\isachardoublequoteopen}kKS\ key\ {\isasymin}\ LocalSecrets\ PQ{\isachardoublequoteclose}\ \isacommand{by}\isamarkupfalse%
\ auto\isanewline
\ \ \ \ \isacommand{show}\isamarkupfalse%
\ {\isacharquery}thesis\ \isanewline
\ \ \ \ \isacommand{proof}\isamarkupfalse%
\ {\isacharparenleft}cases\ {\isachardoublequoteopen}know\ Q\ m{\isachardoublequoteclose}{\isacharparenright}\isanewline
\ \ \ \ \ \ \isacommand{assume}\isamarkupfalse%
\ {\isachardoublequoteopen}know\ Q\ m{\isachardoublequoteclose}\isanewline
\ \ \ \ \ \ \isacommand{from}\isamarkupfalse%
\ this\ \isacommand{show}\isamarkupfalse%
\ {\isacharquery}thesis\ \isacommand{by}\isamarkupfalse%
\ simp\isanewline
\ \ \ \ \isacommand{next}\isamarkupfalse%
\ \isanewline
\ \ \ \ \ \ \isacommand{assume}\isamarkupfalse%
\ not{\isacharunderscore}knowQm{\isacharcolon}{\isachardoublequoteopen}{\isasymnot}\ know\ Q\ m{\isachardoublequoteclose}\isanewline
\ \ \ \ \ \ \isacommand{from}\isamarkupfalse%
\ not{\isacharunderscore}knowQm\ \isakeyword{and}\ a{\isadigit{1}}\ \isacommand{have}\isamarkupfalse%
\ sg{\isadigit{3}}a{\isacharcolon}{\isachardoublequoteopen}{\isasymnot}\ ine\ Q\ {\isacharparenleft}kE\ key{\isacharparenright}{\isachardoublequoteclose}\ \isacommand{by}\isamarkupfalse%
\ simp\isanewline
\ \ \ \ \ \ \isacommand{from}\isamarkupfalse%
\ not{\isacharunderscore}knowQm\ \isakeyword{and}\ a{\isadigit{1}}\ \isacommand{have}\isamarkupfalse%
\ sg{\isadigit{3}}b{\isacharcolon}{\isachardoublequoteopen}kKS\ key\ {\isasymnotin}\ LocalSecrets\ Q{\isachardoublequoteclose}\ \isacommand{by}\isamarkupfalse%
\ simp\isanewline
\ \ \ \ \ \ \isacommand{show}\isamarkupfalse%
\ {\isacharquery}thesis\isanewline
\ \ \ \ \ \ \isacommand{proof}\isamarkupfalse%
\ {\isacharparenleft}cases\ {\isachardoublequoteopen}kKS\ key\ {\isasymin}\ LocalSecrets\ P{\isachardoublequoteclose}{\isacharparenright}\isanewline
\ \ \ \ \ \ \ \ \isacommand{assume}\isamarkupfalse%
\ {\isachardoublequoteopen}kKS\ key\ {\isasymin}\ LocalSecrets\ P{\isachardoublequoteclose}\isanewline
\ \ \ \ \ \ \ \ \isacommand{from}\isamarkupfalse%
\ this\ \isakeyword{and}\ a{\isadigit{1}}\ \isacommand{show}\isamarkupfalse%
\ {\isacharquery}thesis\ \isacommand{by}\isamarkupfalse%
\ simp\isanewline
\ \ \ \ \ \ \isacommand{next}\isamarkupfalse%
\isanewline
\ \ \ \ \ \ \ \ \isacommand{assume}\isamarkupfalse%
\ {\isachardoublequoteopen}kKS\ key\ {\isasymnotin}\ LocalSecrets\ P{\isachardoublequoteclose}\isanewline
\ \ \ \ \ \ \ \ \isacommand{from}\isamarkupfalse%
\ sg{\isadigit{2}}\ \isakeyword{and}\ subPQ\ \isakeyword{and}\ cCompLoc\ \isakeyword{and}\ sg{\isadigit{3}}a\ \isakeyword{and}\ this\ \isakeyword{and}\ sg{\isadigit{3}}b\ \isacommand{have}\isamarkupfalse%
\ {\isachardoublequoteopen}ine\ P\ {\isacharparenleft}kE\ key{\isacharparenright}{\isachardoublequoteclose}\isanewline
\ \ \ \ \ \ \ \ \ \ \isacommand{by}\isamarkupfalse%
\ {\isacharparenleft}simp\ add{\isacharcolon}\ LocalSecretsComposition{\isacharunderscore}ine{\isadigit{1}}{\isacharunderscore}k{\isacharparenright}\isanewline
\ \ \ \ \ \ \ \ \isacommand{from}\isamarkupfalse%
\ this\ \isakeyword{and}\ a{\isadigit{1}}\ \isacommand{show}\isamarkupfalse%
\ {\isacharquery}thesis\ \isacommand{by}\isamarkupfalse%
\ simp\isanewline
\ \ \ \ \ \ \isacommand{qed}\isamarkupfalse%
\isanewline
\ \ \ \ \isacommand{qed}\isamarkupfalse%
\isanewline
\ \ \isacommand{qed}\isamarkupfalse%
\isanewline
\isacommand{next}\isamarkupfalse%
\isanewline
\ \ \isacommand{fix}\isamarkupfalse%
\ secret\isanewline
\ \ \isacommand{assume}\isamarkupfalse%
\ a{\isadigit{2}}{\isacharcolon}{\isachardoublequoteopen}m\ {\isacharequal}\ sKS\ secret{\isachardoublequoteclose}\isanewline
\ \ \isacommand{from}\isamarkupfalse%
\ this\ \isacommand{show}\isamarkupfalse%
\ {\isacharquery}thesis\isanewline
\ \ \isacommand{proof}\isamarkupfalse%
\ {\isacharparenleft}cases\ {\isachardoublequoteopen}ine\ PQ\ {\isacharparenleft}sE\ secret{\isacharparenright}{\isachardoublequoteclose}{\isacharparenright}\isanewline
\ \ \ \ \isacommand{assume}\isamarkupfalse%
\ a{\isadigit{2}}{\isadigit{1}}{\isacharcolon}{\isachardoublequoteopen}ine\ PQ\ {\isacharparenleft}sE\ secret{\isacharparenright}{\isachardoublequoteclose}\isanewline
\ \ \ \ \isacommand{from}\isamarkupfalse%
\ this\ \isakeyword{and}\ subPQ\ \isakeyword{and}\ cCompI\ \isakeyword{and}\ a{\isadigit{2}}\ \isacommand{have}\isamarkupfalse%
\isanewline
\ \ \ \ \ {\isachardoublequoteopen}ine\ P\ {\isacharparenleft}sE\ secret{\isacharparenright}\ \ {\isasymor}\ ine\ Q\ {\isacharparenleft}sE\ secret{\isacharparenright}{\isachardoublequoteclose}\isanewline
\ \ \ \ \ \ \isacommand{by}\isamarkupfalse%
\ {\isacharparenleft}simp\ add{\isacharcolon}\ TBtheorem{\isadigit{1}}a{\isacharparenright}\isanewline
\ \ \ \ \isacommand{from}\isamarkupfalse%
\ this\ \isakeyword{and}\ a{\isadigit{2}}\ \isacommand{show}\isamarkupfalse%
\ {\isacharquery}thesis\ \isacommand{by}\isamarkupfalse%
\ auto\isanewline
\ \ \isacommand{next}\isamarkupfalse%
\isanewline
\ \ \ \ \isacommand{assume}\isamarkupfalse%
\ a{\isadigit{2}}{\isadigit{2}}{\isacharcolon}{\isachardoublequoteopen}{\isasymnot}\ ine\ PQ\ {\isacharparenleft}sE\ secret{\isacharparenright}{\isachardoublequoteclose}\isanewline
\ \ \ \ \isacommand{from}\isamarkupfalse%
\ this\ \isakeyword{and}\ knowPQ\ \isakeyword{and}\ a{\isadigit{2}}\ \isacommand{have}\isamarkupfalse%
\ sg{\isadigit{5}}{\isacharcolon}\isanewline
\ \ \ \ \ {\isachardoublequoteopen}sKS\ secret\ {\isasymin}\ LocalSecrets\ PQ{\isachardoublequoteclose}\ \isacommand{by}\isamarkupfalse%
\ auto\isanewline
\ \ \ \ \isacommand{show}\isamarkupfalse%
\ {\isacharquery}thesis\ \isanewline
\ \ \ \ \isacommand{proof}\isamarkupfalse%
\ {\isacharparenleft}cases\ {\isachardoublequoteopen}know\ Q\ m{\isachardoublequoteclose}{\isacharparenright}\isanewline
\ \ \ \ \ \ \isacommand{assume}\isamarkupfalse%
\ {\isachardoublequoteopen}know\ Q\ m{\isachardoublequoteclose}\isanewline
\ \ \ \ \ \ \isacommand{from}\isamarkupfalse%
\ this\ \isacommand{show}\isamarkupfalse%
\ {\isacharquery}thesis\ \isacommand{by}\isamarkupfalse%
\ simp\isanewline
\ \ \ \ \isacommand{next}\isamarkupfalse%
\ \isanewline
\ \ \ \ \ \ \isacommand{assume}\isamarkupfalse%
\ not{\isacharunderscore}knowQm{\isacharcolon}{\isachardoublequoteopen}{\isasymnot}\ know\ Q\ m{\isachardoublequoteclose}\isanewline
\ \ \ \ \ \ \isacommand{from}\isamarkupfalse%
\ not{\isacharunderscore}knowQm\ \isakeyword{and}\ a{\isadigit{2}}\ \isacommand{have}\isamarkupfalse%
\ sg{\isadigit{6}}a{\isacharcolon}{\isachardoublequoteopen}{\isasymnot}\ ine\ Q\ {\isacharparenleft}sE\ secret{\isacharparenright}{\isachardoublequoteclose}\ \isacommand{by}\isamarkupfalse%
\ simp\isanewline
\ \ \ \ \ \ \isacommand{from}\isamarkupfalse%
\ not{\isacharunderscore}knowQm\ \isakeyword{and}\ a{\isadigit{2}}\ \isacommand{have}\isamarkupfalse%
\ sg{\isadigit{6}}b{\isacharcolon}{\isachardoublequoteopen}sKS\ secret\ {\isasymnotin}\ LocalSecrets\ Q{\isachardoublequoteclose}\ \isacommand{by}\isamarkupfalse%
\ simp\isanewline
\ \ \ \ \ \ \isacommand{show}\isamarkupfalse%
\ {\isacharquery}thesis\isanewline
\ \ \ \ \ \ \isacommand{proof}\isamarkupfalse%
\ {\isacharparenleft}cases\ {\isachardoublequoteopen}sKS\ secret\ {\isasymin}\ LocalSecrets\ P{\isachardoublequoteclose}{\isacharparenright}\isanewline
\ \ \ \ \ \ \ \ \isacommand{assume}\isamarkupfalse%
\ {\isachardoublequoteopen}sKS\ secret\ {\isasymin}\ LocalSecrets\ P{\isachardoublequoteclose}\isanewline
\ \ \ \ \ \ \ \ \isacommand{from}\isamarkupfalse%
\ this\ \isakeyword{and}\ a{\isadigit{2}}\ \isacommand{show}\isamarkupfalse%
\ {\isacharquery}thesis\ \isacommand{by}\isamarkupfalse%
\ simp\isanewline
\ \ \ \ \ \ \isacommand{next}\isamarkupfalse%
\isanewline
\ \ \ \ \ \ \ \ \isacommand{assume}\isamarkupfalse%
\ {\isachardoublequoteopen}sKS\ secret\ {\isasymnotin}\ LocalSecrets\ P{\isachardoublequoteclose}\isanewline
\ \ \ \ \ \ \ \ \isacommand{from}\isamarkupfalse%
\ sg{\isadigit{5}}\ \isakeyword{and}\ subPQ\ \isakeyword{and}\ cCompLoc\ \isakeyword{and}\ sg{\isadigit{6}}a\ \isakeyword{and}\ this\ \isakeyword{and}\ sg{\isadigit{6}}b\ \isacommand{have}\isamarkupfalse%
\ \isanewline
\ \ \ \ \ \ \ \ \ {\isachardoublequoteopen}ine\ P\ {\isacharparenleft}sE\ secret{\isacharparenright}{\isachardoublequoteclose}\isanewline
\ \ \ \ \ \ \ \ \ \ \isacommand{by}\isamarkupfalse%
\ {\isacharparenleft}simp\ add{\isacharcolon}\ LocalSecretsComposition{\isacharunderscore}ine{\isadigit{1}}{\isacharunderscore}s{\isacharparenright}\isanewline
\ \ \ \ \ \ \ \ \isacommand{from}\isamarkupfalse%
\ this\ \isakeyword{and}\ a{\isadigit{2}}\ \isacommand{show}\isamarkupfalse%
\ {\isacharquery}thesis\ \isacommand{by}\isamarkupfalse%
\ simp\isanewline
\ \ \ \ \ \ \isacommand{qed}\isamarkupfalse%
\isanewline
\ \ \ \ \isacommand{qed}\isamarkupfalse%
\isanewline
\ \ \isacommand{qed}\isamarkupfalse%
\isanewline
\isacommand{qed}\isamarkupfalse%
\endisatagproof
{\isafoldproof}%
\isadelimproof
\isanewline
\endisadelimproof
\isanewline
\isacommand{lemma}\isamarkupfalse%
\ eout{\isacharunderscore}knows{\isacharunderscore}nonKS{\isacharunderscore}k{\isacharcolon}\isanewline
\ \isakeyword{assumes}\ {\isachardoublequoteopen}m\ {\isasymnotin}\ {\isacharparenleft}specKeys\ A{\isacharparenright}{\isachardoublequoteclose}\isanewline
\ \ \ \ \ \ \ \ \ \isakeyword{and}\ {\isachardoublequoteopen}eout\ A\ {\isacharparenleft}kE\ m{\isacharparenright}{\isachardoublequoteclose}\isanewline
\ \ \ \ \ \ \ \ \ \isakeyword{and}\ {\isachardoublequoteopen}eoutKnowsECorrect\ A\ {\isacharparenleft}kE\ m{\isacharparenright}{\isachardoublequoteclose}\isanewline
\ \ \ \isakeyword{shows}\ {\isachardoublequoteopen}knows\ A\ {\isacharbrackleft}kE\ m{\isacharbrackright}{\isachardoublequoteclose}\isanewline
\isadelimproof
\endisadelimproof
\isatagproof
\isacommand{using}\isamarkupfalse%
\ assms\isanewline
\isacommand{by}\isamarkupfalse%
\ {\isacharparenleft}metis\ Expression{\isachardot}distinct{\isacharparenleft}{\isadigit{1}}{\isacharparenright}\ Expression{\isachardot}inject{\isacharparenleft}{\isadigit{1}}{\isacharparenright}\ eoutKnowsECorrect{\isacharunderscore}L{\isadigit{1}}{\isacharparenright}%
\endisatagproof
{\isafoldproof}%
\isadelimproof
\isanewline
\endisadelimproof
\isanewline
\isacommand{lemma}\isamarkupfalse%
\ eout{\isacharunderscore}knows{\isacharunderscore}nonKS{\isacharunderscore}s{\isacharcolon}\isanewline
\ \isakeyword{assumes}\ h{\isadigit{1}}{\isacharcolon}{\isachardoublequoteopen}m\ {\isasymnotin}\ specSecrets\ A{\isachardoublequoteclose}\isanewline
\ \ \ \ \ \ \ \ \ \isakeyword{and}\ h{\isadigit{2}}{\isacharcolon}{\isachardoublequoteopen}eout\ A\ {\isacharparenleft}sE\ m{\isacharparenright}{\isachardoublequoteclose}\isanewline
\ \ \ \ \ \ \ \ \ \isakeyword{and}\ h{\isadigit{3}}{\isacharcolon}{\isachardoublequoteopen}eoutKnowsECorrect\ A\ {\isacharparenleft}sE\ m{\isacharparenright}{\isachardoublequoteclose}\isanewline
\ \ \ \isakeyword{shows}\ {\isachardoublequoteopen}knows\ A\ {\isacharbrackleft}sE\ m{\isacharbrackright}{\isachardoublequoteclose}\isanewline
\isadelimproof
\endisadelimproof
\isatagproof
\isacommand{using}\isamarkupfalse%
\ assms\isanewline
\isacommand{by}\isamarkupfalse%
\ {\isacharparenleft}metis\ Expression{\isachardot}distinct{\isacharparenleft}{\isadigit{1}}{\isacharparenright}\ Expression{\isachardot}inject{\isacharparenleft}{\isadigit{2}}{\isacharparenright}\ eoutKnowsECorrect{\isacharunderscore}def{\isacharparenright}%
\endisatagproof
{\isafoldproof}%
\isadelimproof
\isanewline
\endisadelimproof
\isanewline
\isacommand{lemma}\isamarkupfalse%
\ not{\isacharunderscore}knows{\isacharunderscore}k{\isacharunderscore}not{\isacharunderscore}ine{\isacharcolon}\isanewline
\isakeyword{assumes}\ {\isachardoublequoteopen}{\isasymnot}\ knows\ A\ {\isacharbrackleft}kE\ m{\isacharbrackright}{\isachardoublequoteclose}\isanewline
\isakeyword{shows}\ \ \ \ {\isachardoublequoteopen}{\isasymnot}\ ine\ A\ {\isacharparenleft}kE\ m{\isacharparenright}{\isachardoublequoteclose}\isanewline
\isadelimproof
\endisadelimproof
\isatagproof
\isacommand{using}\isamarkupfalse%
\ assms\ \isacommand{by}\isamarkupfalse%
\ {\isacharparenleft}metis\ knows{\isadigit{2}}know{\isacharunderscore}neg{\isacharunderscore}k\ not{\isacharunderscore}know{\isacharunderscore}k{\isacharunderscore}not{\isacharunderscore}ine{\isacharparenright}%
\endisatagproof
{\isafoldproof}%
\isadelimproof
\ \isanewline
\endisadelimproof
\isanewline
\isacommand{lemma}\isamarkupfalse%
\ not{\isacharunderscore}knows{\isacharunderscore}s{\isacharunderscore}not{\isacharunderscore}ine{\isacharcolon}\isanewline
\isakeyword{assumes}\ {\isachardoublequoteopen}{\isasymnot}\ knows\ A\ {\isacharbrackleft}sE\ m{\isacharbrackright}{\isachardoublequoteclose}\isanewline
\isakeyword{shows}\ \ \ \ {\isachardoublequoteopen}{\isasymnot}\ ine\ A\ {\isacharparenleft}sE\ m{\isacharparenright}{\isachardoublequoteclose}\isanewline
\isadelimproof
\endisadelimproof
\isatagproof
\isacommand{using}\isamarkupfalse%
\ assms\ \isacommand{by}\isamarkupfalse%
\ {\isacharparenleft}metis\ knows{\isadigit{2}}know{\isacharunderscore}neg{\isacharunderscore}s\ not{\isacharunderscore}know{\isacharunderscore}s{\isacharunderscore}not{\isacharunderscore}ine{\isacharparenright}%
\endisatagproof
{\isafoldproof}%
\isadelimproof
\ \isanewline
\endisadelimproof
\isanewline
\isacommand{lemma}\isamarkupfalse%
\ not{\isacharunderscore}knows{\isacharunderscore}k{\isacharunderscore}not{\isacharunderscore}eout{\isacharcolon}\isanewline
\isakeyword{assumes}\ {\isachardoublequoteopen}m\ {\isasymnotin}\ specKeys\ A{\isachardoublequoteclose}\isanewline
\ \ \ \ \ \ \ \isakeyword{and}\ {\isachardoublequoteopen}{\isasymnot}\ knows\ A\ {\isacharbrackleft}kE\ m{\isacharbrackright}{\isachardoublequoteclose}\isanewline
\ \ \ \ \ \ \ \isakeyword{and}\ {\isachardoublequoteopen}eoutKnowsECorrect\ A\ {\isacharparenleft}kE\ m{\isacharparenright}{\isachardoublequoteclose}\isanewline
\isakeyword{shows}\ \ \ \ {\isachardoublequoteopen}{\isasymnot}\ eout\ A\ {\isacharparenleft}kE\ m{\isacharparenright}{\isachardoublequoteclose}\isanewline
\isadelimproof
\endisadelimproof
\isatagproof
\isacommand{using}\isamarkupfalse%
\ assms\ \isacommand{by}\isamarkupfalse%
\ {\isacharparenleft}metis\ eout{\isacharunderscore}knows{\isacharunderscore}nonKS{\isacharunderscore}k{\isacharparenright}%
\endisatagproof
{\isafoldproof}%
\isadelimproof
\isanewline
\endisadelimproof
\isanewline
\isacommand{lemma}\isamarkupfalse%
\ not{\isacharunderscore}knows{\isacharunderscore}s{\isacharunderscore}not{\isacharunderscore}eout{\isacharcolon}\isanewline
\isakeyword{assumes}\ {\isachardoublequoteopen}m\ {\isasymnotin}\ specSecrets\ A{\isachardoublequoteclose}\isanewline
\ \ \ \ \ \ \ \isakeyword{and}\ {\isachardoublequoteopen}{\isasymnot}\ knows\ A\ {\isacharbrackleft}sE\ m{\isacharbrackright}{\isachardoublequoteclose}\isanewline
\ \ \ \ \ \ \ \isakeyword{and}\ {\isachardoublequoteopen}eoutKnowsECorrect\ A\ {\isacharparenleft}sE\ m{\isacharparenright}{\isachardoublequoteclose}\isanewline
\isakeyword{shows}\ \ \ \ {\isachardoublequoteopen}{\isasymnot}\ eout\ A\ {\isacharparenleft}sE\ m{\isacharparenright}{\isachardoublequoteclose}\isanewline
\isadelimproof
\endisadelimproof
\isatagproof
\isacommand{using}\isamarkupfalse%
\ assms\ \isacommand{by}\isamarkupfalse%
\ {\isacharparenleft}metis\ eout{\isacharunderscore}knows{\isacharunderscore}nonKS{\isacharunderscore}s{\isacharparenright}%
\endisatagproof
{\isafoldproof}%
\isadelimproof
\ \isanewline
\endisadelimproof
\isanewline
\isacommand{lemma}\isamarkupfalse%
\ \ adv{\isacharunderscore}not{\isacharunderscore}knows{\isadigit{1}}{\isacharcolon}\isanewline
\isakeyword{assumes}\ {\isachardoublequoteopen}out\ P\ {\isasymsubseteq}\ ins\ A{\isachardoublequoteclose}\isanewline
\ \ \ \ \ \ \ \isakeyword{and}\ {\isachardoublequoteopen}{\isasymnot}\ knows\ A\ {\isacharbrackleft}kE\ m{\isacharbrackright}{\isachardoublequoteclose}\isanewline
\isakeyword{shows}\ \ \ \ {\isachardoublequoteopen}{\isasymnot}\ eout\ P\ {\isacharparenleft}kE\ m{\isacharparenright}{\isachardoublequoteclose}\isanewline
\isadelimproof
\endisadelimproof
\isatagproof
\isacommand{using}\isamarkupfalse%
\ assms\ \isacommand{by}\isamarkupfalse%
\ {\isacharparenleft}metis\ adv{\isacharunderscore}not{\isacharunderscore}know{\isadigit{1}}\ knows{\isadigit{2}}know{\isacharunderscore}neg{\isacharunderscore}k{\isacharparenright}%
\endisatagproof
{\isafoldproof}%
\isadelimproof
\isanewline
\endisadelimproof
\isanewline
\isacommand{lemma}\isamarkupfalse%
\ adv{\isacharunderscore}not{\isacharunderscore}knows{\isadigit{2}}{\isacharcolon}\isanewline
\isakeyword{assumes}\ {\isachardoublequoteopen}out\ P\ \ {\isasymsubseteq}\ ins\ A{\isachardoublequoteclose}\isanewline
\ \ \ \ \ \ \ \ \isakeyword{and}\ {\isachardoublequoteopen}{\isasymnot}\ knows\ A\ {\isacharbrackleft}sE\ m{\isacharbrackright}{\isachardoublequoteclose}\ \isanewline
\isakeyword{shows}\ \ \ \ {\isachardoublequoteopen}{\isasymnot}\ eout\ P\ {\isacharparenleft}sE\ m{\isacharparenright}{\isachardoublequoteclose}\isanewline
\isadelimproof
\endisadelimproof
\isatagproof
\isacommand{using}\isamarkupfalse%
\ assms\ \isacommand{by}\isamarkupfalse%
\ {\isacharparenleft}metis\ adv{\isacharunderscore}not{\isacharunderscore}know{\isadigit{2}}\ knows{\isadigit{2}}know{\isacharunderscore}neg{\isacharunderscore}s{\isacharparenright}%
\endisatagproof
{\isafoldproof}%
\isadelimproof
\isanewline
\endisadelimproof
\isanewline
\isacommand{lemma}\isamarkupfalse%
\ knows{\isacharunderscore}decomposition{\isacharunderscore}{\isadigit{1}}{\isacharunderscore}k{\isacharcolon}\isanewline
\isakeyword{assumes}\ {\isachardoublequoteopen}kKS\ a\ {\isasymnotin}\ specKeysSecrets\ P{\isachardoublequoteclose}\isanewline
\ \ \ \ \ \ \ \isakeyword{and}\ {\isachardoublequoteopen}kKS\ a\ {\isasymnotin}\ specKeysSecrets\ Q{\isachardoublequoteclose}\isanewline
\ \ \ \ \ \ \ \isakeyword{and}\ {\isachardoublequoteopen}subcomponents\ PQ\ {\isacharequal}\ {\isacharbraceleft}P{\isacharcomma}\ Q{\isacharbraceright}{\isachardoublequoteclose}\isanewline
\ \ \ \ \ \ \ \isakeyword{and}\ {\isachardoublequoteopen}knows\ PQ\ {\isacharbrackleft}kE\ a{\isacharbrackright}{\isachardoublequoteclose}\isanewline
\ \ \ \ \ \ \ \isakeyword{and}\ {\isachardoublequoteopen}correctCompositionIn\ PQ{\isachardoublequoteclose}\isanewline
\ \ \ \ \ \ \ \isakeyword{and}\ {\isachardoublequoteopen}correctCompositionLoc\ PQ{\isachardoublequoteclose}\isanewline
\isakeyword{shows}\ {\isachardoublequoteopen}knows\ P\ {\isacharbrackleft}kE\ a{\isacharbrackright}\ {\isasymor}\ knows\ Q\ {\isacharbrackleft}kE\ a{\isacharbrackright}{\isachardoublequoteclose}\isanewline
\isadelimproof
\endisadelimproof
\isatagproof
\isacommand{using}\isamarkupfalse%
\ assms\ \isacommand{by}\isamarkupfalse%
\ {\isacharparenleft}metis\ know{\isacharunderscore}decomposition\ knows{\isadigit{1}}k{\isacharparenright}%
\endisatagproof
{\isafoldproof}%
\isadelimproof
\isanewline
\endisadelimproof
\isanewline
\isacommand{lemma}\isamarkupfalse%
\ knows{\isacharunderscore}decomposition{\isacharunderscore}{\isadigit{1}}{\isacharunderscore}s{\isacharcolon}\isanewline
\isakeyword{assumes}\ {\isachardoublequoteopen}sKS\ a\ {\isasymnotin}\ specKeysSecrets\ P{\isachardoublequoteclose}\isanewline
\ \ \ \ \ \ \ \isakeyword{and}\ {\isachardoublequoteopen}sKS\ a\ {\isasymnotin}\ specKeysSecrets\ Q{\isachardoublequoteclose}\isanewline
\ \ \ \ \ \ \ \isakeyword{and}\ {\isachardoublequoteopen}subcomponents\ PQ\ {\isacharequal}\ {\isacharbraceleft}P{\isacharcomma}\ Q{\isacharbraceright}{\isachardoublequoteclose}\isanewline
\ \ \ \ \ \ \ \isakeyword{and}\ {\isachardoublequoteopen}knows\ PQ\ {\isacharbrackleft}sE\ a{\isacharbrackright}{\isachardoublequoteclose}\isanewline
\ \ \ \ \ \ \ \isakeyword{and}\ {\isachardoublequoteopen}correctCompositionIn\ PQ{\isachardoublequoteclose}\isanewline
\ \ \ \ \ \ \ \isakeyword{and}\ {\isachardoublequoteopen}correctCompositionLoc\ PQ{\isachardoublequoteclose}\isanewline
\isakeyword{shows}\ {\isachardoublequoteopen}knows\ P\ {\isacharbrackleft}sE\ a{\isacharbrackright}\ {\isasymor}\ knows\ Q\ {\isacharbrackleft}sE\ a{\isacharbrackright}{\isachardoublequoteclose}\isanewline
\isadelimproof
\endisadelimproof
\isatagproof
\isacommand{using}\isamarkupfalse%
\ assms\ \isacommand{by}\isamarkupfalse%
\ {\isacharparenleft}metis\ know{\isacharunderscore}decomposition\ knows{\isadigit{1}}s{\isacharparenright}%
\endisatagproof
{\isafoldproof}%
\isadelimproof
\isanewline
\endisadelimproof
\isanewline
\isacommand{lemma}\isamarkupfalse%
\ knows{\isacharunderscore}decomposition{\isacharunderscore}{\isadigit{1}}{\isacharcolon}\isanewline
\isakeyword{assumes}\ {\isachardoublequoteopen}subcomponents\ PQ\ {\isacharequal}\ {\isacharbraceleft}P{\isacharcomma}\ Q{\isacharbraceright}{\isachardoublequoteclose}\isanewline
\ \ \ \ \ \ \ \isakeyword{and}\ {\isachardoublequoteopen}knows\ PQ\ {\isacharbrackleft}a{\isacharbrackright}{\isachardoublequoteclose}\isanewline
\ \ \ \ \ \ \ \isakeyword{and}\ {\isachardoublequoteopen}correctCompositionIn\ PQ{\isachardoublequoteclose}\isanewline
\ \ \ \ \ \ \ \isakeyword{and}\ {\isachardoublequoteopen}correctCompositionLoc\ PQ{\isachardoublequoteclose}\isanewline
\ \ \ \ \ \ \ \isakeyword{and}\ {\isachardoublequoteopen}{\isacharparenleft}{\isasymexists}\ z{\isachardot}\ a\ {\isacharequal}\ kE\ z{\isacharparenright}\ {\isasymor}\ {\isacharparenleft}{\isasymexists}\ z{\isachardot}\ a\ {\isacharequal}\ sE\ z{\isacharparenright}{\isachardoublequoteclose}\isanewline
\ \ \ \ \ \ \ \isakeyword{and}\ {\isachardoublequoteopen}{\isasymforall}\ z{\isachardot}\ a\ {\isacharequal}\ kE\ z\ {\isasymlongrightarrow}\ \isanewline
\ \ \ \ \ \ \ \ \ kKS\ z\ {\isasymnotin}\ specKeysSecrets\ P\ {\isasymand}\ kKS\ z\ {\isasymnotin}\ specKeysSecrets\ Q{\isachardoublequoteclose}\isanewline
\ \ \ \ \ \ \ \isakeyword{and}\ h{\isadigit{7}}{\isacharcolon}{\isachardoublequoteopen}{\isasymforall}\ z{\isachardot}\ a\ {\isacharequal}\ sE\ z\ {\isasymlongrightarrow}\ \isanewline
\ \ \ \ \ \ \ \ \ sKS\ z\ {\isasymnotin}\ specKeysSecrets\ P\ {\isasymand}\ sKS\ z\ {\isasymnotin}\ specKeysSecrets\ Q{\isachardoublequoteclose}\isanewline
\isakeyword{shows}\ {\isachardoublequoteopen}knows\ P\ {\isacharbrackleft}a{\isacharbrackright}\ {\isasymor}\ knows\ Q\ {\isacharbrackleft}a{\isacharbrackright}{\isachardoublequoteclose}\isanewline
\isadelimproof
\endisadelimproof
\isatagproof
\isacommand{using}\isamarkupfalse%
\ assms\isanewline
\isacommand{by}\isamarkupfalse%
\ {\isacharparenleft}metis\ knows{\isacharunderscore}decomposition{\isacharunderscore}{\isadigit{1}}{\isacharunderscore}k\ knows{\isacharunderscore}decomposition{\isacharunderscore}{\isadigit{1}}{\isacharunderscore}s{\isacharparenright}%
\endisatagproof
{\isafoldproof}%
\isadelimproof
\ \isanewline
\endisadelimproof
\isanewline
\isacommand{lemma}\isamarkupfalse%
\ knows{\isacharunderscore}composition{\isadigit{1}}{\isacharunderscore}k{\isacharcolon}\isanewline
\isakeyword{assumes}\ {\isachardoublequoteopen}{\isacharparenleft}kKS\ m{\isacharparenright}\ {\isasymnotin}\ specKeysSecrets\ P{\isachardoublequoteclose}\isanewline
\ \ \ \ \ \ \ \isakeyword{and}\ {\isachardoublequoteopen}{\isacharparenleft}kKS\ m{\isacharparenright}\ {\isasymnotin}\ specKeysSecrets\ Q{\isachardoublequoteclose}\isanewline
\ \ \ \ \ \ \ \isakeyword{and}\ {\isachardoublequoteopen}knows\ P\ {\isacharbrackleft}kE\ m{\isacharbrackright}{\isachardoublequoteclose}\isanewline
\ \ \ \ \ \ \ \isakeyword{and}\ {\isachardoublequoteopen}subcomponents\ PQ\ {\isacharequal}\ {\isacharbraceleft}P{\isacharcomma}Q{\isacharbraceright}{\isachardoublequoteclose}\isanewline
\ \ \ \ \ \ \ \isakeyword{and}\ {\isachardoublequoteopen}correctCompositionIn\ PQ{\isachardoublequoteclose}\ \isanewline
\ \ \ \ \ \ \ \isakeyword{and}\ {\isachardoublequoteopen}correctCompositionKS\ PQ{\isachardoublequoteclose}\isanewline
\isakeyword{shows}\ {\isachardoublequoteopen}knows\ PQ\ {\isacharbrackleft}kE\ m{\isacharbrackright}{\isachardoublequoteclose}\isanewline
\isadelimproof
\endisadelimproof
\isatagproof
\isacommand{using}\isamarkupfalse%
\ assms\ \isacommand{by}\isamarkupfalse%
\ {\isacharparenleft}metis\ know{\isacharunderscore}composition\ knows{\isadigit{1}}k{\isacharparenright}%
\endisatagproof
{\isafoldproof}%
\isadelimproof
\isanewline
\endisadelimproof
\isanewline
\isacommand{lemma}\isamarkupfalse%
\ knows{\isacharunderscore}composition{\isadigit{1}}{\isacharunderscore}s{\isacharcolon}\isanewline
\isakeyword{assumes}\ {\isachardoublequoteopen}{\isacharparenleft}sKS\ m{\isacharparenright}\ {\isasymnotin}\ specKeysSecrets\ P{\isachardoublequoteclose}\isanewline
\ \ \ \ \ \ \ \isakeyword{and}\ {\isachardoublequoteopen}{\isacharparenleft}sKS\ m{\isacharparenright}\ {\isasymnotin}\ specKeysSecrets\ Q{\isachardoublequoteclose}\isanewline
\ \ \ \ \ \ \ \isakeyword{and}\ {\isachardoublequoteopen}knows\ P\ {\isacharbrackleft}sE\ m{\isacharbrackright}{\isachardoublequoteclose}\isanewline
\ \ \ \ \ \ \ \isakeyword{and}\ {\isachardoublequoteopen}subcomponents\ PQ\ {\isacharequal}\ {\isacharbraceleft}P{\isacharcomma}Q{\isacharbraceright}{\isachardoublequoteclose}\ \isanewline
\ \ \ \ \ \ \ \isakeyword{and}\ {\isachardoublequoteopen}correctCompositionIn\ PQ{\isachardoublequoteclose}\isanewline
\ \ \ \ \ \ \ \isakeyword{and}\ {\isachardoublequoteopen}correctCompositionKS\ PQ{\isachardoublequoteclose}\isanewline
\isakeyword{shows}\ {\isachardoublequoteopen}knows\ PQ\ {\isacharbrackleft}sE\ m{\isacharbrackright}{\isachardoublequoteclose}\isanewline
\isadelimproof
\endisadelimproof
\isatagproof
\isacommand{using}\isamarkupfalse%
\ assms\ \isacommand{by}\isamarkupfalse%
\ {\isacharparenleft}metis\ know{\isacharunderscore}composition\ knows{\isadigit{1}}s{\isacharparenright}%
\endisatagproof
{\isafoldproof}%
\isadelimproof
\isanewline
\endisadelimproof
\isanewline
\isacommand{lemma}\isamarkupfalse%
\ knows{\isacharunderscore}composition{\isadigit{2}}{\isacharunderscore}k{\isacharcolon}\isanewline
\isakeyword{assumes}\ {\isachardoublequoteopen}{\isacharparenleft}kKS\ m{\isacharparenright}\ {\isasymnotin}\ specKeysSecrets\ P{\isachardoublequoteclose}\isanewline
\ \ \ \ \ \ \ \isakeyword{and}\ {\isachardoublequoteopen}{\isacharparenleft}kKS\ m{\isacharparenright}\ {\isasymnotin}\ specKeysSecrets\ Q{\isachardoublequoteclose}\isanewline
\ \ \ \ \ \ \ \isakeyword{and}\ {\isachardoublequoteopen}knows\ Q\ {\isacharbrackleft}kE\ m{\isacharbrackright}{\isachardoublequoteclose}\isanewline
\ \ \ \ \ \ \ \isakeyword{and}\ {\isachardoublequoteopen}subcomponents\ PQ\ {\isacharequal}\ {\isacharbraceleft}P{\isacharcomma}Q{\isacharbraceright}{\isachardoublequoteclose}\isanewline
\ \ \ \ \ \ \ \isakeyword{and}\ {\isachardoublequoteopen}correctCompositionIn\ PQ{\isachardoublequoteclose}\ \isanewline
\ \ \ \ \ \ \ \isakeyword{and}\ {\isachardoublequoteopen}correctCompositionKS\ PQ{\isachardoublequoteclose}\isanewline
\isakeyword{shows}\ {\isachardoublequoteopen}knows\ PQ\ {\isacharbrackleft}kE\ m{\isacharbrackright}{\isachardoublequoteclose}\isanewline
\isadelimproof
\endisadelimproof
\isatagproof
\isacommand{using}\isamarkupfalse%
\ assms\isanewline
\isacommand{by}\isamarkupfalse%
\ {\isacharparenleft}metis\ know{\isadigit{2}}knowsPQ{\isacharunderscore}k\ know{\isacharunderscore}composition\ knows{\isadigit{2}}know{\isacharunderscore}k{\isacharparenright}%
\endisatagproof
{\isafoldproof}%
\isadelimproof
\isanewline
\endisadelimproof
\isanewline
\isacommand{lemma}\isamarkupfalse%
\ knows{\isacharunderscore}composition{\isadigit{2}}{\isacharunderscore}s{\isacharcolon}\isanewline
\isakeyword{assumes}\ {\isachardoublequoteopen}{\isacharparenleft}sKS\ m{\isacharparenright}\ {\isasymnotin}\ specKeysSecrets\ P{\isachardoublequoteclose}\isanewline
\ \ \ \ \ \ \ \isakeyword{and}\ {\isachardoublequoteopen}{\isacharparenleft}sKS\ m{\isacharparenright}\ {\isasymnotin}\ specKeysSecrets\ Q{\isachardoublequoteclose}\isanewline
\ \ \ \ \ \ \ \isakeyword{and}\ {\isachardoublequoteopen}knows\ Q\ {\isacharbrackleft}sE\ m{\isacharbrackright}{\isachardoublequoteclose}\isanewline
\ \ \ \ \ \ \ \isakeyword{and}\ {\isachardoublequoteopen}subcomponents\ PQ\ {\isacharequal}\ {\isacharbraceleft}P{\isacharcomma}Q{\isacharbraceright}{\isachardoublequoteclose}\isanewline
\ \ \ \ \ \ \ \isakeyword{and}\ {\isachardoublequoteopen}correctCompositionIn\ PQ{\isachardoublequoteclose}\isanewline
\ \ \ \ \ \ \ \isakeyword{and}\ {\isachardoublequoteopen}correctCompositionKS\ PQ{\isachardoublequoteclose}\ \isanewline
\isakeyword{shows}\ {\isachardoublequoteopen}knows\ PQ\ {\isacharbrackleft}sE\ m{\isacharbrackright}{\isachardoublequoteclose}\isanewline
\isadelimproof
\endisadelimproof
\isatagproof
\isacommand{using}\isamarkupfalse%
\ assms\isanewline
\isacommand{by}\isamarkupfalse%
\ {\isacharparenleft}metis\ know{\isadigit{2}}knowsPQ{\isacharunderscore}s\ know{\isacharunderscore}composition\ knows{\isadigit{2}}know{\isacharunderscore}s{\isacharparenright}%
\endisatagproof
{\isafoldproof}%
\isadelimproof
\isanewline
\endisadelimproof
\isanewline
\isacommand{lemma}\isamarkupfalse%
\ knows{\isacharunderscore}composition{\isacharunderscore}neg{\isadigit{1}}{\isacharunderscore}k{\isacharcolon}\isanewline
\isakeyword{assumes}\ {\isachardoublequoteopen}kKS\ m\ {\isasymnotin}\ specKeysSecrets\ P{\isachardoublequoteclose}\isanewline
\ \ \ \ \ \ \ \isakeyword{and}\ {\isachardoublequoteopen}kKS\ m\ {\isasymnotin}\ specKeysSecrets\ Q{\isachardoublequoteclose}\isanewline
\ \ \ \ \ \ \ \isakeyword{and}\ {\isachardoublequoteopen}{\isasymnot}\ knows\ P\ {\isacharbrackleft}kE\ m{\isacharbrackright}{\isachardoublequoteclose}\isanewline
\ \ \ \ \ \ \ \isakeyword{and}\ {\isachardoublequoteopen}{\isasymnot}\ knows\ Q\ {\isacharbrackleft}kE\ m{\isacharbrackright}{\isachardoublequoteclose}\ \isanewline
\ \ \ \ \ \ \ \isakeyword{and}\ {\isachardoublequoteopen}subcomponents\ PQ\ {\isacharequal}\ {\isacharbraceleft}P{\isacharcomma}Q{\isacharbraceright}{\isachardoublequoteclose}\isanewline
\ \ \ \ \ \ \ \isakeyword{and}\ {\isachardoublequoteopen}correctCompositionLoc\ PQ{\isachardoublequoteclose}\isanewline
\ \ \ \ \ \ \ \isakeyword{and}\ {\isachardoublequoteopen}correctCompositionIn\ PQ{\isachardoublequoteclose}\isanewline
\ \ \ \ \ \ \ \isakeyword{and}\ {\isachardoublequoteopen}correctCompositionKS\ PQ{\isachardoublequoteclose}\ \isanewline
\isakeyword{shows}\ {\isachardoublequoteopen}{\isasymnot}\ knows\ PQ\ {\isacharbrackleft}kE\ m{\isacharbrackright}{\isachardoublequoteclose}\isanewline
\isadelimproof
\endisadelimproof
\isatagproof
\isacommand{using}\isamarkupfalse%
\ assms\ \isacommand{by}\isamarkupfalse%
\ {\isacharparenleft}metis\ know{\isacharunderscore}decomposition\ knows{\isadigit{1}}k{\isacharparenright}%
\endisatagproof
{\isafoldproof}%
\isadelimproof
\isanewline
\endisadelimproof
\isanewline
\isacommand{lemma}\isamarkupfalse%
\ knows{\isacharunderscore}composition{\isacharunderscore}neg{\isadigit{1}}{\isacharunderscore}s{\isacharcolon}\isanewline
\isakeyword{assumes}\ {\isachardoublequoteopen}sKS\ m\ {\isasymnotin}\ specKeysSecrets\ P{\isachardoublequoteclose}\isanewline
\ \ \ \ \ \ \ \isakeyword{and}\ {\isachardoublequoteopen}sKS\ m\ {\isasymnotin}\ specKeysSecrets\ Q{\isachardoublequoteclose}\isanewline
\ \ \ \ \ \ \ \isakeyword{and}\ {\isachardoublequoteopen}{\isasymnot}\ knows\ P\ {\isacharbrackleft}sE\ m{\isacharbrackright}{\isachardoublequoteclose}\isanewline
\ \ \ \ \ \ \ \isakeyword{and}\ {\isachardoublequoteopen}{\isasymnot}\ knows\ Q\ {\isacharbrackleft}sE\ m{\isacharbrackright}{\isachardoublequoteclose}\ \isanewline
\ \ \ \ \ \ \ \isakeyword{and}\ {\isachardoublequoteopen}subcomponents\ PQ\ {\isacharequal}\ {\isacharbraceleft}P{\isacharcomma}Q{\isacharbraceright}{\isachardoublequoteclose}\isanewline
\ \ \ \ \ \ \ \isakeyword{and}\ {\isachardoublequoteopen}correctCompositionLoc\ PQ{\isachardoublequoteclose}\isanewline
\ \ \ \ \ \ \ \isakeyword{and}\ {\isachardoublequoteopen}correctCompositionIn\ PQ{\isachardoublequoteclose}\isanewline
\ \ \ \ \ \ \ \isakeyword{and}\ {\isachardoublequoteopen}correctCompositionKS\ PQ{\isachardoublequoteclose}\ \isanewline
\isakeyword{shows}\ {\isachardoublequoteopen}{\isasymnot}\ knows\ PQ\ {\isacharbrackleft}sE\ m{\isacharbrackright}{\isachardoublequoteclose}\isanewline
\isadelimproof
\endisadelimproof
\isatagproof
\isacommand{using}\isamarkupfalse%
\ assms\ \isacommand{by}\isamarkupfalse%
\ {\isacharparenleft}metis\ knows{\isacharunderscore}decomposition{\isacharunderscore}{\isadigit{1}}{\isacharunderscore}s{\isacharparenright}%
\endisatagproof
{\isafoldproof}%
\isadelimproof
\isanewline
\endisadelimproof
\isanewline
\isacommand{lemma}\isamarkupfalse%
\ knows{\isacharunderscore}concat{\isacharunderscore}{\isadigit{1}}{\isacharcolon}\isanewline
\isakeyword{assumes}\ {\isachardoublequoteopen}knows\ P\ {\isacharparenleft}a\ {\isacharhash}\ e{\isacharparenright}{\isachardoublequoteclose}\isanewline
\isakeyword{shows}\ \ \ \ {\isachardoublequoteopen}knows\ P\ {\isacharbrackleft}a{\isacharbrackright}{\isachardoublequoteclose}\isanewline
\isadelimproof
\endisadelimproof
\isatagproof
\isacommand{using}\isamarkupfalse%
\ assms\ \isacommand{by}\isamarkupfalse%
\ {\isacharparenleft}metis\ append{\isacharunderscore}Cons\ append{\isacharunderscore}Nil\ knows{\isadigit{2}}{\isacharparenright}%
\endisatagproof
{\isafoldproof}%
\isadelimproof
\isanewline
\endisadelimproof
\isanewline
\isacommand{lemma}\isamarkupfalse%
\ knows{\isacharunderscore}concat{\isacharunderscore}{\isadigit{2}}{\isacharcolon}\isanewline
\isakeyword{assumes}\ {\isachardoublequoteopen}knows\ P\ {\isacharparenleft}a\ {\isacharhash}\ e{\isacharparenright}{\isachardoublequoteclose}\isanewline
\isakeyword{shows}\ \ \ \ {\isachardoublequoteopen}knows\ P\ e{\isachardoublequoteclose}\isanewline
\isadelimproof
\endisadelimproof
\isatagproof
\isacommand{using}\isamarkupfalse%
\ assms\ \isacommand{by}\isamarkupfalse%
\ {\isacharparenleft}metis\ append{\isacharunderscore}Cons\ append{\isacharunderscore}Nil\ knows{\isadigit{2}}a{\isacharparenright}%
\endisatagproof
{\isafoldproof}%
\isadelimproof
\isanewline
\endisadelimproof
\isanewline
\isacommand{lemma}\isamarkupfalse%
\ knows{\isacharunderscore}concat{\isacharunderscore}{\isadigit{3}}{\isacharcolon}\isanewline
\isakeyword{assumes}\ {\isachardoublequoteopen}knows\ P\ {\isacharbrackleft}a{\isacharbrackright}{\isachardoublequoteclose}\isanewline
\ \ \ \ \ \ \ \isakeyword{and}\ {\isachardoublequoteopen}knows\ P\ e{\isachardoublequoteclose}\isanewline
\isakeyword{shows}\ {\isachardoublequoteopen}knows\ P\ {\isacharparenleft}a\ {\isacharhash}\ e{\isacharparenright}{\isachardoublequoteclose}\isanewline
\isadelimproof
\endisadelimproof
\isatagproof
\isacommand{using}\isamarkupfalse%
\ assms\ \isacommand{by}\isamarkupfalse%
\ {\isacharparenleft}metis\ append{\isacharunderscore}Cons\ append{\isacharunderscore}Nil\ knows{\isadigit{3}}{\isacharparenright}%
\endisatagproof
{\isafoldproof}%
\isadelimproof
\isanewline
\endisadelimproof
\isanewline
\isacommand{lemma}\isamarkupfalse%
\ not{\isacharunderscore}knows{\isacharunderscore}conc{\isacharunderscore}knows{\isacharunderscore}elem{\isacharunderscore}not{\isacharunderscore}knows{\isacharunderscore}tail{\isacharcolon}\isanewline
\isakeyword{assumes}\ {\isachardoublequoteopen}{\isasymnot}\ knows\ P\ {\isacharparenleft}a\ {\isacharhash}\ e{\isacharparenright}{\isachardoublequoteclose}\isanewline
\ \ \ \ \ \ \ \isakeyword{and}\ {\isachardoublequoteopen}knows\ P\ {\isacharbrackleft}a{\isacharbrackright}{\isachardoublequoteclose}\isanewline
\isakeyword{shows}\ {\isachardoublequoteopen}{\isasymnot}\ knows\ P\ e{\isachardoublequoteclose}\isanewline
\isadelimproof
\endisadelimproof
\isatagproof
\isacommand{using}\isamarkupfalse%
\ assms\ \isacommand{by}\isamarkupfalse%
\ {\isacharparenleft}metis\ knows{\isacharunderscore}concat{\isacharunderscore}{\isadigit{3}}{\isacharparenright}%
\endisatagproof
{\isafoldproof}%
\isadelimproof
\isanewline
\endisadelimproof
\ \ \ \ \isanewline
\isacommand{lemma}\isamarkupfalse%
\ not{\isacharunderscore}knows{\isacharunderscore}conc{\isacharunderscore}not{\isacharunderscore}knows{\isacharunderscore}elem{\isacharunderscore}tail{\isacharcolon}\isanewline
\isakeyword{assumes}\ {\isachardoublequoteopen}{\isasymnot}\ knows\ P\ {\isacharparenleft}a{\isacharhash}e{\isacharparenright}{\isachardoublequoteclose}\isanewline
\isakeyword{shows}\ \ \ \ {\isachardoublequoteopen}{\isasymnot}\ knows\ P\ {\isacharbrackleft}a{\isacharbrackright}\ {\isasymor}\ {\isasymnot}\ knows\ P\ e{\isachardoublequoteclose}\isanewline
\isadelimproof
\endisadelimproof
\isatagproof
\isacommand{using}\isamarkupfalse%
\ assms\ \isacommand{by}\isamarkupfalse%
\ {\isacharparenleft}metis\ append{\isacharunderscore}Cons\ append{\isacharunderscore}Nil\ knows{\isadigit{3}}{\isacharparenright}%
\endisatagproof
{\isafoldproof}%
\isadelimproof
\isanewline
\endisadelimproof
\isanewline
\isacommand{lemma}\isamarkupfalse%
\ not{\isacharunderscore}knows{\isacharunderscore}elem{\isacharunderscore}not{\isacharunderscore}knows{\isacharunderscore}conc{\isacharcolon}\isanewline
\isakeyword{assumes}\ {\isachardoublequoteopen}{\isasymnot}\ knows\ P\ {\isacharbrackleft}a{\isacharbrackright}{\isachardoublequoteclose}\isanewline
\isakeyword{shows}\ \ \ \ {\isachardoublequoteopen}{\isasymnot}\ knows\ P\ {\isacharparenleft}a\ {\isacharhash}\ e{\isacharparenright}{\isachardoublequoteclose}\isanewline
\isadelimproof
\endisadelimproof
\isatagproof
\isacommand{using}\isamarkupfalse%
\ assms\ \isacommand{by}\isamarkupfalse%
\ {\isacharparenleft}metis\ knows{\isacharunderscore}concat{\isacharunderscore}{\isadigit{1}}{\isacharparenright}%
\endisatagproof
{\isafoldproof}%
\isadelimproof
\isanewline
\endisadelimproof
\isanewline
\isacommand{lemma}\isamarkupfalse%
\ not{\isacharunderscore}knows{\isacharunderscore}tail{\isacharunderscore}not{\isacharunderscore}knows{\isacharunderscore}conc{\isacharcolon}\isanewline
\isakeyword{assumes}\ {\isachardoublequoteopen}{\isasymnot}\ knows\ P\ e{\isachardoublequoteclose}\isanewline
\isakeyword{shows}\ \ \ \ {\isachardoublequoteopen}{\isasymnot}\ knows\ P\ {\isacharparenleft}a\ {\isacharhash}\ e{\isacharparenright}{\isachardoublequoteclose}\isanewline
\isadelimproof
\endisadelimproof
\isatagproof
\isacommand{using}\isamarkupfalse%
\ assms\ \isacommand{by}\isamarkupfalse%
\ {\isacharparenleft}metis\ knows{\isacharunderscore}concat{\isacharunderscore}{\isadigit{2}}{\isacharparenright}%
\endisatagproof
{\isafoldproof}%
\isadelimproof
\isanewline
\endisadelimproof
\isanewline
\isacommand{lemma}\isamarkupfalse%
\ knows{\isacharunderscore}composition{\isadigit{3}}{\isacharcolon}\isanewline
\ \isakeyword{fixes}\ e{\isacharcolon}{\isacharcolon}{\isachardoublequoteopen}Expression\ list{\isachardoublequoteclose}\isanewline
\ \isakeyword{assumes}\ {\isachardoublequoteopen}knows\ P\ e{\isachardoublequoteclose}\isanewline
\ \ \ \ \ \isakeyword{and}\ subPQ{\isacharcolon}{\isachardoublequoteopen}subcomponents\ PQ\ {\isacharequal}\ {\isacharbraceleft}P{\isacharcomma}Q{\isacharbraceright}{\isachardoublequoteclose}\isanewline
\ \ \ \ \ \isakeyword{and}\ cCompI{\isacharcolon}{\isachardoublequoteopen}correctCompositionIn\ PQ{\isachardoublequoteclose}\isanewline
\ \ \ \ \ \isakeyword{and}\ cCompKS{\isacharcolon}{\isachardoublequoteopen}correctCompositionKS\ PQ{\isachardoublequoteclose}\isanewline
\ \ \ \ \ \isakeyword{and}\ {\isachardoublequoteopen}{\isasymforall}\ {\isacharparenleft}m{\isacharcolon}{\isacharcolon}Expression{\isacharparenright}{\isachardot}\ {\isacharparenleft}{\isacharparenleft}m\ mem\ e{\isacharparenright}\ {\isasymlongrightarrow}\ \isanewline
\ \ \ \ \ \ \ \ \ \ \ \ {\isacharparenleft}{\isacharparenleft}{\isasymexists}\ z{\isadigit{1}}{\isachardot}\ m\ {\isacharequal}\ {\isacharparenleft}kE\ z{\isadigit{1}}{\isacharparenright}{\isacharparenright}\ {\isasymor}\ {\isacharparenleft}{\isasymexists}\ z{\isadigit{2}}{\isachardot}\ m\ {\isacharequal}\ {\isacharparenleft}sE\ z{\isadigit{2}}{\isacharparenright}{\isacharparenright}{\isacharparenright}{\isacharparenright}{\isachardoublequoteclose}\isanewline
\ \ \ \ \ \isakeyword{and}\ {\isachardoublequoteopen}notSpecKeysSecretsExpr\ P\ e{\isachardoublequoteclose}\isanewline
\ \ \ \ \ \isakeyword{and}\ {\isachardoublequoteopen}notSpecKeysSecretsExpr\ Q\ e{\isachardoublequoteclose}\ \isanewline
\ \isakeyword{shows}\ {\isachardoublequoteopen}knows\ PQ\ e{\isachardoublequoteclose}\isanewline
\isadelimproof
\endisadelimproof
\isatagproof
\isacommand{using}\isamarkupfalse%
\ assms\isanewline
\isacommand{proof}\isamarkupfalse%
\ {\isacharparenleft}induct\ e{\isacharparenright}\isanewline
\ \ \isacommand{case}\isamarkupfalse%
\ Nil\ \isanewline
\ \ \isacommand{from}\isamarkupfalse%
\ this\ \isacommand{show}\isamarkupfalse%
\ {\isacharquery}case\ \isacommand{by}\isamarkupfalse%
\ {\isacharparenleft}simp\ only{\isacharcolon}\ knows{\isacharunderscore}emptyexpression{\isacharparenright}\isanewline
\isacommand{next}\isamarkupfalse%
\isanewline
\ \ \isacommand{fix}\isamarkupfalse%
\ a\ l\ \isanewline
\ \ \isacommand{case}\isamarkupfalse%
\ {\isacharparenleft}Cons\ a\ l{\isacharparenright}\isanewline
\ \ \isacommand{from}\isamarkupfalse%
\ Cons\ \isacommand{have}\isamarkupfalse%
\ sg{\isadigit{1}}{\isacharcolon}{\isachardoublequoteopen}knows\ P\ {\isacharbrackleft}a{\isacharbrackright}{\isachardoublequoteclose}\ \isacommand{by}\isamarkupfalse%
\ {\isacharparenleft}simp\ add{\isacharcolon}\ knows{\isacharunderscore}concat{\isacharunderscore}{\isadigit{1}}{\isacharparenright}\isanewline
\ \ \isacommand{from}\isamarkupfalse%
\ Cons\ \isacommand{have}\isamarkupfalse%
\ sg{\isadigit{2}}{\isacharcolon}{\isachardoublequoteopen}knows\ P\ l{\isachardoublequoteclose}\ \isacommand{by}\isamarkupfalse%
\ {\isacharparenleft}simp\ only{\isacharcolon}\ knows{\isacharunderscore}concat{\isacharunderscore}{\isadigit{2}}{\isacharparenright}\isanewline
\ \ \isacommand{from}\isamarkupfalse%
\ sg{\isadigit{1}}\ \isacommand{have}\isamarkupfalse%
\ sg{\isadigit{3}}{\isacharcolon}{\isachardoublequoteopen}a\ mem\ {\isacharparenleft}a\ {\isacharhash}\ l{\isacharparenright}{\isachardoublequoteclose}\ \isacommand{by}\isamarkupfalse%
\ simp\isanewline
\ \ \isacommand{from}\isamarkupfalse%
\ Cons\ \isakeyword{and}\ sg{\isadigit{2}}\ \isacommand{have}\isamarkupfalse%
\ sg{\isadigit{2}}a{\isacharcolon}{\isachardoublequoteopen}knows\ PQ\ l{\isachardoublequoteclose}\ \isanewline
\ \ \ \ \isacommand{by}\isamarkupfalse%
\ {\isacharparenleft}simp\ add{\isacharcolon}\ notSpecKeysSecretsExpr{\isacharunderscore}L{\isadigit{2}}{\isacharparenright}\isanewline
\ \ \isacommand{from}\isamarkupfalse%
\ Cons\ \isakeyword{and}\ sg{\isadigit{1}}\ \isakeyword{and}\ sg{\isadigit{2}}\ \isakeyword{and}\ sg{\isadigit{3}}\ \isacommand{show}\isamarkupfalse%
\ {\isacharquery}case\isanewline
\ \ \isacommand{proof}\isamarkupfalse%
\ {\isacharparenleft}cases\ {\isachardoublequoteopen}{\isasymexists}\ z{\isadigit{1}}{\isachardot}\ a\ {\isacharequal}\ kE\ z{\isadigit{1}}{\isachardoublequoteclose}{\isacharparenright}\isanewline
\ \ \ \ \isacommand{assume}\isamarkupfalse%
\ {\isachardoublequoteopen}{\isasymexists}\ z{\isadigit{1}}{\isachardot}\ a\ {\isacharequal}\ {\isacharparenleft}kE\ z{\isadigit{1}}{\isacharparenright}{\isachardoublequoteclose}\isanewline
\ \ \ \ \isacommand{from}\isamarkupfalse%
\ this\ \isacommand{obtain}\isamarkupfalse%
\ z\ \isakeyword{where}\ a{\isadigit{1}}{\isacharcolon}{\isachardoublequoteopen}a\ {\isacharequal}\ {\isacharparenleft}kE\ z{\isacharparenright}{\isachardoublequoteclose}\ \isacommand{by}\isamarkupfalse%
\ auto\isanewline
\ \ \ \ \isacommand{from}\isamarkupfalse%
\ a{\isadigit{1}}\ \isakeyword{and}\ Cons\ \isacommand{have}\isamarkupfalse%
\ sg{\isadigit{4}}{\isacharcolon}{\isachardoublequoteopen}{\isacharparenleft}kKS\ z{\isacharparenright}\ {\isasymnotin}\ specKeysSecrets\ P{\isachardoublequoteclose}\isanewline
\ \ \ \ \ \ \isacommand{by}\isamarkupfalse%
\ {\isacharparenleft}simp\ add{\isacharcolon}\ notSpecKeysSecretsExpr{\isacharunderscore}def{\isacharparenright}\isanewline
\ \ \ \ \isacommand{from}\isamarkupfalse%
\ a{\isadigit{1}}\ \isakeyword{and}\ Cons\ \isacommand{have}\isamarkupfalse%
\ sg{\isadigit{5}}{\isacharcolon}{\isachardoublequoteopen}{\isacharparenleft}kKS\ z{\isacharparenright}\ {\isasymnotin}\ specKeysSecrets\ Q{\isachardoublequoteclose}\ \isanewline
\ \ \ \ \ \ \isacommand{by}\isamarkupfalse%
\ {\isacharparenleft}simp\ add{\isacharcolon}\ notSpecKeysSecretsExpr{\isacharunderscore}def{\isacharparenright}\isanewline
\ \ \ \ \isacommand{from}\isamarkupfalse%
\ sg{\isadigit{1}}\ \isakeyword{and}\ a{\isadigit{1}}\ \isacommand{have}\isamarkupfalse%
\ sg{\isadigit{6}}{\isacharcolon}{\isachardoublequoteopen}knows\ P\ {\isacharbrackleft}kE\ z{\isacharbrackright}{\isachardoublequoteclose}\ \isacommand{by}\isamarkupfalse%
\ simp\isanewline
\ \ \ \ \isacommand{from}\isamarkupfalse%
\ sg{\isadigit{4}}\ \isakeyword{and}\ sg{\isadigit{5}}\ \isakeyword{and}\ sg{\isadigit{6}}\ \isakeyword{and}\ subPQ\ \isakeyword{and}\ cCompI\ \isakeyword{and}\ cCompKS\ \ \isanewline
\ \ \ \ \ \ \isacommand{have}\isamarkupfalse%
\ {\isachardoublequoteopen}knows\ PQ\ {\isacharbrackleft}kE\ z{\isacharbrackright}{\isachardoublequoteclose}\ \isanewline
\ \ \ \ \ \ \isacommand{by}\isamarkupfalse%
\ {\isacharparenleft}rule\ knows{\isacharunderscore}composition{\isadigit{1}}{\isacharunderscore}k{\isacharparenright}\isanewline
\ \ \ \ \isacommand{from}\isamarkupfalse%
\ this\ \isakeyword{and}\ sg{\isadigit{2}}a\ \isakeyword{and}\ a{\isadigit{1}}\ \isacommand{show}\isamarkupfalse%
\ {\isacharquery}case\ \isacommand{by}\isamarkupfalse%
\ {\isacharparenleft}simp\ add{\isacharcolon}\ knows{\isacharunderscore}concat{\isacharunderscore}{\isadigit{3}}{\isacharparenright}\isanewline
\ \ \isacommand{next}\isamarkupfalse%
\ \isanewline
\ \ \ \ \isacommand{assume}\isamarkupfalse%
\ {\isachardoublequoteopen}{\isasymnot}\ {\isacharparenleft}{\isasymexists}z{\isadigit{1}}{\isachardot}\ a\ {\isacharequal}\ kE\ z{\isadigit{1}}{\isacharparenright}{\isachardoublequoteclose}\isanewline
\ \ \ \ \isacommand{from}\isamarkupfalse%
\ this\ \isakeyword{and}\ Cons\ \isakeyword{and}\ sg{\isadigit{3}}\ \isacommand{have}\isamarkupfalse%
\ {\isachardoublequoteopen}{\isasymexists}\ z{\isadigit{2}}{\isachardot}\ a\ {\isacharequal}\ {\isacharparenleft}sE\ z{\isadigit{2}}{\isacharparenright}{\isachardoublequoteclose}\ \isacommand{by}\isamarkupfalse%
\ auto\isanewline
\ \ \ \ \isacommand{from}\isamarkupfalse%
\ this\ \isacommand{obtain}\isamarkupfalse%
\ z\ \isakeyword{where}\ a{\isadigit{2}}{\isacharcolon}{\isachardoublequoteopen}a\ {\isacharequal}\ {\isacharparenleft}sE\ z{\isacharparenright}{\isachardoublequoteclose}\ \isacommand{by}\isamarkupfalse%
\ auto\isanewline
\ \ \ \ \isacommand{from}\isamarkupfalse%
\ a{\isadigit{2}}\ \isakeyword{and}\ Cons\ \isacommand{have}\isamarkupfalse%
\ sg{\isadigit{8}}{\isacharcolon}{\isachardoublequoteopen}{\isacharparenleft}sKS\ z{\isacharparenright}\ {\isasymnotin}\ specKeysSecrets\ P{\isachardoublequoteclose}\ \isanewline
\ \ \ \ \ \ \isacommand{by}\isamarkupfalse%
\ {\isacharparenleft}simp\ add{\isacharcolon}\ notSpecKeysSecretsExpr{\isacharunderscore}def{\isacharparenright}\isanewline
\ \ \ \ \isacommand{from}\isamarkupfalse%
\ a{\isadigit{2}}\ \isakeyword{and}\ Cons\ \isacommand{have}\isamarkupfalse%
\ sg{\isadigit{9}}{\isacharcolon}{\isachardoublequoteopen}{\isacharparenleft}sKS\ z{\isacharparenright}\ {\isasymnotin}\ specKeysSecrets\ Q{\isachardoublequoteclose}\isanewline
\ \ \ \ \ \ \isacommand{by}\isamarkupfalse%
\ {\isacharparenleft}simp\ add{\isacharcolon}\ notSpecKeysSecretsExpr{\isacharunderscore}def{\isacharparenright}\isanewline
\ \ \ \ \isacommand{from}\isamarkupfalse%
\ sg{\isadigit{1}}\ \isakeyword{and}\ a{\isadigit{2}}\ \isacommand{have}\isamarkupfalse%
\ sg{\isadigit{1}}{\isadigit{0}}{\isacharcolon}{\isachardoublequoteopen}knows\ P\ {\isacharbrackleft}sE\ z{\isacharbrackright}{\isachardoublequoteclose}\ \isacommand{by}\isamarkupfalse%
\ simp\ \isanewline
\ \ \ \ \isacommand{from}\isamarkupfalse%
\ sg{\isadigit{8}}\ \isakeyword{and}\ sg{\isadigit{9}}\ \isakeyword{and}\ sg{\isadigit{1}}{\isadigit{0}}\ \isakeyword{and}\ subPQ\ \isakeyword{and}\ cCompI\ \isakeyword{and}\ cCompKS\ \ \isanewline
\ \ \ \ \ \ \isacommand{have}\isamarkupfalse%
\ {\isachardoublequoteopen}knows\ PQ\ {\isacharbrackleft}sE\ z{\isacharbrackright}{\isachardoublequoteclose}\ \isanewline
\ \ \ \ \ \ \isacommand{by}\isamarkupfalse%
\ {\isacharparenleft}rule\ knows{\isacharunderscore}composition{\isadigit{1}}{\isacharunderscore}s{\isacharparenright}\isanewline
\ \ \ \ \isacommand{from}\isamarkupfalse%
\ this\ \isakeyword{and}\ sg{\isadigit{2}}a\ \isakeyword{and}\ a{\isadigit{2}}\ \isacommand{show}\isamarkupfalse%
\ {\isacharquery}case\ \isacommand{by}\isamarkupfalse%
\ {\isacharparenleft}simp\ add{\isacharcolon}\ knows{\isacharunderscore}concat{\isacharunderscore}{\isadigit{3}}{\isacharparenright}\isanewline
\ \ \isacommand{qed}\isamarkupfalse%
\ \isanewline
\isacommand{qed}\isamarkupfalse%
\endisatagproof
{\isafoldproof}%
\isadelimproof
\ \isanewline
\endisadelimproof
\ \isanewline
\isacommand{lemma}\isamarkupfalse%
\ knows{\isacharunderscore}composition{\isadigit{4}}{\isacharcolon}\isanewline
\ \isakeyword{assumes}\ {\isachardoublequoteopen}knows\ Q\ e{\isachardoublequoteclose}\isanewline
\ \ \ \ \ \isakeyword{and}\ subPQ{\isacharcolon}{\isachardoublequoteopen}subcomponents\ PQ\ {\isacharequal}\ {\isacharbraceleft}P{\isacharcomma}Q{\isacharbraceright}{\isachardoublequoteclose}\ \isanewline
\ \ \ \ \ \isakeyword{and}\ cCompI{\isacharcolon}{\isachardoublequoteopen}correctCompositionIn\ PQ{\isachardoublequoteclose}\isanewline
\ \ \ \ \ \isakeyword{and}\ cCompKS{\isacharcolon}{\isachardoublequoteopen}correctCompositionKS\ PQ{\isachardoublequoteclose}\isanewline
\ \ \ \ \ \isakeyword{and}\ {\isachardoublequoteopen}{\isasymforall}\ m{\isachardot}\ m\ mem\ e\ {\isasymlongrightarrow}\ {\isacharparenleft}{\isacharparenleft}{\isasymexists}\ z{\isachardot}\ m\ {\isacharequal}\ kE\ z{\isacharparenright}\ {\isasymor}\ {\isacharparenleft}{\isasymexists}\ z{\isachardot}\ m\ {\isacharequal}\ sE\ z{\isacharparenright}{\isacharparenright}{\isachardoublequoteclose}\isanewline
\ \ \ \ \ \isakeyword{and}\ {\isachardoublequoteopen}notSpecKeysSecretsExpr\ P\ e{\isachardoublequoteclose}\isanewline
\ \ \ \ \ \isakeyword{and}\ {\isachardoublequoteopen}notSpecKeysSecretsExpr\ Q\ e{\isachardoublequoteclose}\ \isanewline
\ \isakeyword{shows}\ {\isachardoublequoteopen}knows\ PQ\ e{\isachardoublequoteclose}\isanewline
\isadelimproof
\endisadelimproof
\isatagproof
\isacommand{using}\isamarkupfalse%
\ assms\isanewline
\isacommand{proof}\isamarkupfalse%
\ {\isacharparenleft}induct\ e{\isacharparenright}\isanewline
\ \ \isacommand{case}\isamarkupfalse%
\ Nil\ \isanewline
\ \ \isacommand{from}\isamarkupfalse%
\ this\ \isacommand{show}\isamarkupfalse%
\ {\isacharquery}case\ \isacommand{by}\isamarkupfalse%
\ {\isacharparenleft}simp\ only{\isacharcolon}\ knows{\isacharunderscore}emptyexpression{\isacharparenright}\isanewline
\isacommand{next}\isamarkupfalse%
\isanewline
\ \ \isacommand{fix}\isamarkupfalse%
\ a\ l\ \isanewline
\ \ \isacommand{case}\isamarkupfalse%
\ {\isacharparenleft}Cons\ a\ l{\isacharparenright}\isanewline
\ \ \isacommand{from}\isamarkupfalse%
\ Cons\ \isacommand{have}\isamarkupfalse%
\ sg{\isadigit{1}}{\isacharcolon}{\isachardoublequoteopen}knows\ Q\ {\isacharbrackleft}a{\isacharbrackright}{\isachardoublequoteclose}\ \isacommand{by}\isamarkupfalse%
\ {\isacharparenleft}simp\ add{\isacharcolon}\ knows{\isacharunderscore}concat{\isacharunderscore}{\isadigit{1}}{\isacharparenright}\isanewline
\ \ \isacommand{from}\isamarkupfalse%
\ Cons\ \isacommand{have}\isamarkupfalse%
\ sg{\isadigit{2}}{\isacharcolon}{\isachardoublequoteopen}knows\ Q\ l{\isachardoublequoteclose}\ \isacommand{by}\isamarkupfalse%
\ {\isacharparenleft}simp\ only{\isacharcolon}\ knows{\isacharunderscore}concat{\isacharunderscore}{\isadigit{2}}{\isacharparenright}\isanewline
\ \ \isacommand{from}\isamarkupfalse%
\ sg{\isadigit{1}}\ \isacommand{have}\isamarkupfalse%
\ sg{\isadigit{3}}{\isacharcolon}{\isachardoublequoteopen}a\ mem\ {\isacharparenleft}a\ {\isacharhash}\ l{\isacharparenright}{\isachardoublequoteclose}\ \isacommand{by}\isamarkupfalse%
\ simp\isanewline
\ \ \isacommand{from}\isamarkupfalse%
\ Cons\ \isakeyword{and}\ sg{\isadigit{2}}\ \isacommand{have}\isamarkupfalse%
\ sg{\isadigit{2}}a{\isacharcolon}{\isachardoublequoteopen}knows\ PQ\ l{\isachardoublequoteclose}\ \isanewline
\ \ \ \ \isacommand{by}\isamarkupfalse%
\ {\isacharparenleft}simp\ add{\isacharcolon}\ notSpecKeysSecretsExpr{\isacharunderscore}L{\isadigit{2}}{\isacharparenright}\isanewline
\ \ \isacommand{from}\isamarkupfalse%
\ Cons\ \isakeyword{and}\ sg{\isadigit{1}}\ \isakeyword{and}\ sg{\isadigit{2}}\ \isakeyword{and}\ sg{\isadigit{3}}\ \isacommand{show}\isamarkupfalse%
\ {\isacharquery}case\isanewline
\ \ \isacommand{proof}\isamarkupfalse%
\ {\isacharparenleft}cases\ {\isachardoublequoteopen}{\isasymexists}\ z{\isadigit{1}}{\isachardot}\ a\ {\isacharequal}\ kE\ z{\isadigit{1}}{\isachardoublequoteclose}{\isacharparenright}\isanewline
\ \ \ \ \isacommand{assume}\isamarkupfalse%
\ {\isachardoublequoteopen}{\isasymexists}\ z{\isadigit{1}}{\isachardot}\ a\ {\isacharequal}\ {\isacharparenleft}kE\ z{\isadigit{1}}{\isacharparenright}{\isachardoublequoteclose}\isanewline
\ \ \ \ \isacommand{from}\isamarkupfalse%
\ this\ \isacommand{obtain}\isamarkupfalse%
\ z\ \isakeyword{where}\ a{\isadigit{1}}{\isacharcolon}{\isachardoublequoteopen}a\ {\isacharequal}\ {\isacharparenleft}kE\ z{\isacharparenright}{\isachardoublequoteclose}\ \isacommand{by}\isamarkupfalse%
\ auto\isanewline
\ \ \ \ \isacommand{from}\isamarkupfalse%
\ a{\isadigit{1}}\ \isakeyword{and}\ Cons\ \isacommand{have}\isamarkupfalse%
\ sg{\isadigit{4}}{\isacharcolon}{\isachardoublequoteopen}{\isacharparenleft}kKS\ z{\isacharparenright}\ {\isasymnotin}\ specKeysSecrets\ P{\isachardoublequoteclose}\isanewline
\ \ \ \ \ \ \isacommand{by}\isamarkupfalse%
\ {\isacharparenleft}simp\ add{\isacharcolon}\ notSpecKeysSecretsExpr{\isacharunderscore}def{\isacharparenright}\isanewline
\ \ \ \ \isacommand{from}\isamarkupfalse%
\ a{\isadigit{1}}\ \isakeyword{and}\ Cons\ \isacommand{have}\isamarkupfalse%
\ sg{\isadigit{5}}{\isacharcolon}{\isachardoublequoteopen}{\isacharparenleft}kKS\ z{\isacharparenright}\ {\isasymnotin}\ specKeysSecrets\ Q{\isachardoublequoteclose}\isanewline
\ \ \ \ \ \ \isacommand{by}\isamarkupfalse%
\ {\isacharparenleft}simp\ add{\isacharcolon}\ notSpecKeysSecretsExpr{\isacharunderscore}def{\isacharparenright}\isanewline
\ \ \ \ \isacommand{from}\isamarkupfalse%
\ sg{\isadigit{1}}\ \isakeyword{and}\ a{\isadigit{1}}\ \isacommand{have}\isamarkupfalse%
\ sg{\isadigit{6}}{\isacharcolon}{\isachardoublequoteopen}knows\ Q\ {\isacharbrackleft}kE\ z{\isacharbrackright}{\isachardoublequoteclose}\ \isacommand{by}\isamarkupfalse%
\ simp\isanewline
\ \ \ \ \isacommand{from}\isamarkupfalse%
\ sg{\isadigit{4}}\ \isakeyword{and}\ sg{\isadigit{5}}\ \isakeyword{and}\ sg{\isadigit{6}}\ \isakeyword{and}\ subPQ\ \isakeyword{and}\ cCompI\ \isakeyword{and}\ cCompKS\ \isanewline
\ \ \ \ \ \ \isacommand{have}\isamarkupfalse%
\ {\isachardoublequoteopen}knows\ PQ\ {\isacharbrackleft}kE\ z{\isacharbrackright}{\isachardoublequoteclose}\ \isanewline
\ \ \ \ \ \ \isacommand{by}\isamarkupfalse%
\ {\isacharparenleft}rule\ knows{\isacharunderscore}composition{\isadigit{2}}{\isacharunderscore}k{\isacharparenright}\isanewline
\ \ \ \ \isacommand{from}\isamarkupfalse%
\ this\ \isakeyword{and}\ sg{\isadigit{2}}a\ \isakeyword{and}\ a{\isadigit{1}}\ \isacommand{show}\isamarkupfalse%
\ {\isacharquery}case\ \isacommand{by}\isamarkupfalse%
\ {\isacharparenleft}simp\ add{\isacharcolon}\ knows{\isacharunderscore}concat{\isacharunderscore}{\isadigit{3}}{\isacharparenright}\isanewline
\ \ \isacommand{next}\isamarkupfalse%
\ \isanewline
\ \ \ \ \isacommand{assume}\isamarkupfalse%
\ {\isachardoublequoteopen}{\isasymnot}\ {\isacharparenleft}{\isasymexists}z{\isadigit{1}}{\isachardot}\ a\ {\isacharequal}\ kE\ z{\isadigit{1}}{\isacharparenright}{\isachardoublequoteclose}\isanewline
\ \ \ \ \isacommand{from}\isamarkupfalse%
\ this\ \isakeyword{and}\ Cons\ \isakeyword{and}\ sg{\isadigit{3}}\ \isacommand{have}\isamarkupfalse%
\ {\isachardoublequoteopen}{\isasymexists}\ z{\isadigit{2}}{\isachardot}\ a\ {\isacharequal}\ {\isacharparenleft}sE\ z{\isadigit{2}}{\isacharparenright}{\isachardoublequoteclose}\ \isacommand{by}\isamarkupfalse%
\ auto\isanewline
\ \ \ \ \isacommand{from}\isamarkupfalse%
\ this\ \isacommand{obtain}\isamarkupfalse%
\ z\ \isakeyword{where}\ a{\isadigit{2}}{\isacharcolon}{\isachardoublequoteopen}a\ {\isacharequal}\ {\isacharparenleft}sE\ z{\isacharparenright}{\isachardoublequoteclose}\ \isacommand{by}\isamarkupfalse%
\ auto\isanewline
\ \ \ \ \isacommand{from}\isamarkupfalse%
\ a{\isadigit{2}}\ \isakeyword{and}\ Cons\ \isacommand{have}\isamarkupfalse%
\ sg{\isadigit{8}}{\isacharcolon}{\isachardoublequoteopen}{\isacharparenleft}sKS\ z{\isacharparenright}\ {\isasymnotin}\ specKeysSecrets\ P{\isachardoublequoteclose}\isanewline
\ \ \ \ \ \ \isacommand{by}\isamarkupfalse%
\ {\isacharparenleft}simp\ add{\isacharcolon}\ notSpecKeysSecretsExpr{\isacharunderscore}def{\isacharparenright}\isanewline
\ \ \ \ \isacommand{from}\isamarkupfalse%
\ a{\isadigit{2}}\ \isakeyword{and}\ Cons\ \isacommand{have}\isamarkupfalse%
\ sg{\isadigit{9}}{\isacharcolon}{\isachardoublequoteopen}{\isacharparenleft}sKS\ z{\isacharparenright}\ {\isasymnotin}\ specKeysSecrets\ Q{\isachardoublequoteclose}\isanewline
\ \ \ \ \ \ \isacommand{by}\isamarkupfalse%
\ {\isacharparenleft}simp\ add{\isacharcolon}\ notSpecKeysSecretsExpr{\isacharunderscore}def{\isacharparenright}\isanewline
\ \ \ \ \isacommand{from}\isamarkupfalse%
\ sg{\isadigit{1}}\ \isakeyword{and}\ a{\isadigit{2}}\ \isacommand{have}\isamarkupfalse%
\ sg{\isadigit{1}}{\isadigit{0}}{\isacharcolon}{\isachardoublequoteopen}knows\ Q\ {\isacharbrackleft}sE\ z{\isacharbrackright}{\isachardoublequoteclose}\ \isacommand{by}\isamarkupfalse%
\ simp\ \isanewline
\ \ \ \ \isacommand{from}\isamarkupfalse%
\ sg{\isadigit{8}}\ \isakeyword{and}\ sg{\isadigit{9}}\ \isakeyword{and}\ sg{\isadigit{1}}{\isadigit{0}}\ \isakeyword{and}\ subPQ\ \isakeyword{and}\ cCompI\ \isakeyword{and}\ cCompKS\ \ \isanewline
\ \ \ \ \ \ \isacommand{have}\isamarkupfalse%
\ {\isachardoublequoteopen}knows\ PQ\ {\isacharbrackleft}sE\ z{\isacharbrackright}{\isachardoublequoteclose}\ \isanewline
\ \ \ \ \ \ \isacommand{by}\isamarkupfalse%
\ {\isacharparenleft}rule\ knows{\isacharunderscore}composition{\isadigit{2}}{\isacharunderscore}s{\isacharparenright}\isanewline
\ \ \ \ \isacommand{from}\isamarkupfalse%
\ this\ \isakeyword{and}\ sg{\isadigit{2}}a\ \isakeyword{and}\ a{\isadigit{2}}\ \isacommand{show}\isamarkupfalse%
\ {\isacharquery}case\ \isacommand{by}\isamarkupfalse%
\ {\isacharparenleft}simp\ add{\isacharcolon}\ knows{\isacharunderscore}concat{\isacharunderscore}{\isadigit{3}}{\isacharparenright}\isanewline
\ \ \isacommand{qed}\isamarkupfalse%
\ \isanewline
\isacommand{qed}\isamarkupfalse%
\endisatagproof
{\isafoldproof}%
\isadelimproof
\isanewline
\endisadelimproof
\isanewline
\isacommand{lemma}\isamarkupfalse%
\ knows{\isacharunderscore}composition{\isadigit{5}}{\isacharcolon}\isanewline
\isakeyword{assumes}\ {\isachardoublequoteopen}knows\ P\ e\ {\isasymor}\ knows\ Q\ e{\isachardoublequoteclose}\ \isanewline
\ \ \ \ \ \ \ \isakeyword{and}\ {\isachardoublequoteopen}subcomponents\ PQ\ {\isacharequal}\ {\isacharbraceleft}P{\isacharcomma}Q{\isacharbraceright}{\isachardoublequoteclose}\isanewline
\ \ \ \ \ \ \ \isakeyword{and}\ {\isachardoublequoteopen}correctCompositionIn\ PQ{\isachardoublequoteclose}\isanewline
\ \ \ \ \ \ \ \isakeyword{and}\ {\isachardoublequoteopen}correctCompositionKS\ PQ{\isachardoublequoteclose}\isanewline
\ \ \ \ \ \ \ \isakeyword{and}\ {\isachardoublequoteopen}{\isasymforall}\ m{\isachardot}\ m\ mem\ e\ {\isasymlongrightarrow}\ {\isacharparenleft}{\isacharparenleft}{\isasymexists}\ z{\isachardot}\ m\ {\isacharequal}\ kE\ z{\isacharparenright}\ {\isasymor}\ {\isacharparenleft}{\isasymexists}\ z{\isachardot}\ m\ {\isacharequal}\ sE\ z{\isacharparenright}{\isacharparenright}{\isachardoublequoteclose}\isanewline
\ \ \ \ \ \ \ \isakeyword{and}\ {\isachardoublequoteopen}notSpecKeysSecretsExpr\ P\ e{\isachardoublequoteclose}\isanewline
\ \ \ \ \ \ \ \isakeyword{and}\ {\isachardoublequoteopen}notSpecKeysSecretsExpr\ Q\ e{\isachardoublequoteclose}\ \isanewline
\isakeyword{shows}\ {\isachardoublequoteopen}knows\ PQ\ e{\isachardoublequoteclose}\isanewline
\isadelimproof
\endisadelimproof
\isatagproof
\isacommand{using}\isamarkupfalse%
\ assms\ \isacommand{by}\isamarkupfalse%
\ {\isacharparenleft}metis\ knows{\isacharunderscore}composition{\isadigit{3}}\ knows{\isacharunderscore}composition{\isadigit{4}}{\isacharparenright}%
\endisatagproof
{\isafoldproof}%
\isadelimproof
\isanewline
\endisadelimproof
\isadelimtheory
\isanewline
\endisadelimtheory
\isatagtheory
\isacommand{end}\isamarkupfalse%
\endisatagtheory
{\isafoldtheory}%
\isadelimtheory
\endisadelimtheory
\ \end{isabellebody}%

\bibliographystyle{abbrv}

\begin{thebibliography}{1}

\bibitem{npw}
T.~Nipkow, L.~C. Paulson, and M.~Wenzel.
\newblock {\em {Isabelle/HOL -- A Proof Assistant for Higher-Order Logic}}.
\newblock LNCS. Springer, 2013.

\bibitem{FocusStreamsCaseStudies-AFP}
M.~Spichkova.
\newblock {Stream Processing Components: Isabelle/HOL Formalisation and Case
  Studies}.
\newblock {\em Archive of Formal Proofs}, Nov. 2013.

\bibitem{sj_TB08}
M.~Spichkova and J.~J\"urjens.
\newblock {Formal Specification of Cryptographic Protocols and Their
  Composition Properties: FOCUS-oriented approach}.
\newblock Technical report, Technische Universit\"at M\"unchen, 2008.

\bibitem{IsabelleManual}
M.~Wenzel.
\newblock {\em The Isabelle/Isar Reference Manual}.
\newblock TU M\"unchen, 2013.
\end{thebibliography}

\end{document}